\documentclass[12pt,twoside, a4paper]{article}
\pdfoutput=1

\def\mc{\mathcal}

\usepackage[dvips]{graphicx}
\usepackage{amssymb}
\usepackage{amssymb,amsmath}
\usepackage{graphicx}
\usepackage{dsfont}
\usepackage{caption}
\usepackage{subcaption}
\usepackage{mathtools}
\usepackage{verbatim}
\usepackage{graphicx}
\usepackage{multirow}
\usepackage[outline]{contour}
\usepackage{xcolor,colortbl}
\usepackage{pdflscape}
\input{epsf.sty}  
\contourlength{.1pt}
\graphicspath{ {./Figures/} }
\numberwithin{equation}{section}

\begin{document}
\begin{center}
\LARGE{\textbf{Twisted compactifications of 6D field theories from maximal 7D gauged supergravity}}
\end{center}
\vspace{1 cm}
\begin{center}
\large{\textbf{Parinya Karndumri}$^a$ and \textbf{Patharadanai Nuchino}$^b$}
\end{center}
\begin{center}
String Theory and Supergravity Group, Department
of Physics, Faculty of Science, Chulalongkorn University, 254 Phayathai Road, Pathumwan, Bangkok 10330, Thailand
\end{center}
E-mail: $^a$parinya.ka@hotmail.com \\
E-mail: $^b$danai.nuchino@hotmail.com \vspace{1 cm}\\
\begin{abstract}
We study supersymmetric $AdS_n\times \Sigma^{7-n}$, $n=2,3,4,5$ solutions in seven-dimensional maximal gauged supergravity with $CSO(p,q,5-p-q)$ and $CSO(p,q,4-p-q)$ gauge groups. These gauged supergravities are consistent truncations of eleven-dimensional supergravity and type IIB theory on $H^{p,q}\circ T^{5-p-q}$ and $H^{p,q}\circ T^{4-p-q}$, respectively. Apart from recovering the previously known $AdS_n\times \Sigma^{7-n}$ solutions in $SO(5)$ gauge group, we find novel classes of $AdS_5\times S^2$, $AdS_3\times S^2\times \Sigma^2$ and $AdS_3\times CP^2$ solutions in non-compact $SO(3,2)$ gauge group together with a class of $AdS_3\times CP^2$ solutions in $SO(4,1)$ gauge group. In $SO(5)$ gauge group, we extensively study holographic RG flow solutions interpolating from the $SO(5)$ supersymmetric $AdS_7$ vacuum to the $AdS_n\times \Sigma^{7-n}$ fixed points and singular geometries in the form of curved domain walls with $Mkw_{n-1}\times \Sigma^{7-n}$ slices. In many cases, the singularities are physically acceptable and can be interpreted as non-conformal phases of $(n-1)$-dimensional SCFTs obtained from twisted compactifications of $N=(2,0)$ SCFT in six dimensions. In $SO(3,2)$ and $SO(4,1)$ gauge groups, we give a large number of RG flows between the new $AdS_n\times \Sigma^{7-n}$ fixed points and curved domain walls while, in $CSO(p,q,4-p-q)$ gauge group, RG flows interpolating between asymptotically locally flat domain walls and curved domain walls are given. 
\end{abstract}
\newpage
\section{Introduction}
Wrapped branes play an important role in the study of the AdS/CFT correspondence \cite{maldacena,Gubser_AdS_CFT,Witten_AdS_CFT} and its generalization to non-conformal field theories DW/QFT correspondence \cite{DW_QFT1,DW_QFT2,DW_QFT3}. In particular, these brane configurations describe RG flows across dimensions from supersymmetric field theories on the worldvolume of the unwrapped branes to lower-dimensional field theories on the worldvolume of the branes wrapped on internal compact manifolds. For supersymmetric theories, the latter are obtained from twisted compactifications of the former on the internal manifolds. Some amount of supersymmetry is preserved by performing a topological twist along the internal manifolds \cite{Witten_twist}.
\\
\indent In this paper, we are interested in solutions describing wrapped $5$-branes in string/M-theory. Rather than searching directly for wrapped brane solutions in string/M-theory, finding supersymmetric solutions of seven-dimensional gauged supergravities in the form of domain walls interpolating between $AdS_7$ and $AdS_n\times \Sigma^{7-n}$ geometries, with $\Sigma^{7-n}$ being a $(7-n)$-dimensional compact manifold, is a more traceable task. In many cases, the resulting solutions can be embedded in ten or eleven dimensions by using consistent truncation ansatze. Solutions of this type in the maximal $N=4$ gauged supergravity with $SO(5)$ gauge group in seven dimensions have been extensively studied in previous works \cite{Maldacena_nogo,Kim_AdS_factor,Cucu_AdSD-2,Gauntlett1,Gauntlett2,4D_SCFT_from_M5,2D_Bobev,Wraped_M5}, see also \cite{Bobby_wrapped_M5,7D_noncompact,7D_twist,AdS3_7D_N2} for similar solutions in $N=2$ gauged supergravity. For solutions in other dimensions, see \cite{Wraped_D3,3D_CFT_from_LS_point,flow_acrossD_bobev,BBC,N3_AdS2,flow_across_Betti,AdS2_trisasakian,6D_twist,
5D_N4_flow,5Dtwist,BH_microstate_6D2,Minwoo_6D_BH2,Calos_6D_flow1,Kim_wrapped_F4} for an incomplete list. 
\\
\indent We will study this type of solutions within the maximal $N=4$ gauged supergravity constructed in \cite{N4_7D_Henning} using the embedding tensor formalism, see also \cite{7D_Pernici} and \cite{Noncompact_7D} for an earlier construction. Unlike the previously known results mentioned above, we will consider more general gauge groups of the form $CSO(p,q,5-p-q)$ and $CSO(p,q,4-p-q)$ obtained respectively from the embedding tensor in $\mathbf{15}$ and $\overline{\mathbf{40}}$ representations of the global symmetry $SL(5)$. Gauged supergravities with these gauge groups can be obtained from consistent truncations of eleven-dimensional supergravity and type IIB theory, respectively, see \cite{Henning_Hohm1} and \cite{Henning_Emanuel}. To the best of our knowledge, supersymmetric $AdS_n\times \Sigma^{7-n}$ solutions in $N=4$ gauged supergravity with non-compact and non-semisimple gauge groups have not been considered in the previous studies.    
\\
\indent For the aforementioned gaugings of $N=4$ supergravity, only $SO(5)$ gauge group admits a fully supersymmetric $AdS_7$ vacuum dual to $N=(2,0)$ superconformal field theory (SCFT) in six dimensions. The $AdS_n\times \Sigma^{7-n}$ solutions describe conformal fixed points in $n-1$ dimensions. In this case, these fixed points correspond to $(n-1)$-dimensional SCFTs obtained from twisted compactifications of $N=(2,0)$ SCFT in six dimensions on $\Sigma^{7-n}$. For all other gauge groups, the vacua are given by half-supersymmetric domain walls dual to six-dimensional $N=(2,0)$ non-conformal field theories. We accordingly interpret the resulting $AdS_n\times \Sigma^{7-n}$ solutions as conformal fixed points in lower-dimensions of these $N=(2,0)$ non-conformal field theories. We will study various possible RG flows from both conformal and non-conformal field theories in six dimensions to these lower-dimensional SCFTs as well as to non-conformal field theories.    
\\
\indent The paper is organized as follows. In section \ref{N4_7D_SUGRA}, we briefly review the
maximal gauged supergravity in seven dimensions. The study of supersymmetric $AdS_n\times \Sigma^{7-n}$ solutions in gauged supergravities with $CSO(p,q,5-p-q)$ and $CSO(p,q,4-p-q)$ gauge groups is presented in sections \ref{Y_gauging} and \ref{Z_gauging}, respectively. Conclusions and comments on the results are given in section \ref{conclusion}. For convenience, we also collect all bosonic field equations of the maximal seven-dimensional gauged supergravity in the appendix.

\section{$N=4$ gauged supergravity in seven dimensions}\label{N4_7D_SUGRA}
In this section, we briefly review seven-dimensional $N = 4$ gauged supergravity in the embedding tensor formalism constructed in \cite{N4_7D_Henning}. We will omit all the detail and only collect relevant fomulae involving the bosonic Lagrangian and fermionic supersymmetry transformations which are essential for finding supersymmetric solutions. The reader is referred to \cite{N4_7D_Henning} for more detail.
\\
\indent The only $N=4$ supermultiplet in seven dimensions is the supergravity multiplet with the field content
\begin{equation}\label{fieldcon}
(e^{\hat{\mu}}_\mu, \psi^a_\mu, A^{MN}_\mu, B_{\mu\nu M}, \chi^{abc}, {\mathcal{V}_M}^A).
\end{equation}
The component fields are given by the graviton $e^{\hat{\mu}}_\mu$, four gravitini $\psi^a_\mu$, ten vectors $A^{MN}_\mu=A^{[MN]}_\mu$, five two-form fields $B_{\mu\nu M}$, sixteen spin-$\frac{1}{2}$ fermions $\chi^{abc}=\chi^{[ab]c}$ and fourteen scalar fields described by the $SL(5)/SO(5)$ coset representative ${\mathcal{V}_M}^A$. In this paper, curved and flat space-time indices are denoted by $\mu$, $\nu,\ldots$ and $\hat{\mu}$, $\hat{\nu},\ldots$, respectively. Lower (upper) $M,N=1,...,5$ indices refer to the (anti-) fundamental representation \textbf{5} ($\overline{\mathbf{5}}$) of the global $SL(5)$ symmetry. 
\\
\indent Fermionic fields are described by symplectic Majorana spinors subject to the conditions
\begin{equation}
\bar{\psi}^T_{\mu a}=\Omega_{ab}C\psi_\mu^b\qquad \textrm{and}\qquad \bar{\chi}^T_{abc}=\Omega_{ad}\Omega_{be}\Omega_{cf}C\chi^{def}
\end{equation}
where $C$ denotes the charge conjugation matrix obeying
\begin{equation}
C=C^T=-C^{-1}=-C^\dagger\, .
\end{equation}
These fermionic fields transform in representations of the local $SO(5)\sim USp(4)$ R-symmetry with $USp(4)$ fundamental or $SO(5)$ spinor indices $a,b,\ldots =1,...,4$. Accordingly, the four gravitini $\psi^a_\mu$ and the spin-$\frac{1}{2}$ fields $\chi^{abc}$ transform as $\mathbf{4}$ and $\mathbf{16}$, respectively. $\chi^{abc}$ satisfy the following conditions
\begin{equation}
\chi^{[abc]}=0\qquad \textrm{and}\qquad \Omega_{ab}\chi^{abc}=0
\end{equation}
with $\Omega_{ab}=\Omega_{[ab]}$ being the $USp(4)$ symplectic form satisfying
\begin{equation}
(\Omega_{ab})^*=\Omega^{ab}\qquad \textrm{and}\qquad \Omega_{ac}\Omega^{bc}=\delta^b_a\, .
\end{equation}
Raising and lowering of $USp(4)$ indices by $\Omega^{ab}$ and $\Omega_{ab}$ correspond to complex conjugation.
\\
\indent The fourteen scalars are described by the $SL(5)/SO(5)$ coset representative ${\mc{V}_M}^A$, transforming under the global $SL(5)$ and local $SO(5)$ symmetries by left and right multiplications. Indices $M=1,2,\ldots, 5$ and $A=1,2,\ldots ,5$ are accordingly $SL(5)$ and $SO(5)$ fundamental indices, respectively. The $SO(5)$ vector indices of ${\mc{V}_M}^A$ can be described by a pair of antisymmetric $USp(4)$ fundamental indices as ${\mc{V}_M}^{ab}={\mc{V}_M}^{[ab]}$ satisfying the relation
\begin{equation}
{\mathcal{V}_M}^{ab}\Omega_{ab}=0\, .
\end{equation}
The inverse of ${\mathcal{V}_M}^A$ denoted by ${\mc{V}_A}^M$ will be written as ${\mathcal{V}_{ab}}^M$ with 
\begin{equation}
{\mathcal{V}_M}^{ab}{\mathcal{V}_{ab}}^N=\delta^N_M \qquad \textrm{and} \qquad {\mathcal{V}_{ab}}^M{\mathcal{V}_M}^{cd}=\delta^{[c}_{a}\delta^{d]}_{b}-\frac{1}{4}\Omega_{ab}\Omega^{cd}\, .
\end{equation}
\indent The bosonic Lagrangian of the $N=4$ seven-dimensional gauged supergravity can be written as
\begin{eqnarray}
e^{-1}\mathcal{L}&=&\frac{1}{2}R-\mathcal{M}_{MP}\mathcal{M}_{NQ}\mathcal{H}_{\mu\nu}^{(2)MN}\mathcal{H}^{(2)PQ\mu\nu }-\frac{1}{6}\mathcal{M}^{MN}\mathcal{H}_{\mu\nu\rho M}^{(3)}{\mathcal{H}^{(3)\mu\nu\rho}}_ N\nonumber \\
& &+\frac{1}{8}(D_\mu\mathcal{M}_{MN})(D^\mu\mathcal{M}^{MN})-e^{-1}\mathcal{L}_{VT}-\mathbf{V}\label{BosLag}
\end{eqnarray}
while the supersymmetry transformations of fermions read
\begin{eqnarray}
\delta\psi^a_\mu&=& D_\mu\epsilon^a-g\gamma_\mu A^{ab}_1\Omega_{bc}\epsilon^c+\frac{1}{15}\mathcal{H}_{\nu\rho\lambda M}^{(3)}({\gamma_\mu}^{\nu\rho\lambda}-\frac{9}{2}\delta^\nu_\mu\gamma^{\rho\lambda})\Omega^{ab}{\mathcal{V}_{bc}}^M\epsilon^c\nonumber \\ & &+\frac{1}{5}\mathcal{H}_{\nu\rho}^{(2)MN}({\gamma_\mu}^{\nu\rho}-8\delta^\nu_\mu\gamma^{\rho}){\mathcal{V}_M}^{ad}\Omega_{de}{\mathcal{V}_N}^{eb}\Omega_{bc}\epsilon^c,\\
\delta\chi^{abc}&=& 2\Omega^{cd}{P_{\mu de}}^{ab}\gamma^\mu\epsilon^e+gA^{d,abc}_2\Omega_{de}\epsilon^e\nonumber \\ & &
+2\mathcal{H}_{\mu\nu}^{(2)MN}\gamma^{\mu\nu}\Omega_{de}\left[{\mathcal{V}_M}^{cd}{\mathcal{V}_N}^{e[a}\epsilon^{b]}-\frac{1}{5}(\Omega^{ab}\delta^c_g-\Omega^{c[a}\delta^{b]}_g){\mathcal{V}_M}^{gf}\Omega_{fh}{\mathcal{V}_N}^{hd}\epsilon^{e}\right] \nonumber \\ & &
-\frac{1}{6}\mathcal{H}_{\mu\nu\rho M}^{(3)}\gamma^{\mu\nu\rho}{\mathcal{V}_{fe}}^M\left[\Omega^{af}\Omega^{be}\epsilon^c-\frac{1}{5}(\Omega^{ab}\Omega^{cf}+4\Omega^{c[a}\Omega^{b]f})\epsilon^e\right].
\end{eqnarray}
The covariant derivative of the supersymmetry parameters is defined by
\begin{equation}
D_\mu\epsilon^a=\nabla_\mu\epsilon^a-{Q_{\mu b}}^a\epsilon^b
\end{equation}
with $\nabla_\mu$ being the space-time covariant derivative. The composite connection ${Q_{\mu a}}^b$ and the vielbein on the $SL(5)/SO(5)$ coset ${P_{\mu ab}}^{cd}$ are obtained from
\begin{equation}
{P_{\mu ab}}^{cd}+2{Q_{\mu [a}}^{[c}\delta^{d]}_{b]}= {\mathcal{V}_{ab}}^M(\partial_\mu{\mathcal{V}_M}^{cd}-gA^{PQ}_\mu{X_{PQ,M}}^N{\mathcal{V}_N}^{cd}).
\end{equation}
The gauge generators in the representation $\mathbf{5}$ of $SL(5)$ can be written in term of the embedding tensor as
\begin{equation}
{X_{MN,P}}^Q= {\Theta_{MN,P}}^Q=\delta^Q_{[M}Y_{N]P}-2\epsilon_{MNPRS}Z^{RS,Q}\, .
\end{equation}
Supersymmetry requires that the embedding tensor can have only two components given by the tensors $Y_{MN}$ and $Z^{MN,P}$ with $Y_{MN}=Y_{(MN)}$, $Z^{MN,P}=Z^{[MN],P}$ and $Z^{[MN,P]}=0$ corresponding to representations $\mathbf{15}$ and $\overline{\mathbf{40}}$ of $SL(5)$, respectively.
\\
\indent The fermion shift matrices $A_1$ and $A_2$ are given by
\begin{eqnarray}
A^{ab}_1&=& -\frac{1}{4\sqrt{2}}\left(\frac{1}{4}B\Omega^{ab}+\frac{1}{5}C^{ab}\right),\label{A1}\\
A^{d,abc}_2&=&\frac{1}{2\sqrt{2}}\left[\Omega^{ec}\Omega^{fd}({C^{ab}}_{ef}-{B^{ab}}_{ef})\right. \nonumber \\
& &\left.+\frac{1}{4}(C^{ab}\Omega^{cd}+\frac{1}{5}\Omega^{ab}C^{cd}+\frac{4}{5}\Omega^{c[a}C^{b]d})\right]\label{A2}
\end{eqnarray}
with 
\begin{eqnarray}\label{BCfunctions}
B&=&\frac{\sqrt{2}}{5}\Omega^{ac}\Omega^{bd}Y_{ab,cd},\\
{{B^{ab}}_{cd}}&=&\sqrt{2}\left[\Omega^{ae}\Omega^{bf}\delta^{[g}_c \delta^{h]}_d-\frac{1}{5}(\delta^{[a}_c \delta^{b]}_d-\frac{1}{4}\Omega^{ab}\Omega_{cd})\Omega^{eg}\Omega^{fh}\right]Y_{ef,gh},\\
C^{ab}&=&8\Omega_{cd}Z^{(ac)[bd]},\\
{C^{ab}}_{cd}&=&8\left(-\Omega_{ce}\Omega_{df}\delta^{[a}_g \delta^{b]}_h+\Omega_{g(c}\delta^{[a}_{d)} \delta^{b]}_e\Omega_{fh}\right)Z^{(ef)[gh]}\, .
\end{eqnarray}
The ``dressed'' components of the embedding tensor are defined by
\begin{eqnarray}
Y_{ab,cd}&=&{\mathcal{V}_{ab}}^M{\mathcal{V}_{cd}}^NY_{MN},\\
\textrm{and}\qquad Z^{(ac)[ef]}&=&\sqrt{2}{\mathcal{V}_M}^{ab}{\mathcal{V}_N}^{cd}{\mathcal{V}_P}^{ef}\Omega_{bd}Z^{MN,P}\, .
\end{eqnarray}
\indent A unimodular symmetric matrix $\mc{M}_{MN}$ describing $SL(5)/SO(5)$ scalars in a manifestly $SO(5)$ invariant manner is defined by
\begin{equation}\label{fullM}
\mathcal{M}_{MN}={\mathcal{V}_M}^{ab}{\mathcal{V}_N}^{cd}\Omega_{ac}\Omega_{bd}
\end{equation}
together with its inverse
\begin{equation}
\mathcal{M}^{MN}={\mathcal{V}_{ab}}^M{\mathcal{V}_{cd}}^N\Omega^{ac}\Omega^{bd}\, .
\end{equation}
The scalar potential is given by
\begin{eqnarray}\label{scalarPot}
\mathbf{V}&=& \frac{g^2}{64}\left[2\mathcal{M}^{MN}Y_{NP}\mathcal{M}^{PQ}Y_{QM}-(\mathcal{M}^{MN}Y_{MN})^2\right]\nonumber \\
& &+g^2Z^{MN,P}Z^{QR,S}\left(\mathcal{M}_{MQ}\mathcal{M}_{NR}\mathcal{M}_{PS}-\mathcal{M}_{MQ}\mathcal{M}_{NP}\mathcal{M}_{RS}\right)
\nonumber \\
&=&-15A_1^{ab}A_{1 ab}+ \frac{1}{8}A_2^{a,bcd}A_{2 a,bcd}
\end{eqnarray}
\indent Unlike in the ungauged supergravity in which all three-form fields can be dualized to two-form fields, the field content of the gauged supergravity can incorporate massive two- and three-form fields. The degrees of freedom in the vector and tensor fields of the ungauged theory will be redistributed among massless and massive vector, two-form and three-form fields after gaugings. In general, with a proper gauge fixing of various tensor gauge transformations, there can be $t$ self-dual massive three-form and $s$ massive two-form fields for $s\equiv\text{rank}\ Z$ and $t\equiv\text{rank}\ Y$. In addition, there are $10-s$ massless vectors and $5-s-t$ massless two-form fields. It should be noted that $t+s\leq 5$ by the quadratic constraint which ensures that the embedding tensor leads to gauge generators for a closed subalgebra of $SL(5)$. Furthermore, more massive gauge fields can arise from broken gauge symmetry.
\\
\indent The field strength tensors of vector and two-form fields are defined by
\begin{eqnarray}
\mathcal{H}_{\mu\nu}^{(2)MN}&=&F^{MN}_{\mu\nu}+gZ^{MN,P}B_{\mu\nu P},\label{ModTensor1}\\
\mathcal{H}_{\mu\nu\rho M}^{(3)}&=&gY_{MN}S^N_{\mu\nu\rho}+3D_{[\mu}B_{\nu\rho]M}\nonumber\\
&&+6\epsilon_{MNPQR}A^{NP}_{[\mu}(\partial_\nu A^{QR}_{\rho]}+\frac{2}{3}g {X_{ST,U}}^QA^{RU}_\nu A^{ST}_{\rho]})\label{ModTensor2}
\end{eqnarray}
with the usual non-abelian gauge field strength
\begin{equation}\label{Ful2Form}
F^{MN}_{\mu\nu}=2\partial_{[\mu}A^{MN}_{\nu]}+g{(X_{PQ})_{RS}}^{MN}A^{PQ}_{[\mu}A^{RS}_{\nu]}\, .
\end{equation}
These field strengths satisfy the following Bianchi identities
\begin{eqnarray}
D_{[\mu}\mathcal{H}^{(2)MN}_{\nu\rho]}&=&\frac{1}{3}gZ^{MN,P}\mathcal{H}^{(3)}_{\mu\nu\rho P}, \label{DefBianchi1}\\
D_{[\mu}\mathcal{H}^{(3)}_{\nu\rho\lambda] M}&=&\frac{3}{2}\epsilon_{MNPQR}\mathcal{H}^{(2)NP}_{[\mu\nu}\mathcal{H}^{(2)QR}_{\rho\lambda]}+\frac{1}{4}gY_{MN}\mathcal{H}^{(4)N}_{\mu\nu\rho\lambda}\label{DefBianchi2}
\end{eqnarray}
where the covariant field strengths of the three-form fields are given by
\begin{eqnarray}\label{4-form}
Y_{MN}\mathcal{H}^{(4)N}_{\mu\nu\rho\lambda}&=&Y_{MN}\left[4D_{[\mu}S^N_{\nu\rho\lambda]}+6F^{NP}_{[\mu\nu}B_{\rho\lambda] P}+3gZ^{NP,Q}B_{[\mu\nu P}B_{\rho\lambda] Q}\right.\nonumber \\
& &+4g\epsilon_{PQRVW}{X_{ST,U}}^VA^{NP}_{[\mu}A^{QR}_\nu A^{ST}_\rho A^{UW}_{\lambda]}\nonumber \\
& &\left.+8\epsilon_{PQRST}A^{NP}_{[\mu}A^{QR}_\nu\partial_\rho A^{ST}_{\lambda]}\right].
\end{eqnarray}
All of these fields interact with each other via the vector-tensor topological term $\mathcal{L}_{VT}$ whose explcit form can be found in \cite{N4_7D_Henning}. 

\section{Solutions from gaugings in $\mathbf{15}$ representation}\label{Y_gauging}
We now consider supersymmetric $AdS_n\times \Sigma^{7-n}$ solutions with gauge group $CSO(p,q,5-p-q)$ obtained from gaugings in $\mathbf{15}$ representation. In this case, non-vanishing components of the embedding tensor can be written as
\begin{equation}
Y_{MN}=\text{diag}(\underbrace{1,..,1}_p,\underbrace{-1,..,-1}_q,\underbrace{0,..,0}_r),\qquad p+q+r=5\, .
\end{equation} 
We will use the following choice of $SO(5)$ gamma matrices to convert an $SO(5)$ vector index to a pair of antisymmetric spinor indices
\begin{eqnarray}
\Gamma_1&=&-\sigma_2\otimes\sigma_2, \qquad  \Gamma_2=\mathbf{I}_2\otimes\sigma_1,\qquad \Gamma_3=\mathbf{I}_2\otimes\sigma_3,\nonumber  \\
\Gamma_4&=&\sigma_1\otimes\sigma_2, \qquad \Gamma_5=\sigma_3\otimes\sigma_2
\end{eqnarray}
where $\sigma_i$ are the usual Pauli matrices. $\Gamma_A$ satisfy the following relations
\begin{eqnarray}
\{\Gamma_A,\Gamma_B\}&=&2\delta_{AB}\mathbf{I}_4,\qquad  (\Gamma_A)^{ab}=-(\Gamma_A)^{ba}, \nonumber \\
\Omega_{ab}(\Gamma_A)^{ab}&=&0, \qquad ((\Gamma_A)^{ab})^*=\Omega_{ac}\Omega_{bd}(\Gamma_A)^{cd},
\end{eqnarray}
and the symplectic form of $USp(4)$ is chosen to be
\begin{equation}
\Omega_{ab}=\Omega^{ab}=\mathbf{I}_2\otimes i\sigma_2\, .
\end{equation}
The coset representative of the form ${\mc{V}_M}^{ab}$ and the inverse ${\mc{V}_{ab}}^M$ are then given by
\begin{equation}
{\mathcal{V}_M}^{ab}=\frac{1}{2} {\mathcal{V}_M}^A(\Gamma_A)^{ab}\qquad \textrm{and}\qquad
{\mathcal{V}_{ab}}^M=\frac{1}{2} {\mathcal{V}_A}^M(\Gamma^A)_{ab}\, .
\end{equation}

\subsection{Supersymmetric $AdS_5\times \Sigma^2$ solutions with $SO(2)\times SO(2)$ symmetry}\label{YAdS5section}
\begin{figure}
  \centering
  \begin{subfigure}[b]{0.4\linewidth}
    \includegraphics[width=\linewidth]{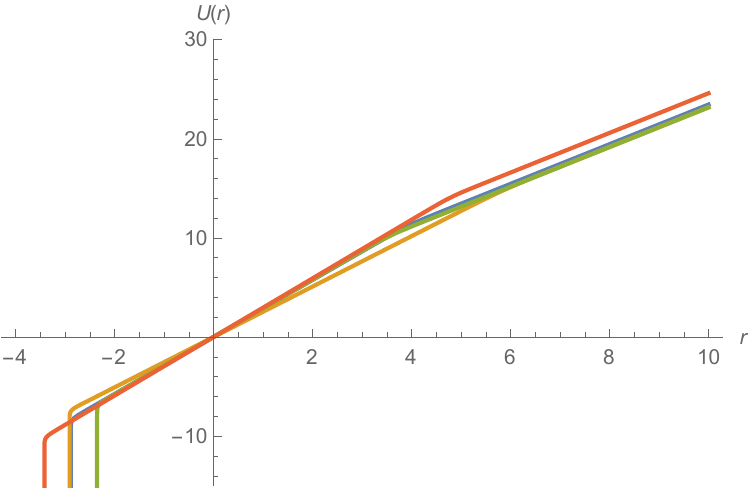}
  \caption{$U$ solution}
  \end{subfigure}
  \begin{subfigure}[b]{0.4\linewidth}
    \includegraphics[width=\linewidth]{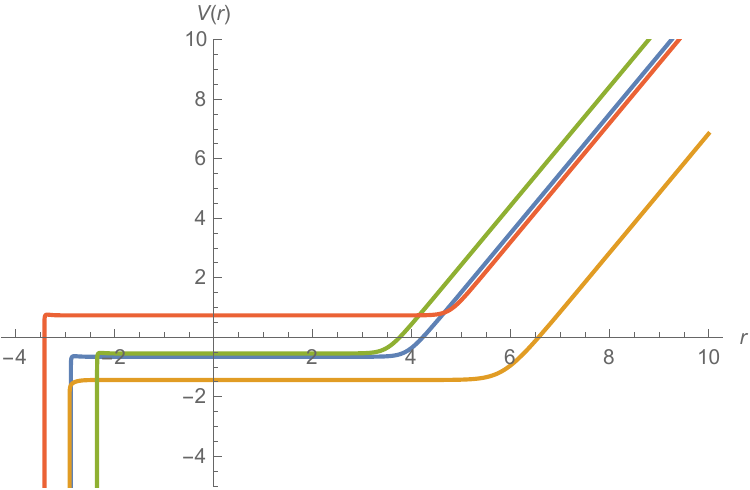}
  \caption{$V$ solution}
  \end{subfigure}
  \begin{subfigure}[b]{0.4\linewidth}
    \includegraphics[width=\linewidth]{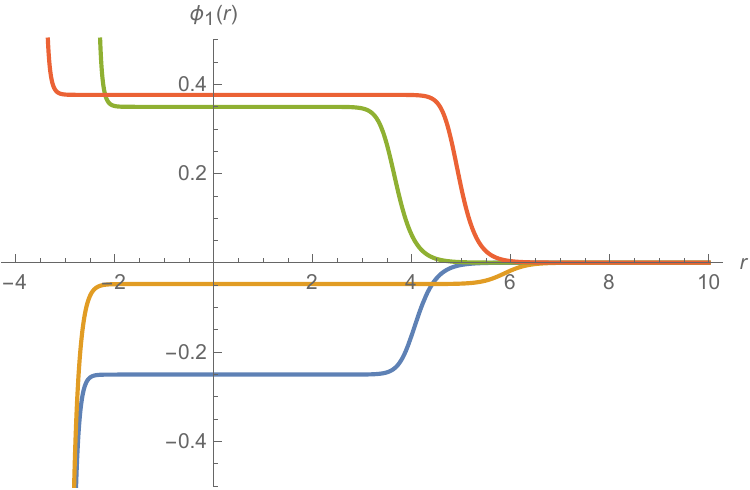}
  \caption{$\phi_1$ solution}
  \end{subfigure}
  \begin{subfigure}[b]{0.4\linewidth}
    \includegraphics[width=\linewidth]{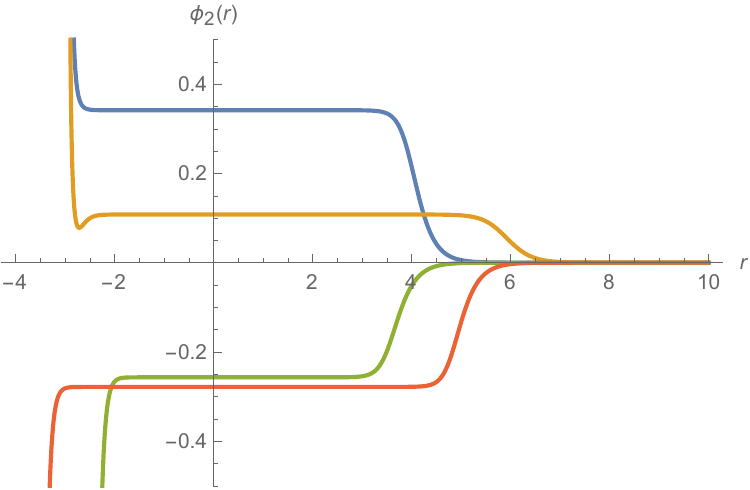}
  \caption{$\phi_2$ solution}
  \end{subfigure}
  \caption{RG flows from the $N=4$ $AdS_7$ critical point as $r\rightarrow\infty$ to $AdS_5\times H^2$ fixed points and curved domain walls for $SO(2)\times SO(2)$ twist in $SO(5)$ gauge group. The blue, orange, green and red curves refer to $p_2=-\frac{1}{4}, -\frac{1}{24}, \frac{1}{4},4$, respectively.}
  \label{Y_AdS5xH2flows}
\end{figure}

\begin{figure}
  \centering
  \begin{subfigure}[b]{0.35\linewidth}
    \includegraphics[width=\linewidth]{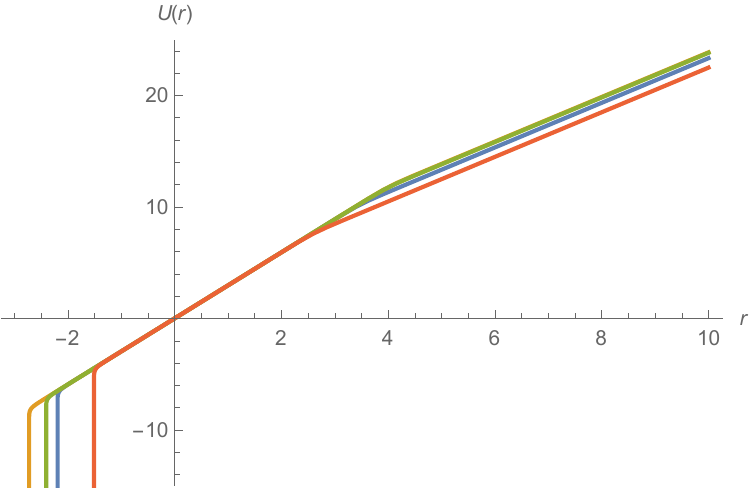}
  \caption{$U$ solution}
  \end{subfigure}
  \begin{subfigure}[b]{0.35\linewidth}
    \includegraphics[width=\linewidth]{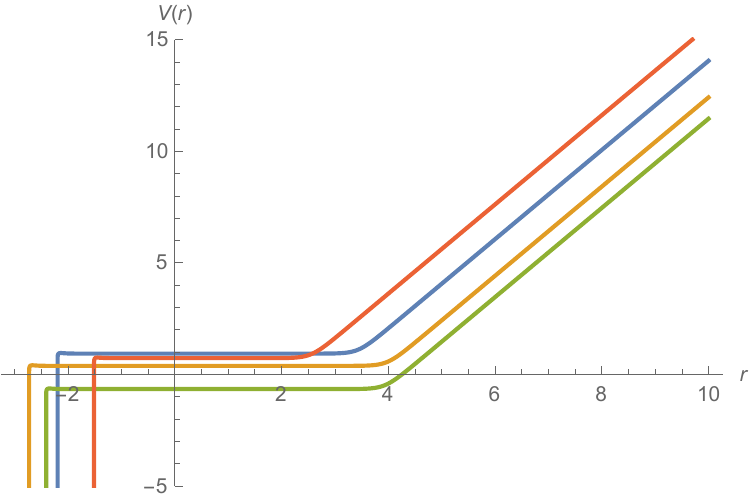}
  \caption{$V$ solution}
  \end{subfigure}\\
  \begin{subfigure}[b]{0.35\linewidth}
    \includegraphics[width=\linewidth]{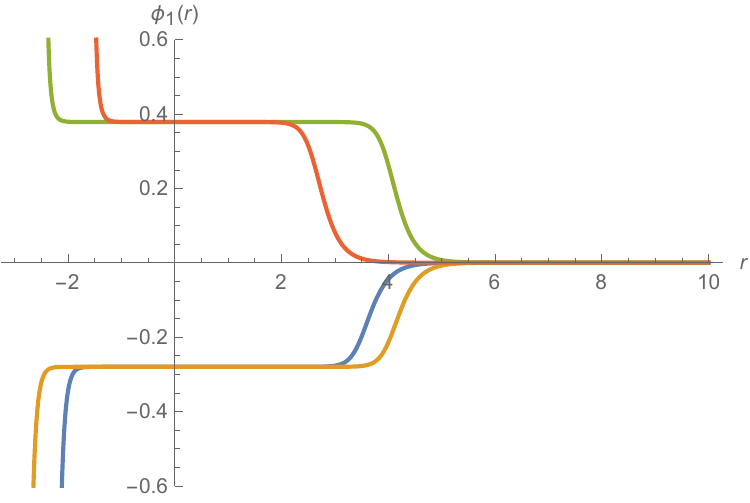}
  \caption{$\phi_1$ solution}
  \end{subfigure}
  \begin{subfigure}[b]{0.35\linewidth}
    \includegraphics[width=\linewidth]{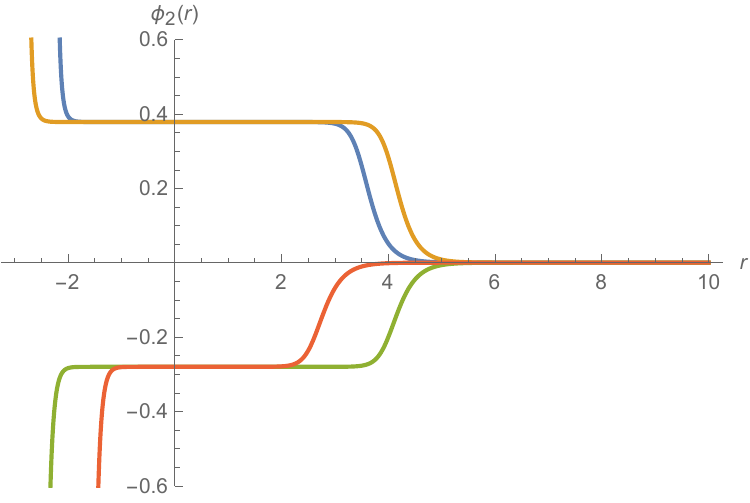}
  \caption{$\phi_2$ solution}
  \end{subfigure}
  \caption{RG flows from the $N=4$ $AdS_7$ critical point as $r\rightarrow\infty$ to $AdS_5\times \mathbb{R}^2$ fixed points and curved domain walls for $SO(2)\times SO(2)$ twist in $SO(5)$ gauge group. The blue, orange, green and red curves refer to $p_2=-6, -2, \frac{1}{4},4$, respectively.}
  \label{Y_AdS5xR2flows}
\end{figure}

\begin{figure}
  \centering
  \begin{subfigure}[b]{0.35\linewidth}
    \includegraphics[width=\linewidth]{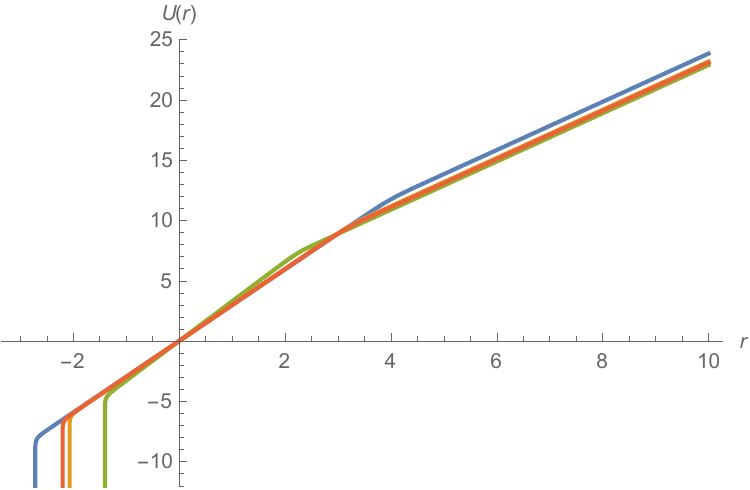}
  \caption{$U$ solution}
  \end{subfigure}
  \begin{subfigure}[b]{0.35\linewidth}
    \includegraphics[width=\linewidth]{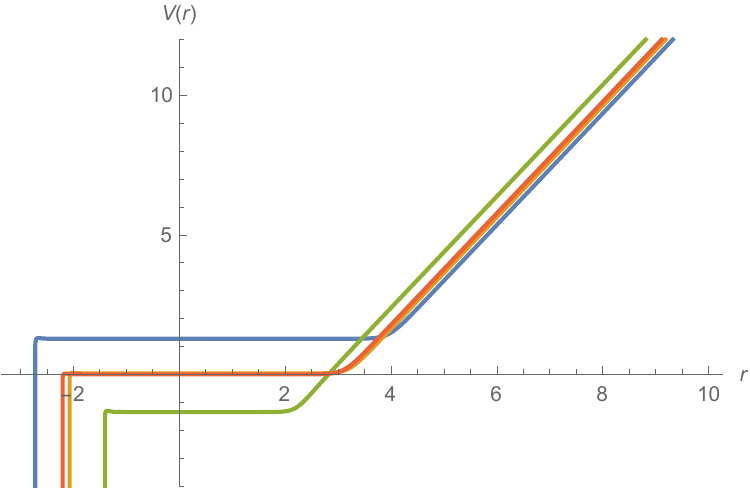}
  \caption{$V$ solution}
  \end{subfigure}
  \begin{subfigure}[b]{0.35\linewidth}
    \includegraphics[width=\linewidth]{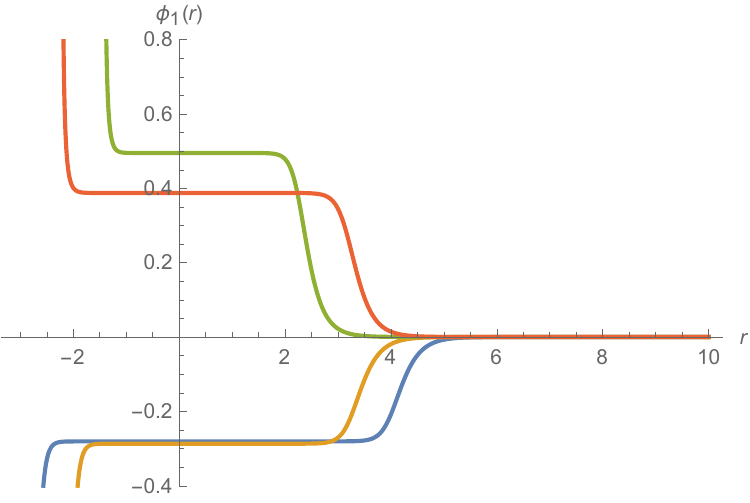}
  \caption{$\phi_1$ solution}
  \end{subfigure}
  \begin{subfigure}[b]{0.35\linewidth}
    \includegraphics[width=\linewidth]{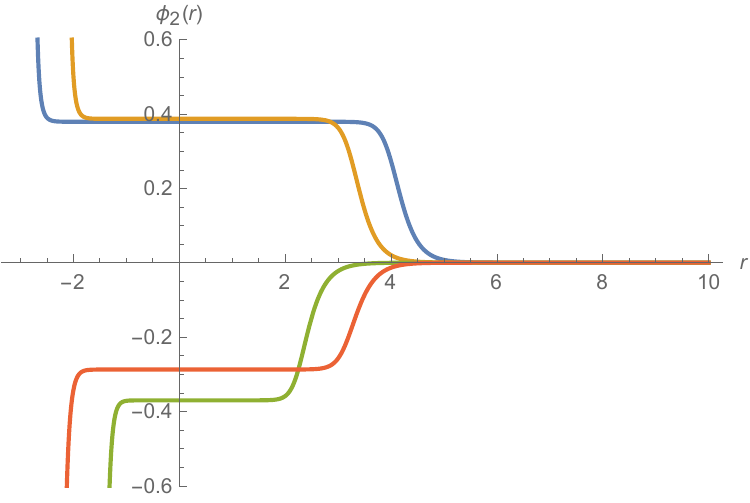}
  \caption{$\phi_2$ solution}
  \end{subfigure}
  \caption{RG flows from the $N=4$ $AdS_7$ critical point as $r\rightarrow\infty$ to $AdS_5\times S^2$ fixed points and curved domain walls for $SO(2)\times SO(2)$ twist in $SO(5)$ gauge group. The blue, orange, green and red curves refer to $p_2=-12, -1, \frac{1}{8},1$, respectively.}
  \label{Y_AdS5xS2flows}
\end{figure}

We begin with solutions of the form $AdS_5\times\Sigma^2$ with the metric ansatz given by
\begin{equation}\label{YAdS57Dmetric}
ds_7^2=e^{2U(r)}dx^2_{1,3}+dr^2+e^{2V(r)}ds^2_{\Sigma^{2}_{k}}\, .
\end{equation}
$\Sigma^2_k$ is a Riemann surface with the metric given by
\begin{equation}\label{Sigma2metric}
 ds^2_{\Sigma^2_k}=d\theta^2+f_k(\theta)^2d\varphi^2,
\end{equation}
and $dx^2_{1,3}=\eta_{mn}dx^{m} dx^{n}$ with $m,n=0, ..,3$ is the flat metric on the four-dimensional Minkowski space $Mkw_4$. The function $f_k(\theta)$ is defined by
\begin{equation}\label{fFn}
f_k(\theta)=\begin{cases}
                        	\sin{\theta}, \ \  \quad k=+1 \\
                       	\theta, \ \ \ \ \ \ \quad k=0\\
			\sinh{\theta}, \quad k=-1
                    \end{cases}
\end{equation}
with $k=1,0,-1$ corresponding to $S^2$, $\mathbb{R}^2$ and $H^2$, respectively.
\\
\indent With the following choice of vielbein
\begin{eqnarray}
e^{\hat{m}}&=&e^{U}dx^m, \qquad e^{\hat{r}}=dr,\nonumber \\
e^{\hat{\theta}}&=& e^{V}d\theta, \qquad e^{\hat{\varphi}}=e^{V}f_{k}(\theta)d\varphi,\label{AdS5xSigma2bein}
\end{eqnarray}
we find the following non-vanishing components of the spin connection
\begin{equation}\label{AdS4xSigma3SpinCon}
\omega_{(1)}^{\hat{m}\hat{r}}=U'e^{\hat{m}}, \qquad \omega_{(1)}^{\hat{i}\hat{r}}= V'e^{\hat{i}},\qquad
\omega_{(1)}^{\hat{\varphi}\hat{\theta}}=\frac{f'_{k}(\theta)}{f_{k}(\theta)}e^{-V}e^{\hat{\varphi}}\, .
\end{equation}
The index $\hat{i}=\hat{\theta}, \hat{\varphi}$ is a flat index on $\Sigma^2_{k}$, and $f'_k(\theta)=\frac{df_k(\theta)}{d\theta}$. The $r$-derivatives will be denoted by $'$ while a $'$ on any function with an explicit argument refers to the derivative of the function with respect to that argument. 
\\
\indent We are interested in solutions with $SO(2)\times SO(2)$ symmetry. Among the fourteen scalars in $SL(5)/SO(5)$ coset, there are two $SO(2)\times SO(2)$ singlet scalars corresponding to the following $SL(5)$ non-compact generators 
\begin{equation}
\hat{Y}_1=e_{1,1}+e_{2,2}-2e_{5,5}\qquad \textrm{and}\qquad
\hat{Y}_2=e_{3,3}+e_{4,4}-2e_{5,5}\, .
\end{equation}
We have introduced $GL(5)$ matrices defined by
\begin{equation}
(e_{IJ})_{MN}=\delta_{IM}\delta_{JN}\, .
\end{equation}
The $SL(5)/SO(5)$ coset representative is then given by 
\begin{equation}\label{YSO(2)xSO(2)Ys}
\mathcal{V}=e^{\phi_1\hat{Y}_1+\phi_2\hat{Y}_2}\, .
\end{equation}
\indent A general form of the embedding tensor for gauge groups with an $SO(2)\times SO(2)$ subgroup can be written as
\begin{equation}\label{SO(2)xSO(2)Ytensor}
Y_{MN}=\text{diag}(+1,+1,\sigma,\sigma,\rho).
\end{equation}
This gives rise to $SO(5)$ ($\rho=\sigma=1$), $SO(4,1)$ ($-\rho=\sigma=1$), $SO(3,2)$ ($\rho=-\sigma=1$), $CSO(4,0,1)$ ($\rho=0$, $\sigma=1$) and $CSO(2,2,1)$ ($\rho=0$, $\sigma=-1$) gauge groups. 
\\
\indent With all these, we can straightforwardly compute the scalar potential
\begin{equation}\label{YSO(2)xSO(2)Pot}
\mathbf{V}=-\frac{1}{64}g^2e^{-2(\phi_1+\phi_2)}\left[8\sigma-\rho^2e^{10(\phi_1+\phi_2)}+4\rho(e^{4\phi_1+6\phi_2}+\sigma e^{6\phi_1+4\phi_2})\right].
\end{equation}
For $SO(5)$ gauge group, this potential admits two $AdS_7$ critical points given by 
\begin{equation}
\phi_1=\phi_2=0,\qquad \mathbf{V}_0=-\frac{15}{64}g^2 \label{SO(5)Cripoint}
\end{equation}
and 
\begin{equation}
\phi_1=\phi_2=\frac{1}{10}\ln{2},\qquad \mathbf{V}_0=-\frac{5g^2}{16\times2^{2/5}} \, . \label{SO(4)Cripoint}
\end{equation}
The former preserves $N=4$ supersymmetry with $SO(5)$ symmetry while the latter is a non-supersymmetric $AdS_7$ vacuum with $SO(4)$ symmetry. We note here that for $\phi_1=\phi_2$ the $SO(2)\times SO(2)$ symmetry is enhanced to $SO(4)$. These two $AdS_7$ vacua have been identified long ago in \cite{Noncompact_7D}.
\\
\indent To perform a topological twist on $\Sigma^2_k$, we turn on the following $SO(2)\times SO(2)$ gauge fields
\begin{equation}\label{YAdS5gaugeAnt}
A^{12}_{(1)}=-e^{-V}\frac{p_{1}}{k}\frac{f'_{k}(\theta)}{f_{k}(\theta)}e^{\hat{\varphi}}\qquad \textrm{and}\qquad
A^{34}_{(1)}=-e^{-V}\frac{p_{2}}{k}\frac{f'_{k}(\theta)}{f_{k}(\theta)}e^{\hat{\varphi}}
\end{equation} 
and set all the other fields to zero. By imposing the twist condition 
\begin{equation}\label{YSO(2)xSO(2)on1QYM}
g(p_1+\sigma p_2)=k
\end{equation}
together with the following projection conditions
\begin{equation}\label{SO(2)xSO(2)Projcon}
\gamma^{\hat{\varphi}\hat{\theta}}\epsilon^a=-{(\Gamma_{12})^a}_b\epsilon^b=-{(\Gamma_{34})^a}_b\epsilon^b
\end{equation}
and
\begin{equation}\label{pureYProj}
\gamma_r\epsilon^a=\epsilon^a,
\end{equation}
we can derive the BPS equations
\begin{eqnarray}
U'&\hspace{-0.3cm}=&\hspace{-0.3cm}\frac{g}{40}(2e^{-2\phi_1}+\rho e^{4(\phi_1+\phi_2)}+2\sigma e^{-2\phi_2})-\frac{2}{5}e^{-2V}(e^{2\phi_1}p_1+e^{2\phi_2}p_2),\\
V'&\hspace{-0.3cm}=&\hspace{-0.3cm}\frac{g}{40}(2e^{-2\phi_1}+\rho e^{4(\phi_1+\phi_2)}+2\sigma e^{-2\phi_2})+\frac{8}{5}e^{-2V}(e^{2\phi_1}p_1+e^{2\phi_2}p_2),\\
\phi_1'&\hspace{-0.3cm}=&\hspace{-0.3cm}\frac{g}{20}(3e^{-2\phi_1}-\rho e^{4(\phi_1+\phi_2)}-2\sigma e^{-2\phi_2})-\frac{2}{5}e^{-2V}(3e^{2\phi_1}p_1-2e^{2\phi_2}p_2),\\
\phi_2'&\hspace{-0.3cm}=&\hspace{-0.3cm}\frac{g}{20}(3\sigma e^{-2\phi_2}-\rho e^{4(\phi_1+\phi_2)}-2 e^{-2\phi_1})+\frac{2}{5}e^{-2V}(2e^{2\phi_1}p_1-3e^{2\phi_2}p_2).\hspace{0.8cm}
\end{eqnarray}
It can be readily verified that these BPS equations together with the twist condition \eqref{YSO(2)xSO(2)on1QYM} imply the second-ordered field equations. The radial component of the gravitino variations $\delta\psi^a_r$ gives the usual solution for the Killing spinors
\begin{equation}
\epsilon^a=e^{\frac{U}{2}}\epsilon^a_{(0)}
\end{equation}
in which $\epsilon^a_{(0)}$ are constant spinors.    
\\
\indent By imposing the conditions $V'=\phi_1'=\phi_2'=0$ and $U'=\frac{1}{L_{\textrm{AdS}_5}}$ on the BPS equations, we find a class of $AdS_5$ fixed point solutions given by
\begin{eqnarray}
e^{2V}&=&-\frac{8(e^{4\phi_1}p_1+2e^{2(\phi_1+\phi_2)}p_2)}{g},\\
e^{10\phi_1}&=&\frac{12p_1^2-24\sigma p_1^2p_2+22\sigma^2 p_1p_2^2-8\sigma^3p_2^3 -2K}{3p_1^3\rho\sigma^2},\\
e^{2\phi_2}&=&\frac{6p_1^3-15\sigma p_1^2p_2+13\sigma^2 p_1p_2^2-4\sigma^3p_2^3 +K}{p_2(9\sigma p_1p_2-6p_1^2-4\sigma^2 p_2^2)}e^{2\phi_1},\\
 L_{\text{AdS}_5}&=&\frac{4(e^{4\phi_1}p_1+2e^{2(\phi_1+\phi_2)}p_2)}{g\left(e^{2\phi_1}p_1+e^{2\phi_2}p_2\right)}
\end{eqnarray}
where
\begin{equation}
K=(6p_1^2-9\sigma p_1p_2+4p_2^2\sigma^2)\sqrt{(p_1^2-\sigma p_1p_2+\sigma^2p_2^2)}\, .
\end{equation}
The $AdS_5$ fixed points do not exist in the case of $\rho=0$. Furthermore, it turns out that good $AdS_5$ fixed points exist only in $SO(5)$ and $SO(3,2)$ gauge groups with $\rho=\sigma=1$ and $\rho=-\sigma=1$, respectively. 
\\
\indent For $SO(5)$ gauge group, there exist $AdS_5$ fixed points when 
\begin{equation}
gp_2\neq -1,0,\qquad gp_2\neq0\qquad \textrm{and}\qquad  gp_2<0\ \cup\ gp_2>1, 
\end{equation}
for $\Sigma^2=H^2, \mathbb{R}^2, S^2$, respectively. In deriving the above conditions, we have chosen $g>0$ for convenience. We emphasize here that the $AdS_5\times\mathbb{R}^2$ fixed point preserves sixteen supercharges while the $AdS_5\times H^2$ and $AdS_5\times S^2$ preserve only eight supercharges. This is due to the fact that no spin connection on $\mathbb{R}^2$ needs to be cancelled by performing a twist. In this case, the projector involving $\gamma_{\hat{\theta}\hat{\varphi}}$ is not needed. All these $AdS_5\times \Sigma^2$ fixed points, dual to four-dimensional SCFTs from M5-branes, together with the corresponding RG flows from the supersymmetric $AdS_7$ vacuum have recently been discussed in \cite{4D_SCFT_from_M5}.  
\\
\indent In this work, we will extend the study of these RG flows by considering more general RG flows from the $N=4$ $AdS_7$ critical point to $AdS_5$ fixed points and then to singular geometries in the form of curved domain walls with $Mkw_4\times \Sigma^2$ slices. According to the usual holographic interpretation, these geometries should be dual to non-conformal field theories in four dimensions arising from the RG flows from four-dimensional SCFTs dual to $AdS_5\times \Sigma^2$ fixed points. The latter are in turn obtained from twisted compactification of $N=(2,0)$ SCFT in six dimensions dual to the $N=4$ $AdS_7$ vacuum. Examples of these RG flows are given in figures \ref{Y_AdS5xH2flows}, \ref{Y_AdS5xR2flows}, and \ref{Y_AdS5xS2flows} for the cases of $AdS_5\times H^2$, $AdS_5\times\mathbb{R}^2$ and $AdS_5\times S^2$ fixed points, respectively. In these solutions, we have chosen the position of the $AdS_5\times \Sigma^2$ fixed points to be $r=0$ and set $g=16$. We note here that on of the magnetic charges is fixed by the twist condition \eqref{YSO(2)xSO(2)on1QYM}. In the numerical solutions, we have chosen $p_2$ to be the independent parameter.

\begin{figure}
  \centering
  \begin{subfigure}[b]{0.3\linewidth}
    \includegraphics[width=\linewidth]{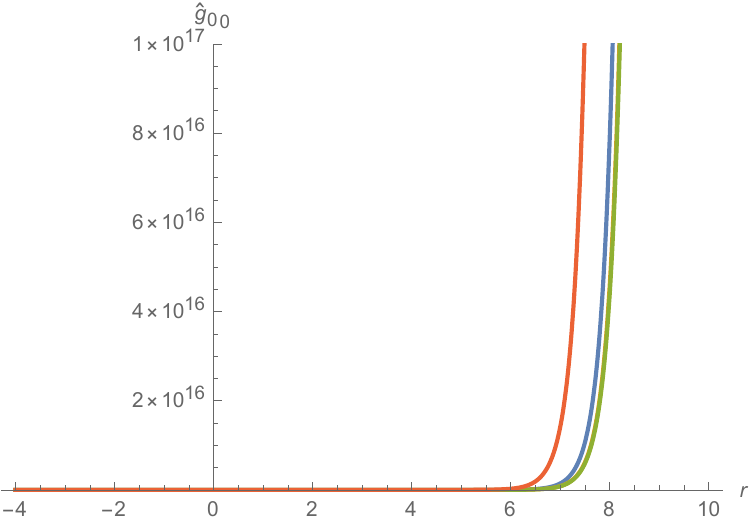}
  \caption{$\Sigma^2=H^2$}
  \end{subfigure}
  \begin{subfigure}[b]{0.3\linewidth}
    \includegraphics[width=\linewidth]{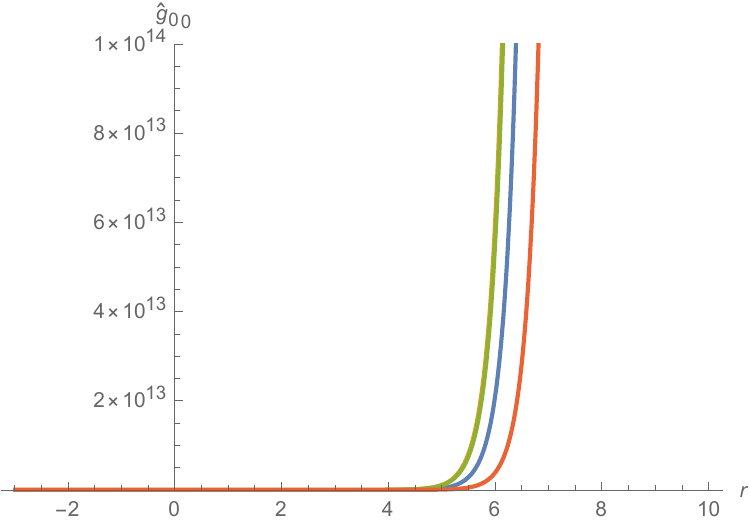}
  \caption{$\Sigma^2=\mathbb{R}^2$}
  \end{subfigure}
  \begin{subfigure}[b]{0.3\linewidth}
    \includegraphics[width=\linewidth]{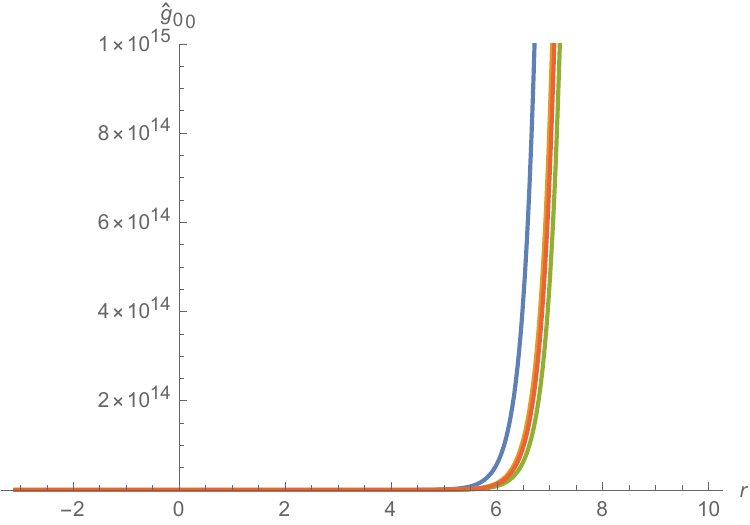}
  \caption{$\Sigma^2=S^2$}
  \end{subfigure}
  \caption{The behavior of $\hat{g}_{00}$ along the RG flows from the $N=4$ $AdS_7$ critical point as $r\rightarrow\infty$ to $AdS_5\times \Sigma^2$ fixed points and curved domain walls with $SO(2)\times SO(2)$ symmetry for $\Sigma^2=H^2, \mathbb{R}^2, S^2$ in $SO(5)$ gauge group.}
  \label{SO5_singularity_AdS5}
\end{figure}

\indent We can use the explicit uplift formulae given in \cite{11D_to_7D_Nastase1, 11D_to_7D_Nastase2} to determine whether these IR singularities are physical by considering the $(00)$-component of the eleven-dimensional metric given by
\begin{equation}
\hat{g}_{00}=\Delta^{\frac{1}{3}}g_{00}\label{11D_g00}\, .
\end{equation}
The warped factor $\Delta$ is defined by 
\begin{equation}
\Delta=\mc{M}^{MN}\delta_{MP}\delta_{NQ}\mu^P\mu^Q
\end{equation}
with $\mu^M$, $M=1,2,\ldots, 5$, being the coordinates on $S^4$ and satisfying $\mu^M\mu^M=1$. Using the coset representative given in \eqref{YSO(2)xSO(2)Ys} and the $S^4$ coordinates
\begin{eqnarray}
\mu^M&=&\left(\cos\xi,\sin\xi\cos\psi\cos\alpha,\sin\xi\cos\psi\sin\alpha,\right.\nonumber \\
& &\left.\sin\xi\sin\psi\cos\beta,\sin\xi\sin\psi\sin\beta\right),
\end{eqnarray}
we find the behavior of $\hat{g}_{00}$ along the flows as shown in figure \ref{SO5_singularity_AdS5}. As can be seen from the figure, $\hat{g}_{00}\rightarrow 0$ near the singularities. These singularities are then physical according to the criterion given in \cite{Maldacena_nogo}. Therefore, the singularities can be interpreted as holographic duals of non-conformal phases of the four-dimensional SCFTs obtained from twisted compactifications of six-dimensional $N=(2,0)$ SCFT on $\Sigma^2$. 

\begin{figure}
  \centering
  \begin{subfigure}[b]{0.35\linewidth}
    \includegraphics[width=\linewidth]{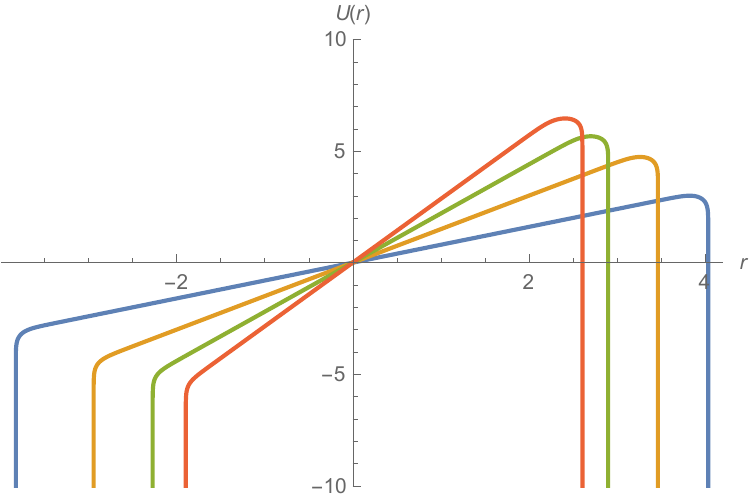}
  \caption{$U$ solution}
  \end{subfigure}
  \begin{subfigure}[b]{0.35\linewidth}
    \includegraphics[width=\linewidth]{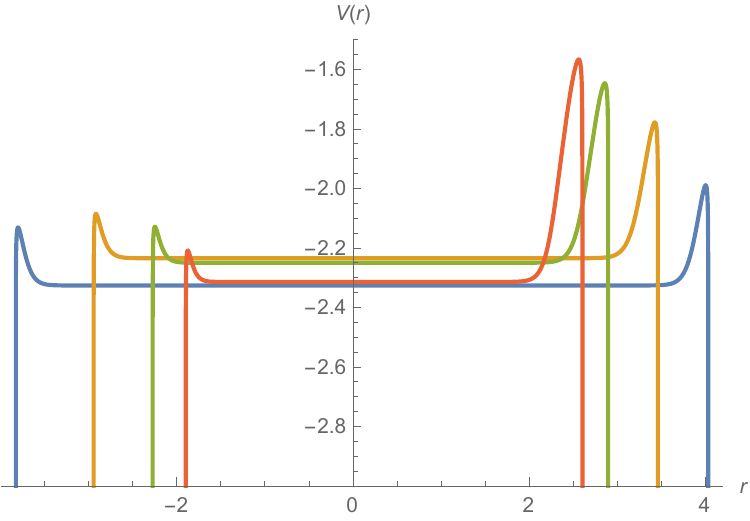}
  \caption{$V$ solution}
  \end{subfigure}\\
  \begin{subfigure}[b]{0.35\linewidth}
    \includegraphics[width=\linewidth]{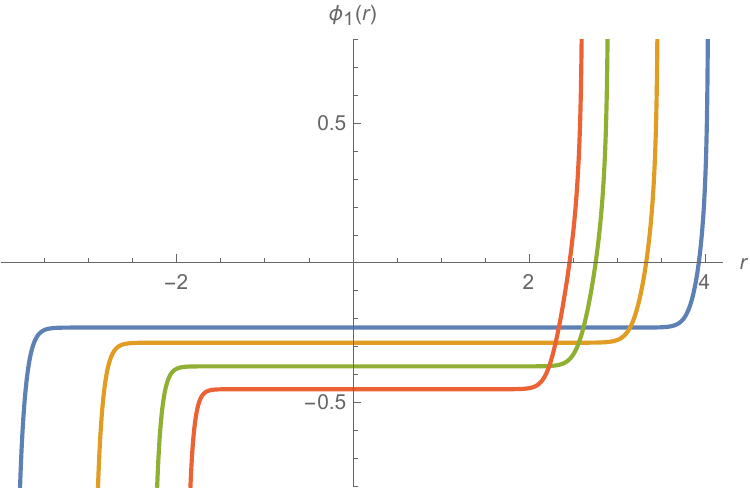}
  \caption{$\phi_1$ solution}
  \end{subfigure}
  \begin{subfigure}[b]{0.35\linewidth}
    \includegraphics[width=\linewidth]{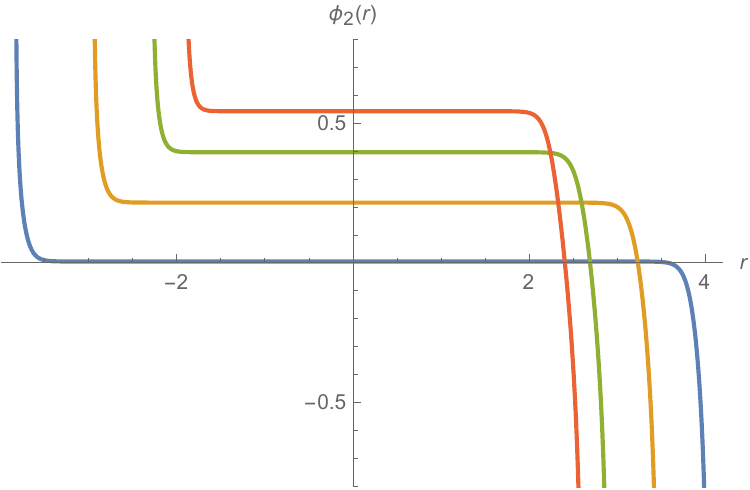}
  \caption{$\phi_2$ solution}
  \end{subfigure}
  \caption{RG flows between $AdS_5\times S^2$ fixed points and curved domain walls for $SO(2)\times SO(2)$ twist in $SO(3,2)$ gauge group. The blue, orange, green and red curves refer to $p_2=-\frac{1}{36}, -\frac{1}{48}, -\frac{1}{64},-\frac{1}{86}$, respectively.}
  \label{YSO2xSO232flows}
\end{figure}

\begin{figure}
\centering
    \includegraphics[width=0.5\linewidth]{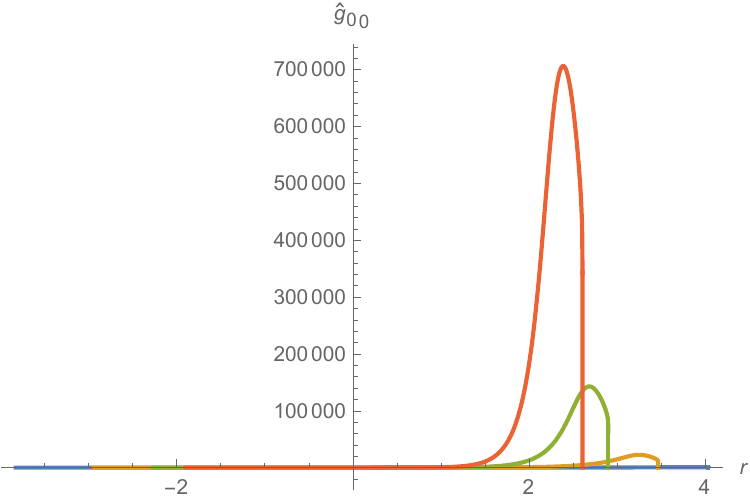}
  \caption{Profiles of $\hat{g}_{00}$ for RG flows between $N=1$ SCFTs and $N=1$ non-conformal field theories in four dimensions from $SO(3,2)$ gauge group}\label{YSO2xSO232g00}
  \end{figure}

We now consider $SO(3,2)$ gauge group. In this case, we find new $AdS_5\times S^2$ fixed points in a small range, with $g>0$,
\begin{equation}
-\frac{1}{2}<g p_2<0\, .
\end{equation}
As in the previous case, these $AdS_5\times S^2$ solutions also preserve eight supercharges and are dual to $N=1$ SCFTs in four dimensions. In constrast to $SO(5)$ gauge group, the vacuum solution in this case is given by a half-supersymmetric domain wall, see the solutions given in \cite{our_7D_DW}. According to the DW/QFT correspondence, these solutions are expected to describe $N=(2,0)$ non-conformal field theories in six dimensions. The above $AdS_5\times S^2$ fixed points can be regarded as conformal fixed points in four dimensions arising from twisted compactifications of the $N=(2,0)$ field theories in six dimensions on $S^2$. In figure \ref{YSO2xSO232flows}, we give examples of RG flows between the $AdS_5\times S^2$ fixed points and curved domain walls with the worldvolume given by $Mkw_4\times S^2$. The latter should describe non-conformal phases of the $N=1$ SCFTs in four dimensions. The two ends of the flows represent two possible non-conformal phases with $(\phi_1\rightarrow \infty,\phi_2\rightarrow -\infty)$ and $(\phi_1\rightarrow -\infty,\phi_2\rightarrow \infty)$. In all of these flow solutions, we have set $g=16$.
\\
\indent In figure \ref{YSO2xSO232g00}, we give the behavior of the eleven-dimensional metric component $\hat{g}_{00}$ along the flows. This is obtained by using the consistent truncation of eleven-dimensional supergravity on $H^{p,q}$ given in \cite{Henning_Hohm1}. The explicit form of $\hat{g}_{00}$ is similar to that given in \eqref{11D_g00} but with the warped factor $\Delta$ given by
\begin{equation}
\Delta=\mc{M}^{MN}\eta_{MP}\eta_{NQ}\mu^P\mu^Q\, .
\end{equation}
The tensor $\eta_{MN}=\textrm{diag}(1,\ldots ,1,-1,\ldots,-1)$ is the $SO(p,q)$ invariant tensor, and $\mu^P$ are coordinates on $H^{p,q}$ satsifying $\mu^P\mu^Q\eta_{PQ}=1$. For $SO(3,2)$, we have a truncation of eleven-dimensional supergravity on $H^{3,2}$ with $\eta_{MN}=\textrm{diag}(1,1,1,-1,-1)$. From figure \ref{YSO2xSO232g00}, we see that $\hat{g}_{00}\rightarrow 0$ on both sides of the flows. Therefore, all of these singularities are physically acceptable. We accordingly interpret these solutions as RG flows between $N=1$ SCFTs and non-conformal field theories in four dimensions obtained from twisted compactifications of $N=(2,0)$ field theory on $S^2$.    

\subsection{Supersymmetric $AdS_4\times \Sigma^3$ solutions with $SO(3)$ symmetry}\label{YAdS4section}
We now carry out a similar analysis for supersymmetric solutions of the form $AdS_4\times \Sigma^3$ with $\Sigma^3$ being a $3$-manifold with constant curvature. The ansatz for the metric takes the form of
\begin{equation}\label{SO(3)7Dmetric}
ds_7^2=e^{2U(r)}dx_{1,2}^2+dr^2+e^{2V(r)}ds^2_{\Sigma^{3}_{k}}
\end{equation}
where $dx^2_{1,2}=\eta_{mn}dx^{m} dx^{n}$, $m,n=0,1,2$ is the metric on the three-dimensional Minkowski space. The metric on $\Sigma^3_{k}$ is given by
\begin{equation}
ds^2_{\Sigma^3_{k}}=d\psi^2+f_{k}(\psi)^2(d\theta^2+\sin^2{\theta}d\varphi^2)\label{Sigma3metric}
\end{equation}
with the function $f_k$ defined in \eqref{fFn}.
\\
\indent Using the vielbein
\begin{eqnarray}
e^{\hat{m}}&=& e^{U}dx^m, \qquad e^{\hat{r}}=dr,\qquad 
e^{\hat{\psi}}= e^{V}d\psi, \nonumber \\ 
e^{\hat{\theta}}&=&\ e^{V}f_{k}(\psi)d\theta, \qquad e^{\hat{\varphi}}=e^{V}f_{k}(\psi)\sin{\theta}d\varphi,\label{AdS4xSigma3bein}
\end{eqnarray}
we find non-vanishing components of the spin connection as follow
\begin{eqnarray}
\omega_{(1)}^{\hat{m}\hat{r}}&=& U'e^{\hat{m}}, \qquad \omega_{(1)}^{\hat{i}\hat{r}}= V'e^{\hat{i}},\qquad
\omega_{(1)}^{\hat{\theta}\hat{\psi}}=\frac{f'_{k}(\psi)}{f_{k}(\psi)}e^{-V}e^{\hat{\theta}},\nonumber \\ \omega_{(1)}^{\hat{\varphi}\hat{\psi}}&=& \frac{f'_{k}(\psi)}{f_{k}(\psi)}e^{-V}e^{\hat{\varphi}}, \qquad \omega_{(1)}^{\hat{\varphi}\hat{\theta}}=\frac{\cot\theta}{f_{k}(\psi)}e^{-V}e^{\hat{\varphi}}\label{AdS4xSigma3SpinCon}
\end{eqnarray}
where $\hat{i}=\hat{\psi}, \hat{\theta}, \hat{\varphi}$ is a flat index on $\Sigma^3_{k}$. We will perform the twist by turning on $SO(3)\subset SO(3)\times SO(2)\subset SO(5)_R$ and $SO(3)_+\subset SO(3)_+\times SO(3)_-\sim SO(4)\subset SO(5)_R$ gauge fields with $SO(5)_R$ denoting the R-symmetry.

\subsubsection{Solutions with $SO(3)$ twists}\label{YSO(3)section}
\begin{figure}
  \centering
  \begin{subfigure}[b]{0.32\linewidth}
    \includegraphics[width=\linewidth]{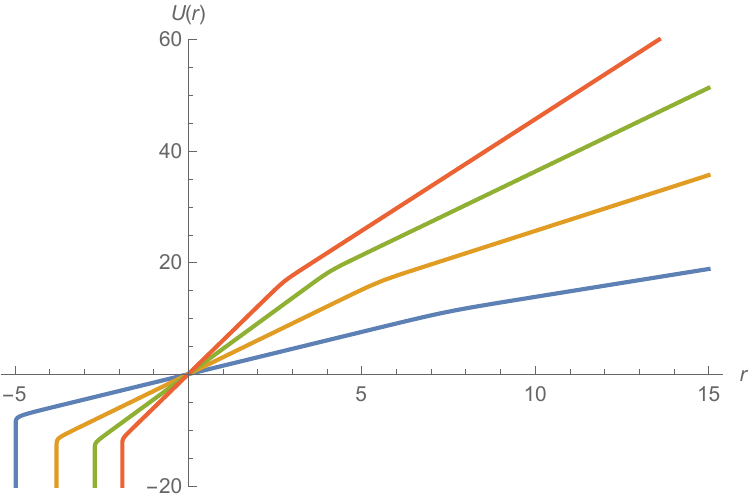}
  \caption{$U$ solution}
  \end{subfigure}
  \begin{subfigure}[b]{0.32\linewidth}
    \includegraphics[width=\linewidth]{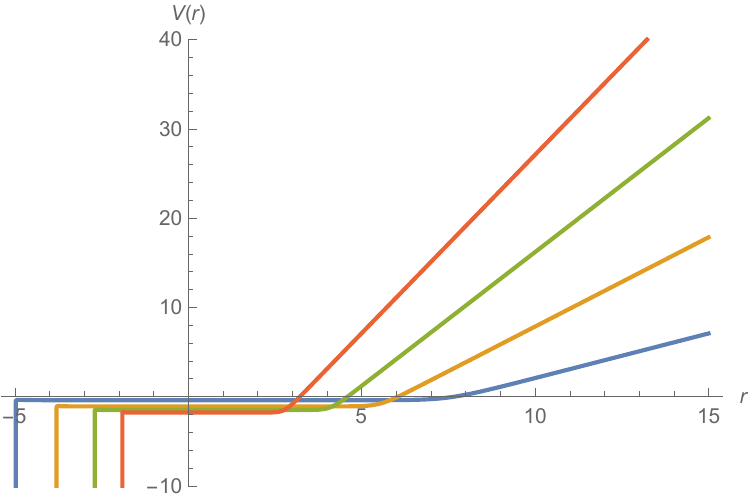}
  \caption{$V$ solution}
  \end{subfigure}
  \begin{subfigure}[b]{0.32\linewidth}
    \includegraphics[width=\linewidth]{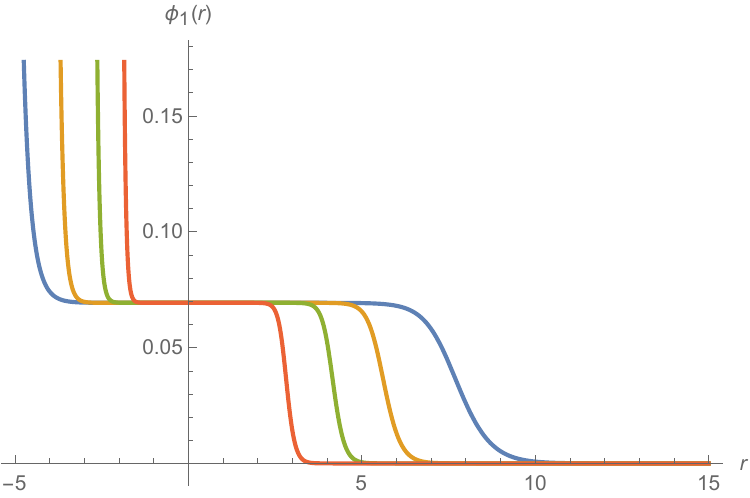}
  \caption{$\phi_1$ solution}
  \end{subfigure}
  \caption{RG flows from the $N=4$ $AdS_7$ critical point to the $AdS_4\times H^3$ fixed point and curved domain walls for $SO(3)$ twists with $\phi_1\rightarrow \infty$. The blue, orange, green and red curves refer to $g=8,16,24,32$, respectively.}
  \label{YSO3flow1}
\end{figure}

\begin{figure}
  \centering
  \begin{subfigure}[b]{0.32\linewidth}
    \includegraphics[width=\linewidth]{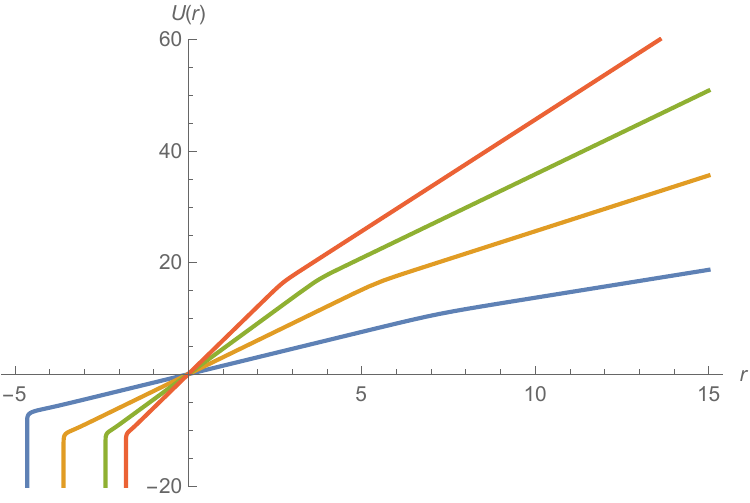}
  \caption{$U$ solution}
  \end{subfigure}
  \begin{subfigure}[b]{0.32\linewidth}
    \includegraphics[width=\linewidth]{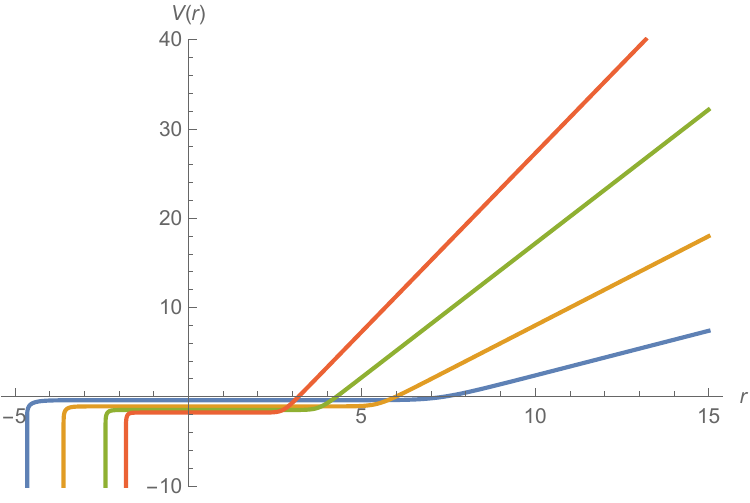}
  \caption{$V$ solution}
  \end{subfigure}
  \begin{subfigure}[b]{0.32\linewidth}
    \includegraphics[width=\linewidth]{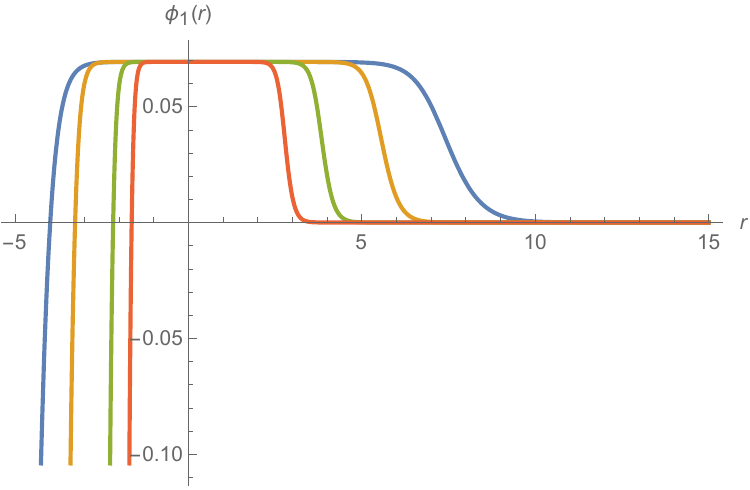}
  \caption{$\phi_1$ solution}
  \end{subfigure}
  \caption{RG flows from the $N=4$ $AdS_7$ critical point to the $AdS_4\times H^3$ fixed point and curved domain walls for $SO(3)$ twists with $\phi_1\rightarrow -\infty$. The blue, orange, green and red curves refer to $g=8,16,24,32$, respectively.}
  \label{YSO3flow2}
\end{figure}

\begin{figure}
  \centering
  \begin{subfigure}[b]{0.38\linewidth}
    \includegraphics[width=\linewidth]{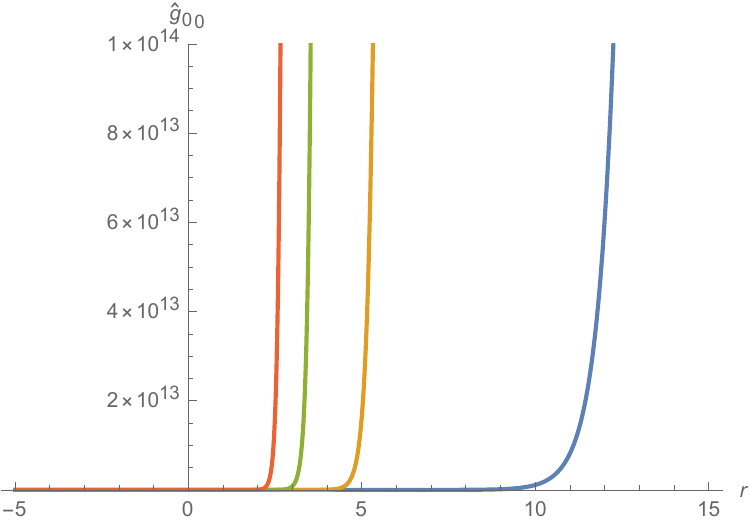}
  \caption{The profiles of $\hat{g}_{00}$ for RG flows with $\phi_1\rightarrow \infty$}
  \end{subfigure}
  \begin{subfigure}[b]{0.38\linewidth}
    \includegraphics[width=\linewidth]{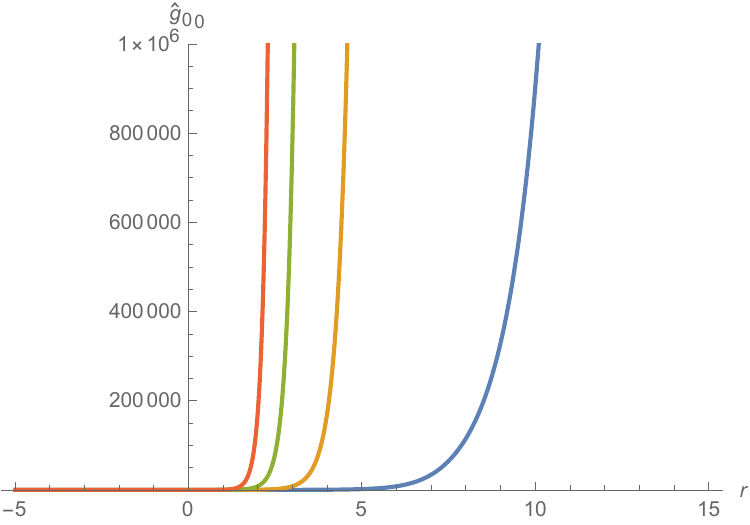}
  \caption{The profiles of $\hat{g}_{00}$ for RG flows with $\phi_1\rightarrow -\infty$}
  \end{subfigure}
  \caption{Behaviors of the $(00)$-component of the eleven-dimensional metric for the RG flows given in figures \ref{YSO3flow1} and \ref{YSO3flow2}.}
  \label{g00_SO3_phi1}
\end{figure}

We first consider solutions with $SO(3)\subset SO(5)_R$ twists by turning on the following $SO(3)$ gauge fields 
\begin{equation}\label{YSO(3)gaugeAnt}
A^{12}_{(1)}=-e^{-V}\frac{p}{k}\frac{f'_{k}(\psi)}{f_{k}(\psi)}e^{\hat{\theta}},\ \ \ A^{13}_{(1)}=-e^{-V}\frac{p}{k}\frac{f'_{k}(\psi)}{f_{k}(\psi)}e^{\hat{\varphi}}, \ \ \ A^{23}_{(1)}=-e^{-V}\frac{p}{k}\frac{\cot\theta}{f_{k}(\psi)}e^{\hat{\varphi}}\, .
\end{equation}
There are three $SO(3)$ singlet scalars corresponding to the $SL(5)$ noncompact generators
\begin{eqnarray}
\tilde{Y}_1&=&2e_{1,1}+2e_{2,2}+2e_{3,3}-3e_{4,4}-3e_{5,5},\nonumber \\ 
\tilde{Y}_2&=&e_{4,5}+e_{5,4},\nonumber \\ 
\tilde{Y}_3&=&e_{4,4}-e_{5,5}\, .\label{YSO(3)Ys}
\end{eqnarray}
The $SL(5)/SO(5)$ coset representative is then given by 
\begin{equation}\label{YSO(3)coset}
\mathcal{V}=e^{\phi_1\tilde{Y}_1+\phi_2\tilde{Y}_2+\phi_3\tilde{Y}_3}\, .
\end{equation}
\indent We will consider gauge groups with an $SO(3)$ subgroup characterized by the embedding tensor of the form
\begin{equation}
Y_{MN}=\text{diag}(+1,+1,+1,\sigma,\rho).
\end{equation}
There are six possible gauge groups given by $SO(5)$ ($\rho=\sigma=1$), $SO(4,1)$ ($-\rho=\sigma=1$), $SO(3,2)$ ($\rho=\sigma=-1$), $CSO(4,0,1)$ ($\rho=0$, $\sigma=1$), $CSO(3,1,1)$ ($\rho=0$, $\sigma=-1$), and $CSO(3,0,2)$ ($\rho=\sigma=0$). With this embedding tensor and the coset representative \eqref{YSO(3)coset}, the scalar potential reads
\begin{eqnarray}
\mathbf{V}&=&-\frac{g^2}{64}\left[3e^{-8\phi_1}+6e^{2\phi_1}\left[(\rho+\sigma)\cosh{2\phi_2}\cosh{2\phi_3}+(\rho-\sigma)\sinh{2\phi_3}\right]\right.\nonumber \\
& &+\frac{1}{4}e^{12\phi_1}\left[-(\rho+\sigma)^2\cosh{4\phi_2}(1+\cosh{4\phi_3})-4(\rho^2-\sigma^2)\cosh{2\phi_2}\sinh{4\phi_3}\right. \nonumber\\
& &\left.\left.+\rho^2+10\rho\sigma+\sigma^2-(3\rho^2-2\rho\sigma+3\sigma^2)\cosh{4\phi_3}\right]\right]
\end{eqnarray}
which admits supersymmetric $N=4$ and non-supersymmetric $AdS_7$ critical points given in \eqref{SO(5)Cripoint} and \eqref{SO(4)Cripoint} at $\phi_1=\phi_2=\phi_3=0$ and $\phi_1=\frac{1}{20}\ln{2}$, $\phi_2=\pm\frac{1}{4}\ln{2}$, and $\phi_3=0$, respectively.
\\
\indent We now impose a simple twist condition 
\begin{equation}
gp=k\label{GenQYM} 
\end{equation}
and the following projectors on the Killing spinors
\begin{equation}\label{SO(3)Projcon}
\gamma^{\hat{4}\hat{5}}\epsilon^a=-{(\Gamma_{12})^a}_b\epsilon^b\qquad \textrm{and} \qquad \gamma^{\hat{5}\hat{6}}\epsilon^a=-{(\Gamma_{23})^a}_b\epsilon^b\, .
\end{equation}
With all other fields vanishing and the $\gamma_r$ projector given in \eqref{pureYProj}, we obtain the BPS equations
\begin{eqnarray}
U'&=&\frac{g}{40}e^{6\phi_1}\left[(\rho+\sigma)\cosh{2\phi_2}\cosh{2\phi_3}+(\rho-\sigma)\sinh{2\phi_3}\right]\nonumber\\&&+\frac{3g}{40}e^{-4\phi_1}-\frac{6}{5}e^{-2(V-2\phi_1)}p,\\
V'&=&\frac{g}{40}e^{6\phi_1}\left[(\rho+\sigma)\cosh{2\phi_2}\cosh{2\phi_3}+(\rho-\sigma)\sinh{2\phi_3}\right]\nonumber\\&&+\frac{3g}{40}e^{-4\phi_1}+\frac{14}{5}e^{-2(V-2\phi_1)}p,\\
\phi_1'&=&\frac{g}{40}e^{6\phi_1}\left[(\rho-\sigma)\sinh{2\phi_3}-(\rho+\sigma)\cosh{2\phi_2}\cosh{2\phi_3}\right]\nonumber\\&&\frac{g}{20}e^{-4\phi_1}-\frac{4}{5}e^{-2(V-2\phi_1)}p,\\
\phi_2'&=&-\frac{g}{8}e^{6\phi_1}(\rho+\sigma)\sinh{2\phi_2}\text{sech}{2\phi_3},\\
\phi_3'&=&-\frac{g}{8}e^{6\phi_1}\left((\rho+\sigma)\cosh{2\phi_2}\sinh{2\phi_3}+(\rho-\sigma)\cosh{2\phi_3}\right).
\end{eqnarray}
\indent From these BPS equations, we find an $AdS_4\times H^3$ fixed point only for $SO(5)$ gauge group given by
\begin{eqnarray}
\phi_1&=&\frac{1}{10}\ln2,\qquad \phi_2= \phi_3=0,\nonumber  \\
V&=&\ln\left[\frac{16^{3/5}}{g}\right],\qquad L_{AdS_4}=\frac{4\times2^{2/5}}{g}\, .
\end{eqnarray}
This is the $AdS_4\times H^3$ solution studied in \cite{Gauntlett1}. The solution preserves eight supercharges and corresponds to $N=2$ SCFT in three dimensions. As in the prevouse case, in addition to the holographic RG flows from the supersymmetric $N=4$ $AdS_7$ vacuum to this $AdS_4\times H^3$ geometry, we also consider more general RG flows from the $AdS_4\times H^3$ fixed point to curved domain walls with a $Mkw_3\times H^3$ slice dual to $N=2$ non-conformal field theories in three dimensions. 
\\
\indent There are many possible RG flows of this type. The simplest possibility is given by RG flows with $\phi_2=\phi_3=0$ along the entire flows. Examples of these RG flows are given in figures \ref{YSO3flow1} and \ref{YSO3flow2} in which $\phi_1\rightarrow \infty$ and $\phi_1\rightarrow -\infty$, respectively. Both types of the singularities are physically acceptable as can be seen from the behavior of the $(00)$-component of the eleven-dimensional metric $\hat{g}_{00}$ given in figure \ref{g00_SO3_phi1}. These singular geometries are then dual to $N=2$ non-conformal field theories in three dimensions obtained from twisted compactifications of the six-dimensional $N=(2,0)$ SCFT on $H^3$.

\begin{figure}
  \centering
  \begin{subfigure}[b]{0.32\linewidth}
    \includegraphics[width=\linewidth]{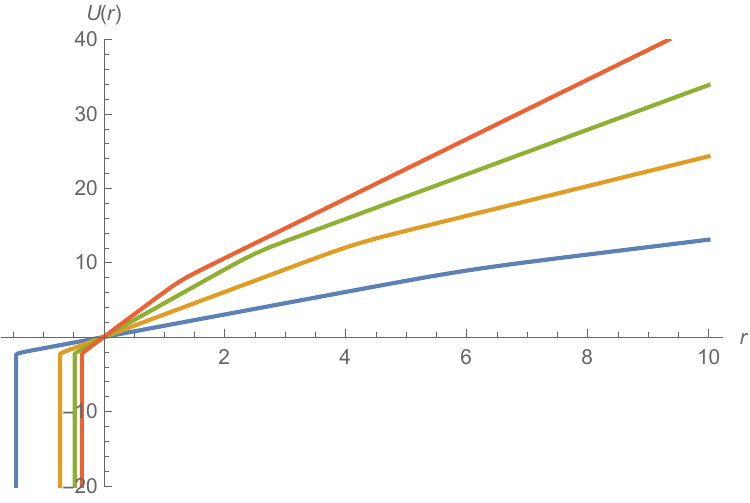}
  \caption{$U$ solution}
  \end{subfigure}
  \begin{subfigure}[b]{0.32\linewidth}
    \includegraphics[width=\linewidth]{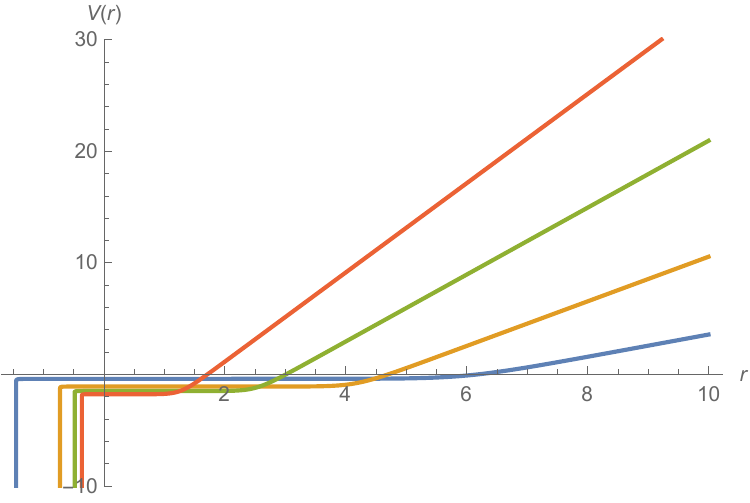}
  \caption{$V$ solution}
  \end{subfigure}
  \begin{subfigure}[b]{0.32\linewidth}
    \includegraphics[width=\linewidth]{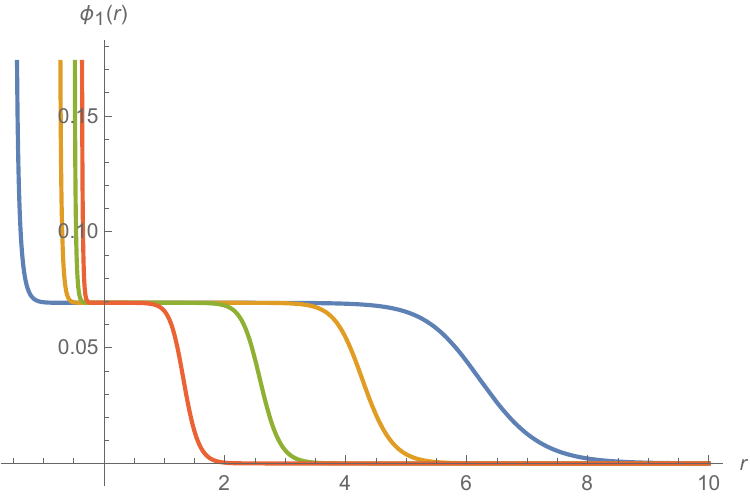}
  \caption{$\phi_1$ solution}
  \end{subfigure}\\
  \begin{subfigure}[b]{0.32\linewidth}
    \includegraphics[width=\linewidth]{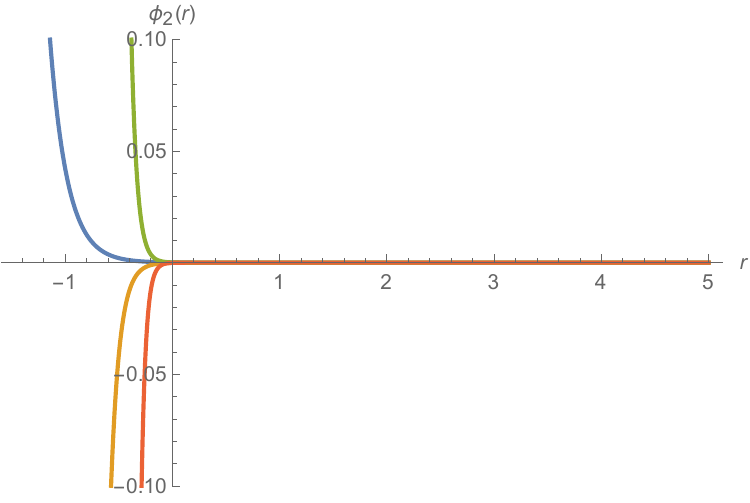}
  \caption{$\phi_2$ solution}
  \end{subfigure}
  \begin{subfigure}[b]{0.32\linewidth}
    \includegraphics[width=\linewidth]{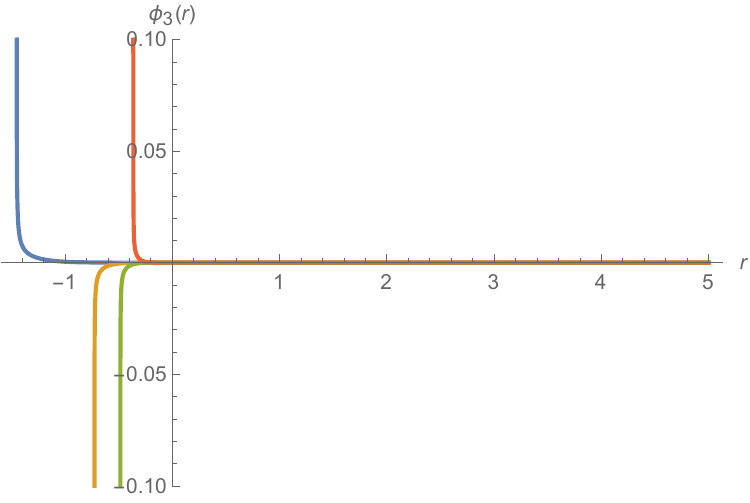}
  \caption{$\phi_3$ solution}
  \end{subfigure}
  \begin{subfigure}[b]{0.32\linewidth}
    \includegraphics[width=\linewidth]{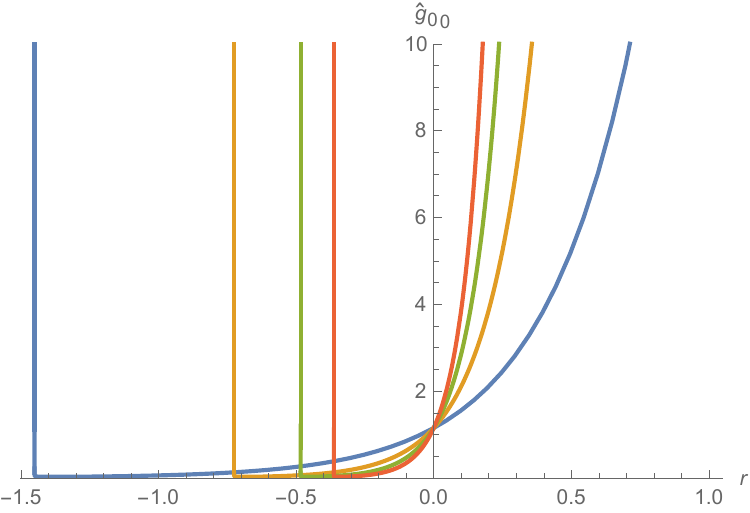}
  \caption{Profiles of $\hat{g}_{00}$}
  \end{subfigure}
  \caption{RG flows from the $N=4$ $AdS_7$ critical point to the $AdS_4\times H^3$ fixed point and curved domain walls for $SO(3)$ twists with $\phi_1$, $\phi_2$ and $\phi_3$ non-vanishing. The blue, orange, green and red curves refer to $g=8,16,24,32$, respectively.}
  \label{YSO3flow3}
\end{figure}

Although $\phi_2$ and $\phi_3$ vahish at both $AdS_7$ and $AdS_4\times H^3$ fixed points, we can consider RG flows to curved domain walls with non-vanishing $\phi_2$ and $\phi_3$. Examples of various possible RG flows are given in figure \ref{YSO3flow3}. The behavior of $\hat{g}_{00}$ near the singularities, $\hat{g}_{00}\rightarrow \infty$, indicates that these singularities are unphysical by the criterion of \cite{Maldacena_nogo}. 

\subsubsection{Solutions with $SO(3)_+$ twists}\label{YSO(3)sdsection}
\begin{figure}
  \centering
  \begin{subfigure}[b]{0.38\linewidth}
    \includegraphics[width=\linewidth]{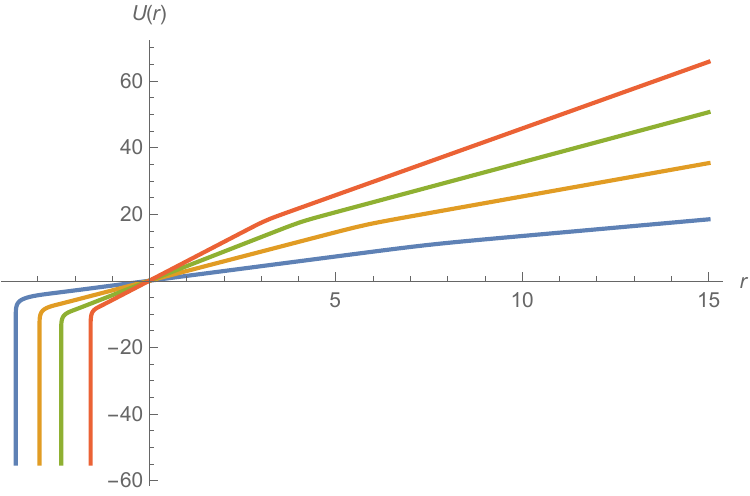}
  \caption{$U$ solution}
  \end{subfigure}
  \begin{subfigure}[b]{0.38\linewidth}
    \includegraphics[width=\linewidth]{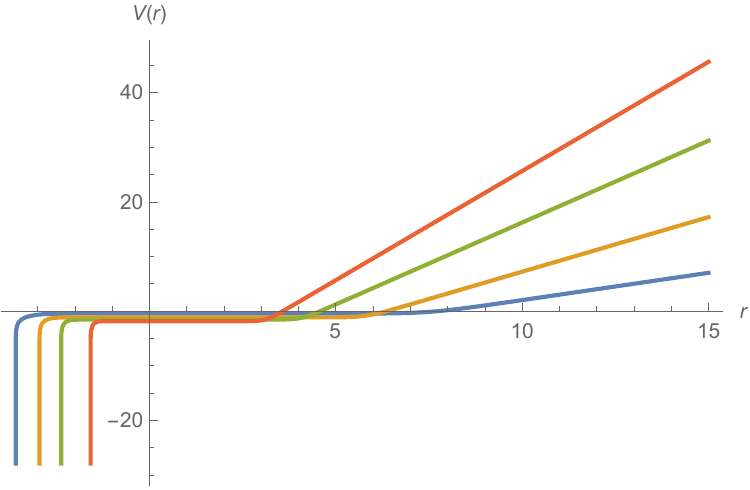}
  \caption{$V$ solution}
  \end{subfigure}\\
  \begin{subfigure}[b]{0.38\linewidth}
    \includegraphics[width=\linewidth]{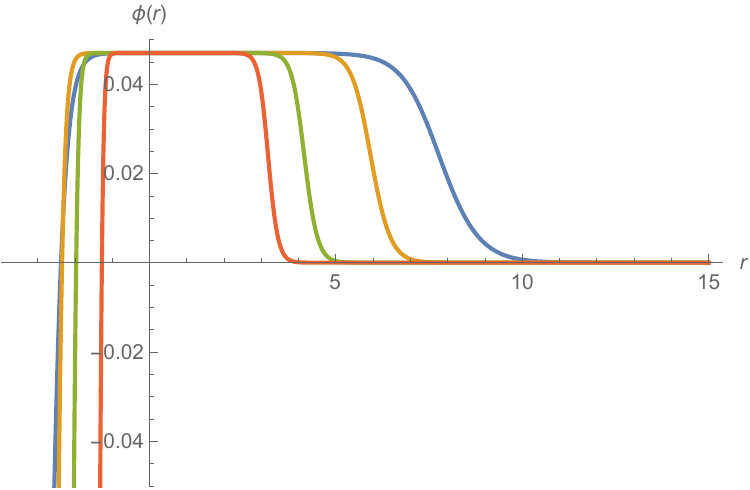}
  \caption{$\phi$ solution}
  \end{subfigure}
    \begin{subfigure}[b]{0.38\linewidth}
    \includegraphics[width=\linewidth]{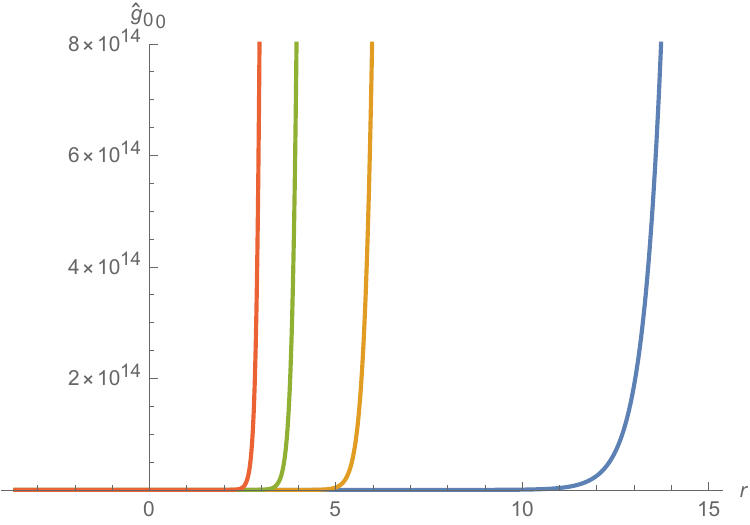}
  \caption{Profiles of $\hat{g}_{00}$}
  \end{subfigure}
  \caption{RG flows from the $N=4$ $AdS_7$ critical point to the $AdS_4\times H^3$ fixed point and curved domain walls for $SO(3)_+$ twists. The blue, orange, green and red curves refer to $g=8,16,24,32$, respectively.}
  \label{YsdSO3flows}
\end{figure}

We now consider another twist given by turning on $SO(3)_+$ gauge fields. In this case, the $SO(3)_+$ is identified with the self-dual $SO(3)$ subgroup of $SO(4)\sim SO(3)_+\times SO(3)_-\subset SO(5)$. We will accordingly turn on the following gauge fields
\begin{eqnarray}
A^{12}_{(1)}=A^{34}_{(1)}&=& -e^{-V}\frac{p}{2k}\frac{f'_{k}(\psi)}{f_{k}(\psi)}e^{\hat{\theta}},\nonumber \\ A^{13}_{(1)}=A^{24}_{(1)}&=& -e^{-V}\frac{p}{2k}\frac{f'_{k}(\psi)}{f_{k}(\psi)}e^{\hat{\varphi}}, \nonumber \\ A^{23}_{(1)}=A^{14}_{(1)}&=& -e^{-V}\frac{p}{2k}\frac{\cot\theta}{f_{k}(\psi)}e^{\hat{\varphi}}\, . \label{YsdSO(3)gaugeAnt}
\end{eqnarray}
\indent The gauge groups containing $SO(3)_+\subset SO(4)$ are given by $SO(5)$, $SO(4,1)$, and $CSO(4,0,1)$. These groups can be gauged altogether by the following embedding tensor
\begin{equation}\label{SO(4)Ytensor}
Y_{MN}=\text{diag}(+1,+1,+1,+1,\rho)
\end{equation}
with $\rho=1,-1,0$, respectively. 
\\
\indent There is only one $SO(3)_+$ singlet scalar corresponding to the $SL(5)$ non-compact generator
\begin{equation}\label{YSO(4)Ys}
\hat{Y}=e_{1,1}+e_{2,2}+e_{3,3}+e_{4,4}-4e_{5,5}\, .
\end{equation}
It should be noted that this generator is invariant under a larger symmetry $SO(4)$. With $SL(5)/SO(5)$ coset representative of the form 
\begin{equation}\label{YSO(4)coset}
\mathcal{V}=e^{\phi\hat{Y}},
\end{equation}
the scalar potential is given by
\begin{equation}\label{YSO(4)Pot}
\mathbf{V}=-\frac{g^2}{64}e^{-4\phi}(8+8\rho e^{10\phi}-\rho^2e^{20\phi})\, .
\end{equation}
As expected, there is an $N=4$ supersymmetric $AdS_7$ critical point at $\phi=0$ and a non-supersymmetric, unstable, $AdS_7$ critical point at $\phi=\frac{1}{10}\ln{2}$.  
\\
\indent To implement the twist, we impose the following projection conditions given in \eqref{SO(3)Projcon}
and
\begin{equation}\label{GammaSDProj}
{(\Gamma_{12})^a}_b\epsilon^b={(\Gamma_{34})^a}_b\epsilon^b\, .
\end{equation}
Together with the twist condition \eqref{GenQYM} and the $\gamma_r$ projection condition \eqref{pureYProj}, we find the following BPS equations 
\begin{eqnarray}
U'&=&\frac{g}{40}(4e^{-2\phi}+\rho e^{8\phi})-\frac{6}{5}e^{-2(V-\phi)}p,\\
V'&=&\frac{g}{40}(4e^{-2\phi}+\rho e^{8\phi})+\frac{14}{5}e^{-2(V-\phi)}p,\\
\phi'&=&\frac{g}{20}(e^{-2\phi}-\rho e^{8\phi})-\frac{3}{5}e^{-2(V-\phi)}p\, .
\end{eqnarray}
As in the $SO(3)$ twist, the BPS equations admit an $AdS_4\times H^3$ fixed point only for $SO(5)$ gauge group. This $AdS_4\times H^3$ vacuum is given by
\begin{equation}
V=\frac{1}{2}\ln\left[\frac{8\times2^{1/5}\times5^{3/5}}{g^2}\right],\quad
\phi=\frac{1}{10}\ln\left[\frac{8}{5}\right],\quad 
L_{\text{AdS}_4}=\frac{2^{3/5}\times5^{4/5}}{g}
\end{equation}
which does not seem to appear in the previously known results.  
\\
\indent Unlike the previous case, this $AdS_4\times H^3$ fixed point preserves only four supercharges and corresponds to $N=1$ SCFT in three dimensions. We can similarly study numerical RG flows from the supersymmetric $AdS_7$ vacuum to this $AdS_4\times H^3$ fixed point and to curved domain walls dual to $N=1$ non-conformal field theories in three dimensions. Examples of these RG flows are given in figure \ref{YsdSO3flows}. It can be seen again that the IR singularities are physical since $\hat{g}_{00}\rightarrow 0$ near the singularities.
\\
\indent For $CSO(4,0,1)$ gauge group, we can analytically solve the BPS equations. The resulting solution is given by
\begin{eqnarray}
\phi&=&C+\frac{g}{160p}(4p\tilde{r}+\tilde{C})^2-\frac{3}{20}\ln(4p\tilde{r}+\tilde{C}),\\
V&=&2\phi+\ln(4p\tilde{r}+\tilde{C}),\\
U&=&V-\ln(4p\tilde{r}+\tilde{C})+C'\, .
\end{eqnarray}
The new radial coordinate $\tilde{r}$ is defined by $\frac{d\tilde{r}}{dr}=e^{-V}$. The integration constants $\tilde{C}$ and $C'$ can be neglected by shifting the coordinate $\tilde{r}$ and rescaling the coordinates $x^m$ on $Mkw_3$. 
\\
\indent Setting $C'=\tilde{C}=0$, we find the leading behavior of the solution at large $\tilde{r}$
\begin{equation}
\phi\sim \tilde{r}^2\qquad \textrm{and} \qquad U\sim V\sim 2\phi\, .
\end{equation} 
In this limit, the contribution from the gauge fields to the BPS equations is highly suppressed due to $V\rightarrow \infty$. The asymptotic behavior is then identified with the standard, flat, domain wall found in \cite{our_7D_DW}. Similar to the case of solutions with an asymptotically locally $AdS_7$ space, we will call this limit an asymptotically locally flat domain wall. 
\\
\indent On the other hand, as $\tilde{r}\rightarrow 0$, we find
\begin{equation}
\phi\sim -\frac{3}{20}\ln (4p\tilde{r}),\qquad V\sim \frac{7}{10}\ln(4p\tilde{r}),\qquad U\sim -\frac{3}{10}\ln(4p\tilde{r}).
\end{equation} 
\indent We also note that, in this case, the complete truncation ansatz in term of type IIA theory on $S^3$ has been constructed in \cite{S3_S4_typeIIA}. Therefore, the solution can be completely embedded in type IIA theory. In this paper, we are only interested in the time component of the ten-dimensional metric given by, see for example \cite{our_7D_DW} for more detail,
\begin{equation} 
\hat{g}_{00}=e^{2U+\frac{3}{2}\phi}\, .
\end{equation}
Using this result, we find that as $\tilde{r}\rightarrow 0$, $\hat{g}_{00}\rightarrow\infty$, so, in this case, the IR singularity is unphysical.

\subsection{Supersymmetric $AdS_3\times \Sigma^4$ solutions}\label{YAdS3section}
In this section, we move on to the analysis of $AdS_3\times \Sigma^4$ solutions. We will consider two types of the internal manifold $\Sigma^4$ namely a Riemannian four-manifold $M^4$ with a constant curvature and a product of two Riemann surfaces $\Sigma^2\times\Sigma^2$ with $SO(4)$ and $SO(2)\times SO(2)$ twists, respectively. 

\subsubsection{$AdS_3\times M^4$ solutions with $SO(4)$ twists}\label{YSO(4)section}
\begin{figure}
  \centering
  \begin{subfigure}[b]{0.38\linewidth}
    \includegraphics[width=\linewidth]{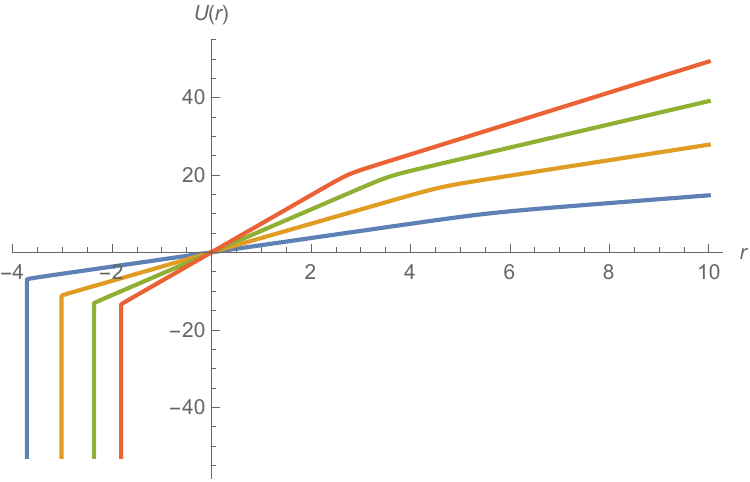}
  \caption{$U$ solution}
  \end{subfigure}
  \begin{subfigure}[b]{0.38\linewidth}
    \includegraphics[width=\linewidth]{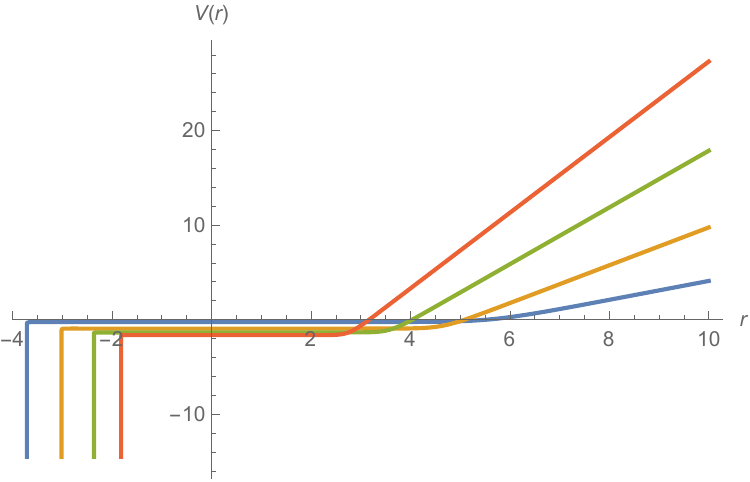}
  \caption{$V$ solution}
  \end{subfigure}\\
  \begin{subfigure}[b]{0.38\linewidth}
    \includegraphics[width=\linewidth]{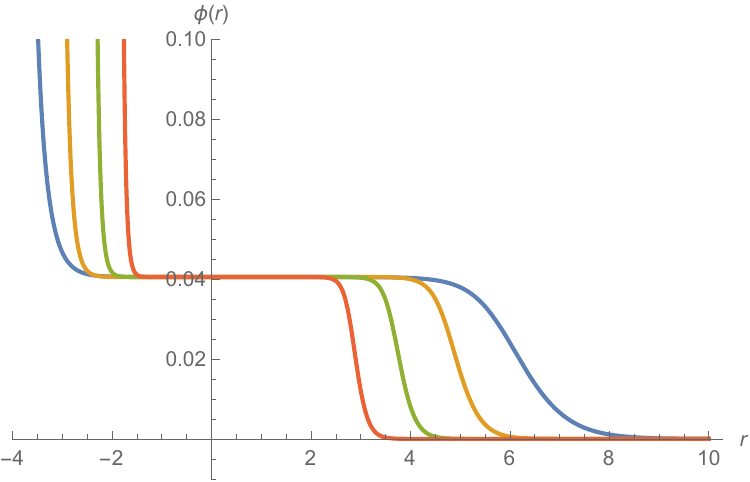}
  \caption{$\phi$ solution}
  \end{subfigure}
    \begin{subfigure}[b]{0.38\linewidth}
    \includegraphics[width=\linewidth]{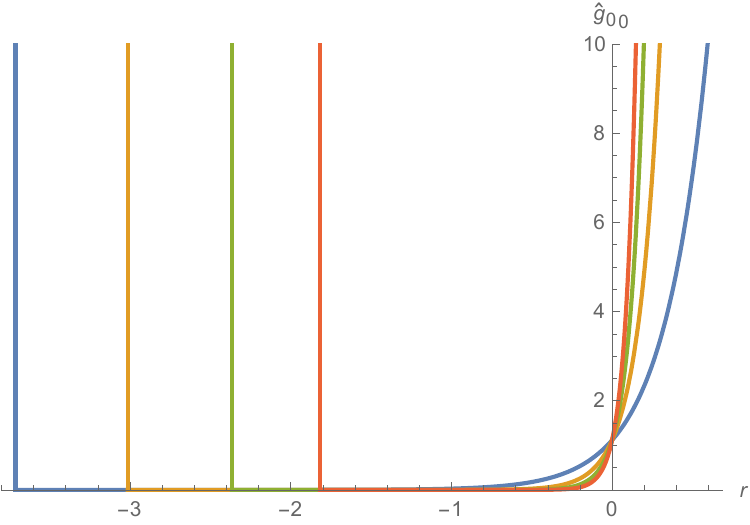}
  \caption{Profiles of $\hat{g}_{00}$}
  \end{subfigure}
  \caption{RG flows from the $N=4$ $AdS_7$ critical point to the $AdS_3\times H^4$ fixed point and curved domain walls for $SO(4)$ twist in $SO(5)$ gauge group. The blue, orange, green and red curves refer to $g=8,16,24,32$, respectively.}
  \label{Y_SO(4)flows}
\end{figure}

As in the previous section, we will consider $SO(4)$ symmetric solutions for $SO(5)$, $SO(4,1)$ and $CSO(4,0,1)$ gauge groups with the embedding tensor given in \eqref{SO(4)Ytensor}. To find $AdS_3\times M^4_k$ solutions, we use the following ansatz for the seven-dimensional metric
\begin{equation}\label{SO(4)7Dmetric}
ds_7^2=e^{2U(r)}dx^2_{1,1}+dr^2+e^{2V(r)}ds^2_{M^{4}_{k}}
\end{equation}
with $dx^2_{1,1}=\eta_{mn}dx^{m} dx^{n}$ for $m,n=0,1$ being the metric on two-dimensional Minkowski space. The explicit form of the metric on $M^4_k$ is given by
\begin{equation}\label{Sigma4metric}
ds^2_{\Sigma^4_k}=d\chi^2+f_k(\chi)^2\left[d\psi^2+\sin^2{\psi}(d\theta^2+\sin^2{\theta}d\varphi^2)\right]
\end{equation}
with $\chi, \psi, \theta \in[0,\frac{\pi}{2}]$, $\varphi\in[0,2\pi]$ and $f_k(\chi)$ is the function defined in \eqref{fFn}. 
\\
\indent With the vielbein basis of the form
\begin{eqnarray}
e^{\hat{m}}&=& e^{U}dx^{m}, \qquad e^{\hat{r}}=dr,  \qquad e^{\hat{\chi}}=e^{V}d\chi, \qquad  e^{\hat{\psi}}= e^{V}f_{k}(\chi)d\psi,\nonumber \\
e^{\hat{\theta}}&=& e^{V}f_{k}(\chi)\sin{\psi}d\theta, \qquad e^{\hat{\varphi}}=e^{V}f_{k}(\chi)\sin{\psi}\sin{\theta}d\varphi,\label{Sigma4bein}
\end{eqnarray}
we obtain the following non-vanishing components of the spin connection
\begin{eqnarray}
\omega_{(1)}^{\hat{m}\hat{r}}&=&  U'e^{\hat{m}}, \qquad \omega_{(1)}^{\hat{i}\hat{r}}= V'e^{\hat{i}}, \nonumber \\
\omega_{(1)}^{\hat{\psi}\hat{\chi}}&=& \frac{f'_{k}(\chi)}{f_{k}(\chi)}e^{-V}e^{\hat{\psi}},\qquad \omega_{(1)}^{\hat{\theta}\hat{\chi}}=\frac{f'_{k}(\chi)}{f_{k}(\chi)}e^{-V}e^{\hat{\theta}},\nonumber \\  \omega_{(1)}^{\hat{\varphi}\hat{\chi}}&=&\frac{f'_{k}(\chi)}{f_{k}(\chi)}e^{-V}e^{\hat{\varphi}},\qquad \omega_{(1)}^{\hat{\theta}\hat{\psi}}= \frac{\cot\psi}{f_{k}(\chi)}e^{-V}e^{\hat{\theta}}, \nonumber  \\ \omega_{(1)}^{\hat{\varphi}\hat{\psi}}&=&\frac{\cot\psi}{f_{k}(\chi)}e^{-V}e^{\hat{\varphi}}, \qquad \omega_{(1)}^{\hat{\varphi}\hat{\theta}}=\frac{\cot\theta}{f_{k}(\chi)\sin{\psi}}e^{-V}e^{\hat{\varphi}}
\label{AdS3xSigma4SpinCon}
\end{eqnarray}
with $\hat{i},\hat{j}=\hat{3},...,\hat{6}=\hat{\chi}, \hat{\psi}, \hat{\theta}, \hat{\varphi}$ being flat indices on $M^4_k$.
\\
\indent We will perform a twist on $M^4_k$ by turning on $SO(4)$ gauge fields to cancel the spin connection as follow
\begin{equation}\label{SO(4)gaugeAnt}
A^{IJ}_{(1)}=-\frac{p}{k}\delta^{I+2}_{[\hat{i}}\delta^{J+2}_{\hat{j}]} \left. \omega^{\hat{i}\hat{j}}_{(1)}\right |_{M^4_k}\, .
\end{equation}
The corresponding two-form field strengths are given by
\begin{equation}\label{SO(4)2form}
\mathcal{H}^{(2)IJ}_{\hat{i}\hat{j}}=F^{IJ}_{\hat{i}\hat{j}}= 2\delta^{I+2}_{[\hat{i}}\delta^{J+2}_{\hat{j}]}e^{-2V}p\, .
\end{equation}
\indent For the $SO(4)$ singlet scalar, we use the same coset representative given in \eqref{YSO(4)coset}. 
However, in this case, the three-form field strengths cannot vanish in order to satisfy the Bianchi's identity for $\mc{H}^{(3)}_{\mu\nu\rho M}$ since the above gauge fields lead to non-vanishing $\epsilon_{MNPQR}\mc{H}^{(2)MN}\wedge \mc{H}^{(2)PQ}$ terms. To preserve the residual $SO(4)$ symmetry, only $\mc{H}^{(3)}_{\mu\nu\rho 5}$ is allowed. We also note that for $SO(5)$ and $SO(4,1)$ gauge groups, the corresponding embedding tensor $Y_{MN}$ is non-degenerate. There are in total $t=\textrm{rank}Y=5$ massive three-form fields, so for these gauge groups, $\mc{H}^{(3)}_{\mu\nu\rho 5}$ is obtained by turning on the massive three-form $S^{(3)}_{\mu\nu\rho 5}$. On the other hand, for $CSO(4,0,1)$ gauge group, we have $Y_{55}=0$, so the contribution to $\mc{H}^{(3)}_{\mu\nu\rho 5}$ comes from a massless two-form field $B_{\mu\nu 5}$. However, we are not able to determine a suitable ansatz for $B_{\mu\nu 5}$ in order to find a consistent set of BPS equations that are compatible with the second-ordered field equations. Accordingly, we will not consider the non-semisimple $CSO(4,0,1)$ gauge group in the following analysis. 
\\
\indent For $SO(5)$ and $SO(4,1)$ gauge groups, the appropriate ansatz for the modified three-form field strength is given by
\begin{equation}\label{SO(3)3form}
\mathcal{H}^{(3)}_{\hat{m}\hat{n}\hat{r} 5}= -\frac{96}{g}\rho e^{-4(V+2\phi)}p^2\varepsilon_{\hat{m}\hat{n}}\, .
\end{equation}
Imposing the twist condition \eqref{GenQYM} and the projector in \eqref{pureYProj} together with additional projectors of the form
\begin{equation}\label{Sigma4ProjCon}
\gamma^{\hat{\chi}\hat{\psi}}\epsilon^a=-{(\Gamma_{12})^a}_b\epsilon^b,\qquad \gamma^{\hat{\psi}\hat{\theta}}\epsilon^a=-{(\Gamma_{23})^a}_b\epsilon^b,\qquad \gamma^{\hat{\theta}\hat{\varphi}}\epsilon^a=-{(\Gamma_{34})^a}_b\epsilon^b,
\end{equation}
we find the BPS equations 
\begin{eqnarray}
U'&=&\frac{g}{40}(4e^{-2\phi}+\rho e^{8\phi})-\frac{12}{5}e^{-2V+2\phi}p+\frac{288}{5g}\rho e^{-4(V+\phi)}p^2,\\
V'&=&\frac{g}{40}(4e^{-2\phi}+\rho e^{8\phi})+\frac{18}{5}e^{-2V+2\phi}p-\frac{192}{5g}\rho e^{-4(V+\phi)}p^2,\\
\phi'&=&\frac{g}{20}(e^{-2\phi}-\rho e^{8\phi})-\frac{6}{5}e^{-2V+2\phi}p-\frac{96}{5g}\rho e^{-4(V+\phi)}p^2\, .
\end{eqnarray}
\indent From these equations, we find an $AdS_3$ fixed point only for $k=-1$ and $\rho=1$. The resulting $AdS_3\times H^4$ solution is given by
\begin{eqnarray}
V&=&\frac{1}{2}\ln\left[\frac{16\times2^{3/5}\times3^{2/5}}{g^2}\right],\qquad
\phi=\frac{1}{10}\ln\left[\frac{3}{2}\right],\nonumber \\
L_{\text{AdS}_3}&=&\frac{2\times2^{4/5}\times3^{1/5}}{g}\, . \label{FinYSO(4)fixedpoint}
\end{eqnarray} 
This is the $AdS_3\times H^4$ fixed point given in \cite{Gauntlett1} for the maximal $SO(5)$ gauged supergravity. The solution preserves four supercharges and corresponds to $N=(1,1)$ SCFT in two dimensions with $SO(4)$ symmetry. As in the previous cases, we will consider RG flows from the supersymmetric $AdS_7$ vacuum to this $AdS_3\times H^4$ fixed point and curved domain walls. Examples of these RG flows are given in figure \ref{Y_SO(4)flows}. Unlike the previous cases, the IR singularities in this case are unphysical due to the behavior $\hat{g}_{00}\rightarrow \infty$.

\subsubsection{$AdS_3\times \Sigma^2\times \Sigma^2$ solutions with $SO(2)\times SO(2)$ twists}\label{YSO(2)xSO(2)section}
\begin{figure}
  \centering
  \begin{subfigure}[b]{0.32\linewidth}
    \includegraphics[width=\linewidth]{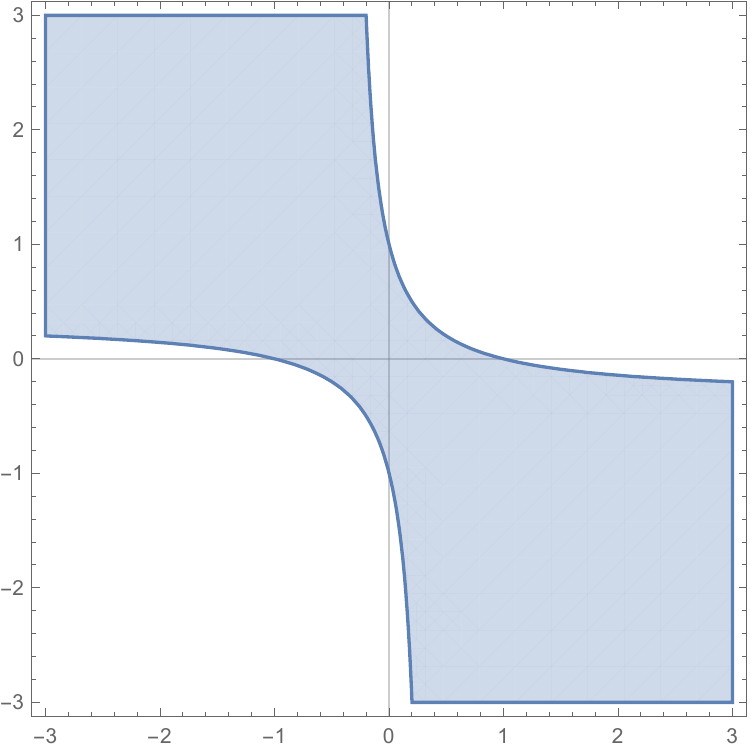}
  \end{subfigure}
  \begin{subfigure}[b]{0.32\linewidth}
    \includegraphics[width=\linewidth]{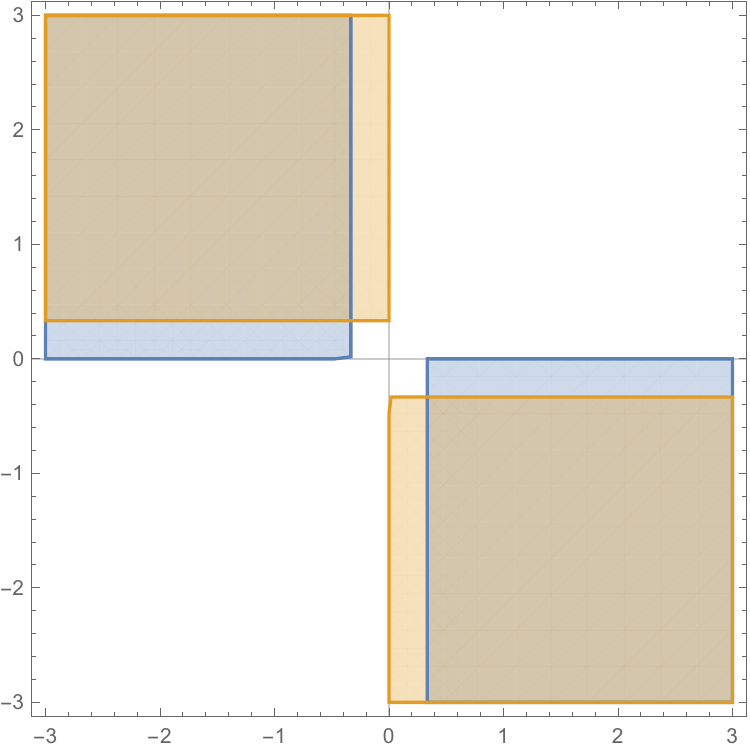}
  \end{subfigure}
  \begin{subfigure}[b]{0.32\linewidth}
    \includegraphics[width=\linewidth]{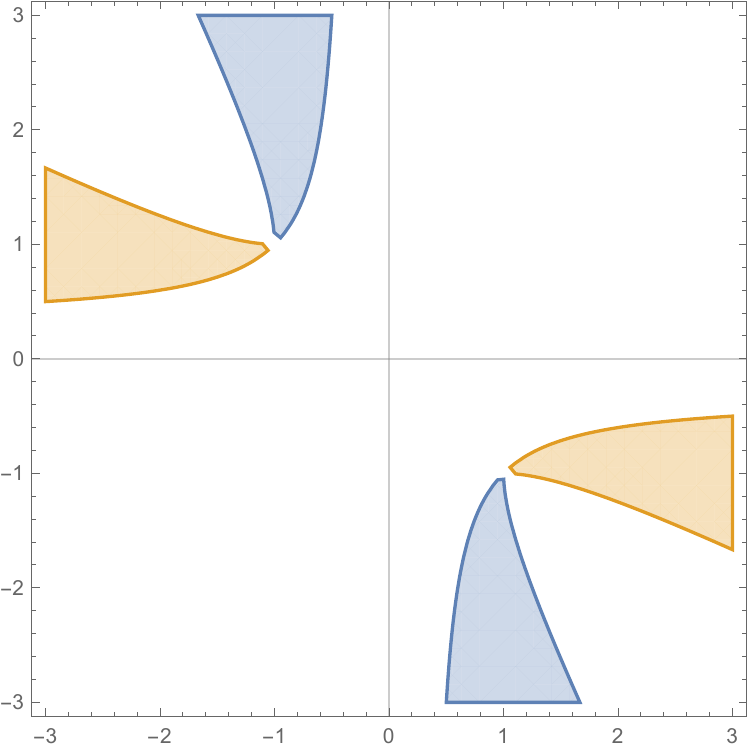}
  \end{subfigure}
  \caption{Regions (blue) in the parameter space $(z_1,z_2)$ where good $AdS_3$ vacua exist in $SO(5)$ gauge group. From left to right, these figures correspond to the cases of $(k_1=k_2=-1)$, $(k_1=-1, k_2=0)$, $(k_1=-k_2=-1)$, respectively. The orange regions are obtained from interchanging $k_1$ and $k_2$.}
  \label{SO(5)RegionPlot}
\end{figure}

\begin{figure}
  \centering
  \begin{subfigure}[b]{0.32\linewidth}
    \includegraphics[width=\linewidth]{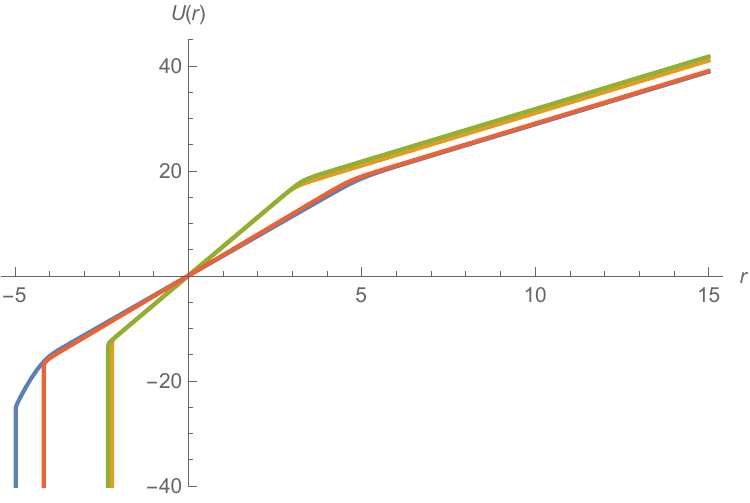}
  \caption{$U$ solution}
  \end{subfigure}
  \begin{subfigure}[b]{0.32\linewidth}
    \includegraphics[width=\linewidth]{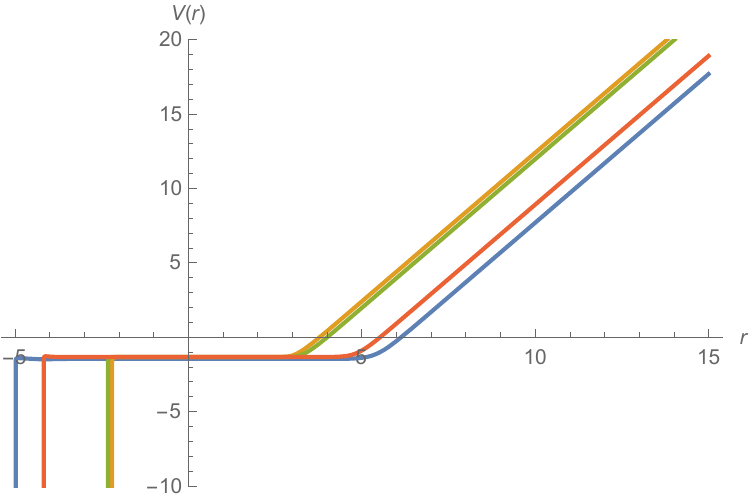}
  \caption{$V$ solution}
  \end{subfigure}
  \begin{subfigure}[b]{0.32\linewidth}
    \includegraphics[width=\linewidth]{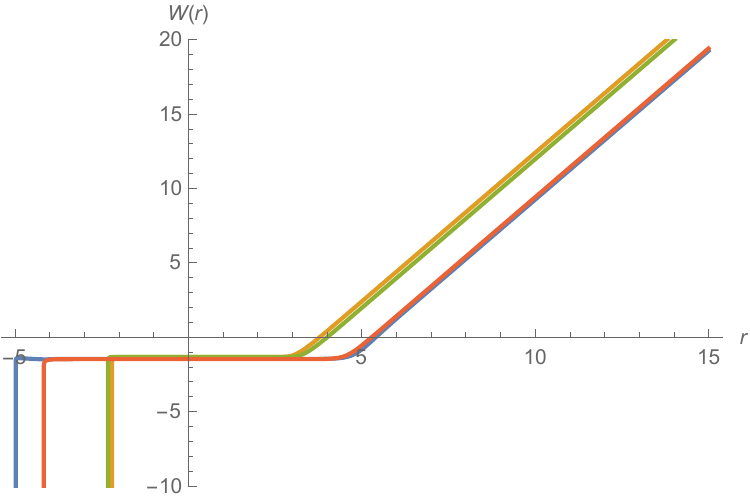}
  \caption{$W$ solution}
  \end{subfigure}\\
  \begin{subfigure}[b]{0.32\linewidth}
    \includegraphics[width=\linewidth]{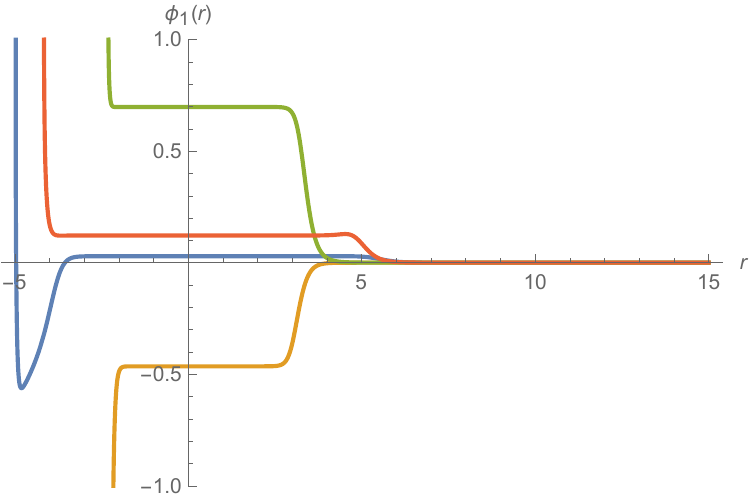}
  \caption{$\phi_1$ solution}
  \end{subfigure}
  \begin{subfigure}[b]{0.32\linewidth}
    \includegraphics[width=\linewidth]{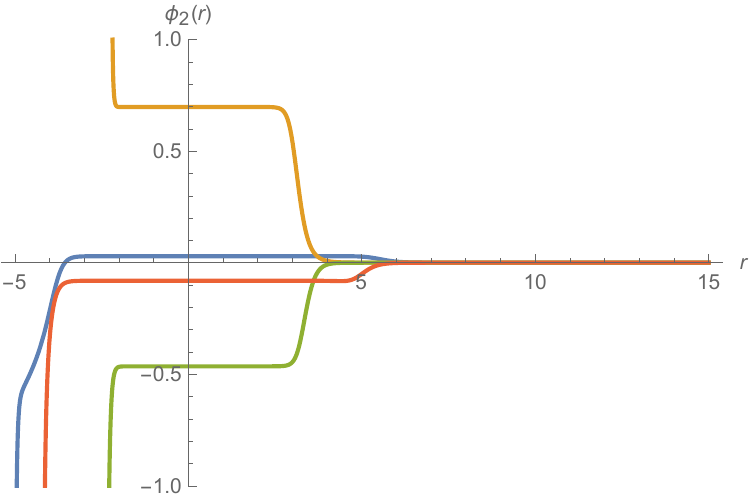}
  \caption{$\phi_2$ solution}
  \end{subfigure}
    \begin{subfigure}[b]{0.32\linewidth}
    \includegraphics[width=\linewidth]{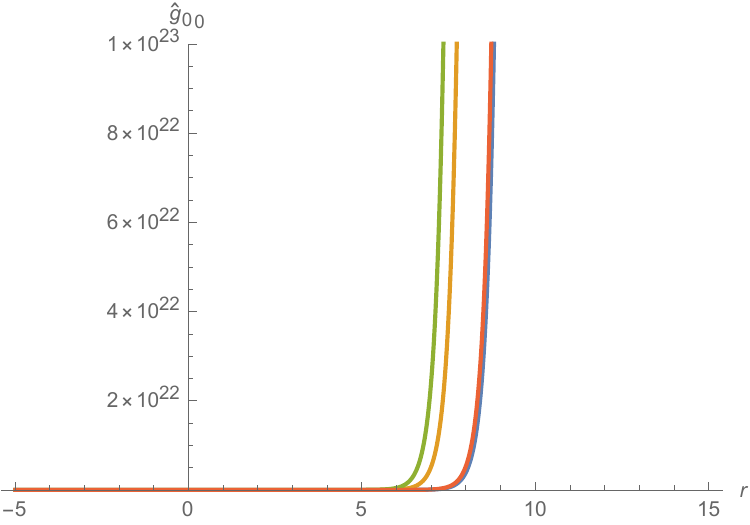}
  \caption{Profiles of $\hat{g}_{00}$}
  \end{subfigure}
  \caption{RG flows from the $N=4$ $AdS_7$ critical point to the $AdS_3\times H^2\times H^2$ fixed points and curved domain walls for $SO(2)\times SO(2)$ twist in $SO(5)$ gauge group. The blue, orange, green and red curves refer to $(z_1,z_2)=(0,0), (0.3,0.3), (-0.3,-0.3), (-1,0.5)$, respectively.}
  \label{AdS3xH2xH2flows}
\end{figure}

\begin{figure}
  \centering
  \begin{subfigure}[b]{0.32\linewidth}
    \includegraphics[width=\linewidth]{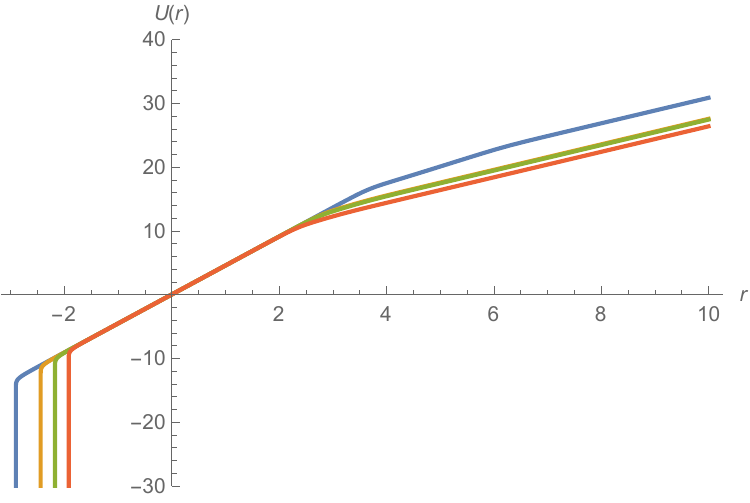}
  \caption{$U$ solution}
  \end{subfigure}
  \begin{subfigure}[b]{0.32\linewidth}
    \includegraphics[width=\linewidth]{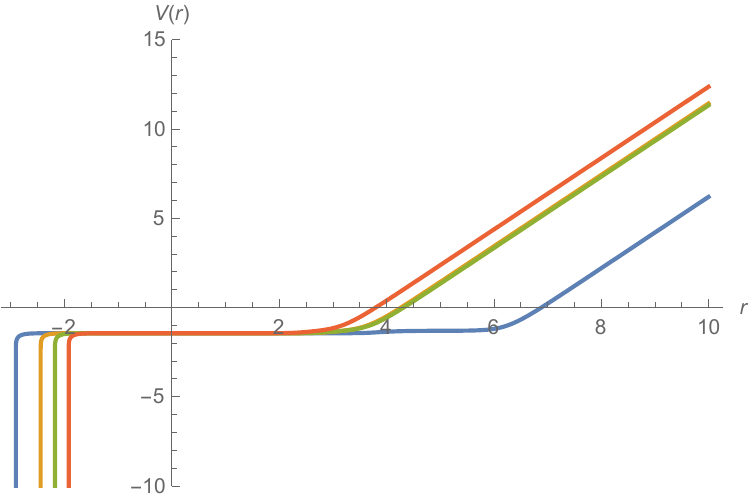}
  \caption{$V$ solution}
  \end{subfigure}
  \begin{subfigure}[b]{0.32\linewidth}
    \includegraphics[width=\linewidth]{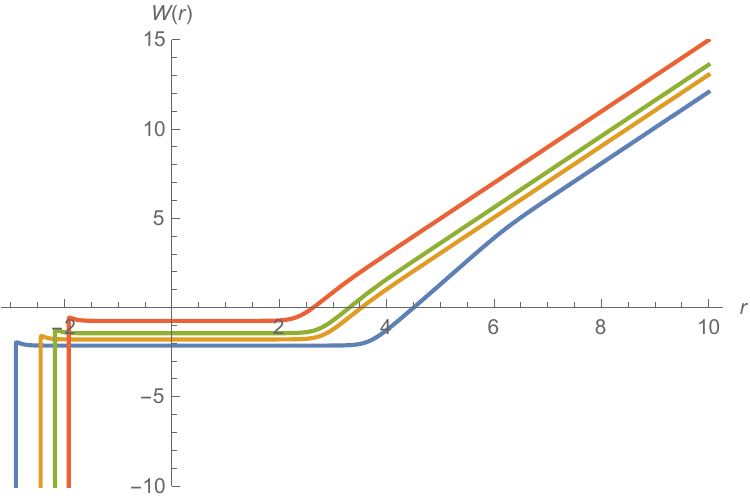}
  \caption{$W$ solution}
  \end{subfigure}
  \begin{subfigure}[b]{0.32\linewidth}
    \includegraphics[width=\linewidth]{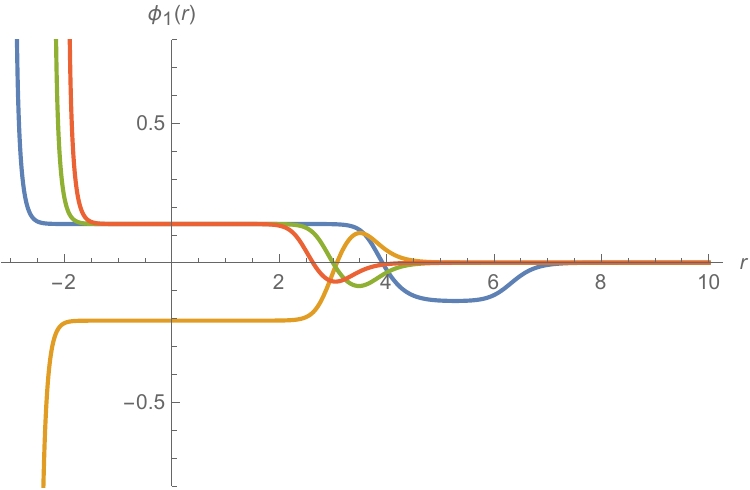}
  \caption{$\phi_1$ solution}
  \end{subfigure}
  \begin{subfigure}[b]{0.32\linewidth}
    \includegraphics[width=\linewidth]{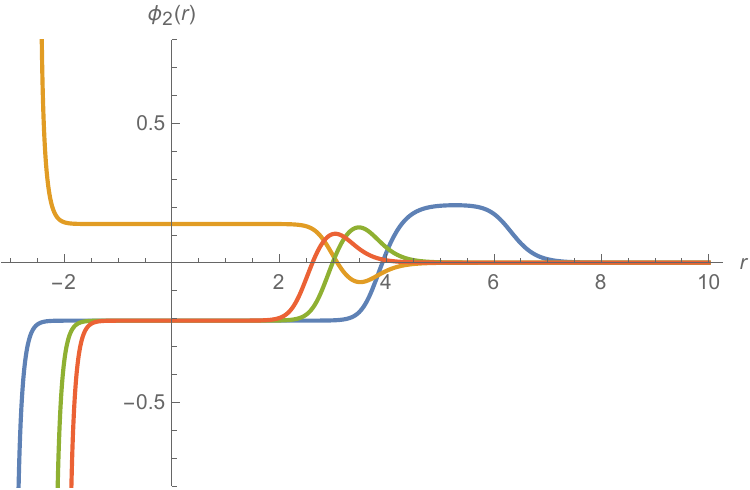}
  \caption{$\phi_2$ solution}
  \end{subfigure}
    \begin{subfigure}[b]{0.32\linewidth}
    \includegraphics[width=\linewidth]{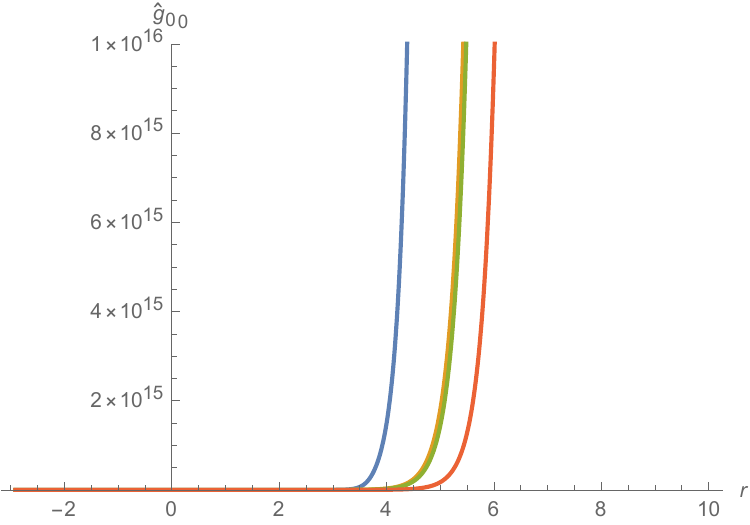}
  \caption{Profiles of $\hat{g}_{00}$}
  \end{subfigure}  
  \caption{RG flows from the $N=4$ $AdS_7$ critical point to the $AdS_3\times H^2\times \mathbb{R}^2$ fixed points and curved domain walls for $SO(2)\times SO(2)$ twist in $SO(5)$ gauge group. The blue, orange, green and red curves refer to $(z_1,z_2)=(1,-0.5), (-1,1),  (1,-2), (-8,1)$, respectively.}
  \label{AdS3xH2xR2flows}
\end{figure}

\begin{figure}
  \centering
  \begin{subfigure}[b]{0.32\linewidth}
    \includegraphics[width=\linewidth]{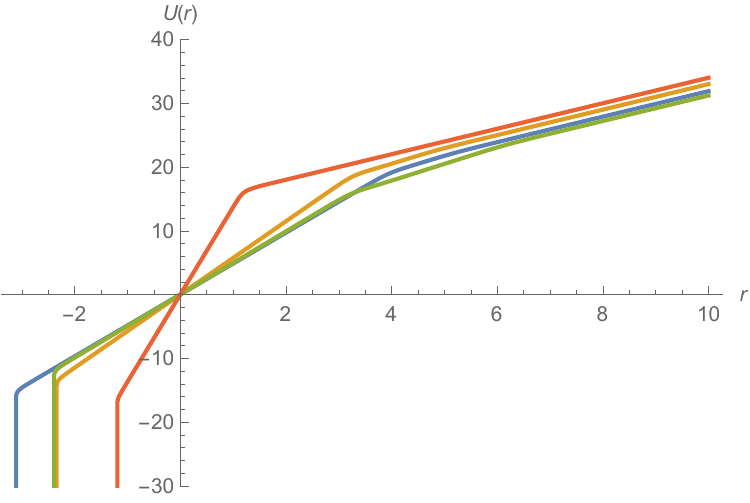}
  \caption{$U$ solution}
  \end{subfigure}
  \begin{subfigure}[b]{0.32\linewidth}
    \includegraphics[width=\linewidth]{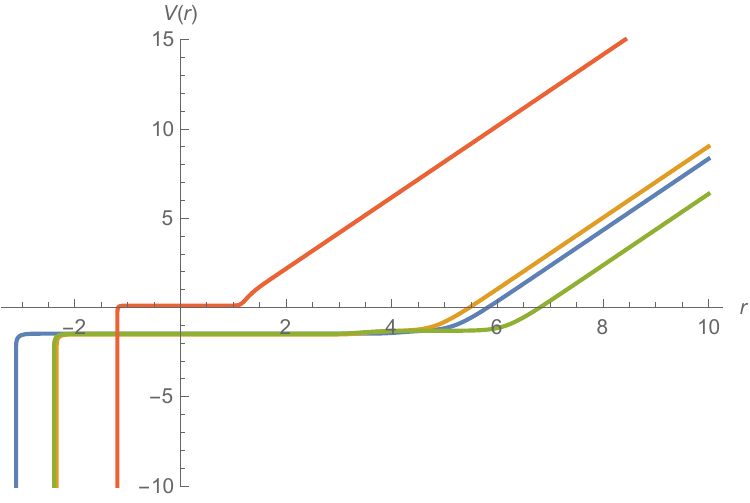}
  \caption{$V$ solution}
  \end{subfigure}
  \begin{subfigure}[b]{0.32\linewidth}
    \includegraphics[width=\linewidth]{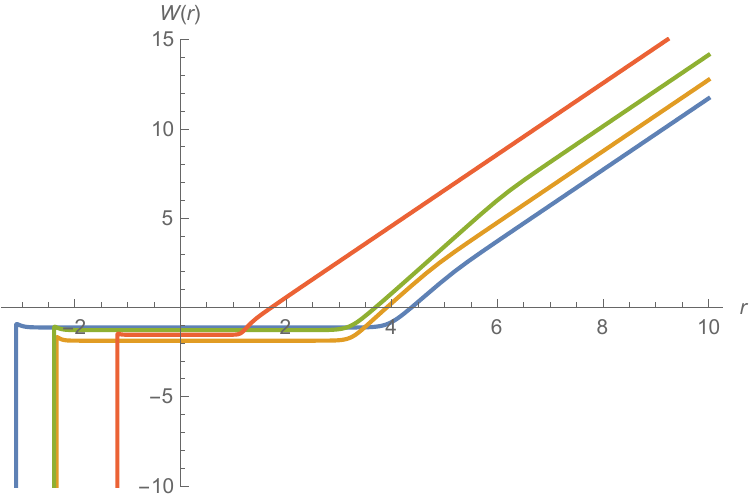}
  \caption{$W$ solution}
  \end{subfigure}\\
  \begin{subfigure}[b]{0.32\linewidth}
    \includegraphics[width=\linewidth]{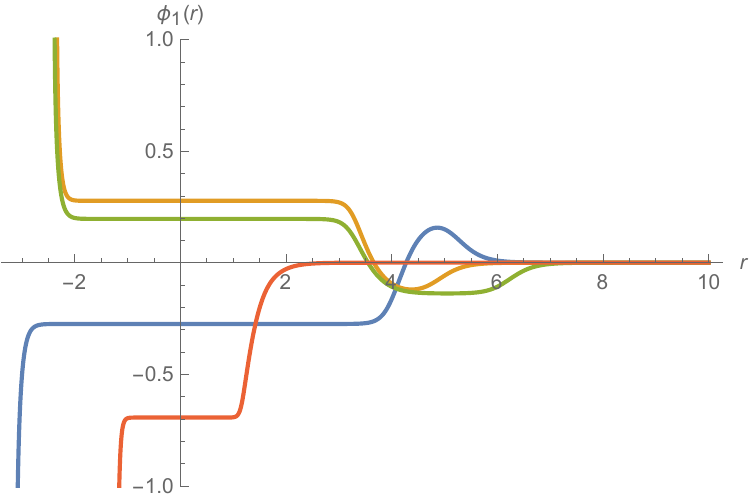}
  \caption{$\phi_1$ solution}
  \end{subfigure}
  \begin{subfigure}[b]{0.32\linewidth}
    \includegraphics[width=\linewidth]{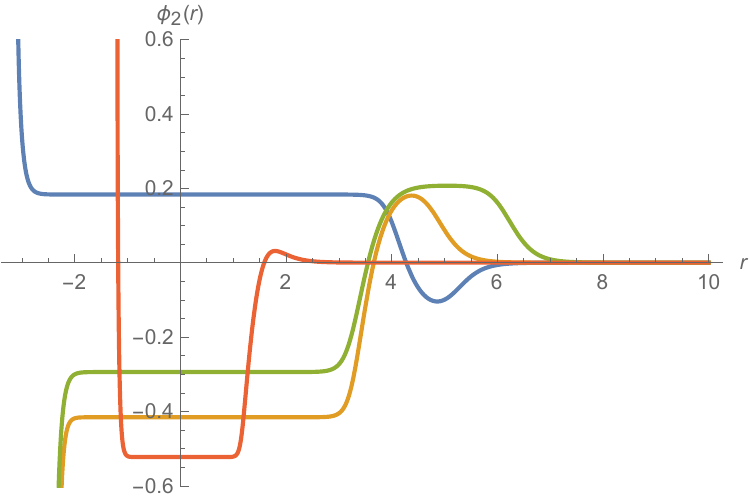}
  \caption{$\phi_2$ solution}
  \end{subfigure}
     \begin{subfigure}[b]{0.32\linewidth}
    \includegraphics[width=\linewidth]{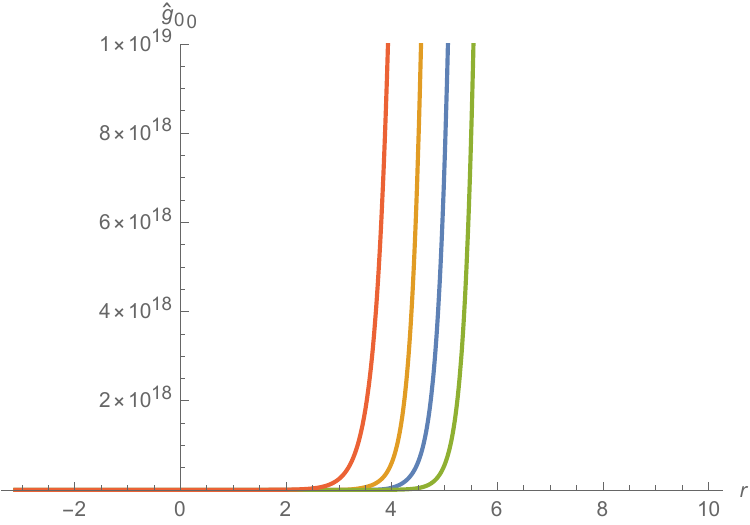}
  \caption{Profiles of $\hat{g}_{00}$}
  \end{subfigure}
  \caption{RG flows from the $N=4$ $AdS_7$ critical point to the $AdS_3\times H^2\times S^2$ fixed points and curved domain walls for $SO(2)\times SO(2)$ twist in $SO(5)$ gauge group. The blue, orange, green and red curves refer to $(z_1,z_2)=(-1,5), (1,-2),  (1,-4), (-3,8)$, respectively.}
  \label{AdS3xH2xS2flows}
\end{figure}

In this section, we consider the manifold $\Sigma^4$ in the form of a product of two Riemann surfaces $\Sigma^2_{k_1}\times\Sigma^2_{k_2}$. The ansatz for the metric takes the form of
\begin{equation}\label{SO(2)xSO(2)7Dmetric}
ds_7^2=e^{2U(r)}dx^2_{1,1}+dr^2+e^{2V(r)}ds^2_{\Sigma^{2}_{k_1}}+e^{2W(r)}ds^2_{\Sigma^{2}_{k_2}}
\end{equation}
in which the metrics on $\Sigma^2_{k_1}$ and $\Sigma^2_{k_2}$ are given in \eqref{Sigma2metric}. 
\\
\indent Using the following choice for the vielbein
\begin{eqnarray}
e^{\hat{m}}&=& e^{U}dx^{m}, \qquad e^{\hat{r}}=dr,\qquad e^{\hat{\theta}_1}=e^{V}d\theta_1,\nonumber \\ e^{\hat{\varphi}_1}&=&e^{V}f_{k_1}(\theta_1)d\varphi_1,\qquad e^{\hat{\theta}_2}=e^{W}d\theta_2, \qquad e^{\hat{\varphi}_2}=e^{W}f_{k_2}(\theta_2)d\varphi_2,\label{Sigma2xSigma2bein}
\end{eqnarray}
we obtain all non-vanishing components of the spin connection as follow
\begin{eqnarray}
\omega_{(1)}^{\hat{m}\hat{r}}&=&  U'e^{\hat{m}}, \qquad \omega_{(1)}^{\hat{i}_1\hat{r}}= V'e^{\hat{i}_1}, \qquad \omega_{(1)}^{\hat{i}_2\hat{r}}= W'e^{\hat{i}_2},\nonumber \\
\omega_{(1)}^{\hat{\varphi}_1\hat{\theta}_1}&=& \frac{f_{k_1}'(\theta_1)}{f_{k_1}(\theta_1)}e^{-V}e^{\hat{\varphi}_1},\hspace{0.4cm} \ \omega_{(1)}^{\hat{\varphi}_2\hat{\theta}_2}=\frac{f_{k_2}'(\theta_2)}{f_{k_2}(\theta_2)}e^{-W}e^{\hat{\varphi}_2}\label{AdS3xSigma2xSigma2SpinCon}
\end{eqnarray}
with $\hat{i}_1=\hat{\theta}_1, \hat{\varphi}_1$ and $\hat{i}_2=\hat{\theta}_2, \hat{\varphi}_2$ being flat indices on $\Sigma^2_{k_1}$ and $\Sigma^2_{k_2}$, respectively.
\\
\indent As in other cases, we consider all gauge groups of the form $CSO(p,q,r)$ with an $SO(2)\times SO(2)$ subgroup. These gauge groups are obtained from the embedding tensor given in \eqref{SO(2)xSO(2)Ytensor}. To perform the twist, we turn on the $SO(2)\times SO(2)$ gauge fields
\begin{eqnarray}
A^{12}_{(1)}&=&-\frac{p_{11}}{k_1}\frac{f_{k_1}'(\theta_1)}{f_{k_1}(\theta_1)}e^{-V}e^{\hat{\varphi}_1}-\frac{p_{12}}{k_2}\frac{f_{k_2}'(\theta_2)}{f_{k_2}(\theta_2)}e^{-W}e^{\hat{\varphi}_2},\nonumber \\
A^{34}_{(1)}&=&-\frac{p_{21}}{k_1}\frac{f_{k_1}'(\theta_1)}{f_{k_1}(\theta_1)}e^{-V}e^{\hat{\varphi}_1}-\frac{p_{22}}{k_2}\frac{f_{k_2}'(\theta_2)}{f_{k_2}(\theta_2)}e^{-W}e^{\hat{\varphi}_2}\, .\label{YAbSO(2)xSO(2)gaugeAnt}
\end{eqnarray} 
The corresponding modified two-form field strengths are given by
\begin{eqnarray}
\mathcal{H}^{(2)12}=F^{12}_{(2)}= e^{-2V}p_{11}e^{\hat{\theta}_1}\wedge e^{\hat{\varphi}_1}+e^{-2W}p_{12}e^{\hat{\theta}_2}\wedge e^{\hat{\varphi}_2},\\ \mathcal{H}^{(2)34}=F^{34}_{(2)}= e^{-2V}p_{21}e^{\hat{\theta}_1}\wedge e^{\hat{\varphi}_1}+e^{-2W}p_{22}e^{\hat{\theta}_2}\wedge e^{\hat{\varphi}_2}\, .
\end{eqnarray}
We also need to turn on the following three-form field strength
\begin{equation}\label{YAbSO(2)xSO(2)3form}
\mathcal{H}^{(3)}_{\hat{m}\hat{n}\hat{r} 5}= \alpha e^{-2(V+W+2\phi_1+2\phi_2)}\varepsilon_{\hat{m}\hat{n}}
\end{equation}
in which $\alpha$ is a constant related to the magnetic charges by the relation 
\begin{equation}\label{YAbSO(2)xSO(2)Con}
g\rho\alpha=-32(p_{12}p_{21}+p_{11}p_{22})\, .
\end{equation}
For $\rho=0$ corresponding to $CSO(2,2,1)$ and $CSO(4,0,1)$ gauge groups, we need to impose a relation on the magnetic charges 
\begin{equation}
p_{12}p_{21}+p_{11}p_{22}=0
\end{equation}
to ensure that the resulting BPS equations are compatible with all the second-ordered field equations.
\\
\indent Using the projection conditions \eqref{pureYProj} and
\begin{equation}\label{AddSO(2)xSO(2)ProjCon}
\gamma_{\hat{\theta_1}\hat{\varphi_1}}\epsilon^a=\gamma_{\hat{\theta_2}\hat{\varphi_2}}\epsilon^a=
-{(\Gamma_{12})^a}_b\epsilon^b=-{(\Gamma_{34})^a}_b\epsilon^b
\end{equation}
together with the twist conditions 
\begin{equation}\label{YSO(2)xSO(2)QYM}
g(p_{11}+\sigma p_{21})=k_1\qquad \textrm{and} \qquad g(p_{12}+\sigma p_{22})=k_2,
\end{equation}
we obtain the following BPS equations 
\begin{eqnarray}\label{YSO(2)xSO(2)BPS}
U'&=&\frac{g}{40}(2e^{-2\phi_1}+\rho e^{4(\phi_1+\phi_2)}+2\sigma e^{-2\phi_2})-\frac{3\alpha}{5g}e^{-2(V+W+\phi_1+\phi_2)}\nonumber\\
&& -\frac{2}{5}\left[e^{-2V}(e^{2\phi_1}p_{11}+e^{2\phi_2}p_{21})+e^{-2W}(e^{2\phi_1}p_{12}+e^{2\phi_2}p_{22})\right],\\
V'&=&\frac{g}{40}(2e^{-2\phi_1}+\rho e^{4(\phi_1+\phi_2)}+2\sigma e^{-2\phi_2})+\frac{2\alpha}{5g}e^{-2(V+W+\phi_1+\phi_2)}\nonumber\\
&& +\frac{2}{5}\left[4e^{-2V}(e^{2\phi_1}p_{11}+e^{2\phi_2}p_{21})-e^{-2W}(e^{2\phi_1}p_{12}+e^{2\phi_2}p_{22})\right],\\
W'&=&\frac{g}{40}(2e^{-2\phi_1}+\rho e^{4(\phi_1+\phi_2)}+2\sigma e^{-2\phi_2})+\frac{2\alpha}{5g}e^{-2(V+W+\phi_1+\phi_2)}\nonumber\\
&& -\frac{2}{5}\left[e^{-2V}(e^{2\phi_1}p_{11}+e^{2\phi_2}p_{21})-4e^{-2W}(e^{2\phi_1}p_{12}+e^{2\phi_2}p_{22})\right],\\
\phi_1'&=&\frac{g}{20}(3e^{-2\phi_1}-\rho e^{4(\phi_1+\phi_2)}-2\sigma e^{-2\phi_2})+\frac{\alpha}{5g}e^{-2(V+W+\phi_1+\phi_2)}\nonumber\\
&& -\frac{2}{5}\left[3e^{2\phi_1}(e^{-2V}p_{11}+e^{-2W}p_{12})-2e^{2\phi_2}(e^{-2V}p_{21}+e^{-2W}p_{22})\right],\\
\phi_2'&=&\frac{g}{20}(3\sigma e^{-2\phi_2}-\rho e^{4(\phi_1+\phi_2)}-2e^{-2\phi_1})+\frac{\alpha}{5g}e^{-2(V+W+\phi_1+\phi_2)}\nonumber\\
&& +\frac{2}{5}\left[2e^{2\phi_1}(e^{-2V}p_{11}+e^{-2W}p_{12})-3e^{2\phi_2}(e^{-2V}p_{21}+e^{-2W}p_{22})\right]\, .\hspace{0.8cm}
\end{eqnarray}
In deriving these equations, we have used the coset representative given in \eqref{YSO(2)xSO(2)Ys} for $SO(2)\times SO(2)$ singlet scalars.
\\
\indent From the BPS equations, we find a class of $AdS_3\times \Sigma^2\times \Sigma^2$ fixed point solutions 
\begin{eqnarray}
e^{2V}&=&-\frac{16(e^{2\phi_1}p_{11}+e^{2\phi_2}p_{21})}{ge^{4(\phi_1+\phi_2)}\rho},\\
e^{2W}&=&-\frac{16(e^{2\phi_1}p_{12}+e^{2\phi_2}p_{22})}{ge^{4(\phi_1+\phi_2)}\rho},\\
e^{10\phi_1}&=&\frac{64\Theta\left[32(p_{12}p_{21}+p_{11}p_{22})-g\rho\alpha-64\sigma p_{21}p_{22}\right]^2}{\left[32\sigma(p_{12}p_{21}+p_{11}p_{22})-g\rho\sigma\alpha-64p_{11}p_{12}\right]^3},\\
e^{10\phi_2}&=&\frac{64\Theta\left[32\sigma(p_{12}p_{21}+p_{11}p_{22})-g\rho\sigma\alpha-64p_{11}p_{12}\right]^2}{\left[32(p_{12}p_{21}+p_{11}p_{22})-g\rho\alpha-64\sigma p_{21}p_{22}\right]^3},\\
L_{\textrm{AdS}_3}&=& \frac{8e^{2(\phi_1+\phi_2)}}{g(e^{2\phi_2}+e^{2\phi_1}\sigma)}
\end{eqnarray}
with
\begin{eqnarray}
\Theta&\hspace{-0.3cm}=&\hspace{-0.3cm}\frac{\Xi\left[32\left(p_{12}p_{22} (p_{11}+p_{21}\sigma)-p_{12}^2 p_{21}-p_{11}p_{22}^2\sigma\right)+g\rho \alpha (p_{12}+p_{22}\sigma)\right]}{\rho\left[1024 p_{12}^2 p_{21}^2+(32 p_{11}p_{22}-g \rho\alpha)^2-64p_{12}p_{21}(32p_{11}p_{22}+g\rho\alpha)\right]},\hspace{0.6cm}\\
\Xi&\hspace{-0.3cm}=&\hspace{-0.3cm}32\left[p_{11}p_{21}(p_{12}+p_{22} \sigma)-p_{11}^2p_{22}-p_{12}p_{21}^2\sigma\right]+g\rho\alpha (p_{11}+ p_{21}\sigma).
\end{eqnarray}
Among all the gauge groups with an $SO(2)\times SO(2)$ subgroup, it turns out that $AdS_3\times \Sigma^2\times \Sigma^2$ solutions are possible only for $SO(5)$ and $SO(3,2)$ gauge groups with $\rho=\sigma=1$ and $\rho=-\sigma=1$, respectively. For $SO(5)$ gauge group, the solutions have been extensively studied in \cite{2D_Bobev}. For $SO(3,2)$ gauge group, all the $AdS_3\times \Sigma^2\times \Sigma^2$ solutions given here are new. 
\\
\indent Following \cite{2D_Bobev}, we define the following two parameters to characterize the possible $AdS_3\times \Sigma^2\times \Sigma^2$ solutions
\begin{equation}\label{zpara}
z_1=g(p_{11}-\sigma p_{21}) \qquad \textrm{ and } \qquad z_2=g(p_{12}-\sigma p_{22})
\end{equation}
where we have set $\rho=1$. In order for the $AdS_3$ fixed points to exist in $SO(5)$ gauge group with $\sigma=1$, one of the Riemann surfaces must be negatively curved, and $AdS_3\times H^2\times \Sigma^2$ solutions can be found within the regions in the parameter space $(z_1, z_2)$ shown in figure \ref{SO(5)RegionPlot}. These regions are the same as those given in \cite{2D_Bobev}. The $AdS_3\times \Sigma^2\times \Sigma^2$ fixed points preserve four supercharges and correspond to $N=(2,0)$ SCFTs in two dimensions with $SO(2)\times SO(2)$ symmetry. Examples of RG flows with $g=16$ from the $N=4$ supersymmetric $AdS_7$ to $AdS_3\times H^2\times \Sigma^2$ fixed points and curved domain walls in the IR are shown in figures \ref{AdS3xH2xH2flows}, \ref{AdS3xH2xR2flows} and \ref{AdS3xH2xS2flows} for $\Sigma^2=H^2,\mathbb{R}^2$ and $S^2$, respectively. All the IR singularities are physical with $\hat{g}_{00}\rightarrow 0$ near the singularities.

\begin{figure}
  \centering
  \begin{subfigure}[b]{0.32\linewidth}
    \includegraphics[width=\linewidth]{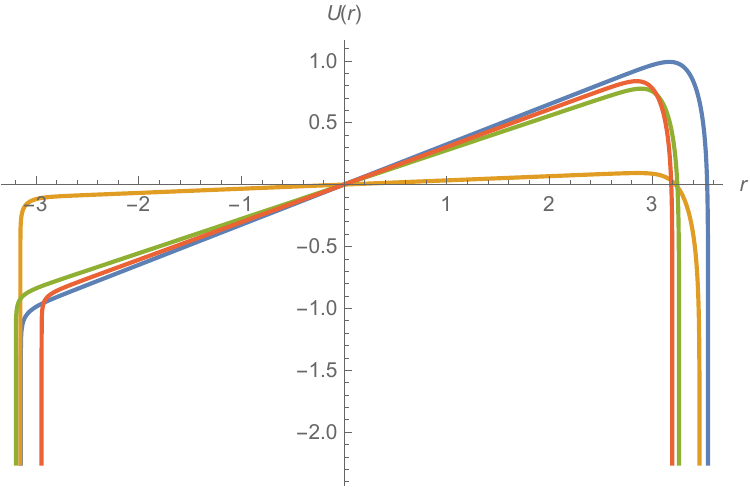}
  \caption{$U$ solution}
  \end{subfigure}
  \begin{subfigure}[b]{0.32\linewidth}
    \includegraphics[width=\linewidth]{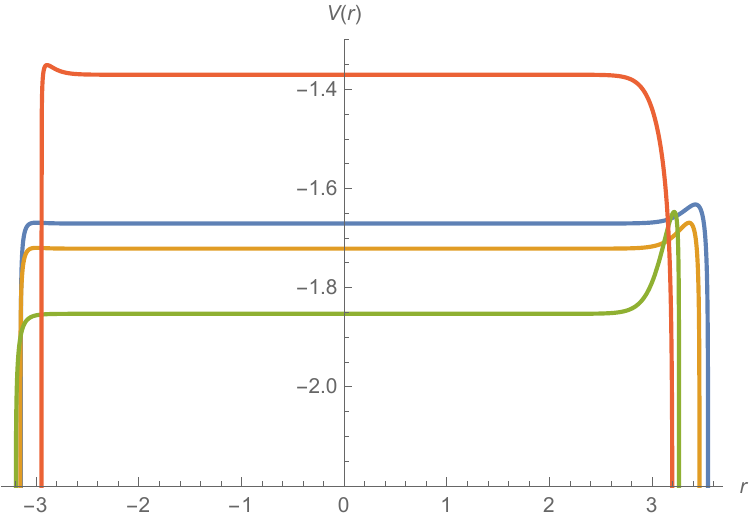}
  \caption{$V$ solution}
  \end{subfigure}
  \begin{subfigure}[b]{0.32\linewidth}
    \includegraphics[width=\linewidth]{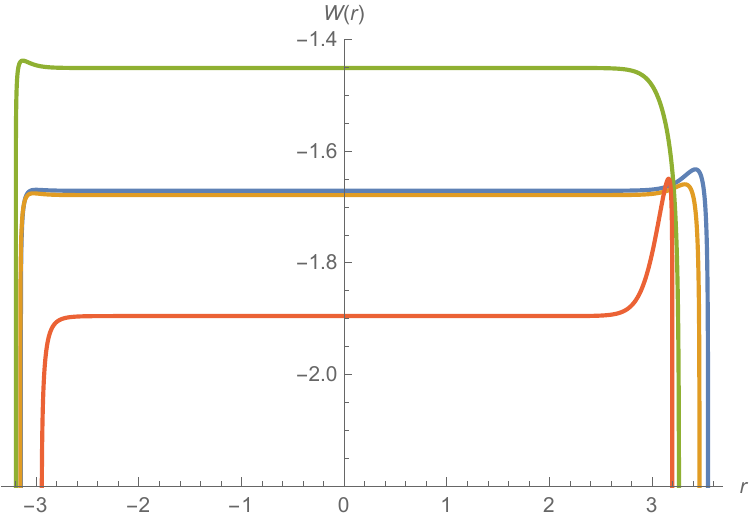}
  \caption{$W$ solution}
  \end{subfigure}\\
  \begin{subfigure}[b]{0.32\linewidth}
    \includegraphics[width=\linewidth]{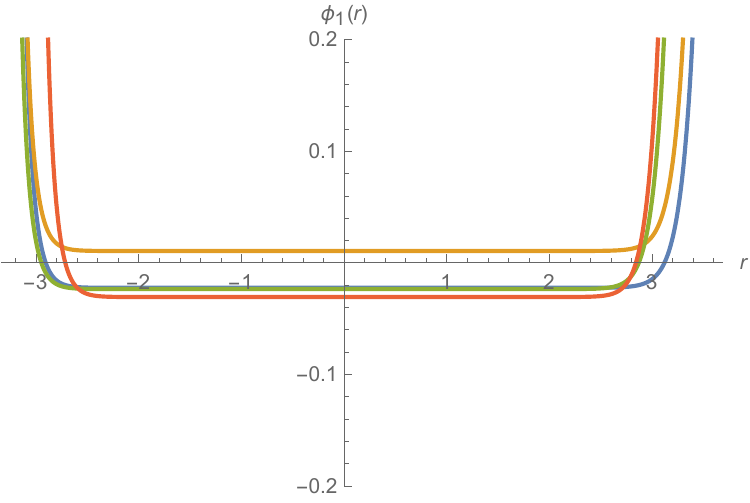}
  \caption{$\phi_1$ solution}
  \end{subfigure}
  \begin{subfigure}[b]{0.32\linewidth}
    \includegraphics[width=\linewidth]{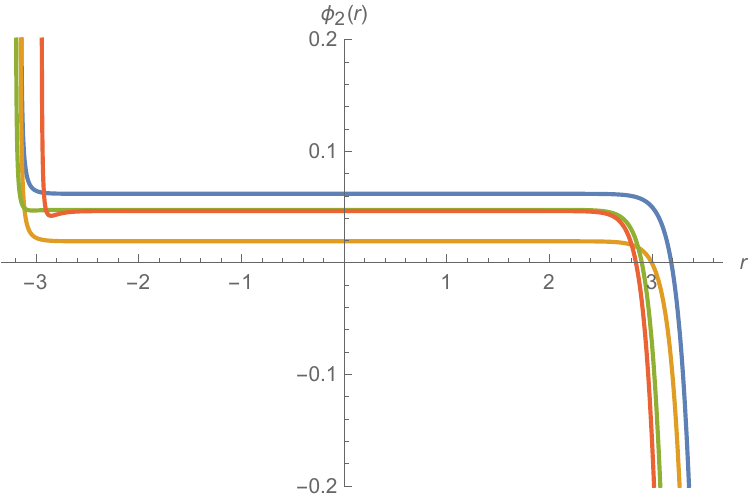}
  \caption{$\phi_2$ solution}
  \end{subfigure}
     \begin{subfigure}[b]{0.32\linewidth}
    \includegraphics[width=\linewidth]{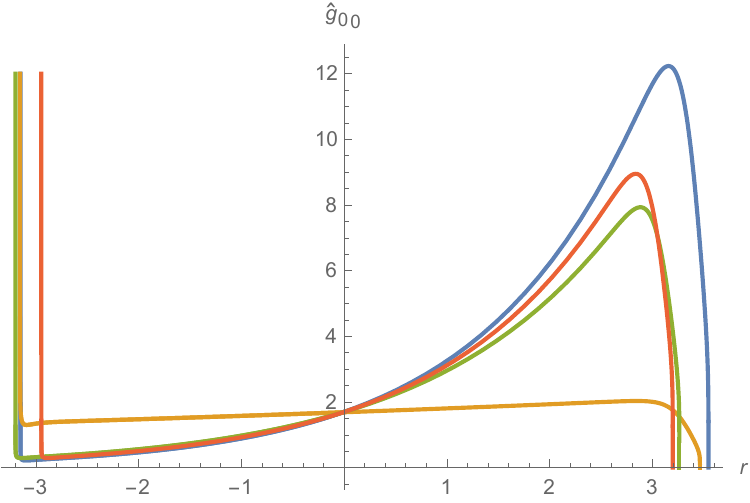}
  \caption{Profiles of $\hat{g}_{00}$}
  \end{subfigure}
  \caption{RG flows between $AdS_3\times S^2\times S^2$ fixed points and curved domain walls for $SO(2)\times SO(2)$ twist in $SO(3,2)$ gauge group. The blue, orange, green and red curves refer to $(z_1,z_2)=(-0.55,-0.55), (-0.55,-0.6),  (-0.35,-0.87), (-1,-0.3)$, respectively.}
  \label{AdS3xS2xS2flows}
\end{figure}

\begin{figure}
  \centering
  \begin{subfigure}[b]{0.32\linewidth}
    \includegraphics[width=\linewidth]{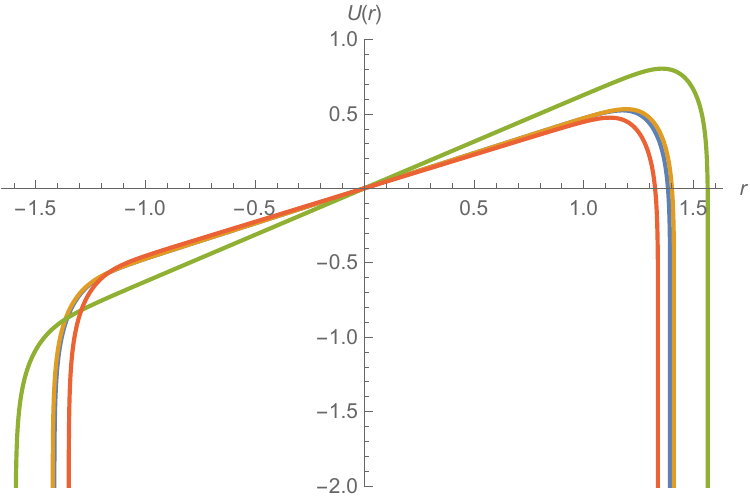}
  \caption{$U$ solution}
  \end{subfigure}
  \begin{subfigure}[b]{0.32\linewidth}
    \includegraphics[width=\linewidth]{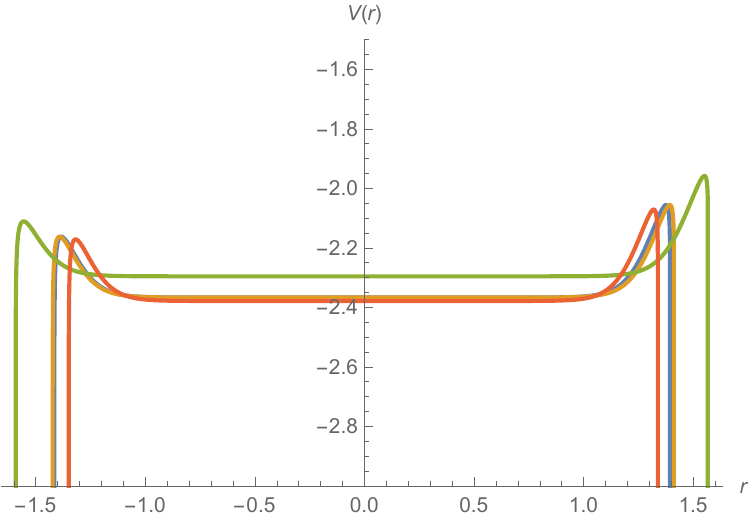}
  \caption{$V$ solution}
  \end{subfigure}
  \begin{subfigure}[b]{0.32\linewidth}
    \includegraphics[width=\linewidth]{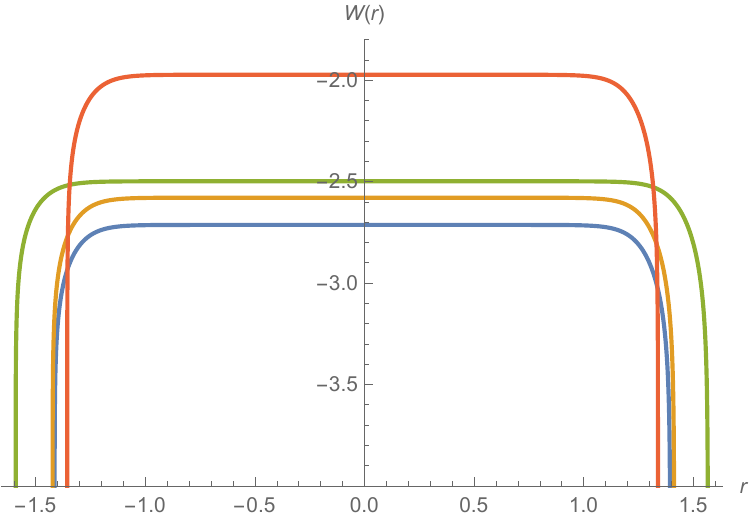}
  \caption{$W$ solution}
  \end{subfigure}\\
  \begin{subfigure}[b]{0.32\linewidth}
    \includegraphics[width=\linewidth]{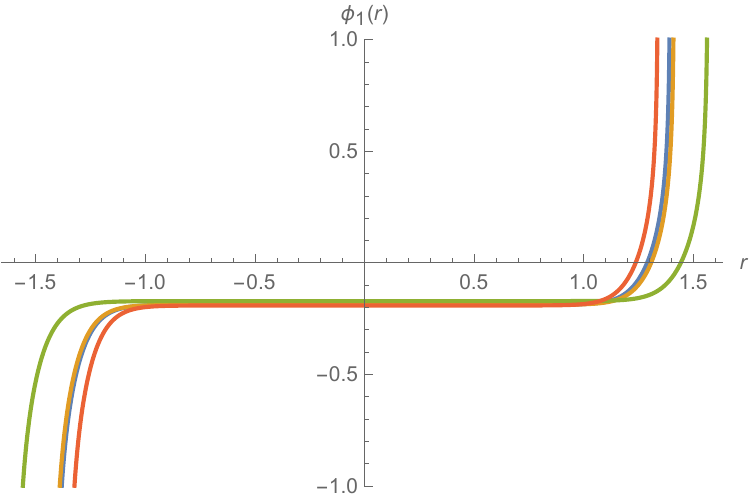}
  \caption{$\phi_1$ solution}
  \end{subfigure}
  \begin{subfigure}[b]{0.32\linewidth}
    \includegraphics[width=\linewidth]{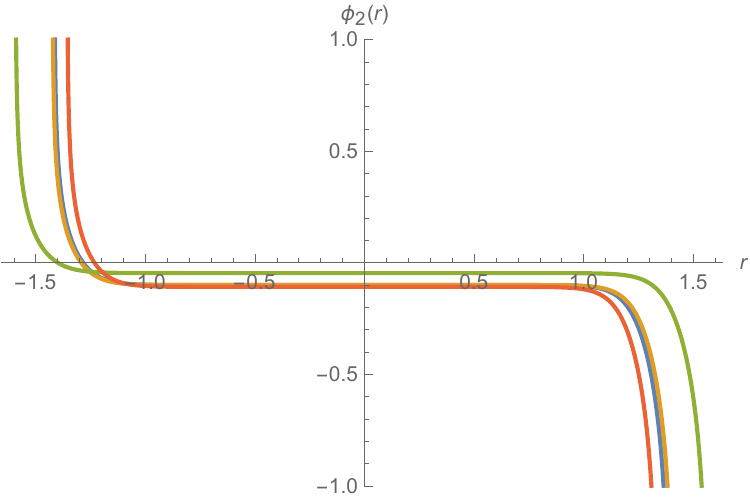}
  \caption{$\phi_2$ solution}
  \end{subfigure}
       \begin{subfigure}[b]{0.32\linewidth}
    \includegraphics[width=\linewidth]{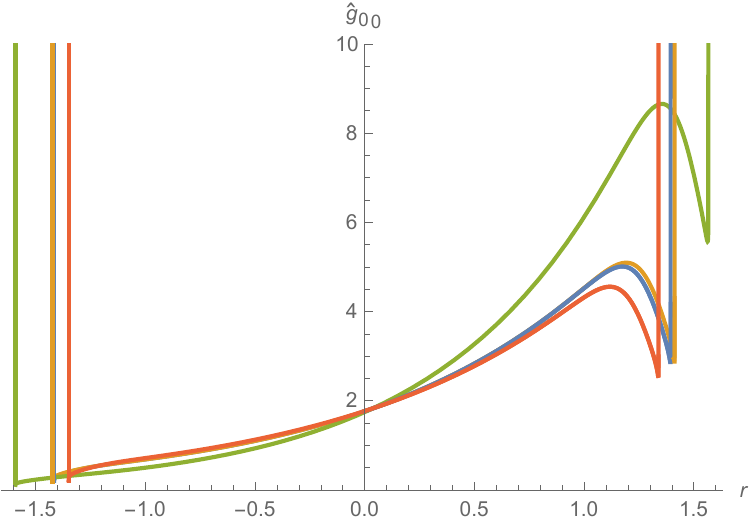}
  \caption{Profiles of $\hat{g}_{00}$}
  \end{subfigure}
  \caption{RG flows between $AdS_3\times S^2\times \mathbb{R}^2$ fixed points and curved domain walls for $SO(2)\times SO(2)$ twist in $SO(3,2)$ gauge group. The blue, orange, green and red curves refer to $(z_1,z_2)=(\frac{1}{34},-\frac{1}{34}), (\frac{1}{34},-\frac{1}{26}),  (\frac{1}{24},-\frac{1}{18}), (\frac{1}{36},-\frac{1}{8})$, respectively.}
  \label{AdS3xS2xR2flows}
\end{figure}

\begin{figure}
  \centering
  \begin{subfigure}[b]{0.32\linewidth}
    \includegraphics[width=\linewidth]{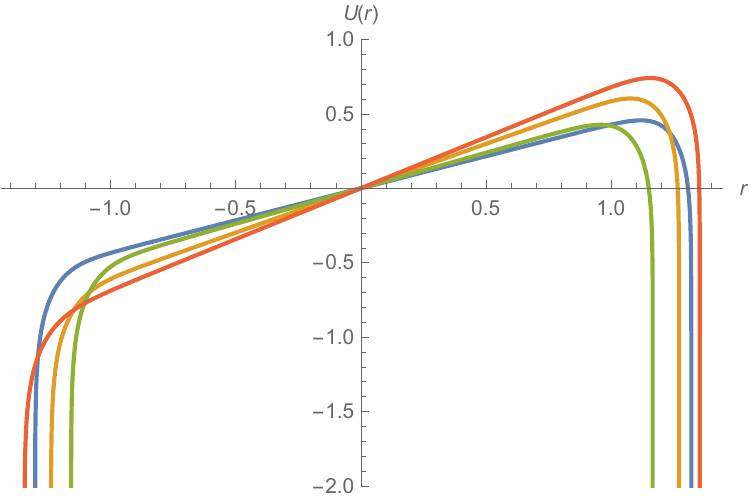}
  \caption{$U$ solution}
  \end{subfigure}
  \begin{subfigure}[b]{0.32\linewidth}
    \includegraphics[width=\linewidth]{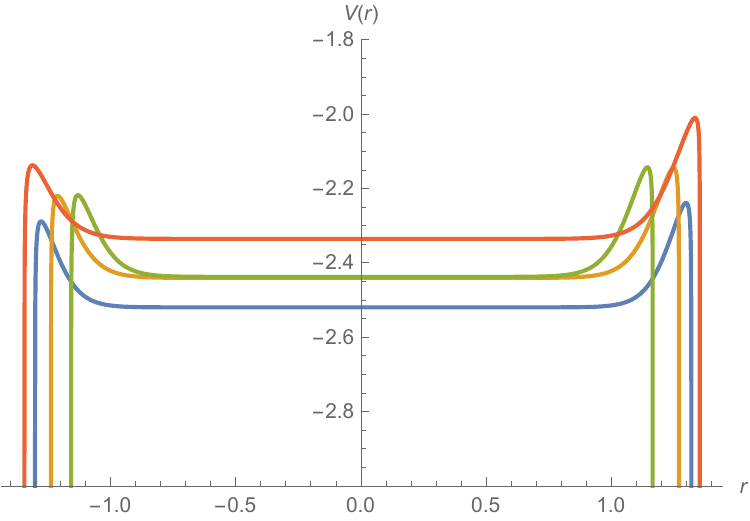}
  \caption{$V$ solution}
  \end{subfigure}
  \begin{subfigure}[b]{0.32\linewidth}
    \includegraphics[width=\linewidth]{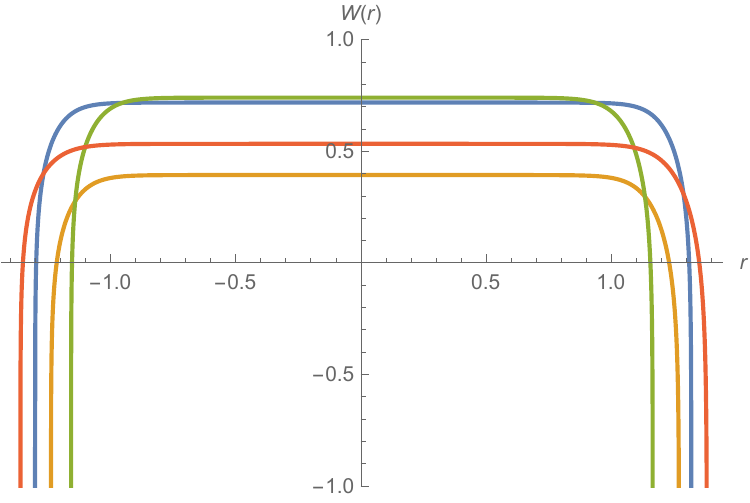}
  \caption{$W$ solution}
  \end{subfigure}
\\
  \begin{subfigure}[b]{0.32\linewidth}
    \includegraphics[width=\linewidth]{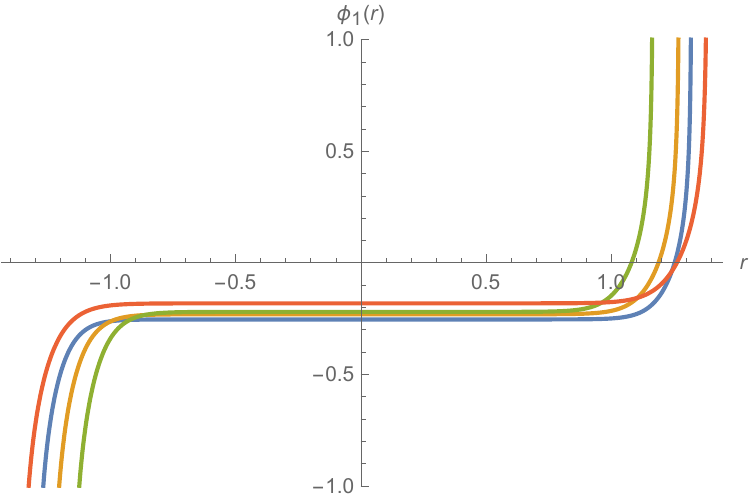}
  \caption{$\phi_1$ solution}
  \end{subfigure}
  \begin{subfigure}[b]{0.32\linewidth}
    \includegraphics[width=\linewidth]{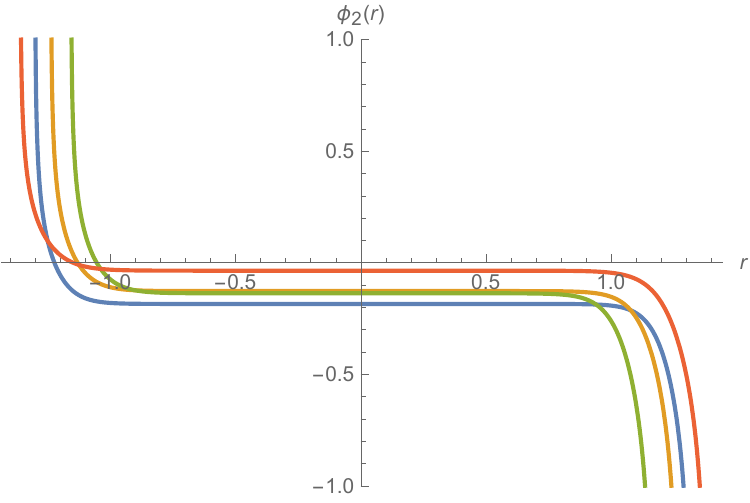}
  \caption{$\phi_2$ solution}
  \end{subfigure}
       \begin{subfigure}[b]{0.32\linewidth}
    \includegraphics[width=\linewidth]{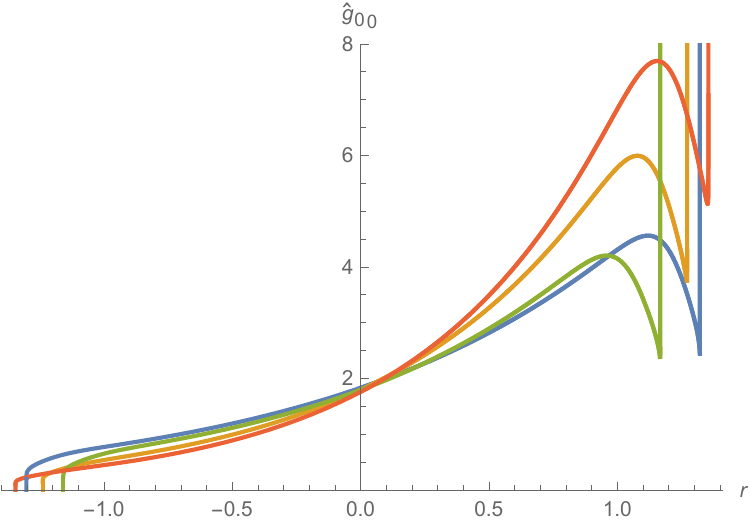}
  \caption{Profiles of $\hat{g}_{00}$}
  \end{subfigure}
  \caption{RG flows between $AdS_3\times S^2\times H^2$ fixed points and curved domain walls for $SO(2)\times SO(2)$ twist in $SO(3,2)$ gauge group. The blue, orange, green and red curves refer to $(z_1,z_2)=(\frac{1}{24},-18), (\frac{1}{16},-12),  (\frac{1}{24},-24), (\frac{1}{16},-24)$, respectively.}
  \label{AdS3xS2xH2flows}
\end{figure}

We now carry out a similar analysis for the case of $SO(3,2)$ gauge group with $\rho=-\sigma=1$. It turns out that in this case, the $AdS_3$ fixed points exist only for at least one of the two Riemann surfaces is positively curved. For definiteness, we will choose $k_1=1$ and $k_2=-1,0,1$ corresponding to $AdS_3\times S^2\times H^2$, $AdS_3\times S^2\times\mathbb{R}^2$ and $AdS_3\times S^2\times S^2$ fixed points, respectively. Using the parameters $z_1$ and $z_2$ defined in \eqref{zpara} with $\sigma=-1$, we find regions in the parameter space $(z_1, z_2)$ for $AdS_3$ vacua to exist in $SO(3,2)$ gauged maximal supergravity as shown in figure \ref{SO(3,2)RegionPlot}. 
\\
\indent For $SO(3,2)$ gauge group, there is no asymptotically locally $AdS_7$ geometry. We will consider RG flows between the $AdS_3\times S^2\times \Sigma^2$ fixed points and curved domain walls with $Mkw_3\times S^2\times \Sigma^2$ slices. These curved domain walls have $SO(2)\times SO(2)$ symmetry and are expected to describe non-conformal field theories in two dimensions obtained from twisted compactifications of $N=(2,0)$ non-conformal field theory in six dimensions. The latter is dual to the half-supersymmetric domain wall of the seven-dimensional gauged supergravity. A number of these RG flows with $g=16$ and different values of $z_1$ and $z_2$ are given in figures \ref{AdS3xS2xS2flows}, \ref{AdS3xS2xR2flows} and \ref{AdS3xS2xH2flows}. We see that all singularities in the flows from $AdS_3\times S^2\times \mathbb{R}^2$ fixed points are unphysical while only the singularities on the right (left) with $\phi_1\rightarrow \infty$ and $\phi_2\rightarrow -\infty$ ($\phi_1\rightarrow -\infty$ and $\phi_2\rightarrow \infty$) in the flows from $AdS_3\times S^2\times S^2$ ($AdS_3\times S^2\times H^2$) fixed points are physical.
\begin{figure}
  \centering
  \begin{subfigure}[b]{0.32\linewidth}
    \includegraphics[width=\linewidth]{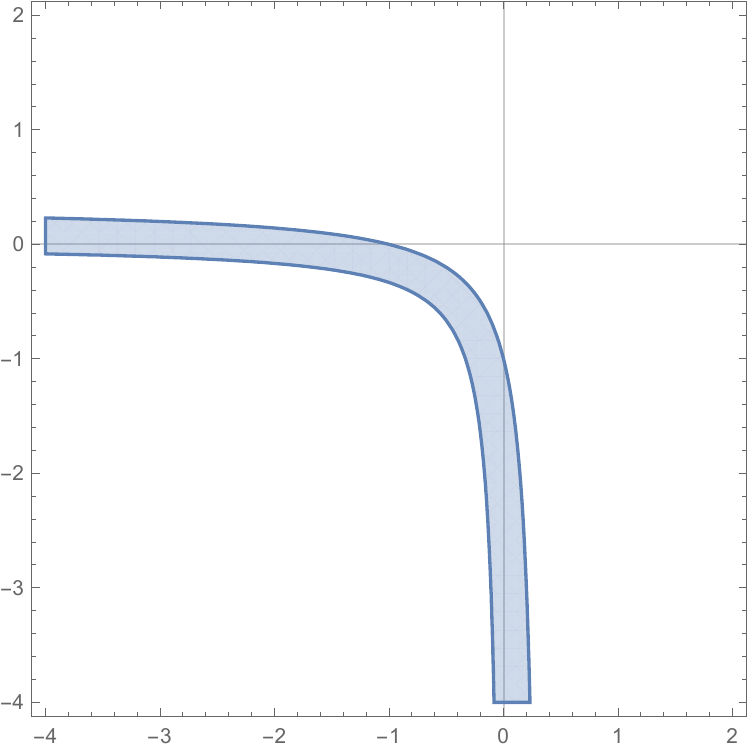}
  \end{subfigure}
  \begin{subfigure}[b]{0.32\linewidth}
    \includegraphics[width=\linewidth]{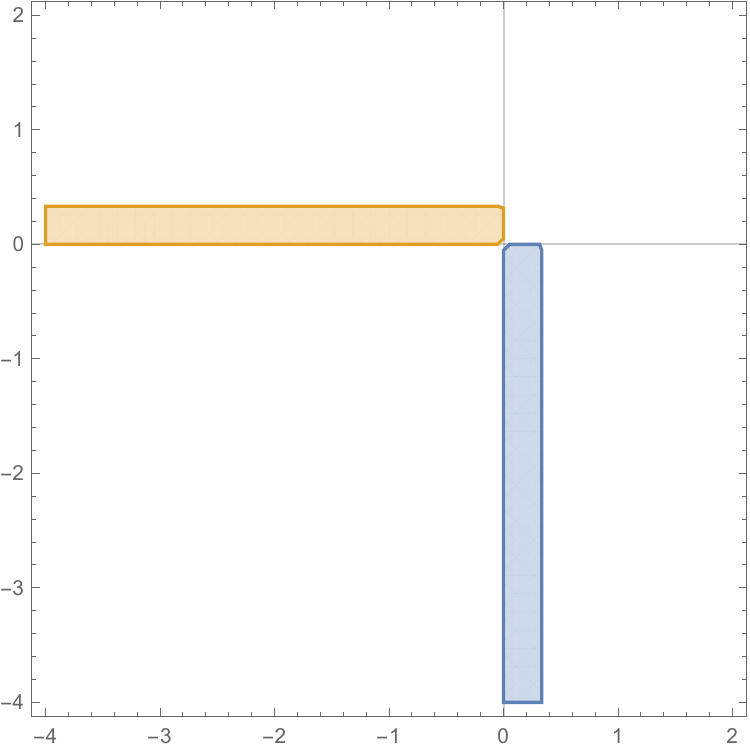}
  \end{subfigure}
  \begin{subfigure}[b]{0.32\linewidth}
    \includegraphics[width=\linewidth]{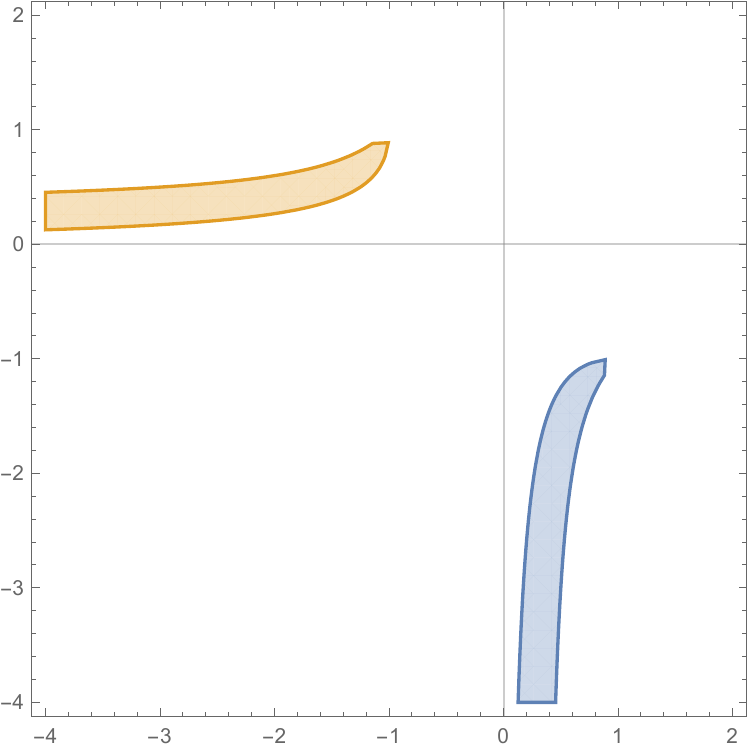}
  \end{subfigure}
  \caption{Regions (blue) in the parameter space ($z_1$,  $z_2$) where good $AdS_3$ vacua exist in $SO(3,2)$ gauge group. From left to right, these figures correspond to the cases of ($k_1=k_2=1$), ($k_1=1$, $k_2=0$), ($k_1=-k_2=1$), respectively. The orange regions are obtained from interchanging $k_1$ and $k_2$.}
  \label{SO(3,2)RegionPlot}
\end{figure}

\subsection{$AdS_3$ vacua from Kahler four-cycles}\label{YKahlersection}
In this section,  we consider twisted compactifications of six-dimensional field theories on Kahler four-cycles $K^4_k$. The constant $k=1,0,-1$ characterizes the constant curvature of $K^4_k$ and corresponds to a two-dimensional complex projective space $CP^2$, a four-dimensional flat space $\mathbb{R}^4$ and a two-dimensional complex hyperbolic space $CH^2$, respectively. The manifold $K^4_k$ has a $U(2)\sim U(1)\times SU(2)$ spin connection. Therefore, we can perform a twist by turning on $SO(2)\sim U(1)$ or $SO(3)\sim SU(2)$ gauge fields to cancel the $U(1)$ or $SU(2)$ parts of the spin connection.
\\
\indent A general ansatz for the seven-dimensional metric takes the form of
\begin{equation}\label{Kahler7Dmetric}
ds_7^2=e^{2U(r)}dx^2_{1,1}+dr^2+e^{2V(r)}ds^2_{K^{4}_{k}}
\end{equation}
in which the explicit form for the metric on the Kahler four-cycle will be specified separately in each case.

\subsubsection{$AdS_3\times K^4_k$ solutions with $SO(3)$ twists}\label{YKahlerSO(3)section}
\begin{figure}
  \centering
  \begin{subfigure}[b]{0.35\linewidth}
    \includegraphics[width=\linewidth]{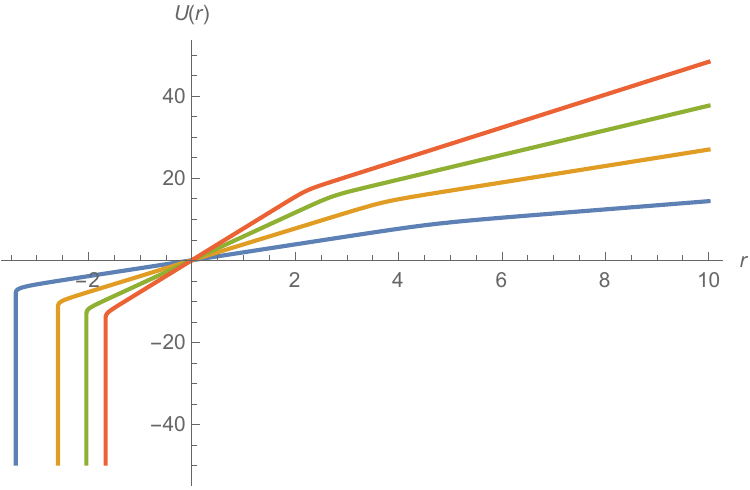}
  \caption{$U$ solution}
  \end{subfigure}
  \begin{subfigure}[b]{0.35\linewidth}
    \includegraphics[width=\linewidth]{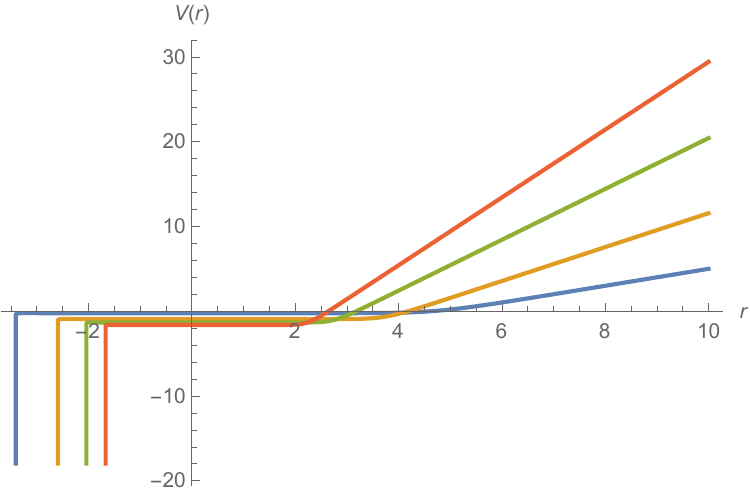}
  \caption{$V$ solution}
  \end{subfigure}\\
  \begin{subfigure}[b]{0.35\linewidth}
    \includegraphics[width=\linewidth]{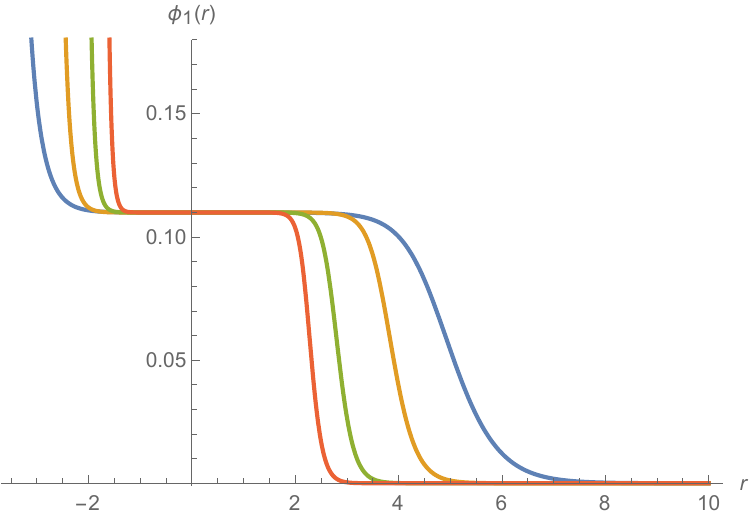}
  \caption{$\phi_1$ solution}
  \end{subfigure}
    \begin{subfigure}[b]{0.35\linewidth}
    \includegraphics[width=\linewidth]{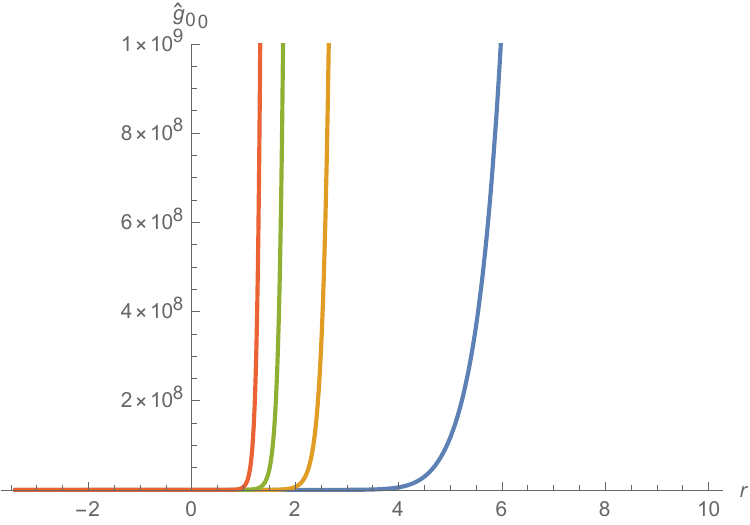}
  \caption{Profiles of $\hat{g}_{00}$}
  \end{subfigure}
  \caption{RG flows from the $N=4$ $AdS_7$ critical point to the $AdS_3\times CH^2$ fixed point and curved domain walls  with $\phi_1\rightarrow \infty$ and $\phi_2=\phi_3=0$ for $SO(3)\sim SU(2)$ twist in $SO(5)$ gauge group. The blue, orange, green and red curves refer to $g=8,16,24,32$, respectively.}
  \label{YSO(3)Kahlerflow1}
\end{figure}

\begin{figure}
  \centering
  \begin{subfigure}[b]{0.35\linewidth}
    \includegraphics[width=\linewidth]{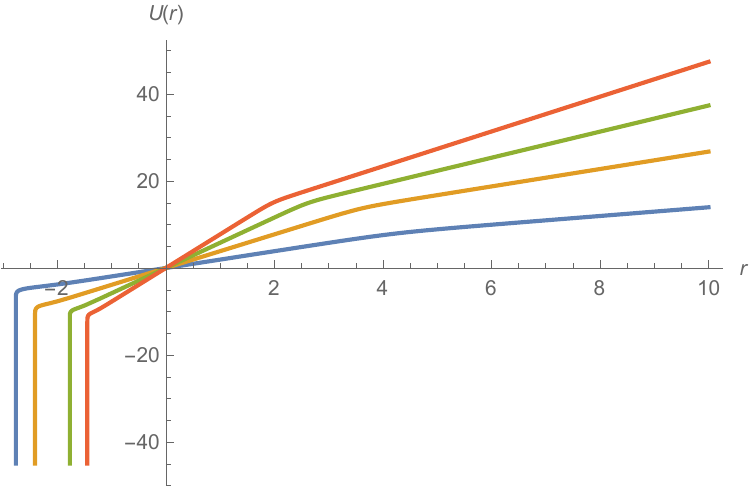}
  \caption{$U$ solution}
  \end{subfigure}
  \begin{subfigure}[b]{0.35\linewidth}
    \includegraphics[width=\linewidth]{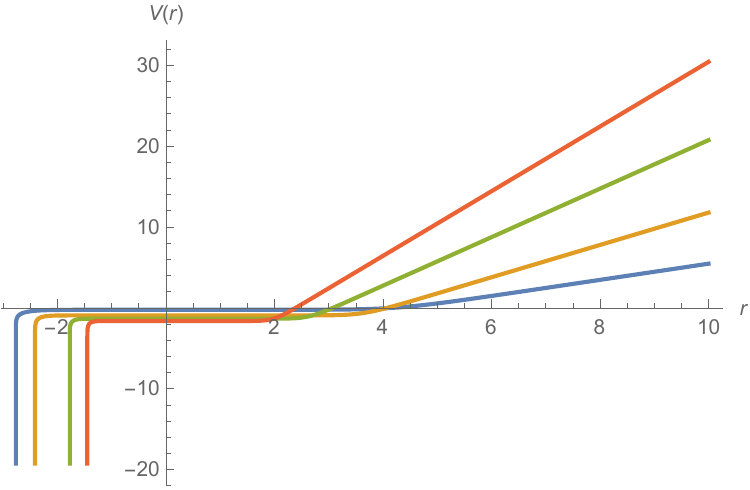}
  \caption{$V$ solution}
  \end{subfigure}\\
  \begin{subfigure}[b]{0.35\linewidth}
    \includegraphics[width=\linewidth]{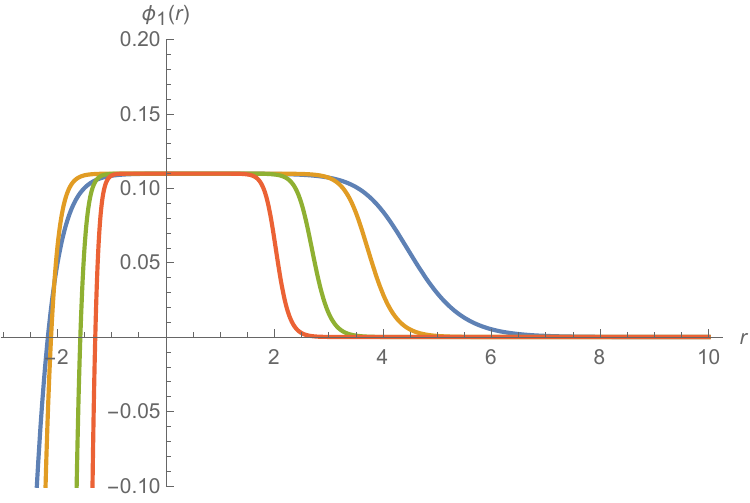}
  \caption{$\phi_1$ solution}
  \end{subfigure}
      \begin{subfigure}[b]{0.35\linewidth}
    \includegraphics[width=\linewidth]{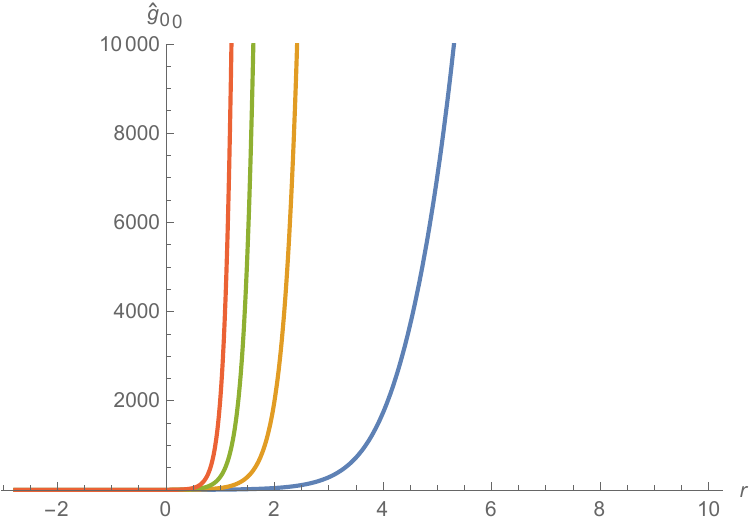}
  \caption{Profiles of $\hat{g}_{00}$}
  \end{subfigure}
  \caption{RG flows from the $N=4$ $AdS_7$ critical point to the $AdS_3\times CH^2$ fixed point and curved domain walls  with $\phi_1\rightarrow -\infty$ and $\phi_2=\phi_3=0$ for $SO(3)\sim SU(2)$ twist in $SO(5)$ gauge group. The blue, orange, green and red curves refer to $g=8,16,24,32$, respectively.}
  \label{YSO(3)Kahlerflow2}
\end{figure}

\begin{figure}
  \centering
  \begin{subfigure}[b]{0.32\linewidth}
    \includegraphics[width=\linewidth]{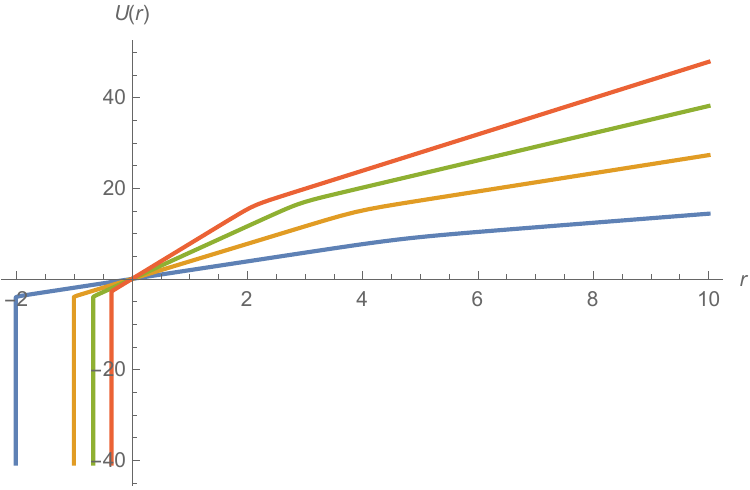}
  \caption{$U$ solution}
  \end{subfigure}
  \begin{subfigure}[b]{0.32\linewidth}
    \includegraphics[width=\linewidth]{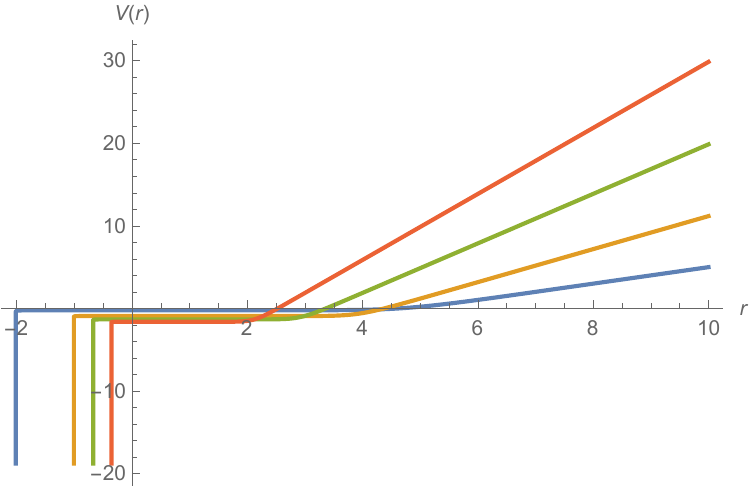}
  \caption{$V$ solution}
  \end{subfigure}
  \begin{subfigure}[b]{0.32\linewidth}
    \includegraphics[width=\linewidth]{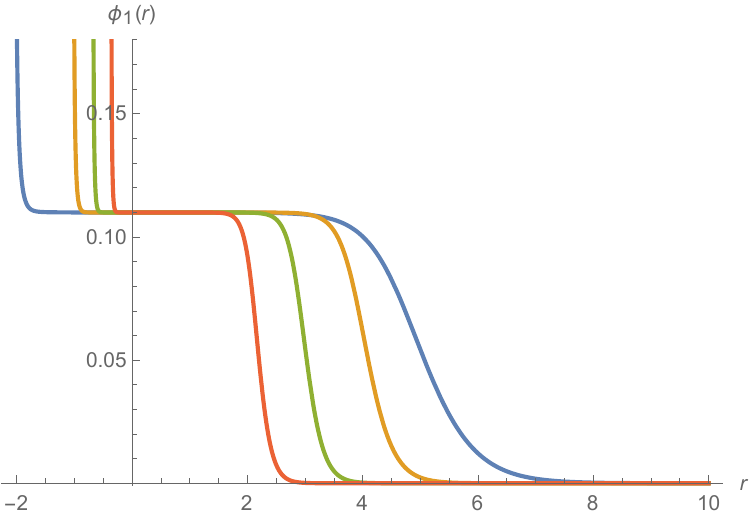}
  \caption{$\phi_1$ solution}
  \end{subfigure}\\
    \begin{subfigure}[b]{0.32\linewidth}
    \includegraphics[width=\linewidth]{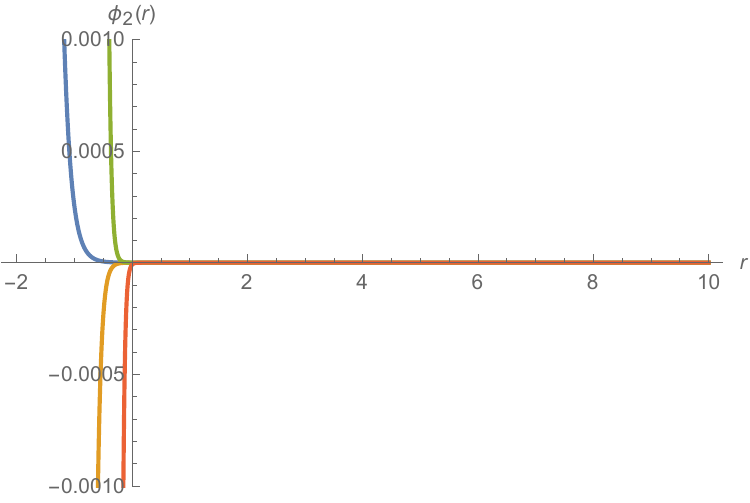}
  \caption{$\phi_2$ solution}
  \end{subfigure}
  \begin{subfigure}[b]{0.32\linewidth}
    \includegraphics[width=\linewidth]{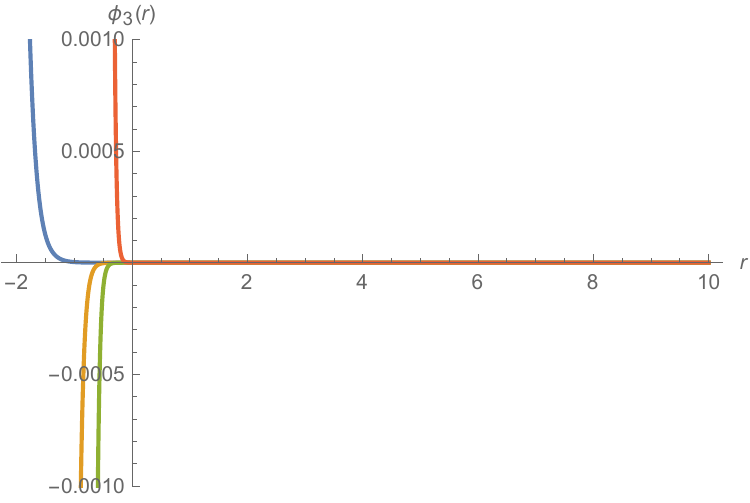}
  \caption{$\phi_3$ solution}
  \end{subfigure}
      \begin{subfigure}[b]{0.32\linewidth}
    \includegraphics[width=\linewidth]{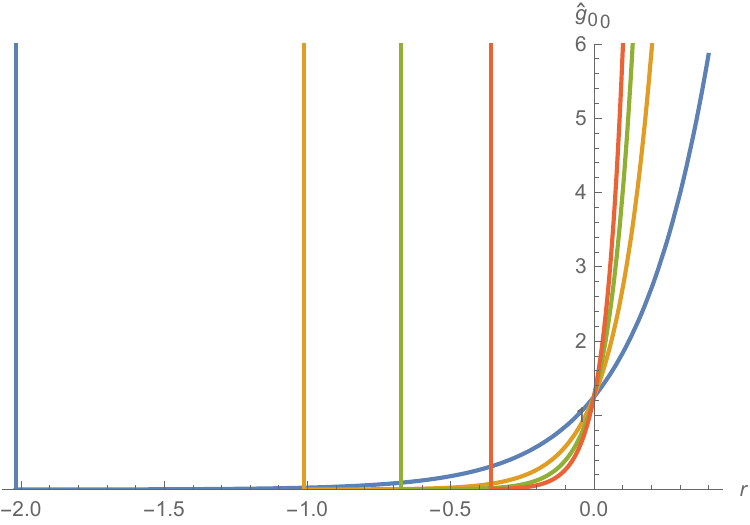}
  \caption{Profiles of $\hat{g}_{00}$}
  \end{subfigure}
  \caption{RG flows from the $N=4$ $AdS_7$ critical point to the $AdS_3\times CH^2$ fixed point and curved domain walls  with $\phi_1$, $\phi_2$ and $\phi_3$ non-vanishing for $SO(3)\sim SU(2)$ twist in $SO(5)$ gauge group. The blue, orange, green and red curves refer to $g=8,16,24,32$, respectively.}
  \label{YSO(3)Kahlerflow3}
\end{figure}

We begin with a twist along the $SU(2)\sim SO(3)$ part of the spin connection and choose the following form of the metric on $K_k^4$
\begin{equation}\label{Kahlermetric}
ds^2_{K^4_k}=d\psi^2+f_k(\psi)^2(\tau_1^2+\tau_2^2+\tau_3^2)
\end{equation}
with $\psi\in[0,\frac{\pi}{2}]$, and $f_k(\psi)$ is the function defined in \eqref{fFn}. $\tau_i$ are $SU(2)$ left-invariant one-forms satisfying 
\begin{equation}
d\tau_i=\frac{1}{2}\epsilon_{ijl}\tau_j\wedge\tau_l, \qquad i,j,l=1,2,3\, . 
\end{equation}
Their explicit forms are given by
\begin{eqnarray}
\tau_1&=&-\sin\vartheta d\theta+\cos\vartheta\sin\theta d \varphi,\nonumber \\
\tau_2&=& \cos\vartheta d\theta+\sin\vartheta\sin\theta d \varphi,\nonumber \\
\tau_3&=& d\vartheta+\cos\theta d \varphi\label{SU(2)Inv1form}
\end{eqnarray}
where the ranges of the variables are $\theta\in[0,\pi]$, $\varphi\in[0,2\pi]$ and $\vartheta\in[0,4\pi]$.
\\
\indent With the following choice of vielbein
\begin{eqnarray}
e^{\hat{m}}&=& e^{U}dx^{m}, \qquad e^{\hat{r}}=dr,  \qquad e^{\hat{3}}=e^{V}f_k(\psi)\tau_1,\nonumber \\ e^{\hat{4}}&=& e^{V}f_k(\psi)\tau_2, \qquad e^{\hat{5}}=e^{V}f_k(\psi)\tau_3, \qquad e^{\hat{6}}=e^{V}d\psi,\label{Kahler4bein}
\end{eqnarray}
we find non-vanishing components of the spin connection
\begin{eqnarray}
\omega_{(1)}^{\hat{m}\hat{r}}&=&  U'e^{\hat{m}}, \qquad \omega_{(1)}^{\hat{i}\hat{r}}= V'e^{\hat{i}},\qquad
\omega_{(1)}^{\hat{3}\hat{6}}=f'_k(\psi)\tau_1,\qquad \omega_{(1)}^{\hat{4}\hat{5}}= \tau_1,\nonumber \\
\omega_{(1)}^{\hat{4}\hat{6}}&=& f'_k(\psi)\tau_2,\qquad \omega_{(1)}^{\hat{5}\hat{3}}=\tau_2,\qquad
\omega_{(1)}^{\hat{5}\hat{6}}=f'_k(\psi)\tau_3,\qquad \omega_{(1)}^{\hat{3}\hat{4}}=\tau_3\label{AdS3xKahler4SpinCon}
\end{eqnarray}
where $\hat{i}=\hat{3}, ..., \hat{6}$ is the flat index on $K^4_k$. $\omega_{(1)}^{\hat{i}\hat{j}}$ are the $SU(2)$ spin connections.
\\
\indent To perform the twist, we turn on $SO(3)$ gauge fields with the following ansatz
\begin{equation}\label{YKahlerSO(3)Ant}
A^{i-2,j-2}_{(1)}=-\frac{p}{k}(f'_k(\psi)-1)\varepsilon^{ijl}\tau_l,\qquad i,j=3,4,5
\end{equation} 
with the modified two-form field strengths given by
\begin{eqnarray}
\mathcal{H}^{12}_{(2)}&=& F^{12}_{(2)}=e^{-2V}p(e^{\hat{3}}\wedge e^{\hat{4}}-e^{\hat{5}}\wedge e^{\hat{6}}),\\
\mathcal{H}^{23}_{(2)}&=& F^{23}_{(2)}=e^{-2V}p(e^{\hat{4}}\wedge e^{\hat{5}}-e^{\hat{3}}\wedge e^{\hat{6}}),\\
\mathcal{H}^{31}_{(2)}&=& F^{31}_{(2)}=e^{-2V}p(e^{\hat{5}}\wedge e^{\hat{3}}-e^{\hat{4}}\wedge e^{\hat{6}}).
\end{eqnarray}
Unlike the previous case, we do not need to turn on the three-form field strengths since, in this case, $\epsilon_{MNPQR}\mc{H}^{(2)NP}\wedge \mc{H}^{(2)QR}=0$.
\\
\indent We then impose the twist condition \eqref{GenQYM} together with the following three projection conditions
\begin{equation}\label{YSO(3)KahlerProjCon}
\gamma^{\hat{3}\hat{4}}\epsilon^a=-{(\Gamma_{12})^a}_b\epsilon^b,\qquad \gamma^{\hat{4}\hat{5}}\epsilon^a=-{(\Gamma_{23})^a}_b\epsilon^b,\qquad \gamma^{\hat{3}\hat{4}}\epsilon^a=-\gamma^{\hat{5}\hat{6}}\epsilon^a\, .
\end{equation}
Using the scalar coset representative \eqref{YSO(3)coset} and the projection \eqref{pureYProj}, we find the following BPS equations
\begin{eqnarray}
U'&=&\frac{g}{40}e^{6\phi_1}\left[(\rho+\sigma)\cosh{2\phi_2}\cosh{2\phi_3}+(\rho-\sigma)\sinh{2\phi_3}\right]\nonumber\\&&+\frac{3g}{40}e^{-4\phi_1}-\frac{12}{5}e^{-2(V-2\phi_1)}p,\\
V'&=&\frac{g}{40}e^{6\phi_1}\left[(\rho+\sigma)\cosh{2\phi_2}\cosh{2\phi_3}+(\rho-\sigma)\sinh{2\phi_3}\right]\nonumber\\&&+\frac{3g}{40}e^{-4\phi_1}+\frac{18}{5}e^{-2(V-2\phi_1)}p,\\
\phi_1'&=&\frac{g}{40}e^{6\phi_1}\left[(\rho-\sigma)\sinh{2\phi_3}-(\rho+\sigma)\cosh{2\phi_2}\cosh{2\phi_3}\right]\nonumber\\&&+\frac{g}{20}e^{-4\phi_1}-\frac{8}{5}e^{-2(V-2\phi_1)}p,\\
\phi_2'&=&-\frac{g}{8}e^{6\phi_1}(\rho+\sigma)\sinh{2\phi_2}\hspace{0.08cm} \text{sech}{2\phi_3},\\
\phi_3'&=&-\frac{g}{8}e^{6\phi_1}\left((\rho+\sigma)\cosh{2\phi_2}\sinh{2\phi_3}+(\rho-\sigma)\cosh{2\phi_3}\right).
\end{eqnarray}
\indent It turns out that only $SO(5)$ gauge group admits an $AdS_3\times CH^2$ fixed point given by 
\begin{eqnarray}
V&=&\frac{1}{2}\ln\left[\frac{16\times3^{4/5}}{g^2}\right],\qquad
\phi_1=\frac{1}{10}\ln3,\nonumber \\
\phi_2&=&\phi_3=0,\qquad
L_{\text{AdS}_3}=\frac{8}{3^{3/5}g}\, .
\end{eqnarray}
This is the $AdS_3\times CH^2$ solution found in \cite{Gauntlett1}. The solution preserves four supercharges and corresponds to $N=(2,0)$ SCFT in two dimensions with $SO(3)\times SO(2)$ symmetry. It should be noted that for $\phi_2=\phi_3=0$, the scalar coset representative is invariant under $SO(3)\times SO(2)\subset SO(5)$. Examples of general RG flows from the supersymmetric $AdS_7$ critical point to the $AdS_3\times CH^2$ fixed point and curved domain walls are shown in figures \ref{YSO(3)Kahlerflow1}, \ref{YSO(3)Kahlerflow2} and \ref{YSO(3)Kahlerflow3}. From these figures, we find that both singularities for $\phi_1\rightarrow \pm \infty$ in the flows with $\phi_2=\phi_3=0$ are physical. On the other hand, the IR singularities in the flows with all $\phi_i$'s non-vanishing are unphysical due to $\hat{g}_{00}\rightarrow \infty$ near the singularities.  

\subsubsection{$AdS_3\times K^4_k$ solutions with $SO(3)_+$ twists}\label{YKahlerSDSO(3)section}
\begin{figure}
  \centering
  \begin{subfigure}[b]{0.38\linewidth}
    \includegraphics[width=\linewidth]{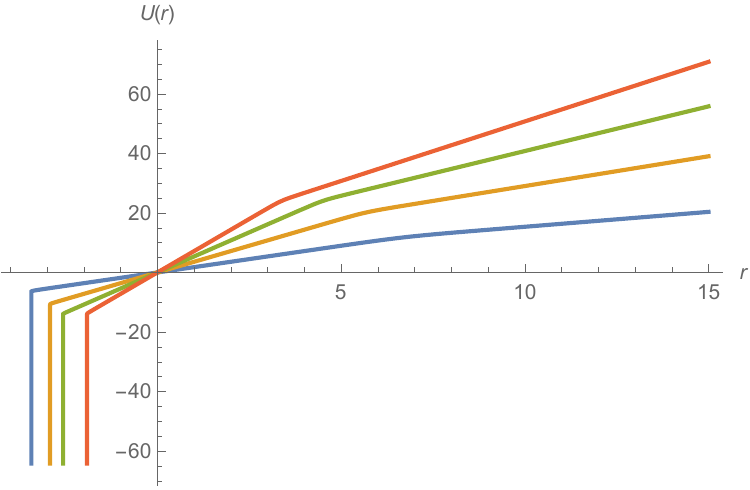}
  \caption{$U$ solution}
  \end{subfigure}
  \begin{subfigure}[b]{0.38\linewidth}
    \includegraphics[width=\linewidth]{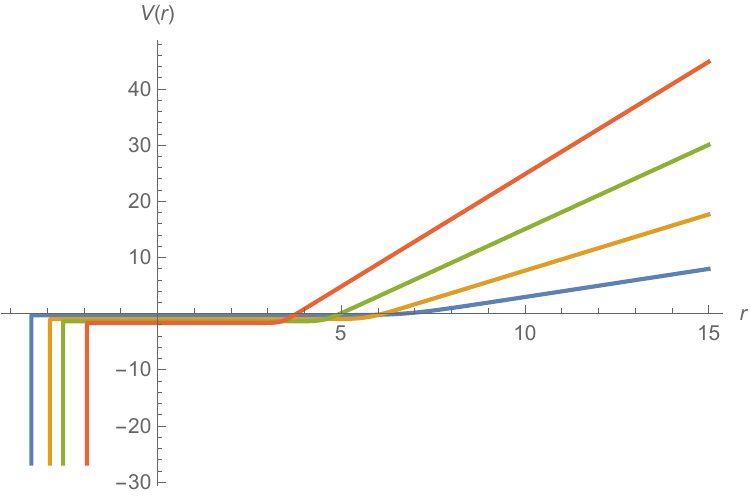}
  \caption{$V$ solution}
  \end{subfigure}\\
  \begin{subfigure}[b]{0.38\linewidth}
    \includegraphics[width=\linewidth]{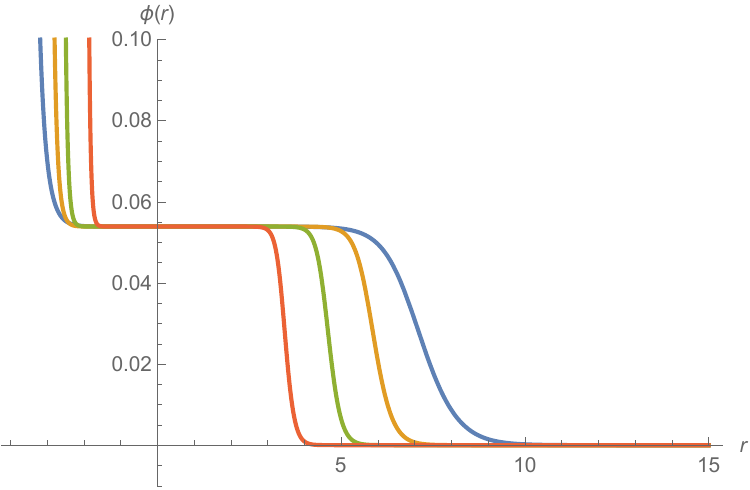}
  \caption{$\phi$ solution}
  \end{subfigure}
      \begin{subfigure}[b]{0.38\linewidth}
    \includegraphics[width=\linewidth]{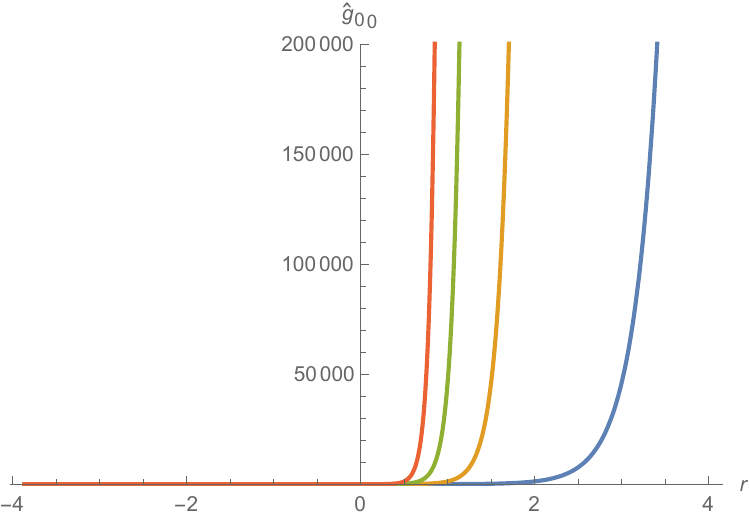}
  \caption{Profiles of $\hat{g}_{00}$}
  \end{subfigure}
  \caption{RG flows from the $N=4$ $AdS_7$ critical point to the $AdS_3\times CH^2$ fixed point and curved domain walls for $SO(3)_+$ twist in $SO(5)$ gauge group. The blue, orange, green and red curves refer to $g=8,16,24,32$, respectively.}
  \label{YSsdO(3)Kahlerflows}
\end{figure}

We now move to $AdS_3\times K^4_k$ solutions with the twist given by identifying the $SU(2)$ part of the spin connection with the self-dual $SO(3)_+\subset SO(3)_+\times SO(3)_-\sim SO(4)$. We begin with the $SO(3)\times SO(3)$ gauge fields of the form
\begin{equation}\label{SO(3)xSO(3)KahlergaugeAnt}
A^{ij}_{(1)}=-\frac{p}{2k}(f'_k(\psi)-1)\epsilon^{ijl}\tau_l\qquad \textrm{and}\qquad
A^{i4}_{(1)}=-\frac{p}{2k}(f'_k(\psi)-1)\delta^{ij}\tau_j\, .
\end{equation} 
The self-dual $SO(3)$ gauge fields can be defined as
\begin{equation}
A^{i}_{(1)}=\frac{1}{2}\epsilon^{ijk}A^{jk}_{(1)}+ A^{i4}_{(1)}=-\frac{p}{k}(f'_k(\psi)-1)\tau_i\, .
\end{equation}
In this case, we perform the twist by imposing the twist condition \eqref{GenQYM} and the three projections given in \eqref{YSO(3)KahlerProjCon} together with an additional projection for the self-duality of $SO(3)$
\begin{equation}\label{GammaSDProj}
{(\Gamma_{12})^a}_b\epsilon^b={(\Gamma_{34})^a}_b\epsilon^b\, .
\end{equation}
Furthermore, by turning on the above $SO(3)$ gauge fields, we need to turn on the the modified three-form field strength of the form
\begin{equation}\label{sdSO(3)3form}
\mathcal{H}^{(3)}_{\hat{m}\hat{n}\hat{r} 5}= -\frac{192}{g}\rho e^{-4(V+2\phi)}p^2\varepsilon_{\hat{m}\hat{n}}\, .
\end{equation}
As in the case of $SO(4)$ symmetric solutions, we will consider only $SO(5)$ and $SO(4,1)$ gauge group with $\rho\neq 0$.
\\
\indent Using the embedding tensor \eqref{SO(4)Ytensor} and the $SO(4)$ invariant coset represenvative \eqref{YSO(4)coset}, we find the following BPS equations
\begin{eqnarray}
U'&=&\frac{g}{40}(4e^{-2\phi}+\rho e^{8\phi})-\frac{12}{5}e^{-2(V-\phi)}p+\frac{144}{5g}\rho e^{-4(V+\phi)}p^2,\\
V'&=&\frac{g}{40}(4e^{-2\phi}+\rho e^{8\phi})+\frac{18}{5}e^{-2(V-\phi)}p-\frac{96}{5g}\rho e^{-4(V+\phi)}p^2,\\
\phi'&=&\frac{g}{20}(e^{-2\phi}-\rho e^{8\phi})-\frac{6}{5}e^{-2(V-\phi)}p-\frac{48}{5g}\rho e^{-4(V+\phi)}p^2
\end{eqnarray}
in which we have also imposed the $\gamma_r$ projection \eqref{pureYProj}. From these equations, an $AdS_3$ fixed point is obtained only in $SO(5)$ gauge group with $\rho=1$ and $k=-1$. This $AdS_3\times CH^2$ solution is given by
\begin{eqnarray}
V&=&\frac{1}{2}\ln\left[\frac{4^{7/5}\times3^{2/5}\times7^{3/5}}{g^2}\right],\nonumber \\
\phi&=&\frac{1}{10}\ln\left[\frac{12}{7}\right],\qquad
L_{\text{AdS}_3}=\frac{4^{6/5}\times3^{1/5}}{g^27^{1/5}} \label{FinYKahlerSDSO(3)fixedpoint}
\end{eqnarray}
which is the $AdS_3\times CH^2$ fixed point found in \cite{Gauntlett1}. The solution preserves two supercharges and corresponds to $N=(1,0)$ SCFT in two dimensions with $SO(3)$ symmetry. Supersymmetric RG flows from the $N=4$ $AdS_7$ vacuum to this $AdS_3\times CH^2$ fixed point and curved domain walls in the IR are given in figure \ref{YSsdO(3)Kahlerflows}. The IR singularities are physically acceptable as indicated by the behavior $\hat{g}_{00}\rightarrow 0$.

\subsubsection{$AdS_3\times K^4_k$ solutions with $SO(2)\times SO(2)$ twists}
\begin{figure}
  \centering
  \begin{subfigure}[b]{0.38\linewidth}
    \includegraphics[width=\linewidth]{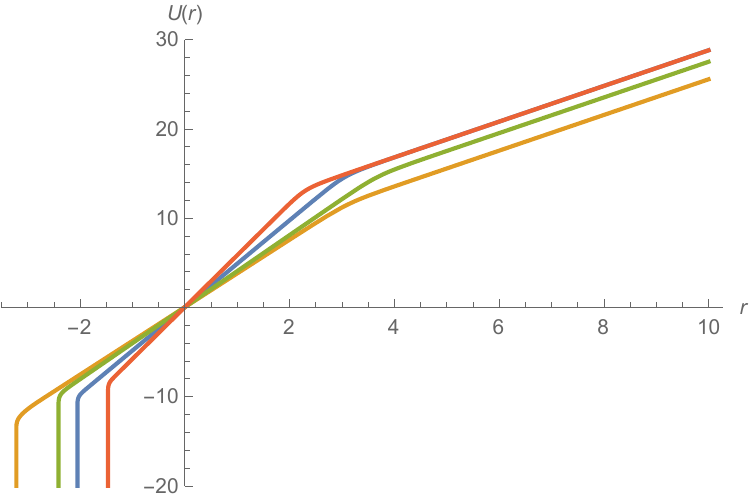}
  \caption{$U$ solution}
  \end{subfigure}
  \begin{subfigure}[b]{0.38\linewidth}
    \includegraphics[width=\linewidth]{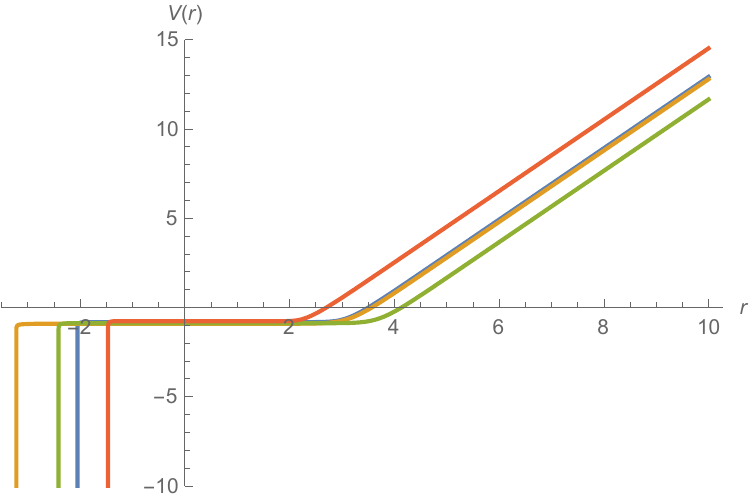}
  \caption{$V$ solution}
  \end{subfigure}
  \begin{subfigure}[b]{0.38\linewidth}
    \includegraphics[width=\linewidth]{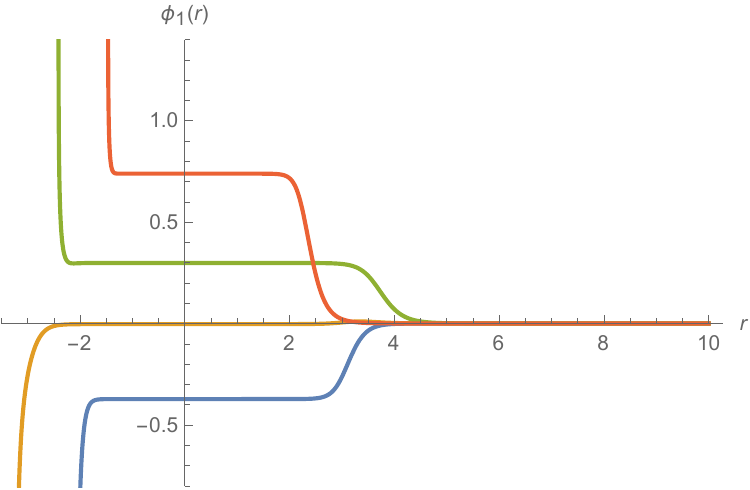}
  \caption{$\phi_1$ solution}
  \end{subfigure}
  \begin{subfigure}[b]{0.38\linewidth}
    \includegraphics[width=\linewidth]{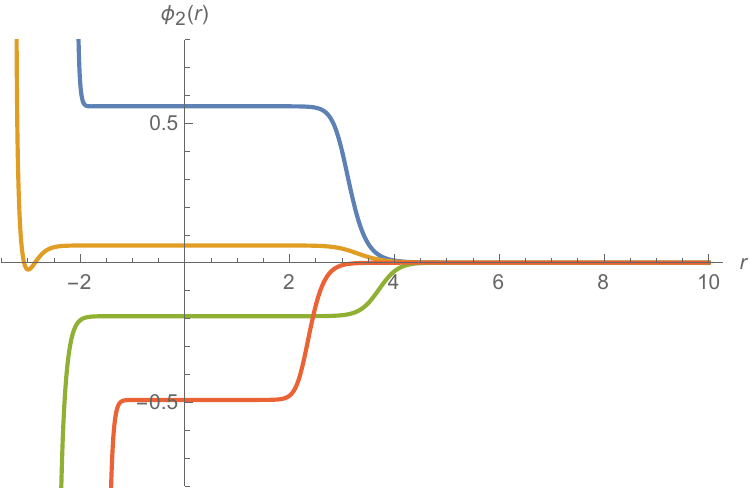}
  \caption{$\phi_2$ solution}
  \end{subfigure}
  \caption{RG flows from the $N=4$ $AdS_7$ critical point to the $AdS_3\times CH^2$ fixed point and curved domain walls for $SO(2)\times SO(2)$ twist in $SO(5)$ gauge group. The blue, orange, green and red curves refer to $p_2=-\frac{1}{25}, -\frac{1}{31}, -\frac{1}{40}, -\frac{1}{46}$, respectively.}
  \label{YSO2xSO2Kahlerflows}
\end{figure}

\begin{figure}
  \centering
    \includegraphics[width=0.5\linewidth]{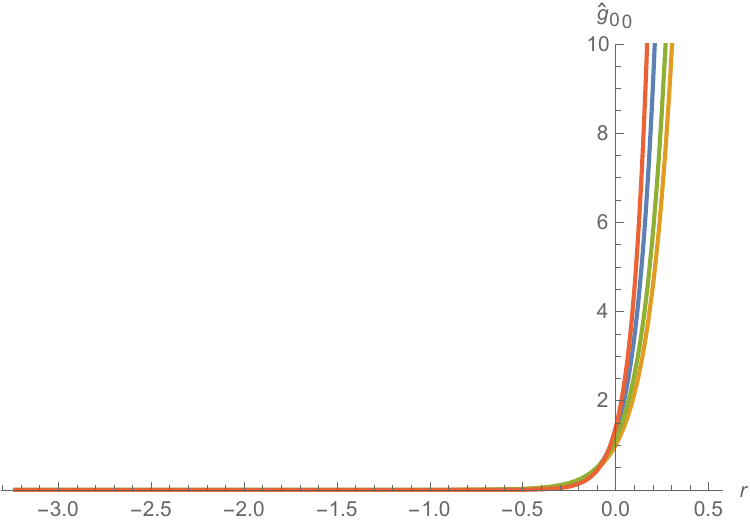}
  \caption{Profiles of the eleven-dimensional metric component $\hat{g}_{00}$ for the RG flows given in figure \ref{YSO2xSO2Kahlerflows}.}
  \label{g00_SO2SO2Kahler}
\end{figure}

As a final case for $AdS_3\times K^4_k$ solutions, we will perform another twist on the Kahler four-cycle by cancelling the $U(1)$ part of the spin connection. To make this $U(1)$ part manifest, we write the metric on $K^4_k$ as
\begin{equation}\label{U(1)Kahlermetric}
ds^2_{K^4_k}=\frac{d\psi^2}{(k\psi^2+1)^2}+\frac{\psi^2\tau_3^2}{(k\psi^2+1)^2}+\frac{\psi^2}{(k\psi^2+1)}(\tau_1^2+\tau_2^2)
\end{equation}
with $\tau_i$ being the $SU(2)$ left-invariant one-forms given in \eqref{SU(2)Inv1form}. The seven-dimensional metric is still given by \eqref{Kahler7Dmetric} with $ds^2_{K^4_k}$ given by \eqref{U(1)Kahlermetric}.
\\
\indent With the following vielbein
\begin{eqnarray}
e^{\hat{m}}&=& e^{U}dx^{m}, \qquad e^{\hat{r}}=dr, \qquad e^{\hat{3}}=\frac{e^{V}\psi}{\sqrt{k\psi^2+1}}\tau_1,\nonumber \\ e^{\hat{4}}&=&\ \frac{e^{V}\psi}{\sqrt{k\psi^2+1}}\tau_2, \qquad e^{\hat{5}}=\frac{e^{V}\psi}{(k\psi^2+1)}\tau_3, \qquad e^{\hat{6}}=\frac{e^{V}d\psi}{(k\psi^2+1)},\label{U(1)Kahler4bein}
\end{eqnarray}
all non-vanishing components of the spin connection are given by
\begin{eqnarray}
\omega_{(1)}^{\hat{m}\hat{r}}&=&  U'e^{\hat{m}}, \qquad \omega_{(1)}^{\hat{i}\hat{r}}= V'e^{\hat{i}},\qquad  \hat{i}=\hat{3}, ..., \hat{6},\nonumber \\
\omega_{(1)}^{\hat{3}\hat{6}}&=& \omega_{(1)}^{\hat{4}\hat{5}}=\frac{\tau_1}{\sqrt{k\psi^2+1}},\qquad \omega^{\hat{3}\hat{4}}=\frac{(2k\psi^2+1)}{(k\psi^2+1)}\tau_3,\nonumber \\
\omega_{(1)}^{\hat{5}\hat{6}}&=& \frac{(1-k\psi^2)}{(k\psi^2+1)}\tau_3,\qquad \omega_{(1)}^{\hat{5}\hat{3}}=\omega_{(1)}^{\hat{4}\hat{6}}=\frac{\tau_2}{\sqrt{k\psi^2+1)}}\, .\label{AdS3xU(1)Kahler4SpinCon}
\end{eqnarray}
\indent To perform the twist, we turn on the $SO(2)\times SO(2)$ gauge fields 
\begin{equation}\label{YKahlerSO(2)xSO(2)gaugeAnt}
A^{12}_{(1)}=p_1\frac{3\psi^2}{\sqrt{k\psi^2+1}}\tau_3\qquad \textrm{and}\qquad A^{34}_{(1)}=p_2\frac{3\psi^2}{\sqrt{k\psi^2+1}}\tau_3
\end{equation}  
and impose the following projection conditions on the Killing spinors
\begin{equation}
\gamma^{\hat{3}\hat{4}}\epsilon^a=-\gamma^{\hat{5}\hat{6}}\epsilon^a=-{(\Gamma_{12})^a}_b\epsilon^b=-{(\Gamma_{34})^a}_b\epsilon^b
\end{equation}
together with the twist condition \eqref{YSO(2)xSO(2)on1QYM}. The associated two-form gauge field strengths are given by
\begin{eqnarray}
\mathcal{H}^{12}_{(2)}=F^{12}_{(2)}=3e^{-2V}p_1J_{(2)}\qquad \textrm{and} \qquad
\mathcal{H}^{34}_{(2)}=F^{34}_{(2)}=3e^{-2V}p_2J_{(2)}
\end{eqnarray}
where $J_{(2)}$ is the Kahler structure defined by
\begin{equation}\label{KahlerSTR}
J_{(2)}=e^{\hat{3}}\wedge e^{\hat{4}}-e^{\hat{5}}\wedge e^{\hat{6}}\, .
\end{equation}
With the above non-vanishing $SO(2)\times SO(2)$ gauge fields, we need to turn on the modified three-form field strength of the form
\begin{equation}\label{YAbSO(2)xSO(2)3form}
\mathcal{H}^{(3)}_{\hat{m}\hat{n}\hat{r} 5}= \frac{576}{g}\rho e^{-4\left(V+\phi_1+\phi_2\right)}p_1p_2  \varepsilon_{\hat{m}\hat{n}}
\end{equation}
with $\rho$ being the parameter in the embedding tensor \eqref{SO(2)xSO(2)Ytensor} for gauge groups with an $SO(2)\times SO(2)$ subgroup. As in the previous cases, the appearance of $\rho$ in \eqref{YAbSO(2)xSO(2)3form} implies that the resulting BPS equations are not compatible with the field equations for the case of $\rho=0$. In subsequent analysis, we will accordingly consider only gauge groups with $\rho\neq0$.
\\
\indent With the $\gamma_r$ projector \eqref{pureYProj} and the scalar coset representative \eqref{YSO(2)xSO(2)Ys}, the corresponding BPS equations read
\begin{eqnarray}
U'&=&\frac{g}{40}(2e^{-2\phi_1}+\rho e^{4(\phi_1+\phi_2)}+2\sigma e^{-2\phi_2})-\frac{12}{5}e^{-2V}(e^{2\phi_1}p_1+e^{2\phi_2}p_2)\nonumber\\&&+\frac{1728}{5g}\rho e^{-2(2V+\phi_1+\phi_2)}p_1p_2,\\
V'&=&\frac{g}{40}(2e^{-2\phi_1}+\rho e^{4(\phi_1+\phi_2)}+2\sigma e^{-2\phi_2})+\frac{18}{5}e^{-2V}(e^{2\phi_1}p_1+e^{2\phi_2}p_2)\nonumber\\&&-\frac{1152}{5g}\rho e^{-2(2V+\phi_1+\phi_2)}p_1p_2,\\
\phi_1'&=&\frac{g}{20}(3e^{-2\phi_1}-\rho e^{4(\phi_1+\phi_2)}-2\sigma e^{-2\phi_2})-\frac{12}{5}e^{-2V}(3e^{2\phi_1}p_1-2e^{2\phi_2}p_2)\nonumber\\&&-\frac{576}{5g}\rho e^{-2(2V+\phi_1+\phi_2)}p_1p_2,\\
\phi_2'&=&\frac{g}{20}(3\sigma e^{-2\phi_2}-\rho e^{4(\phi_1+\phi_2)}-2 e^{-2\phi_1})+\frac{12}{5}e^{-2V}(2e^{2\phi_1}p_1-3e^{2\phi_2}p_2)\nonumber\\&&-\frac{576}{5g}\rho e^{-2(2V+\phi_1+\phi_2)}p_1p_2\, .
\end{eqnarray}
From these equations, we find the following $AdS_3$ fixed point solutions
\begin{eqnarray}
e^{2V}&=&-\frac{48(p_1e^{2\phi_1}+p_2e^{2\phi_2})}{g\rho e^{4(\phi_1+\phi_2)}},\\
e^{10\phi_1}&=&\frac{p_2^2\rho\left((p_1+p_1\rho^2-p_2\sigma)(p_1+p_2\sigma)\right)^2}{p_1^3(2+\rho^2)(p_2(1+\rho^2)\sigma-p_1)^3},\\
e^{10\phi_2}&=&\frac{p_1^2\rho\left(p_1-p_2(1+\rho^2)\sigma)(p_1+p_2\sigma)\right)^2}{p_2^3(2+\rho^2)(p_1+p_1\rho^2-p_2\sigma)^3},\\
L_{\text{AdS}_3}&=&\frac{8e^{-4(\phi_1+\phi_2)}(p_1e^{2\phi_1}+p_2e^{2\phi_2})^2}{g\rho\left(p_1^2e^{4\phi_1}+p_2^2e^{4\phi_2}+2p_1p_2e^{2(\phi_1+\phi_2)}(1+\rho^2)\right)}.
\end{eqnarray}
These solutions preserve four supercharges and are dual to $N=(2,0)$ two-dimensional SCFTs. 
\\
\indent For $SO(5)$ gauge group, there exist $AdS_3\times CH^2$ fixed points in the range
\begin{equation}
-\frac{2}{3}<gp_2<-\frac{1}{3}
\end{equation}
in which we have taken $g>0$ for convenience. Up to some differences in notations, these $AdS_3\times CH^2$ fixed points are the same as the solutions studied in \cite{2D_Bobev}. As in the previous cases, we study RG flows from the supersymmetric $AdS_7$ vacuum to the $AdS_3\times CH^2$ fixed points and curved domain walls in the IR. Some examples of these flows are given in figure \ref{YSO2xSO2Kahlerflows} for $g=16$ and different values of $p_2$. The behaviors of the eleven-dimensional metric component $\hat{g}_{00}$ for these RG flows are shown in figure \ref{g00_SO2SO2Kahler} which indicates that the singularities are physical.  

\begin{figure}
  \centering
  \begin{subfigure}[b]{0.38\linewidth}
    \includegraphics[width=\linewidth]{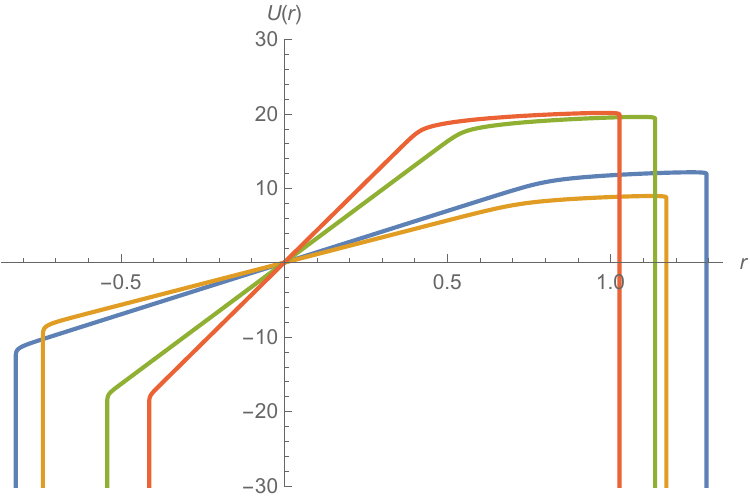}
  \caption{$U$ solution}
  \end{subfigure}
  \begin{subfigure}[b]{0.38\linewidth}
    \includegraphics[width=\linewidth]{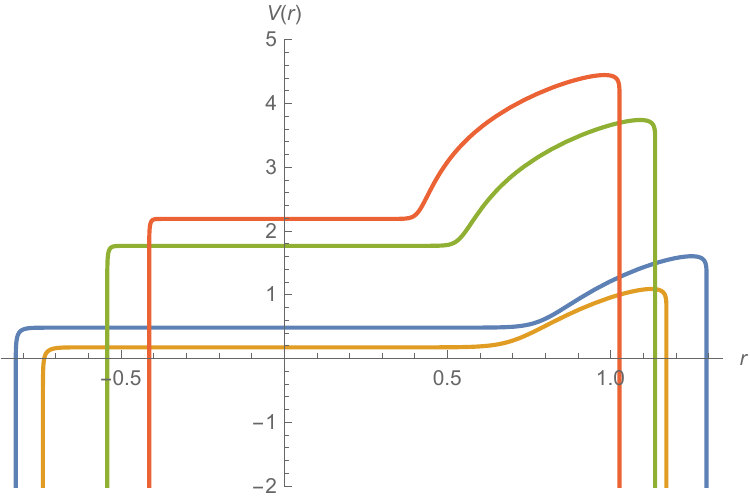}
  \caption{$V$ solution}
  \end{subfigure}
  \begin{subfigure}[b]{0.38\linewidth}
    \includegraphics[width=\linewidth]{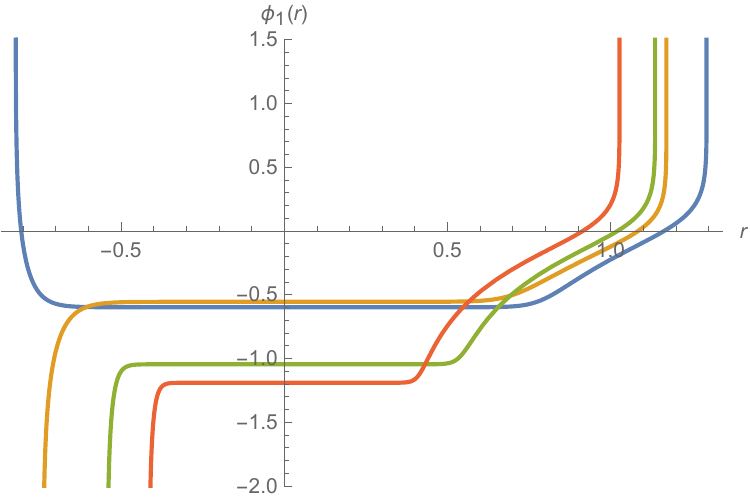}
  \caption{$\phi_1$ solution}
  \end{subfigure}
  \begin{subfigure}[b]{0.38\linewidth}
    \includegraphics[width=\linewidth]{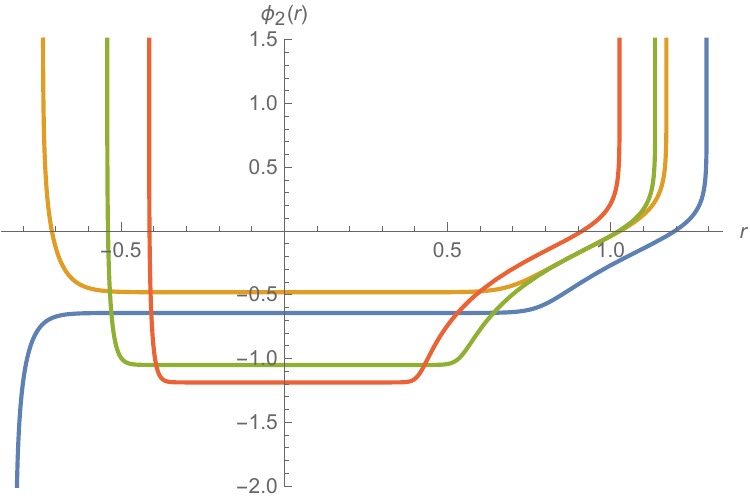}
  \caption{$\phi_2$ solution}
  \end{subfigure}
  \caption{RG flows between $AdS_3\times CP^2$ fixed points and curved domain walls for $SO(2)\times SO(2)$ twist in $SO(4,1)$ gauge group. The blue, orange, green and red curves refer to $p_2=\frac{1}{2}, -\frac{1}{4}, 4, -8, -\frac{1}{47}$, respectively.}
  \label{YSO2xSO2KahlerSO41flows}
\end{figure}

\begin{figure}
  \centering
  \begin{subfigure}[b]{0.38\linewidth}
    \includegraphics[width=\linewidth]{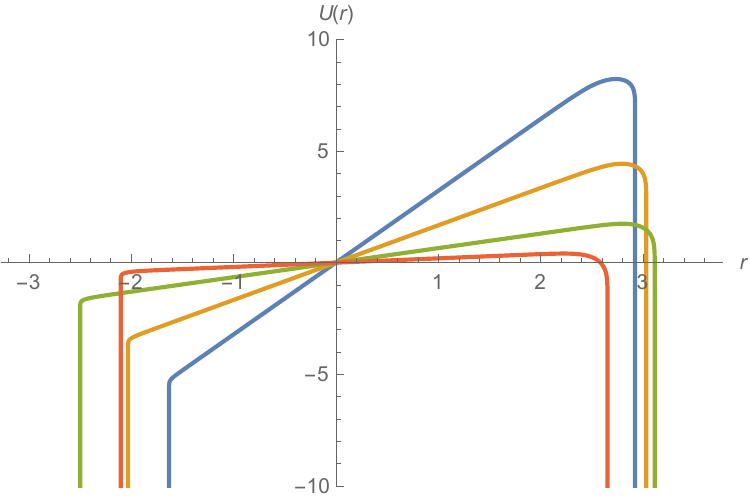}
  \caption{$U$ solution}
  \end{subfigure}
  \begin{subfigure}[b]{0.38\linewidth}
    \includegraphics[width=\linewidth]{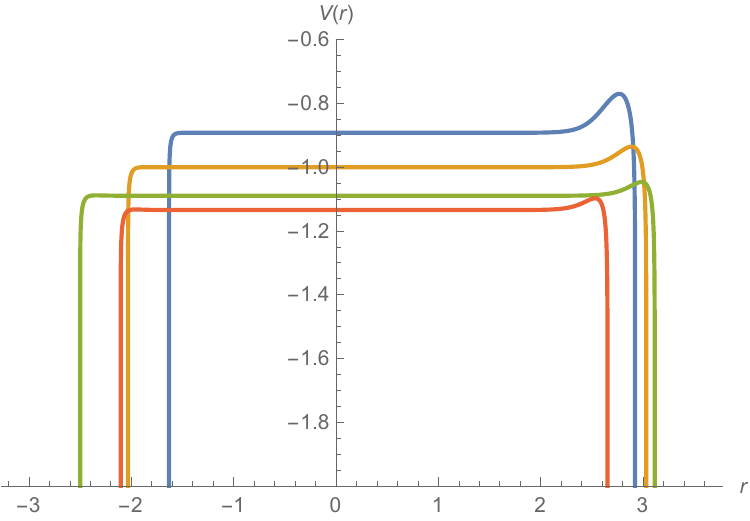}
  \caption{$V$ solution}
  \end{subfigure}
  \begin{subfigure}[b]{0.38\linewidth}
    \includegraphics[width=\linewidth]{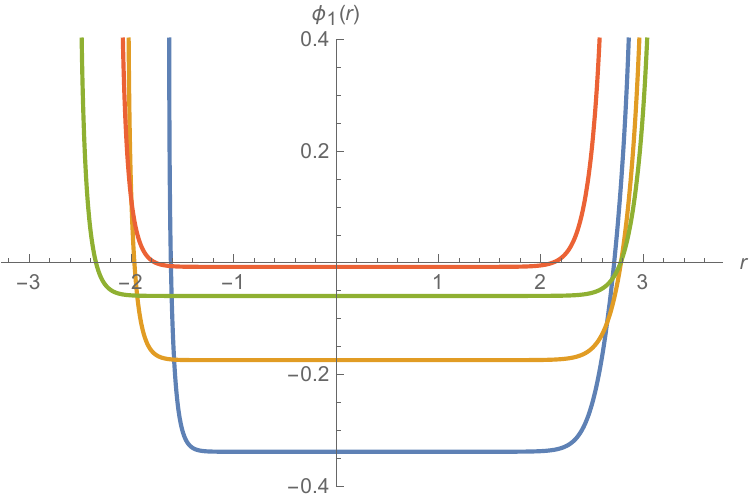}
  \caption{$\phi_1$ solution}
  \end{subfigure}
  \begin{subfigure}[b]{0.38\linewidth}
    \includegraphics[width=\linewidth]{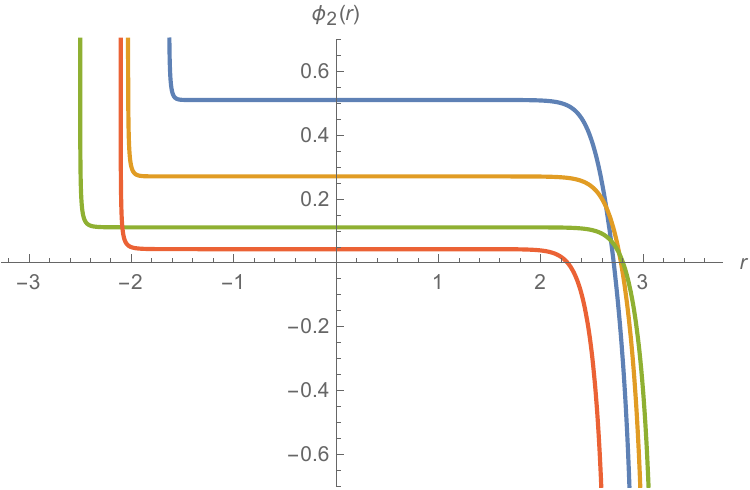}
  \caption{$\phi_2$ solution}
  \end{subfigure}
  \caption{RG flows between $AdS_3\times CP^2$ fixed points and curved domain walls for $SO(2)\times SO(2)$ twist in $SO(3,2)$ gauge group. The blue, orange, green and red curves refer to $p_2=-\frac{1}{23}, -\frac{1}{22}, -\frac{1}{21}, -\frac{2}{41}$, respectively.}
  \label{YSO2xSO2KahlerSO32flows}
\end{figure}

\begin{figure}
  \centering
  \begin{subfigure}[b]{0.38\linewidth}
    \includegraphics[width=\linewidth]{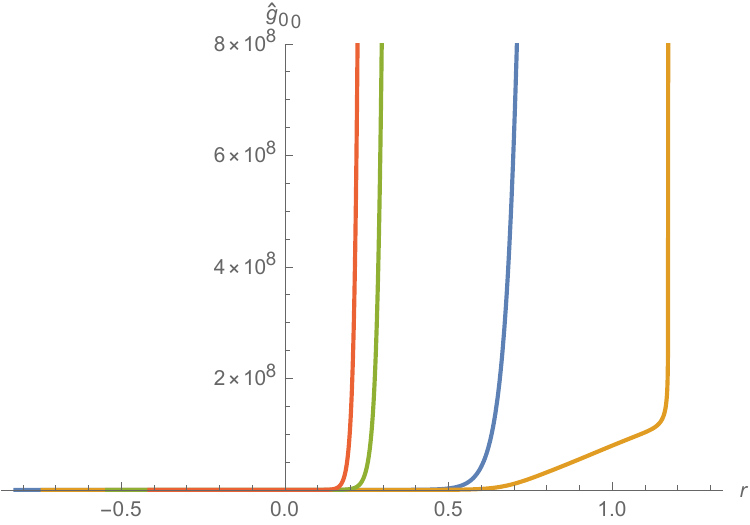}
  \caption{RG flows in $SO(4,1)$ gauge group}
  \end{subfigure}
  \begin{subfigure}[b]{0.38\linewidth}
    \includegraphics[width=\linewidth]{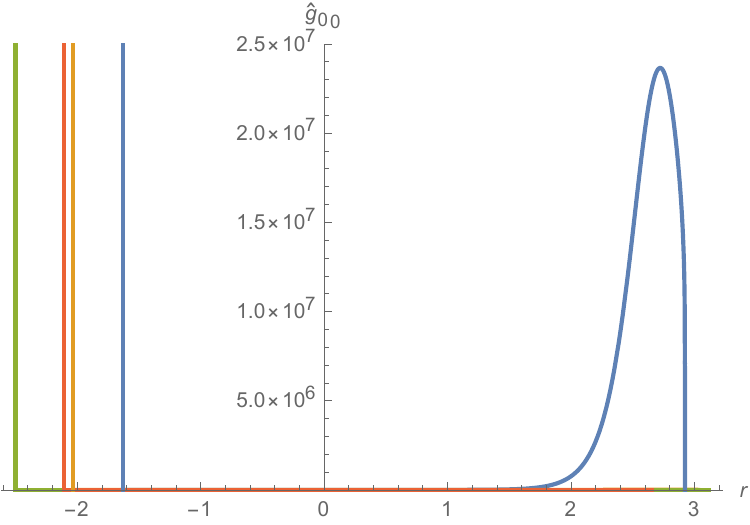}
  \caption{RG flows in $SO(3,2)$ gauge group}
  \end{subfigure}
  \caption{Profiles of the eleven-dimensional metric component $\hat{g}_{00}$ for the RG flows beween $AdS_3\times CP^2$ fixed points and curved domain walls in $SO(4,1)$ and $SO(3,2)$ gauge groups.}
  \label{g00SO2xSO2KahlerCP2}
\end{figure}

Apart from these $AdS_3\times CH^2$ fixed points, we find new $AdS_3\times CP^2$ fixed points in $SO(4,1)$ and $SO(3,2)$ gauge groups respectively in the following ranges, with $g>0$,
\begin{equation}
g p_2<0 \ \cup\ g p_2>1  \qquad \textrm{and} \qquad
-\frac{(3-\sqrt{3})}{6}<g p_2<-\frac{2}{3}\, .
\end{equation}
Since there is no supersymmetric $AdS_7$ critical point for $SO(4,1)$ and $SO(3,2)$ gauge groups, we will study supersymmetric RG flows between these $AdS_3\times CP^2$ fixed points and curved domain walls with $SO(2)\times SO(2)$ symmetry. Examples of these RG flows in $SO(4,1)$ and $SO(3,2)$ gauge groups are shown respectively in figures \ref{YSO2xSO2KahlerSO41flows} and \ref{YSO2xSO2KahlerSO32flows} with $g=16$ and different values of $p_2$. From the behaviors of $\hat{g}_{00}$ in figure \ref{g00SO2xSO2KahlerCP2}, we find that the singularities on the left (right) with $\phi_1\rightarrow \pm \infty$ and $\phi_2\rightarrow \mp \infty$ ($\phi_1\rightarrow \infty$ and $\phi_2\rightarrow - \infty$) of the flows in $SO(4,1)$ ($SO(3,2)$) gauge group are physical.

\subsection{Supersymmetric $AdS_2\times \Sigma^5$ solutions}\label{YAdS2section}
\begin{figure}
  \centering
  \begin{subfigure}[b]{0.32\linewidth}
    \includegraphics[width=\linewidth]{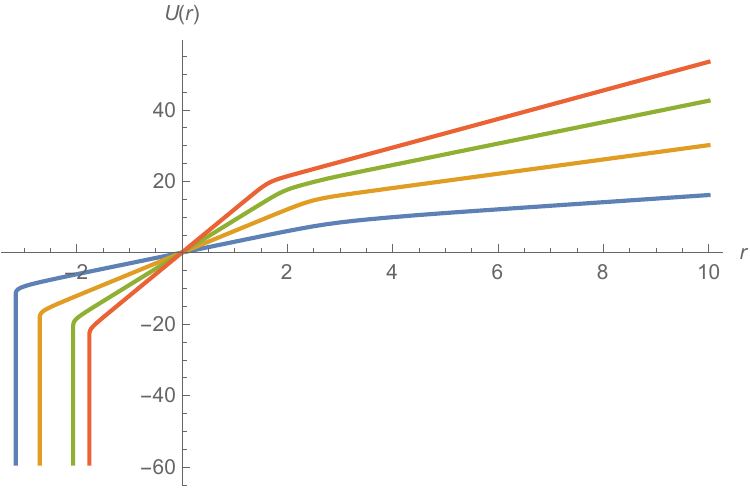}
  \caption{$U$ solution}
  \end{subfigure}
  \begin{subfigure}[b]{0.32\linewidth}
    \includegraphics[width=\linewidth]{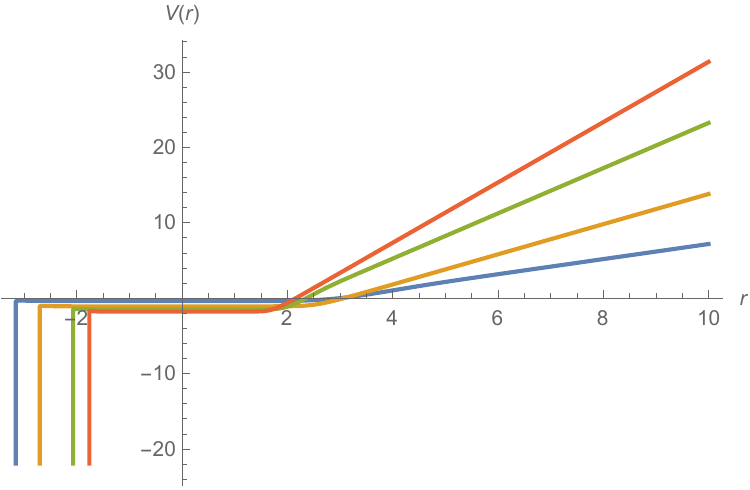}
  \caption{$V$ solution}
  \end{subfigure}
  \begin{subfigure}[b]{0.32\linewidth}
    \includegraphics[width=\linewidth]{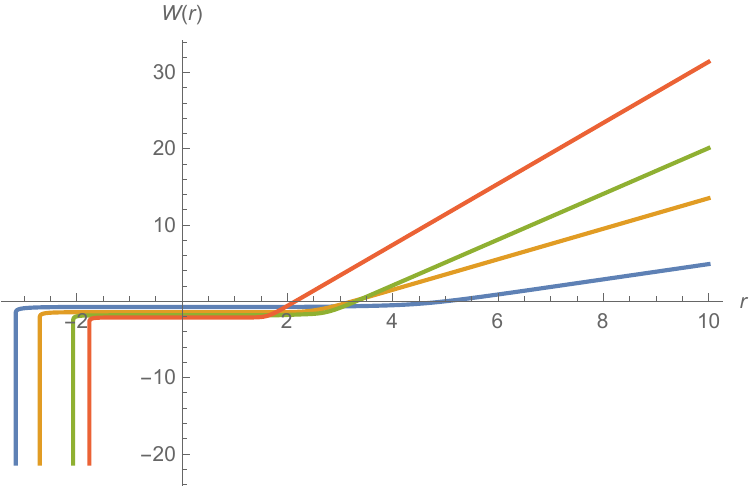}
  \caption{$W$ solution}
  \end{subfigure}\\
  \begin{subfigure}[b]{0.32\linewidth}
    \includegraphics[width=\linewidth]{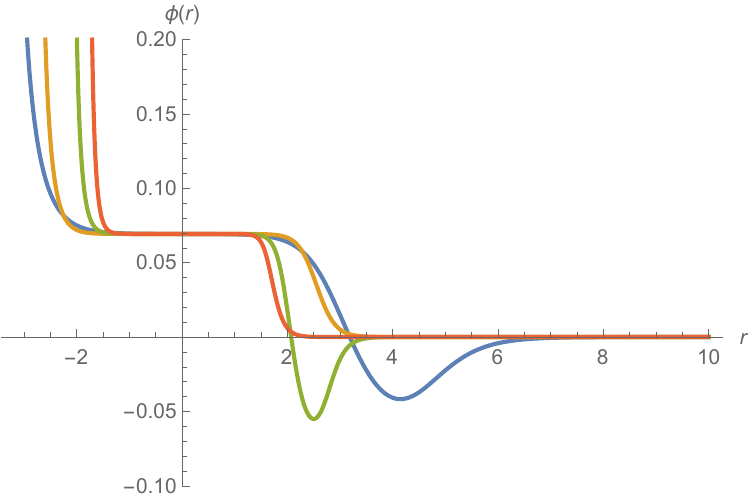}
  \caption{$\phi$ solution}
    \end{subfigure}
     \begin{subfigure}[b]{0.32\linewidth}
    \includegraphics[width=\linewidth]{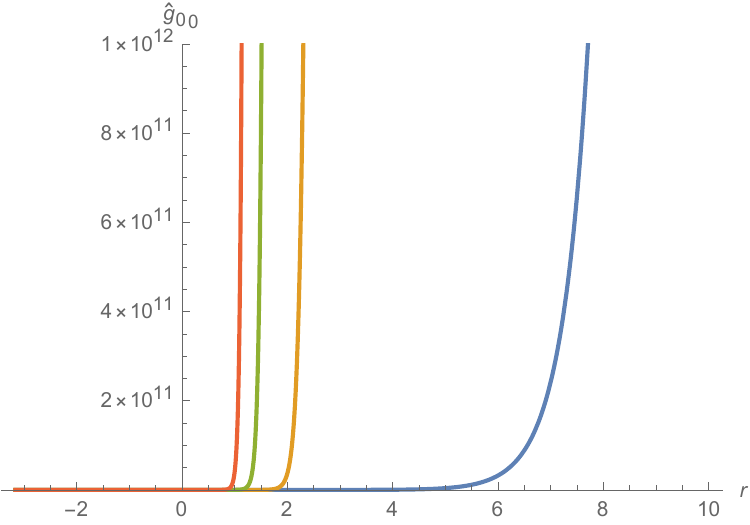}
  \caption{Profiles of $\hat{g}_{00}$}
    \end{subfigure}
  \caption{RG flows from supersymmetric $AdS_7$ vacuum to the $AdS_2\times H^3\times H^2$ fixed point and curved domain walls with $SO(3)\times SO(2)$ twist in $SO(5)$ gauge group. The blue, orange, green and red curves refer to $g=8,16,24,32$, respectively.}
  \label{SO(3)xSO(2)flows}
\end{figure}

We end this section by considering solutions of the form $AdS_2\times \Sigma^5$. $AdS_2\times \Sigma^5$ solutions for the manifold $\Sigma^5$ being $S^5$ or $H^5$ have been given in \cite{Gauntlett1}. The twist is performed by turning on $SO(5)$ gauge fields. This is obviously possible only for $SO(5)$ gauge group. In addition, no scalars in $SL(5)/SO(5)$ coset are singlets under $SO(5)$, so the solutions are given purely in term of the seven-dimensional metric. The corresponding RG flows from the supersymmetric $AdS_7$ vacuum and the $AdS_2\times H^5$ or $AdS_2\times S^5$ fixed points have already been analytically given in \cite{Gauntlett1}. We will not repeat the analysis for this case here. 
\\
\indent However, if we consider $\Sigma^5$ as a product of three- and two-manifolds $\Sigma^3\times \Sigma^2$, it is possible to perform a twist by turning on $SO(3)\times SO(2)$ gauge fields along $\Sigma^3\times \Sigma^2$. In this case, there are two gauge groups with an $SO(3)\times SO(2)$ subgroup namely $SO(5)$ and $SO(3,2)$. The ansatz for the seven-dimensional metric takes the form of
\begin{equation}\label{SO(3)xSO(2)7Dmetric}
ds_7^2=-e^{2U(r)}dt^2+dr^2+e^{2V(r)}ds^2_{\Sigma^{3}_{k_1}}+e^{2W(r)}ds^2_{\Sigma^{2}_{k_2}}\, .
\end{equation}
The explicit form of the metrics on the $\Sigma^{3}_{k_1}$ and $\Sigma^{2}_{k_2}$ are given in \eqref{Sigma3metric} and \eqref{Sigma2metric}, respectively. 
\\
\indent Using the vielbein
\begin{eqnarray}
e^{\hat{t}}&=& e^{U}dt, \quad e^{\hat{r}}=dr,\quad  e^{\hat{\theta}_2}=e^{W}d\theta_2,\quad e^{\hat{\varphi}_2}=e^{W}f_{k_2}(\theta_2)d\varphi_2,\nonumber \\ 
e^{\hat{\psi}_1}&=& e^{V}d\psi_1, \quad e^{\hat{\theta}_1}= e^{V}f_{k_1}(\psi_1)d\theta_1, \quad e^{\hat{\varphi}_1}=e^{V}f_{k_1}(\psi_1)\sin{\theta_1}d\varphi_1, \label{Sigma3xSigma2bein}
\end{eqnarray}
we find non-vanishing components of the spin connection as follow
\begin{eqnarray}
\omega_{(1)}^{\hat{t}\hat{r}}&=&  U'e^{\hat{t}}, \qquad \omega_{(1)}^{\hat{i}_1\hat{r}}= V'e^{\hat{i}_1}, \qquad \omega_{(1)}^{\hat{i}_2\hat{r}}= W'e^{\hat{i}_2},\nonumber \\
\omega_{(1)}^{\hat{\theta}_1\hat{\psi}_1}&=& \frac{f'_{k_1}(\psi_1)}{f_{k_1}(\psi_1)}e^{-V}e^{\hat{\theta}_1},\qquad \omega_{(1)}^{\hat{\varphi}_1\hat{\psi}_1}=\frac{f'_{k_1}(\psi_1)}{f_{k_1}(\psi_1)}e^{-V}e^{\hat{\varphi}_1}, \nonumber \\ \omega_{(1)}^{\hat{\varphi}_1\hat{\theta}_1}&=& \frac{\cot\theta_1}{f_{k_1}(\psi_1)}e^{-V}e^{\hat{\varphi}_1},\qquad
\omega_{(1)}^{\hat{\varphi}_2\hat{\theta}_2}=\frac{f_{k_2}'(\theta_2)}{f_{k_2}(\theta_2)}e^{-W}e^{\hat{\varphi}_2}\label{AdS2xSigma3xSigma2SpinCon}
\end{eqnarray}
where $\hat{i}_1=\hat{\psi}_1, \hat{\theta}_1, \hat{\varphi}_1$ and $\hat{i}_2=\hat{\theta}_2, \hat{\varphi}_2$ are flat indices on $\Sigma^3_{k_1}$ and $\Sigma^2_{k_2}$, respectively.
\\
\indent There is only one $SO(3)\times SO(2)$ invariant scalar field corresponding to the following non-compact generator
\begin{equation}
\mathcal{Y}=2e_{1,1}+2e_{2,2}+2e_{3,3}-3e_{4,4}-3e_{5,5}.
\end{equation}
Therefore, the $SL(5)/SO(5)$ coset representative is parametrized by
\begin{equation}\label{YSO(3)xSO(2)Ys}
\mathcal{V}=e^{\phi\mathcal{Y}}.
\end{equation}
\indent We now turn on the $SO(3)\times SO(2)$ gauge fields of the form
\begin{eqnarray}
A^{12}_{(1)}&=& -\frac{p_1}{k_1}\frac{f'_{k_1}(\psi_1)}{f_{k_1}(\psi_1)}e^{-V}e^{\hat{\theta}_1},\qquad A^{13}_{(1)}=-\frac{p_1}{k_1}\frac{f'_{k_1}(\psi_1)}{f_{k_1}(\psi_1)}e^{-V}e^{\hat{\varphi}_1},\nonumber \\ 
A^{23}_{(1)}&=& -\frac{p_1}{k_1}\frac{\cot(\theta_1)}{f_{k_1}(\psi_1)}e^{-V}e^{\hat{\varphi}_1}, \qquad A^{45}_{(1)}=-\frac{p_2}{k_2}\frac{f_{k_2}'(\theta_2)}{f_{k_2}(\theta_2)}e^{-W}e^{\hat{\varphi}_2}\label{Sigma3xSigma2gaugeAnt}
\end{eqnarray}
with the corresponding two-form field strengths given by
\begin{eqnarray}
\mathcal{H}^{(2)12}_{\hat{\psi}_1\hat{\theta}_1}&=& F^{12}_{\hat{\psi}_1\hat{\theta}_1}=e^{-2V}p_1,\qquad \mathcal{H}^{(2)23}_{\hat{\theta}_1\hat{\varphi}_1}=F^{23}_{\hat{\theta}_1\hat{\varphi}_1}=e^{-2V}p_1,\nonumber \\ \mathcal{H}^{(2)31}_{\hat{\varphi}_1\hat{\psi}_1}&=& F^{31}_{\hat{\varphi}_1\hat{\psi}_1}=e^{-2V}p_1,\qquad \mathcal{H}^{(2)45}_{\hat{\theta}_2\hat{\varphi}_2}=F^{45}_{\hat{\theta}_2\hat{\varphi}_2}= e^{-2W}p_2\, .\label{SO(3)xSO(2)2form}
\end{eqnarray}
With all these gauge fields non-vanishing, we also need to turn on the three-form field strengths 
\begin{equation}\label{SO(3)xSO(2)3form}
\mathcal{H}^{(3)}_{\hat{t}\hat{r}\hat{i}_1 M}= -\frac{32}{g}\delta_{\hat{i}_1M}e^{4\phi-2V-2W}p_1p_2\, .
\end{equation}
\\
\indent We then impose the twist conditions
\begin{equation}\label{YSO(3)xSO(2)QYM}
g p_1=k_1 \qquad \textrm{and} \qquad \sigma g p_2=k_2
\end{equation}
and the following projectors on the Killing spinors
\begin{equation}\label{Sigma3xSigma2ProjCon}
\gamma^{\hat{\psi}_1\hat{\theta}_1}\epsilon^a=-{(\Gamma_{12})^a}_b\epsilon^b, \quad \gamma^{\hat{\theta}_1\hat{\varphi}_1}\epsilon^a=-{(\Gamma_{23})^a}_b\epsilon^b, \quad \gamma^{\hat{\theta}_2\hat{\varphi}_2}\epsilon^a=-{(\Gamma_{45})^a}_b\epsilon^b\, .
\end{equation}
Using the embedding tensor in the form
\begin{equation}
Y_{MN}=\textrm{diag}(+1,+1,+1,\sigma,\sigma)
\end{equation}
for $\sigma=\pm 1$ and the scalar coset representative \eqref{YSO(3)xSO(2)Ys}, we can derive the following BPS equations
\begin{eqnarray}
U'&\hspace{-0.3cm}=&\hspace{-0.3cm}\frac{g}{40}(3e^{-4\phi}+2\sigma e^{6\phi})+\frac{288e^{2\phi}p_1p_2}{5ge^{2(V+W)}}-\frac{2}{5}(3e^{-2V+4\phi}p_1+e^{-2W-6\phi}p_2),\hspace{1cm} \\
V'&\hspace{-0.3cm}=&\hspace{-0.3cm}\frac{g}{40}(3e^{-4\phi}+2\sigma e^{6\phi})-\frac{32e^{2\phi}p_1p_2}{5ge^{2(V+W)}}+\frac{2}{5}(7e^{-2V+4\phi}p_1-e^{-2W-6\phi}p_2), \\
W'&\hspace{-0.3cm}=&\hspace{-0.3cm}\frac{g}{40}(3e^{-4\phi}+2\sigma e^{6\phi})-\frac{192e^{2\phi}p_1p_2}{5ge^{2(V+W)}}-\frac{2}{5}(3e^{-2V+4\phi}p_1-2e^{-2W-6\phi}p_2), \\
\phi'&\hspace{-0.3cm}=&\hspace{-0.3cm}\frac{g}{20}(e^{-4\phi}-\sigma e^{6\phi})+\frac{32e^{2\phi}p_1p_2}{5ge^{2(V+W)}}-\frac{2}{5}(2e^{-2V+4\phi}p_1-e^{-2W-6\phi}p_2)\qquad
\end{eqnarray}
in which we have also used the $\gamma_r$ projector in \eqref{pureYProj}.
\\
\indent From these equations, we find an $AdS_2$ fixed point only for $k_1=k_2=-1$ and $\sigma=1$. The resulting $AdS_2\times H^3\times H^2$ fixed point is given by
\begin{eqnarray}
V&=&\frac{1}{2}\ln\left[\frac{16\times2^{4/5}}{g^2}\right],\qquad
W=\frac{1}{2}\ln\left[\frac{16}{g^22^{1/5}}\right],\nonumber \\
\phi&=&\frac{1}{10}\ln2,\qquad
L_{\textrm{AdS}_2}=\frac{2\times2^{2/5}}{g}\label{YSO(3)xSO(2)fixedpoint}
\end{eqnarray}
which is the solution found in \cite{Gauntlett2}. The three projectors in \eqref{Sigma3xSigma2ProjCon} imply that this solution preserves four supercharges. The solution is dual to superconformal quantum mechanics. Examples of RG flows from the supersymmetric $AdS_7$ vacuum to the $AdS_2\times H^3\times H^2$ fixed point and curved domain walls in the IR are given in figure \ref{SO(3)xSO(2)flows}. From the behavior of the eleven-dimensional metric component $\hat{g}_{00}$, we see that the singularity is physically acceptable. Therefore, this singularity is expected to describe supersymmetric quantum mechanics obtained from a twisted compactification of $N=(2,0)$ SCFT in six dimensions.

\section{Solutions from gaugings in $\overline{\mathbf{40}}$ representation}\label{Z_gauging}
In this section, we repeat the same analysis for gaugings from $\overline{\mathbf{40}}$ representation. These result in gauge groups of the form $CSO(p,q,r)$ with $p+q+r=4$. In this case, the embedding tensor is given by
\begin{equation}\label{pureZ}
Y_{MN}=0\qquad \textrm{and}\qquad Z^{MN,P}=v^{[M}w^{N]P}
\end{equation}
with $w^{MN}=w^{(MN)}$. The $SL(5)$ symmetry can be used to fix $v^M=\delta^M_5$. Following \cite{N4_7D_Henning}, we will split the index $M=(i,5)$ and set $w^{55}=w^{i5}=0$. The remaining $SL(4)\subset SL(5)$ symmetry can be used to diagonalize $w^{ij}$ in the form
\begin{equation}\label{wij}
w^{ij}=\text{diag}(\underbrace{1,..,1}_p,\underbrace{-1,..,-1}_q,\underbrace{0,..,0}_r).
\end{equation}
This generates $CSO(p,q,r)$ gauge groups for $p+q+r=4$ with the corresponding gauge generators given by
\begin{equation}\label{pureZgaugeGen}
{(X_{ij})_k}^l=2\epsilon_{ijkm}w^{ml}\, .
\end{equation}
\indent With the $SL(5)$ index splitting $M=(i,5)$, it is also useful to parametrize the $SL(5)/SO(5)$ coset in term of the $SL(4)/SO(4)$ submanifold as
\begin{equation}\label{t=4decompose}
\mathcal{V}=e^{b_it^i}\widetilde{\mathcal{V}}e^{\phi_0t_0}\, .
\end{equation}
$\widetilde{\mathcal{V}}$ is the $SL(4)/SO(4)$ coset representative, and $t_0$, $t^i$ refer respectively to $SO(1,1)$ and four nilpotent generators in the decomposition $SL(5)\rightarrow SL(4)\times SO(1,1)$. The unimodular matrix $\mathcal{M}_{MN}$ decomposes accordingly
\begin{equation}
\mathcal{M}_{MN}=\begin{pmatrix}e^{-2\phi_0}\widetilde{\mathcal{M}}_{ij}+e^{8\phi_0} & e^{8\phi_0}b_i \\
e^{8\phi_0}b_j& e^{8\phi_0} 
\end{pmatrix}\label{MSL5SL4}
\end{equation}
with $\widetilde{\mathcal{M}}_{ij}=(\widetilde{\mathcal{V}}\widetilde{\mathcal{V}}^T)_{ij}$. The scalar potential for the embedding tensor \eqref{pureZ} reads
\begin{equation}\label{ZwithNilscalarPot}
\mathbf{V}=\frac{g^2}{4}e^{14\phi_0}b_iw^{ij}\widetilde{\mathcal{M}}_{jk}w^{kl}b_l+\frac{g^2}{4}e^{4\phi_0}\left(2\widetilde{\mathcal{M}}_{ij}w^{jk}\widetilde{\mathcal{M}}_{kl}w^{li}-(\widetilde{\mathcal{M}}_{ij}w^{ij})^2\right).
\end{equation}
It can be straightforwardly verified that the nilpotent scalars $b_i$ appear at least quadratically in the Lagrangian. Therefore, these scalars can always be consistently truncated out. In the following analysis, we will consider only supersymmetric solutions with $b_i=0$ for simplicity.
\\
\indent As in the case of gaugings in $\mathbf{15}$ representation, when the compact manifold $\Sigma^d$ has dimension $d>3$, the modified three-form field strengths need to be turned on in order to satisfy the corresponding Bianchi's identity. However, with $Y_{MN}=0$, there are no massive three-form fields. In this case, the contribution to $\mc{H}^{(3)}_{\mu\nu\rho M}$ arises solely from the two-form fields. There are respectively $5-s$ and $s$, for $s=\textrm{rank}\, Z$, massless and massive two-form fields. The latter also appear in the modified gauge field strengths $\mc{H}^{(2)MN}_{\mu\nu}$. In particular, with the embedding tensor given in \eqref{pureZ}, we find
\begin{equation}
\mathcal{H}^{(2)ij}_{\mu\nu}=F^{ij}_{\mu\nu} \qquad \textrm{and} \qquad \mathcal{H}^{(2)5i}_{\mu\nu}=\frac{g}{2}w^{ij}B_{\mu\nu j}
\end{equation}
in which $B_{\mu\nu j}$ are massive two-form fields. However, we are not able to find a consistent set of BPS equations that are compatible with the field equations for non-vanishing massive two-form fields. Therefore, in the following analysis, we will truncate out all the massive two-form fields. Finally, we point out here that the $CSO(p,q,r)$ with $p+q+r=4$ gauge group is not large enough to accomodate $SO(5)$ or $SO(3)\times SO(2)$ subgroups. Consequently, it is not possible to have $AdS_2\times \Sigma^5$ or $AdS_2\times \Sigma^3\times \Sigma^2$ solutions.

\subsection{Solutions with the twists on $\Sigma^2$}
We first look for $AdS_5\times\Sigma^2$ solutions for $\Sigma^2$ being a Riemann surface. The ansatz for the seven-dimensional metric is given in \eqref{YAdS57Dmetric}. We will consider solutions obtained from $SO(2)\times SO(2)$ and $SO(2)$ twists on $\Sigma^2$. The procedure is essentially the same as in the gaugings in $\mathbf{15}$ representation, so we will not give all the details here to avoid a repetition. 

\subsubsection{Solutions with $SO(2)\times SO(2)$ twists}\label{ZAdS5doubleSO(2)section}
\begin{figure}
  \centering
  \begin{subfigure}[b]{0.36\linewidth}
    \includegraphics[width=\linewidth]{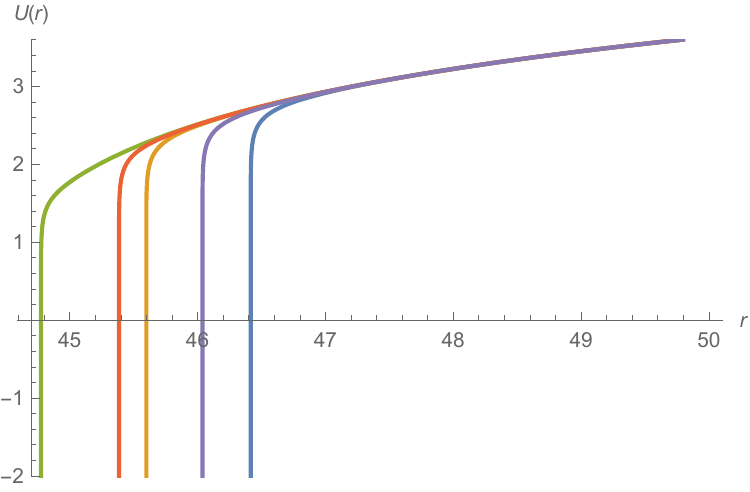}
\caption{$U$ solution}
  \end{subfigure}
  \begin{subfigure}[b]{0.36\linewidth}
    \includegraphics[width=\linewidth]{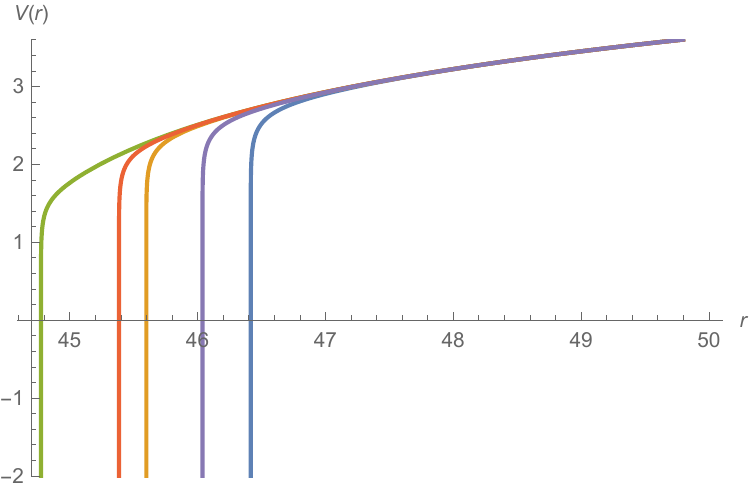}
\caption{$V$ solution}
  \end{subfigure}
  \begin{subfigure}[b]{0.36\linewidth}
    \includegraphics[width=\linewidth]{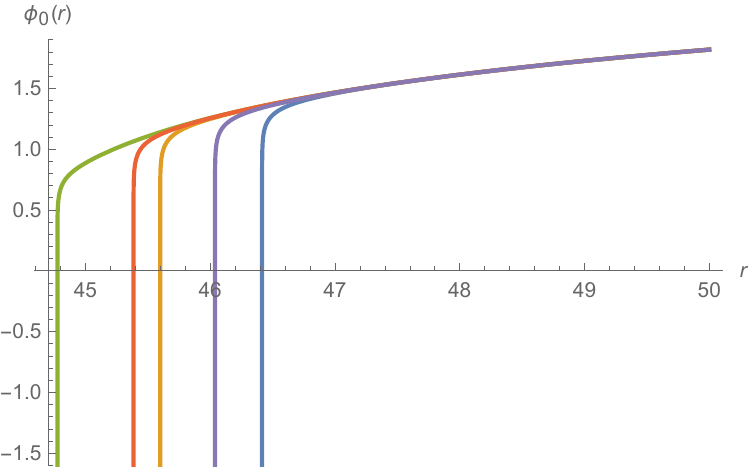}
\caption{$\phi_0$ solution}
  \end{subfigure}
  \begin{subfigure}[b]{0.36\linewidth}
    \includegraphics[width=\linewidth]{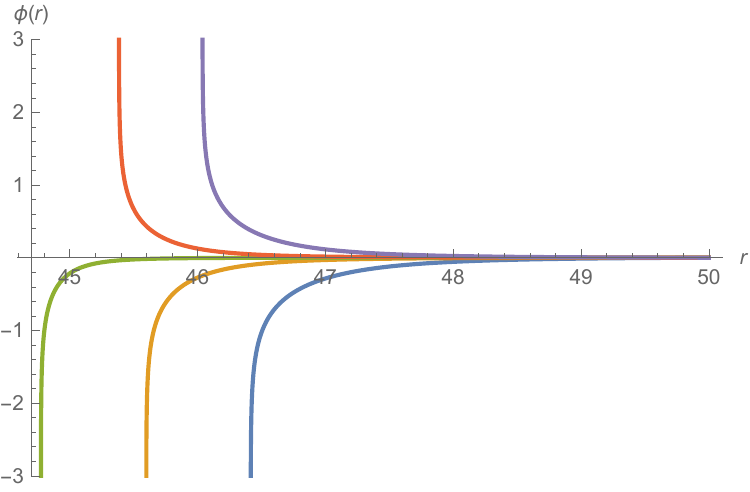}
\caption{$\phi$ solution}
  \end{subfigure}
\caption{Supersymmetric flows interpolating between asymptotically locally flat domain walls and $Mkw_4\times S^2$-sliced curved domain walls for $SO(2)\times SO(2)$ twist in $SO(4)$ gauge group. The blue, orange, green, red and purple curves refer to $p_2=-0.5, -0.03, 0.03, 0.06, 0.25$, respectively.}
\label{Z_AdS5xS2_SO2xSO2_SO4g_flow}
\end{figure}

\begin{figure}
  \centering
  \begin{subfigure}[b]{0.36\linewidth}
    \includegraphics[width=\linewidth]{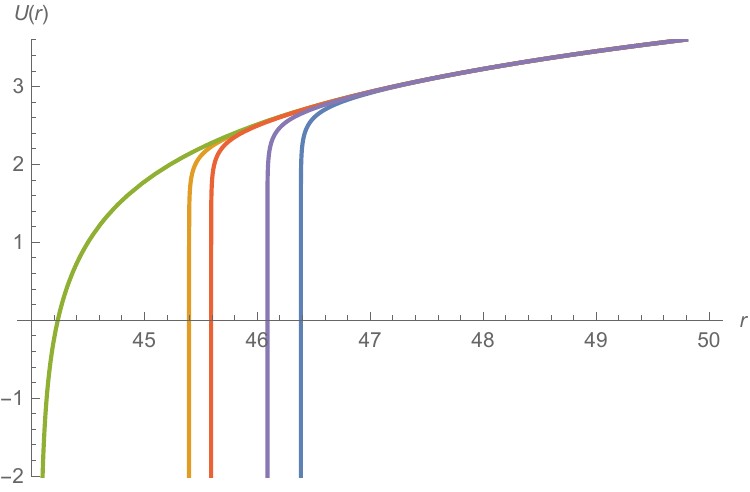}
\caption{$U$ solution}
  \end{subfigure}
  \begin{subfigure}[b]{0.36\linewidth}
    \includegraphics[width=\linewidth]{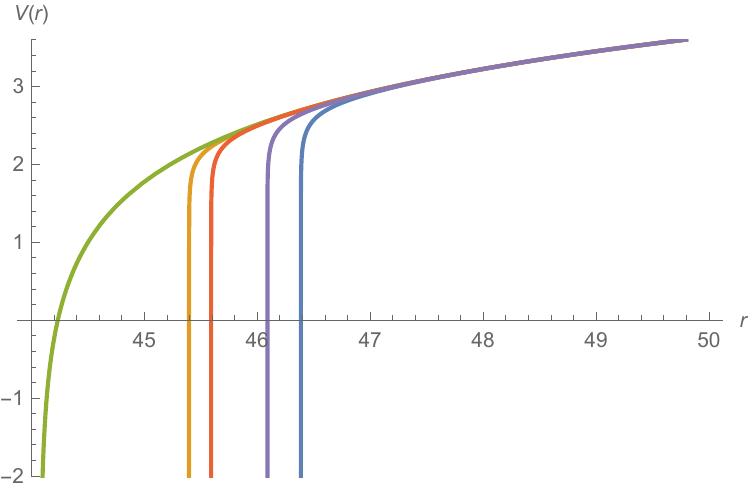}
\caption{$V$ solution}
  \end{subfigure}
  \begin{subfigure}[b]{0.36\linewidth}
    \includegraphics[width=\linewidth]{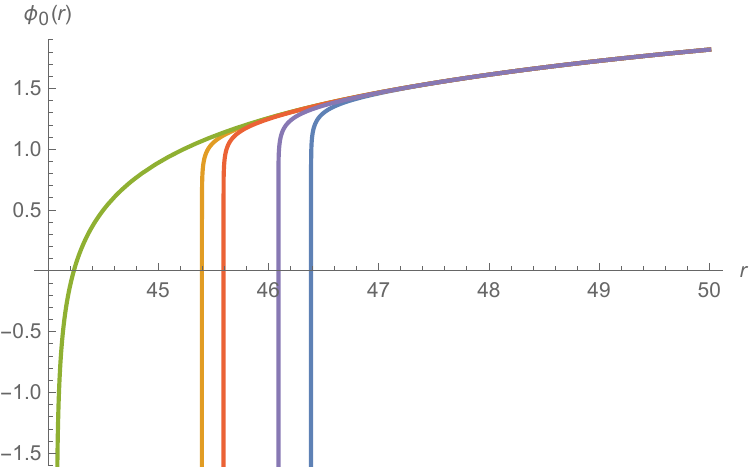}
\caption{$\phi_0$ solution}
  \end{subfigure}
  \begin{subfigure}[b]{0.36\linewidth}
    \includegraphics[width=\linewidth]{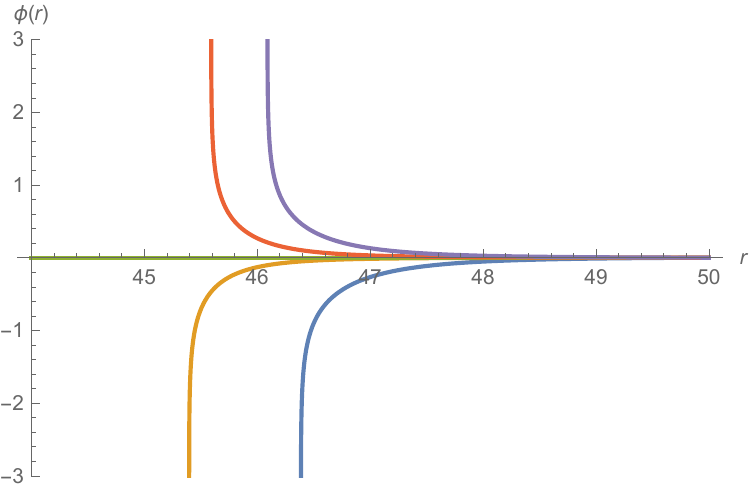}
\caption{$\phi$ solution}
  \end{subfigure}
\caption{Supersymmetric flows interpolating between asymptotically locally flat domain walls and $Mkw_4\times \mathbb{R}^2$-sliced domain walls for $SO(2)\times SO(2)$ twist in $SO(4)$ gauge group. The blue, orange, green, red and purple curves refer to $p_2=-0.5, -0.03, 0, 0.06, 0.25$, respectively.}
\label{Z_AdS5xR2_SO2xSO2_SO4g_flow}
\end{figure}

\begin{figure}
  \centering
  \begin{subfigure}[b]{0.36\linewidth}
    \includegraphics[width=\linewidth]{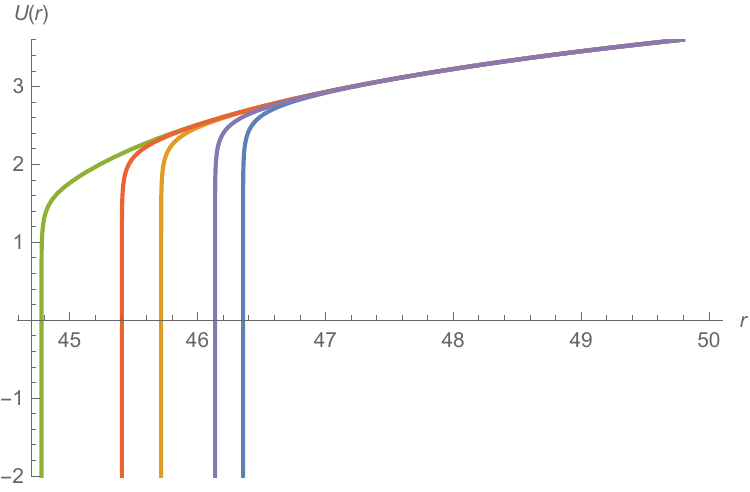}
\caption{$U$ solution}
  \end{subfigure}
  \begin{subfigure}[b]{0.36\linewidth}
    \includegraphics[width=\linewidth]{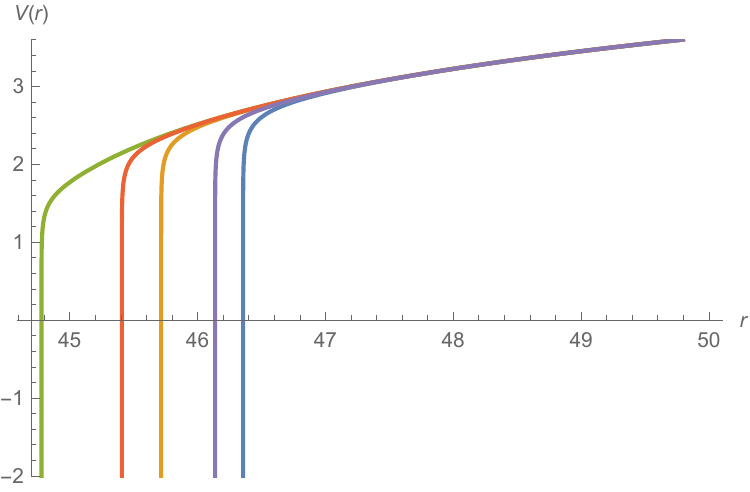}
\caption{$V$ solution}
  \end{subfigure}
  \begin{subfigure}[b]{0.36\linewidth}
    \includegraphics[width=\linewidth]{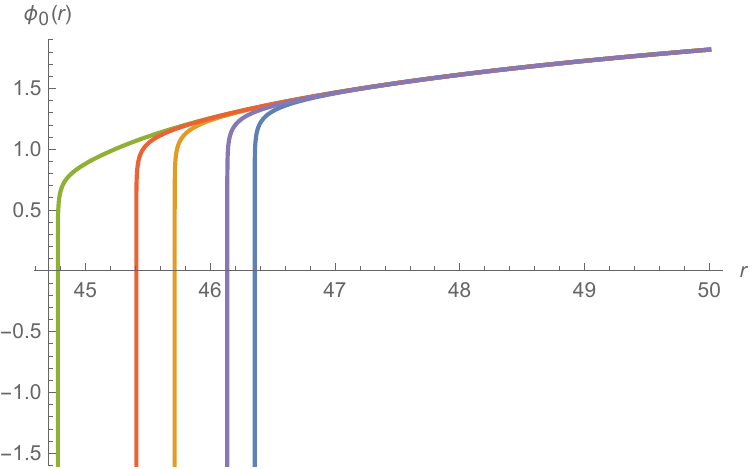}
\caption{$\phi_0$ solution}
  \end{subfigure}
  \begin{subfigure}[b]{0.36\linewidth}
    \includegraphics[width=\linewidth]{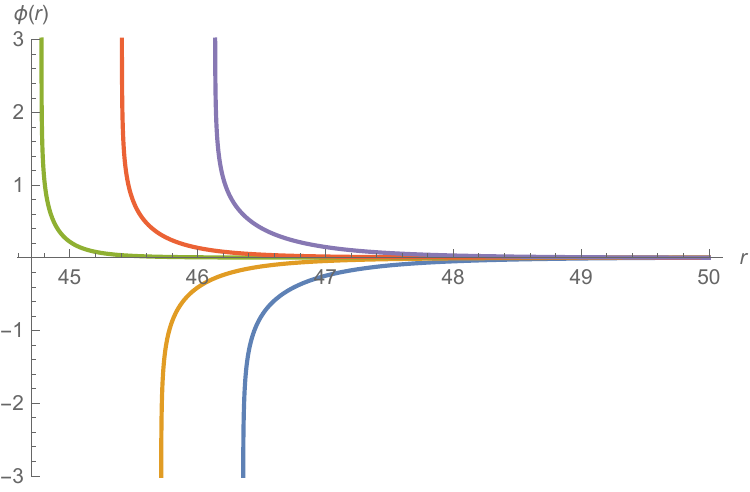}
\caption{$\phi$ solution}
  \end{subfigure}
\caption{Supersymmetric flows interpolating between asymptotically locally flat domain walls and $Mkw_4\times H^2$-sliced curved domain walls for $SO(2)\times SO(2)$ twist in $SO(4)$ gauge group. The blue, orange, green, red and purple curves refer to $p_2=-0.5, -0.12, -0.03, 0, 0.25$, respectively.}
\label{Z_AdS5xH2_SO2xSO2_SO4g_flow}
\end{figure}

\begin{figure}
  \centering
  \begin{subfigure}[b]{0.32\linewidth}
    \includegraphics[width=\linewidth]{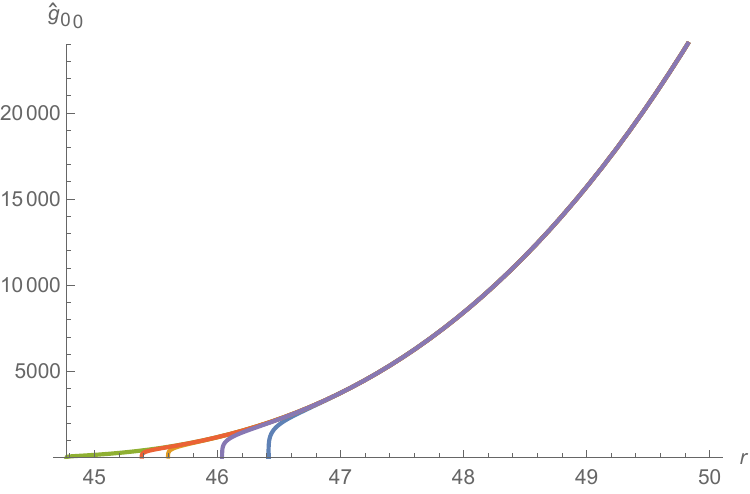}
\caption{Flows with $k=1$}
  \end{subfigure}
  \begin{subfigure}[b]{0.32\linewidth}
    \includegraphics[width=\linewidth]{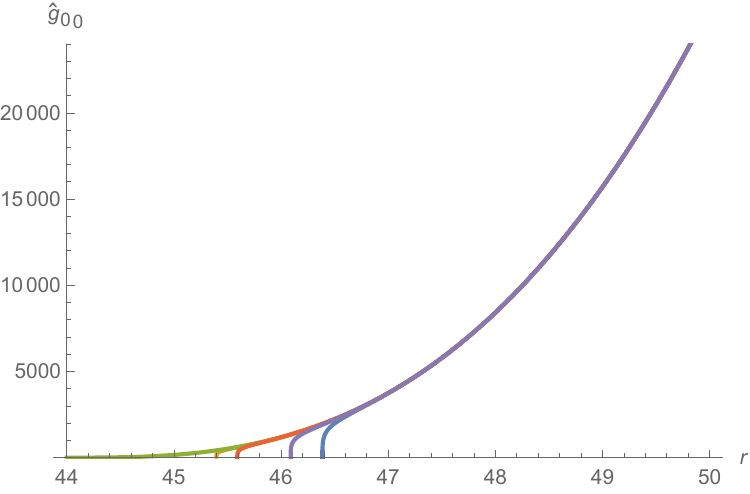}
\caption{Flows with $k=0$}
  \end{subfigure}
  \begin{subfigure}[b]{0.32\linewidth}
    \includegraphics[width=\linewidth]{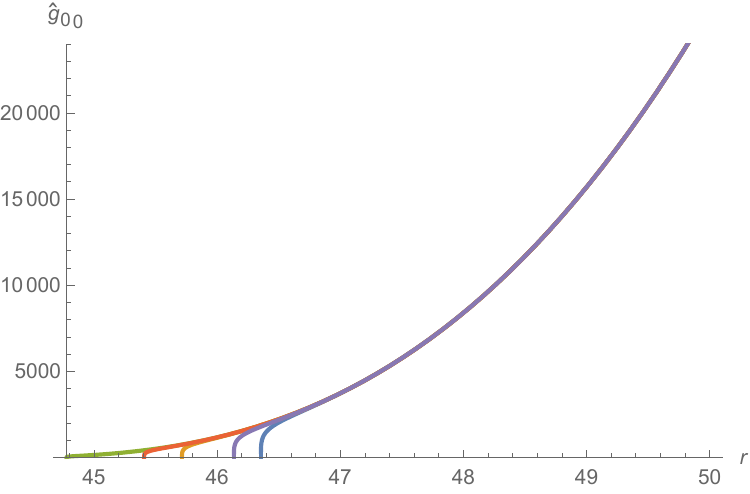}
\caption{Flows with $k=-1$}
  \end{subfigure}
\caption{Profiles of the ten-dimensional metric component $\hat{g}_{00}$ for the supersymmetric flows given in figures \ref{Z_AdS5xS2_SO2xSO2_SO4g_flow}, \ref{Z_AdS5xR2_SO2xSO2_SO4g_flow} and \ref{Z_AdS5xH2_SO2xSO2_SO4g_flow}.}
\label{Z_AdS5xSig2_SO2xSO2_SO4g_g00}
\end{figure}

Gauge groups with an $SO(2)\times SO(2)$ subgroup can be obtained from the embedding tensor of the form
\begin{equation}\label{SO(2)xSO(2)wij}
w^{ij}=\text{diag}(1,1,\sigma,\sigma)
\end{equation}
with the parameter $\sigma=1,-1$ corresponding to $SO(4)$ and $SO(2,2)$ gauge groups, respectively. There is only one $SO(2)\times SO(2)$ singlet scalar from $SL(4)/SO(4)$ coset described by the coset representative 
\begin{equation}\label{ZSO2xSO2coset}
\widetilde{\mathcal{V}}=\text{diag}(e^\phi,e^\phi,e^{-\phi},e^{-\phi}).
\end{equation}
The scalar potential is given by
\begin{equation}\label{ZSO(2)xSO(2)Pot}
\mathbf{V}=-2\sigma e^{-4\phi_0}g^2\, .
\end{equation}
In this case, there is no supersymmetric $AdS_7$ fixed point. The supersymmetric vacuum is given by half-supersymmetric domain walls dual to $N=(2,0)$ non-conformal field theories in six dimensions. 
\\
\indent We now perform the twist by turning on the following $SO(2)\times SO(2)$ gauge fields 
\begin{equation}\label{ZAdS5gaugeAnt}
A^{12}_{(1)}=e^{-V}\frac{p_{2}}{4k}\frac{f'_{k}(\theta)}{f_{k}(\theta)}e^{\hat{\varphi}} \qquad \textrm{and} \qquad A^{34}_{(1)}=e^{-V}\frac{p_{1}}{4k}\frac{f'_{k}(\theta)}{f_{k}(\theta)}e^{\hat{\varphi}}
\end{equation} 
and imposing the projection conditions given in \eqref{SO(2)xSO(2)Projcon} and 
\begin{equation}\label{pureZProj}
\gamma_{\hat{r}}\epsilon^a=-{(\Gamma_{5})^a}_b\epsilon^b
\end{equation}
together with the twist condition \eqref{YSO(2)xSO(2)on1QYM}. 
\\
\indent With all these, we find the following BPS equations
\begin{eqnarray}
U'&\hspace{-0.2cm}=&\hspace{-0.2cm}\frac{g}{5}e^{-2(\phi_0+\phi)}(e^{4\phi}+\sigma)-\frac{1}{10}e^{-2(V-\phi_0)}(e^{-2\phi}p_1+e^{2\phi}p_2),\quad \\
V'&\hspace{-0.2cm}=&\hspace{-0.2cm}\frac{g}{5}e^{-2(\phi_0+\phi)}(e^{4\phi}+\sigma)+\frac{2}{5}e^{-2(V-\phi_0)}(e^{-2\phi}p_1+e^{2\phi}p_2),\quad \\
\phi_0'&\hspace{-0.2cm}=&\hspace{-0.2cm}\frac{g}{10}e^{-2(\phi_0+\phi)}(e^{4\phi}+\sigma)-\frac{1}{20}e^{-2(V-\phi_0)}(e^{-2\phi}p_1+e^{2\phi}p_2),\quad \\
\phi_1'&\hspace{-0.2cm}=&\hspace{-0.2cm}-\frac{g}{2}e^{-2(\phi_0+\phi)}(e^{4\phi}-\sigma)+\frac{1}{4}e^{-2(V-\phi_0)}(e^{-2\phi}p_1-e^{2\phi}p_2).\quad 
\end{eqnarray}
From these equations, there are no fixed point solutions satisfying the conditions $\phi'=\phi_0'=V'=0$ and $U'=\frac{1}{L_{\textrm{AdS}_5}}$. In subsequent analysis, we will consider interpolating solutions between an asymptotically locally flat domain wall and curved domain walls in $SO(4)$ gauge group. Similar solutions can also be found in $SO(2,2)$ gauge group.
\\
\indent For large $V$, the contribution from the gauge fields is highly suppressed. In this limit, we find
\begin{equation}\label{Z_SO2xSO2_asymDW}
\phi\sim\frac{1}{r^5},\qquad \phi_0\sim-\frac{1}{10}\log{\phi}, \qquad U\sim V\sim 2\phi_0
\end{equation}
which implies that $U\sim V\rightarrow \infty$ as $r\rightarrow \infty$. Examples of flow solutions with this asymptotic behavior are given in figures \ref{Z_AdS5xS2_SO2xSO2_SO4g_flow}, \ref{Z_AdS5xR2_SO2xSO2_SO4g_flow} and \ref{Z_AdS5xH2_SO2xSO2_SO4g_flow} for $\Sigma^2=S^2,\mathbb{R}^2,H^2$, respectively. In these solutions, we have set  $g=16$. We note here that the flows to the flat $Mkw_4\times \mathbb{R}^2$-sliced domain walls given in figure \ref{Z_AdS5xR2_SO2xSO2_SO4g_flow} are possible provided that we set $p_2=-p_1$ as required by the twist condition. It should also be pointed out that the green curve in figure \ref{Z_AdS5xR2_SO2xSO2_SO4g_flow} is simply the usual flat domain wall since $k=p_1=p_2=0$. In this case, the solution preserves the full $SO(4)$ gauge symmetry due to the vanishing of the $SO(2)\times SO(2)$ singlet scalar $\phi$. This solution has already been given analytically in \cite{our_7D_DW}. 
\\
\indent As shown in \cite{Henning_Emanuel}, the maximal gauged supergravity in seven dimensions with $CSO(p,q,4-p-q)$ gauge group obtained from the embedding tensor in $\overline{\mathbf{40}}$ representation can be embedded in type IIB theory via a truncation on $H^{p,q}\circ T^{4-p-q}$. For the present discussion, we only need the ten-dimensional metric which, for $SO(p,4-p)$ gauge group, is given by
\begin{equation}
\hat{g}_{\mu\nu}=\kappa^{\frac{3}{4}}\Delta^{\frac{1}{4}}g_{\mu\nu} \label{metric_IIB}
\end{equation}
with
\begin{equation}
\Delta=\mu_i \mu_j\eta^{ik}\eta^{jl}\widetilde{\mc{M}}_{kl}\, .\label{warp_IIB}
\end{equation}
$\eta^{ij}$ is the $SO(p,4-p)$ invariant tensor, and $\mu_i$ are coordinates on $H^{p,q}$ satisfying $\mu_i\mu_j\eta^{ij}=1$. In term of the parametrization \eqref{t=4decompose}, $\kappa$ is identified as follows
\begin{equation}
\kappa=e^{2\phi_0}\, .
\end{equation}
For $CSO(p,q,4-p-q)$ gauge group, we decompose the $SL(4)$ indices $i,j,\ldots$ into $(\hat{i},\tilde{i})$ with $\hat{i}=1,\ldots p+q$ and $\tilde{i}=p+q+1,\ldots, 4$. The ten-dimensional metric and the warped factor are still given by \eqref{metric_IIB} and \eqref{warp_IIB} but with $\eta^{ij}=(\eta^{\hat{i}\hat{j}},\eta^{\tilde{i}\tilde{j}})$ replaced by the $SO(p,q)$ invariant tensor $\eta^{\hat{i}\hat{j}}$ and $\eta^{\tilde{i}\tilde{j}}=0$. In this case, $\mu_{\hat{i}}$ become coordinates on $H^{p,q}$ satisfying $\eta^{\hat{i}\hat{j}}\mu_{\hat{i}}\mu_{\hat{j}}=1$ while $\mu_{\tilde{i}}$ are coordinates on $T^{4-p-q}$.  
\\
\indent For the present case of $SO(4)$ gauge group, we simply have $\eta^{ij}=\delta^{ij}$ for $i,j=1,2,3,4$. The behavior of the time component of the ten-dimensional metric $\hat{g}_{00}$ for the flow solutions in figures \ref{Z_AdS5xS2_SO2xSO2_SO4g_flow}, \ref{Z_AdS5xR2_SO2xSO2_SO4g_flow} and \ref{Z_AdS5xH2_SO2xSO2_SO4g_flow} is shown in figure \ref{Z_AdS5xSig2_SO2xSO2_SO4g_g00}. Near the IR singularities, we find $\hat{g}_{00}\rightarrow 0$, so the singularities are physically acceptable. 

\subsubsection{Solutions with $SO(2)$ twists}
We then move to another twist on $\Sigma^2$ by turning on only an $SO(2)$ gauge field. This can be achieved from the $SO(2)\times SO(2)$ gauge fields given in \eqref{ZAdS5gaugeAnt} by setting $p_2=0$ and $p_1=p$. In this case, the $SL(4)/SO(4)$ coset representative is given by 
\begin{equation}
\widetilde{\mathcal{V}}=e^{\phi_1\widetilde{\mathcal{Y}}_1+\phi_2\widetilde{\mathcal{Y}}_2
+\phi_3\widetilde{\mathcal{Y}}_3}\label{ZSO(2)Ys}
\end{equation}
in which $\widetilde{\mathcal{Y}}_1$, $\widetilde{\mathcal{Y}}_2$, and $\widetilde{\mathcal{Y}}_3$ are non-compact generators commuting with the $SO(2)$ symmetry generated by $X_{12}$. The explicit form of these generators is given by
\begin{equation}
\widetilde{\mathcal{Y}}_1=e_{1,1}+e_{2,2}-e_{3,3}-e_{4,4},\qquad
\widetilde{\mathcal{Y}}_2=e_{3,4}+e_{4,3},\qquad
\widetilde{\mathcal{Y}}_3=e_{3,3}-e_{4,4}\, .
\end{equation}
The embedding tensor is taken to be
\begin{equation}
w^{ij}=\textrm{diag}(1,1,\sigma,\rho)\label{ZSO2_theta}
\end{equation}
corresponding to six gauge groups with an $SO(2)$ subgroup. These are given by $SO(4)$ ($\rho=\sigma=1$), $SO(3,1)$ ($-\rho=\sigma=1$), $SO(2,2)$ ($\rho=\sigma=-1$), $SO(3,0,1)$ ($\rho=0$, $\sigma=1$), $SO(2,1,1)$ ($\rho=0$, $\sigma=-1$), and $CSO(2,0,2)$ ($\rho=\sigma=0$).
\\
\indent Imposing the twist condition \eqref{GenQYM} and the projector \eqref{pureZProj} together with
\begin{equation}\label{SO(2)Projcon}
\gamma^{\hat{\varphi}\hat{\theta}}\epsilon^a=-{(\Gamma_{12})^a}_b\epsilon^b,
\end{equation}
we obtain the BPS equations
\begin{eqnarray}
U'&\hspace{-0.2cm}=&\hspace{-0.2cm}\frac{g}{10}e^{-2(\phi_0+\phi_1)}\left[2e^{4\phi_1}-(\rho-\sigma)\sinh{2\phi_3}+(\rho+\sigma)\cosh{2\phi_3}\cosh{2\phi_2}\right]\nonumber\\&&\hspace{-0.2cm}-\frac{1}{10}pe^{-2(V-\phi_0+\phi_1)},\\
V'&\hspace{-0.2cm}=&\hspace{-0.2cm}\frac{g}{10}e^{-2(\phi_0+\phi_1)}\left[2e^{4\phi_1}-(\rho-\sigma)\sinh{2\phi_3}+(\rho+\sigma)\cosh{2\phi_3}\cosh{2\phi_2}\right]\nonumber\\&&\hspace{-0.2cm}+\frac{2}{5}pe^{-2(V-\phi_0+\phi_1)},\\
\phi_0'&\hspace{-0.2cm}=&\hspace{-0.2cm}\frac{g}{20}e^{-2(\phi_0+\phi_1)}\left[2e^{4\phi_1}-(\rho-\sigma)\sinh{2\phi_3}+(\rho+\sigma)\cosh{2\phi_2}\cosh{2\phi_3}\right]\nonumber\\&&\hspace{-0.2cm}-\frac{1}{20}pe^{-2(V-\phi_0+\phi_1)},\\
\phi_1'&\hspace{-0.2cm}=&\hspace{-0.2cm}-\frac{g}{4}e^{-2(\phi_0+\phi_1)}\left[2e^{4\phi_1}+(\rho-\sigma)\sinh{2\phi_3}-(\rho+\sigma)\cosh{2\phi_2}\cosh{2\phi_3}\right]\nonumber\\&&\hspace{-0.2cm}+\frac{1}{4}pe^{-2(V-\phi_0+\phi_1)},\\
\phi_2'&\hspace{-0.2cm}=&\hspace{-0.2cm}-\frac{g}{2}e^{-2(\phi_0+\phi_1)}(\rho+\sigma)\sinh{2\phi_2}\text{\ sech\ }{2\phi_3},\\
\phi_3'&\hspace{-0.2cm}=&\hspace{-0.2cm}\frac{g}{2}e^{-2(\phi_0+\phi_1)}\left[(\rho-\sigma)\cosh{2\phi_3}-(\rho+\sigma)\cosh{2\phi_2}\sinh{2\phi_3}\right].
\end{eqnarray}
As in the case of $SO(2)\times SO(2)$ twist, there do not exist any $AdS_5\times \Sigma^2$ fixed points. The numerical analysis for finding flow solutions interpolating between locally asymptotically flat domain walls and curved domain walls can also be obtained as in the previous case.

\subsection{Solutions with the twists on $\Sigma^3$}
In this section, we repeat the same analysis for solutions with the twists on $\Sigma^3$. We will consider two different twists by turning on $SO(3)$ and $SO(3)_+$ gauge fields. 

\subsubsection{Solutions with $SO(3)$ twists}
\begin{figure}
  \centering
  \begin{subfigure}[b]{0.36\linewidth}
    \includegraphics[width=\linewidth]{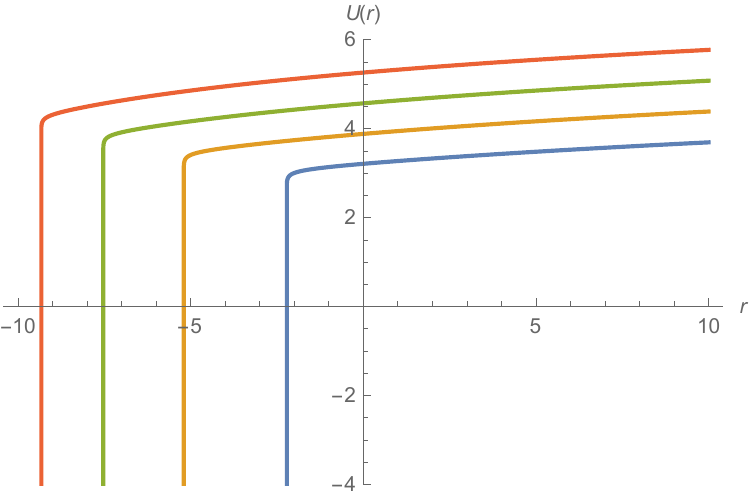}
\caption{$U$ solution}
  \end{subfigure}
  \begin{subfigure}[b]{0.36\linewidth}
    \includegraphics[width=\linewidth]{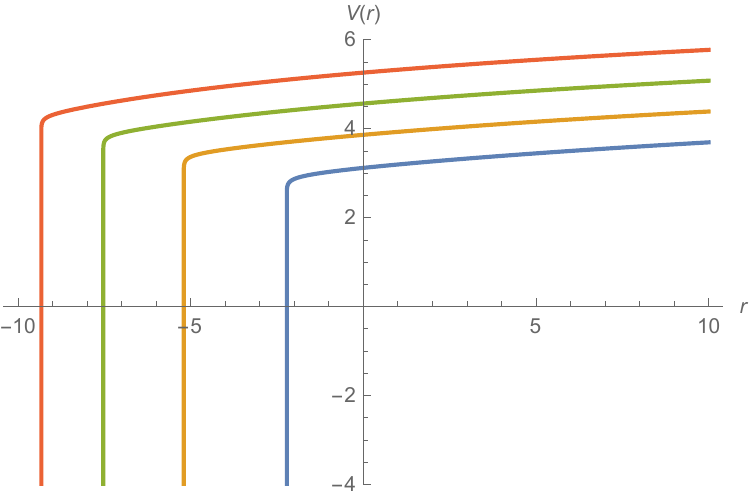}
\caption{$V$ solution}
  \end{subfigure}
  \begin{subfigure}[b]{0.36\linewidth}
    \includegraphics[width=\linewidth]{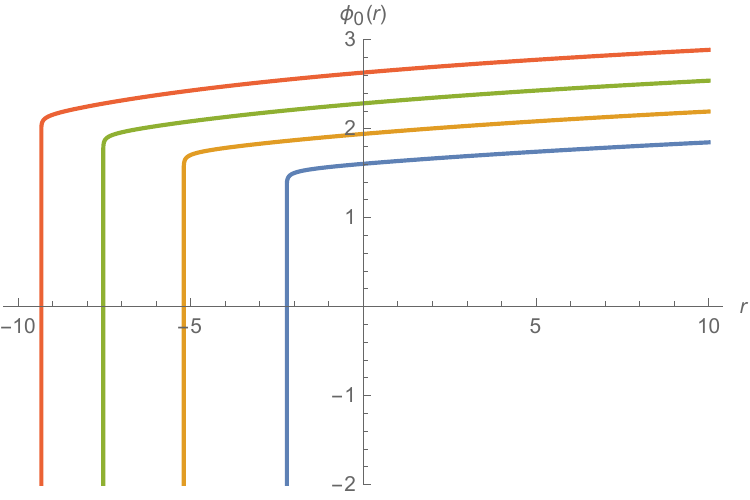}
\caption{$\phi_0$ solution}
  \end{subfigure}
  \begin{subfigure}[b]{0.36\linewidth}
    \includegraphics[width=\linewidth]{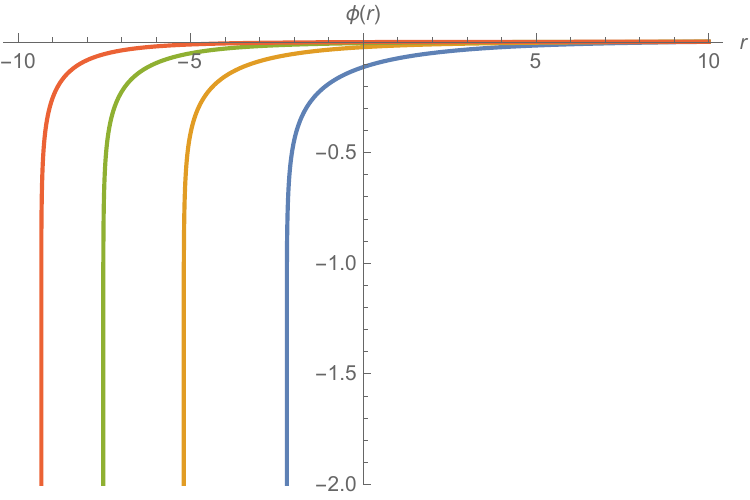}
\caption{$\phi$ solution}
  \end{subfigure}
\caption{Supersymmetric flows interpolating between asymptotically locally flat domain walls and $Mkw_3\times S^3$-sliced curved domain walls for $SO(3)$ twist in $SO(4)$ gauge group. The blue, orange, green and red curves refer to $g=4,8,16,32$, respectively.}
\label{Z_AdS4xS3_SO3_SO4g_flow}
\end{figure}

\begin{figure}
  \centering
  \begin{subfigure}[b]{0.36\linewidth}
    \includegraphics[width=\linewidth]{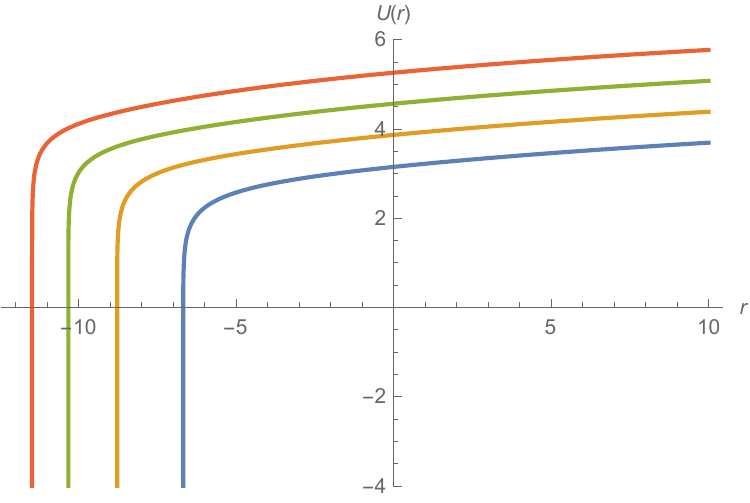}
\caption{$U$ solution}
  \end{subfigure}
  \begin{subfigure}[b]{0.36\linewidth}
    \includegraphics[width=\linewidth]{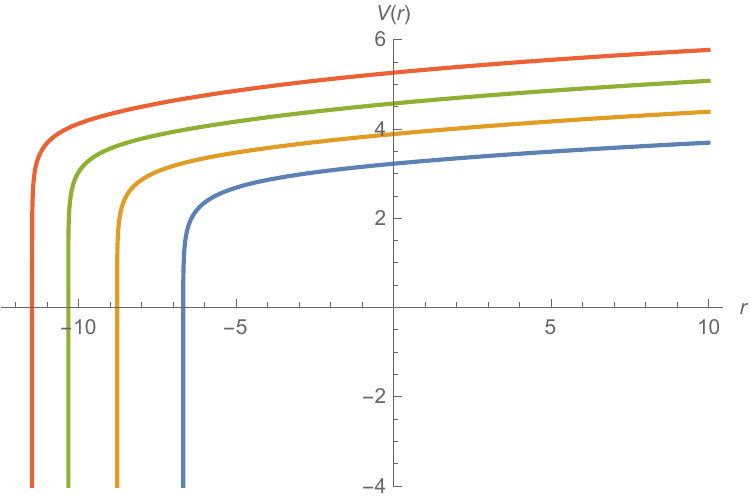}
\caption{$V$ solution}
  \end{subfigure}
  \begin{subfigure}[b]{0.36\linewidth}
    \includegraphics[width=\linewidth]{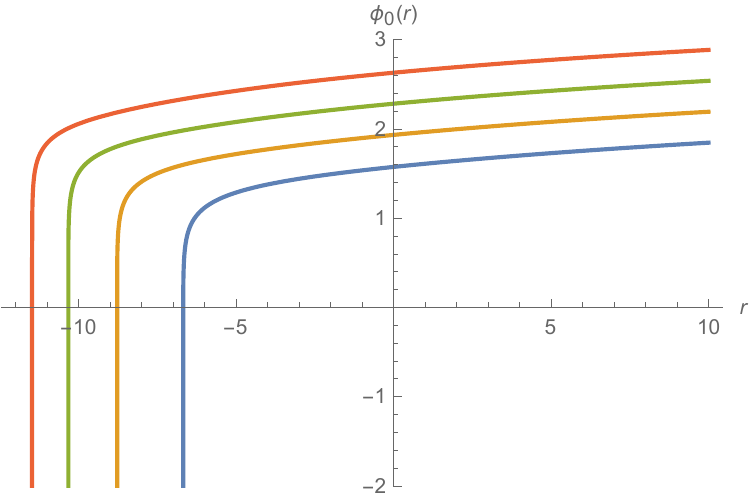}
\caption{$\phi_0$ solution}
  \end{subfigure}
  \begin{subfigure}[b]{0.36\linewidth}
    \includegraphics[width=\linewidth]{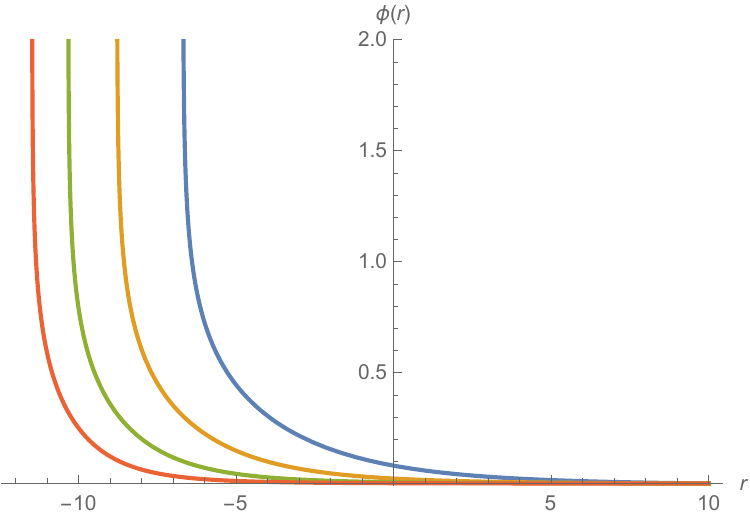}
\caption{$\phi$ solution}
  \end{subfigure}
\caption{Supersymmetric flows interpolating between asymptotically locally flat domain walls and $Mkw_3\times H^3$-sliced curved domain walls for $SO(3)$ twist in $SO(4)$ gauge group. The blue, orange, green and red curves refer to $g=4,8,16,32$, respectively.}
\label{Z_AdS4xH3_SO3_SO4g_flow}
\end{figure}

\begin{figure}
  \centering
  \begin{subfigure}[b]{0.45\linewidth}
    \includegraphics[width=\linewidth]{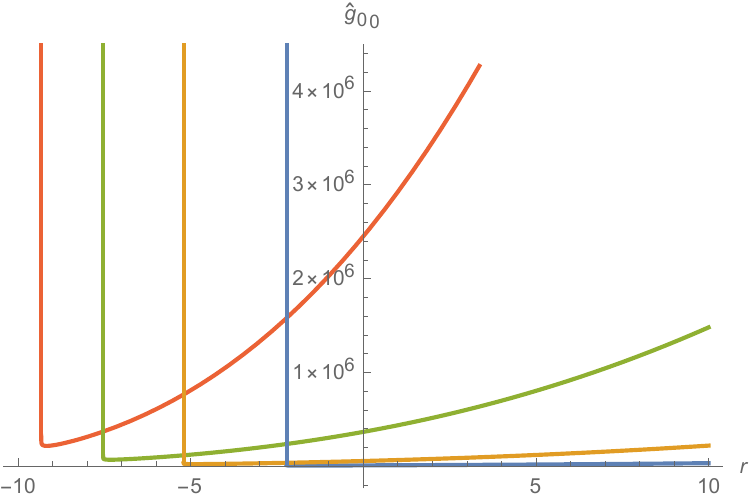}
\caption{$k=1$}
  \end{subfigure}
  \begin{subfigure}[b]{0.45\linewidth}
    \includegraphics[width=\linewidth]{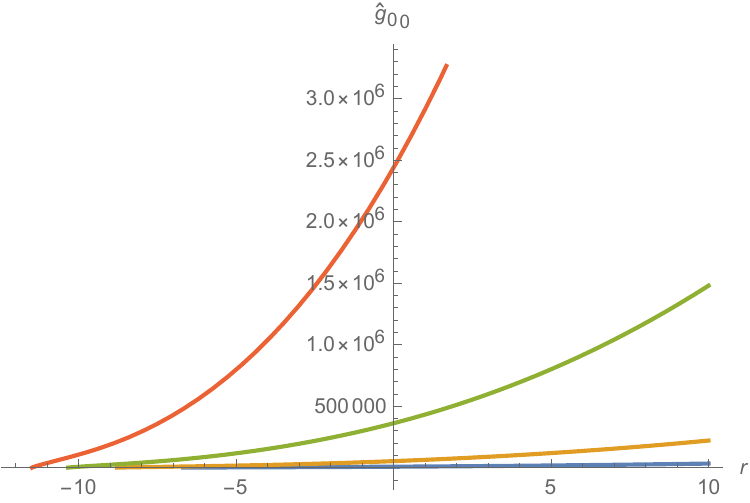}
\caption{$k=-1$}
  \end{subfigure}
\caption{Profiles of the ten-dimensional metric component $\hat{g}_{00}$ for the supersymmetric flows given in figures \ref{Z_AdS4xS3_SO3_SO4g_flow} and \ref{Z_AdS4xH3_SO3_SO4g_flow}.}
\label{Z_AdS4xSig3_SO3_SO4g_g00}
\end{figure}

In this case, the gauge groups with an $SO(3)$ subgroup are described by the embedding tensor 
\begin{equation}
w^{ij}=\textrm{diag}(1,1,1,\rho)\label{Z_SO3_embedding}
\end{equation}
with $\rho=-1,0,1$ corresponding to $SO(3,1)$, $CSO(3,0,1)$ and $SO(4)$ gauge groups, respectively.
\\
\indent There is one $SO(3)$ singlet scalar from the $SL(4)/SO(4)$ coset with the coset representative given by
\begin{equation}\label{ZSO(3)coset}
\widetilde{\mathcal{V}}=\text{diag}(e^\phi,e^\phi,e^\phi,e^{-3\phi})
\end{equation}
leading to the scalar potential
\begin{equation}\label{ZSO(3)Pot}
\mathbf{V}=-\frac{g^2}{4}e^{-4(\phi_0+3\phi)}(3e^{16\phi}+6\rho e^{8\phi}-\rho^2)\, .
\end{equation}
\indent To perform the twist, we turn on the $SO(3)$ gauge fields
\begin{eqnarray}
A^{34}_{(1)}&=&e^{-V}\frac{p}{4k}\frac{f'_{k}(\psi)}{f_{k}(\psi)}e^{\hat{\theta}},\nonumber  \\ 
A^{42}_{(1)}&=&e^{-V}\frac{p}{4k}\frac{f'_{k}(\psi)}{f_{k}(\psi)}e^{\hat{\varphi}}, \nonumber  \\ 
A^{14}_{(1)}&=&e^{-V}\frac{p}{4k}\frac{\cot(\theta)}{f_{k}(\psi)}e^{\hat{\varphi}}\label{ZSO(3)gaugeAnt}
\end{eqnarray}
and impose the projectors \eqref{SO(3)Projcon} on the Killing spinors together with the twist condition \eqref{GenQYM}. With all these and the $\gamma_r$ projector \eqref{pureZProj}, the resulting BPS equations read
\begin{eqnarray}
U'&=&\frac{g}{10}e^{-2(\phi_0+3\phi)}(3e^{8\phi}+\rho)-\frac{3}{10}e^{-2(V-\phi_0+\phi)}p,\\
V'&=&\frac{g}{10}e^{-2(\phi_0+3\phi)}(3e^{8\phi}+\rho)+\frac{7}{10}e^{-2(V-\phi_0+\phi)}p,\\
\phi_0'&=&\frac{g}{20}e^{-2(\phi_0+3\phi)}(3e^{8\phi}+\rho)-\frac{3}{20}e^{-2(V-\phi_0+\phi)}p,\\
\phi'&=&-\frac{g}{4}e^{-2(\phi_0+3\phi)}(e^{8\phi}-\rho)+\frac{1}{4}e^{-2(V-\phi_0+\phi)}p.
\end{eqnarray}
As in the previous case, there do not exist any $AdS_4$ fixed points for these equations. We then look for solutions interpolating between asymptotically locally flat domain walls and curved domain walls.
\\
\indent For $CSO(3,0,1)$ gauge group with $\rho=0$, these equations can be analytically solved. First of all, the BPS equations give 
\begin{equation}
U=2\phi_0\, .
\end{equation}
We have neglected an additive integration constant for $U$ which can be absorbed by rescaling the coordinates on $Mkw_3$. With $\rho=0$, we find that $\phi_0'+\frac{3}{5}\phi'=0$ which gives
\begin{equation}
\phi_0=-\frac{3}{5}\phi+C_0
\end{equation}
with an integration constant $C_0$. 
\\
\indent Taking a linear combination $V'+\frac{6}{5}\phi'$ and changing to a new radial coordinate $\tilde{r}$ given by $\frac{d\tilde{r}}{dr}=e^{-\frac{4}{5}\phi}$, we find
\begin{equation}
V=\frac{1}{2}\ln (2p\tilde{r}+C)-\frac{6}{5}\phi\, .
\end{equation}
The integration constant $C$ can also be set to zero by shifting the coordinate $\tilde{r}$. With all these results, the equation for $\phi'$ gives
\begin{equation}
\phi=-\frac{1}{4}\ln \left[\frac{2}{3}g\tilde{r}-\frac{\tilde{C}}{p\sqrt{p\tilde{r}}}\right]
\end{equation}
in which we have set $C=0$ for simplicity, and $\tilde{C}$ is another integration constant. 
\\
\indent As $\tilde{r}\rightarrow 0$, we find that the above solution becomes a locally flat domain wall with $U\sim V\rightarrow \infty$. The asymptotic behavior is given by 
\begin{equation}
\phi\sim \frac{1}{8}\ln \tilde{r},\qquad \phi_0\sim -\frac{3}{40}\ln \tilde{r},\qquad U\sim V\sim -\frac{3}{20}\ln \tilde{r}\, .
\end{equation}
For $\tilde{r}\rightarrow \infty$, we find
\begin{eqnarray}
& &\phi\sim -\frac{1}{4}\ln \tilde{r},\qquad \phi_0\sim \frac{3}{20}\ln \tilde{r},\nonumber \\
& &V\sim \frac{4}{5}\ln \tilde{r},\qquad U\sim \frac{3}{10}\ln \tilde{r}\, .
\end{eqnarray} 
Computing the type IIB metric, we obtain
\begin{equation}
\hat{g}_{00}\sim e^{2U+\frac{3}{2}\phi_0+\frac{1}{2}\phi}\sim \tilde{r}^{\frac{7}{10}}\rightarrow \infty,
\end{equation}
as $\tilde{r}\rightarrow \infty$, which implies that the IR singularity is unphysical.
\\
\indent For $\rho\neq 0$, we can only partially solve the BPS equations. As in the $\rho=0$ case, the BPS equations give $U=2\phi_0$. Taking a linear combination $\phi_0'+\frac{3}{5}\phi'$ and defining a new coordinate $\tilde{r}$ via $\frac{d\tilde{r}}{dr}=e^{-\frac{24}{5}\phi}$, we find
\begin{equation}
\phi_0=\frac{1}{2}\ln\left[\frac{2}{5}(g\rho \tilde{r}+C)\right]-\frac{3}{5}\phi\, .
\end{equation}
The complete solutions can be obtained numerically. As $r\rightarrow \infty$, we find
\begin{equation}\label{Z_SO3_asymDW}
\phi\sim\frac{1}{r^5},\qquad \phi_0\sim-\frac{1}{10}\log{\phi}, \qquad U\sim V\sim 2\phi_0\, . 
\end{equation}
Examples of these solutions for $SO(4)$ gauge group are given in figures \ref{Z_AdS4xS3_SO3_SO4g_flow} and \ref{Z_AdS4xH3_SO3_SO4g_flow} for different values of $g$. The behavior of the ten-dimensional metric $\hat{g}_{00}$ for these flow solutions is shown in figure \ref{Z_AdS4xSig3_SO3_SO4g_g00} from which only the singularities of $Mkw_3\times H^3$-sliced domain walls are physical. For $Mkw_3\times \mathbb{R}^3$-sliced domain walls with $k=0$, the twist condition gives $p=0$ resulting in the usual flat domain walls.  

\subsubsection{Solutions with $SO(3)_+$ twists}\label{ZsdSO(3)Sec}
We now consider another twist by turning on the gauge fields for self-dual $SO(3)_+$ gauge symmetry. Only the scalar field $\phi_0$ is $SO(3)_+$ singlet, so we simply have $\widetilde{\mc{M}}_{ij}=\delta_{ij}$. Furthermore, we consider only $SO(4)$ gauge group since this is the only gauge group that contains the $SO(3)_+$ subgroup. 
\\
\indent The $SO(3)_+$ gauge fileds are given by
\begin{eqnarray}
A^{12}_{(1)}&=&A^{34}_{(1)}=e^{-V}\frac{p}{8k}\frac{f'_{k}(\psi)}{f_{k}(\psi)}e^{\hat{\theta}},\nonumber  \\ 
A^{13}_{(1)}&=&A^{42}_{(1)}=e^{-V}\frac{p}{8k}\frac{f'_{k}(\psi)}{f_{k}(\psi)}e^{\hat{\varphi}}, \nonumber \\ 
A^{23}_{(1)}&=&A^{14}_{(1)}=e^{-V}\frac{p}{8k}\frac{\cot(\theta)}{f_{k}(\psi)}e^{\hat{\varphi}}\, .\label{ZSO(3)xSO(3)gaugeAnt}
\end{eqnarray}
With the projectors \eqref{SO(3)Projcon}, \eqref{GammaSDProj} and \eqref{pureZProj} together with the twist condition \eqref{GenQYM}, the resulting BPS equations are given by
\begin{eqnarray}
U'&=&\frac{2g}{5}e^{-2\phi_0}-\frac{3}{10}e^{-2(V-\phi_0)}p,\\
V'&=&\frac{2g}{5}e^{-2\phi_0}+\frac{7}{10}e^{-2(V-\phi_0)}p,\\
\phi_0'&=&\frac{g}{5}e^{-2\phi_0}-\frac{3}{20}e^{-2(V-\phi_0)}p\, .
\end{eqnarray}
As in the case of $SO(3)$ twist, we do not find any $AdS_4$ fixed points from these equations.
\\
\indent From the BPS equations, we again find that $U=2\phi_0$. By taking a linear combination $V'-2\phi_0'$ and defining a new radial coordinate $\tilde{r}$ by $\frac{d\tilde{r}}{dr}=e^{-V}$, we obtain
\begin{equation}
V=\ln (p\tilde{r}+C)+2\phi_0\, .\label{V_sol1}
\end{equation}
Using $\phi_0$ from \eqref{V_sol1} in equation for $V'$, we find, after changing to the coordinate $\tilde{r}$,
\begin{equation}
V=\frac{g}{5p}(p\tilde{r}+C)^2+\frac{7}{10}\ln(p\tilde{r}+C).
\end{equation}
\indent As $\tilde{r}\rightarrow \infty$, we find, with $C$ set to zero by shifting $\tilde{r}$,
\begin{equation}
U\sim V\sim \frac{1}{5}gp\tilde{r}^2\qquad \textrm{and}\qquad \phi_0\sim \frac{1}{10}gp\tilde{r}^2
\end{equation}
which is identified with a domain wall solution given in \cite{our_7D_DW}. On the other hand, as $\tilde{r}\rightarrow 0$, the solution becomes
\begin{equation}
U\sim -\frac{3}{5}\ln (p\tilde{r}),\qquad V\sim \frac{7}{10}\ln(p\tilde{r}),\qquad \phi_0\sim -\frac{3}{10}\ln(p\tilde{r}).
\end{equation}
This singularity is unphysical since the ten-dimensional metric gives
\begin{equation}
\hat{g}_{00}\sim e^{2U+\frac{3}{2}\phi_0}\rightarrow \infty\, .
\end{equation}
\indent We note that this solution is the same as that given in section \ref{YSO(3)sdsection} for $CSO(4,0,1)$ gauge group. The two gauged supergravities can be obtained from truncations of type IIB and type IIA theories on $S^3$, respectively. In fact, there is a duality between these solutions as pointed out in \cite{Henning_Emanuel}.  

\subsection{Solutions with the twists on $\Sigma^4$}
We finally look for supersymmetric solutions obtained from the twists on a four-manifold $\Sigma^4$. As in the case of gaugings in $\mathbf{15}$ representation, we will consider two types of $\Sigma^4$ in terms of a product of two Riemann surfaces $\Sigma^2\times \Sigma^2$ and a Kahler four-cycle $K^4$.

\subsubsection{Solutions with $SO(2)\times SO(2)$  twists on $\Sigma^2\times \Sigma^2$}\label{firstZSO(2)xSO(2)section}
\begin{figure}
  \centering
  \begin{subfigure}[b]{0.32\linewidth}
    \includegraphics[width=\linewidth]{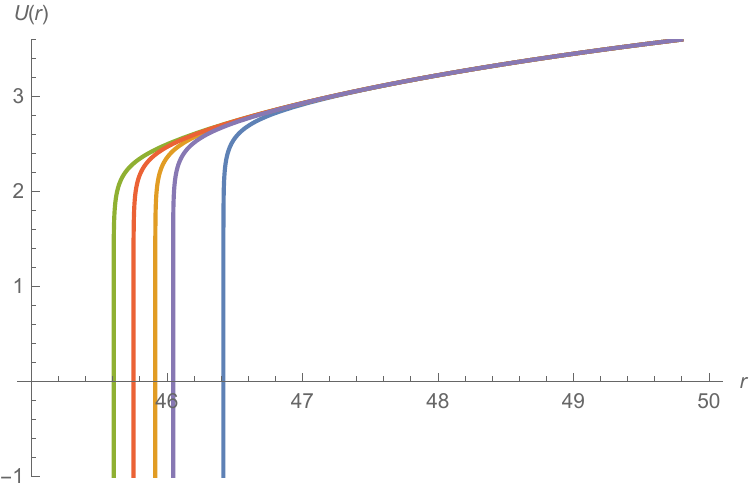}
\caption{$U$ solution}
  \end{subfigure}
  \begin{subfigure}[b]{0.32\linewidth}
    \includegraphics[width=\linewidth]{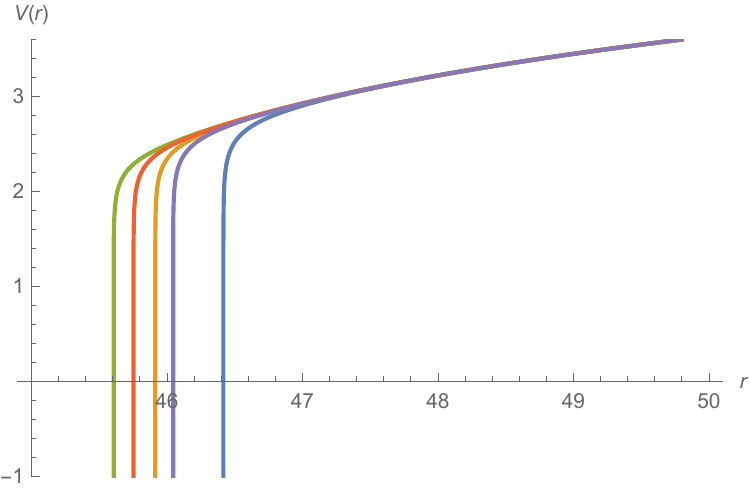}
\caption{$V$ solution}
  \end{subfigure}
  \begin{subfigure}[b]{0.32\linewidth}
    \includegraphics[width=\linewidth]{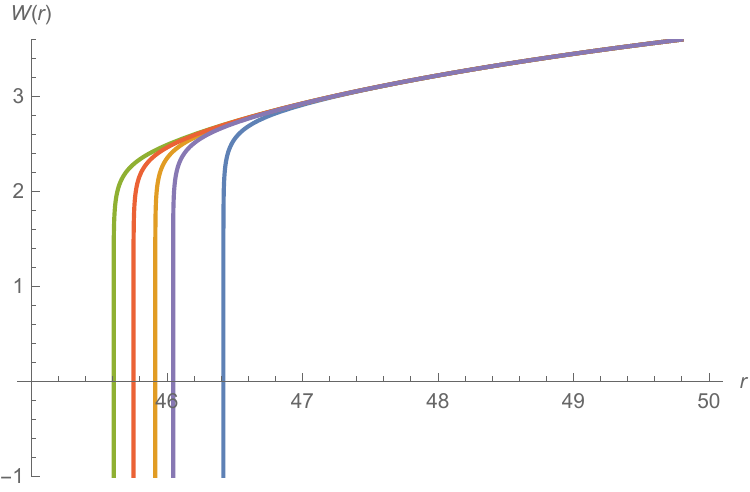}
\caption{$W$ solution}
  \end{subfigure}
  \begin{subfigure}[b]{0.32\linewidth}
    \includegraphics[width=\linewidth]{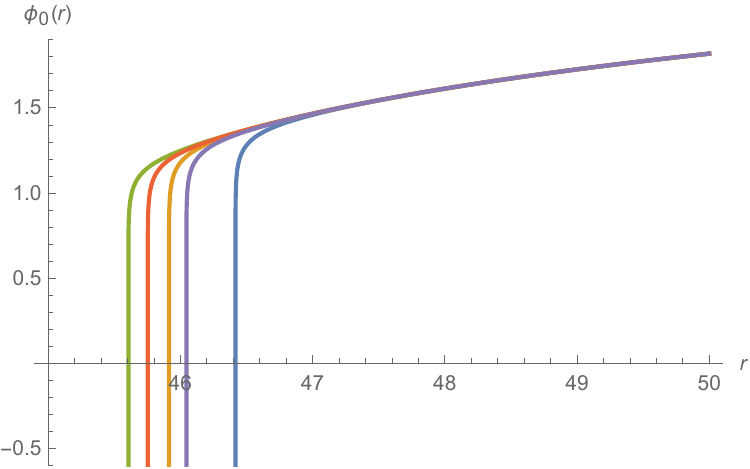}
\caption{$\phi_0$ solution}
  \end{subfigure}
  \begin{subfigure}[b]{0.32\linewidth}
    \includegraphics[width=\linewidth]{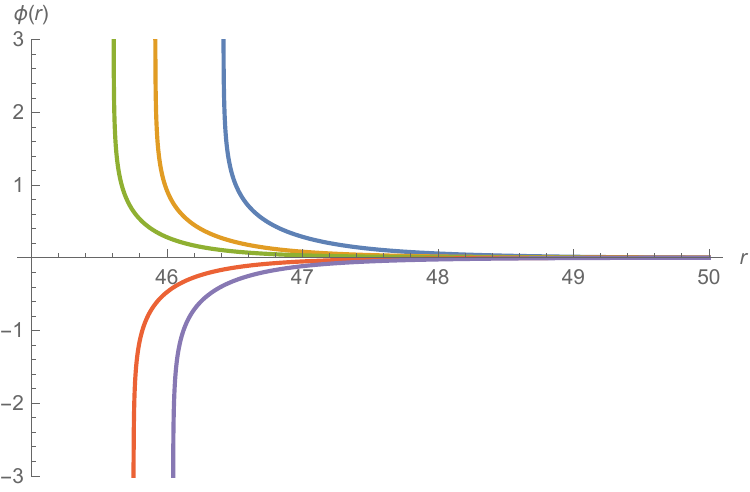}
\caption{$\phi$ solution}
  \end{subfigure}
\caption{Supersymmetric flows interpolating between asymptotically locally flat domain walls and $Mkw_2\times S^2\times S^2$-sliced curved domain walls in $SO(4)$ gauge group. The blue, orange, green, red and purple curves refer to $p_{21}=-0.5, -0.12, 0, 0.12, 0.25$, respectively.}
\label{Z_AdS3xS2xS2_SO2xSO2_SO4g_flow}
\end{figure}

\begin{figure}
  \centering
  \begin{subfigure}[b]{0.32\linewidth}
    \includegraphics[width=\linewidth]{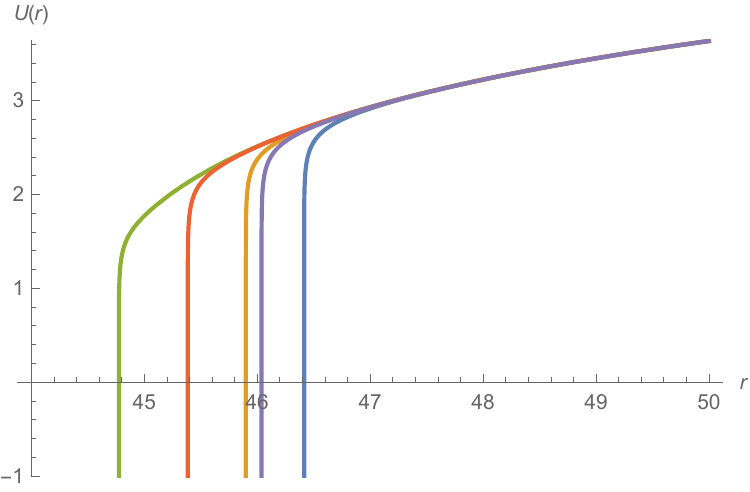}
\caption{$U$ solution}
  \end{subfigure}
  \begin{subfigure}[b]{0.32\linewidth}
    \includegraphics[width=\linewidth]{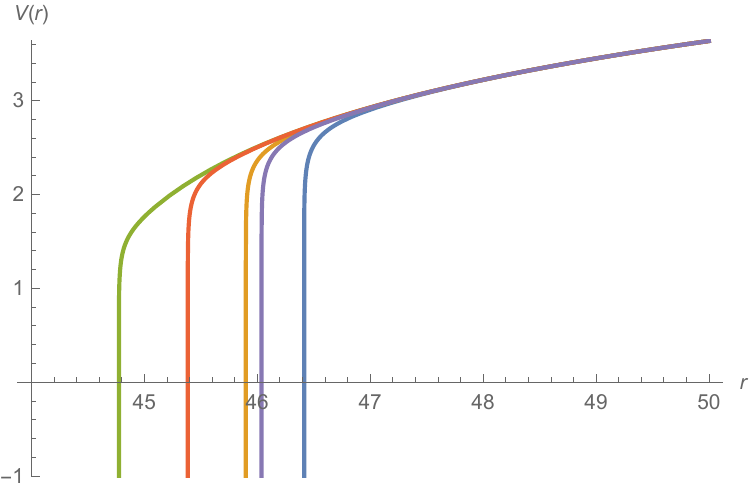}
\caption{$V$ solution}
  \end{subfigure}
  \begin{subfigure}[b]{0.32\linewidth}
    \includegraphics[width=\linewidth]{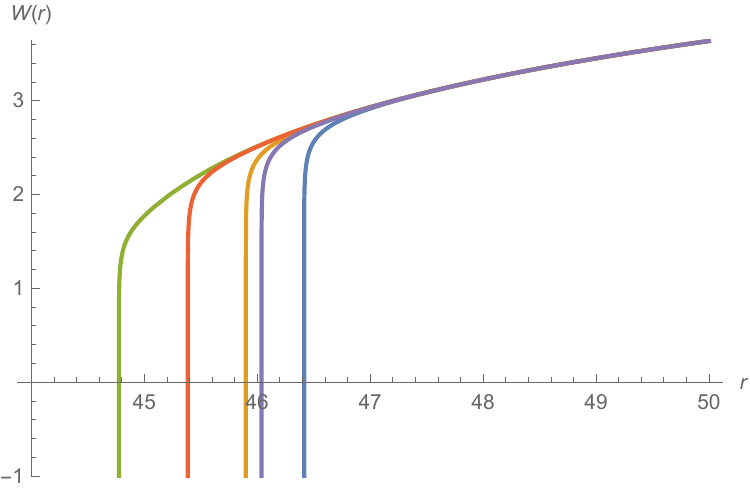}
\caption{$W$ solution}
  \end{subfigure}
  \begin{subfigure}[b]{0.32\linewidth}
    \includegraphics[width=\linewidth]{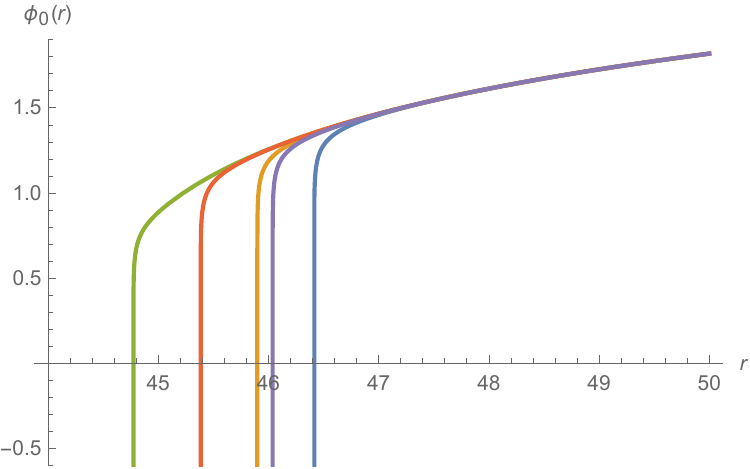}
\caption{$\phi_0$ solution}
  \end{subfigure}
  \begin{subfigure}[b]{0.32\linewidth}
    \includegraphics[width=\linewidth]{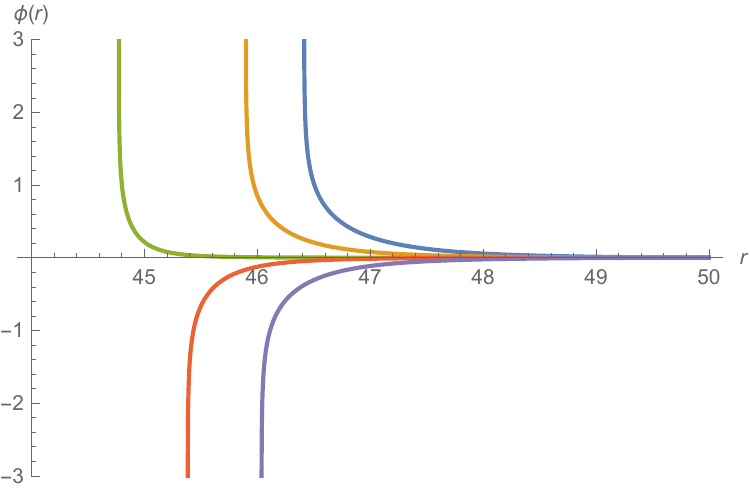}
\caption{$\phi$ solution}
  \end{subfigure}
\caption{Supersymmetric flows interpolating between asymptotically locally flat domain walls and $Mkw_2\times S^2 \times \mathbb{R}^2$-sliced curved domain walls in $SO(4)$ gauge group. The blue, orange, green, red and purple curves refer to $p_{21}=-0.5, -0.12, 0.03, 0.06, 0.25$, respectively.}
\label{Z_AdS3xS2xR2_SO2xSO2_SO4g_flow}
\end{figure}

\begin{figure}
  \centering
  \begin{subfigure}[b]{0.32\linewidth}
    \includegraphics[width=\linewidth]{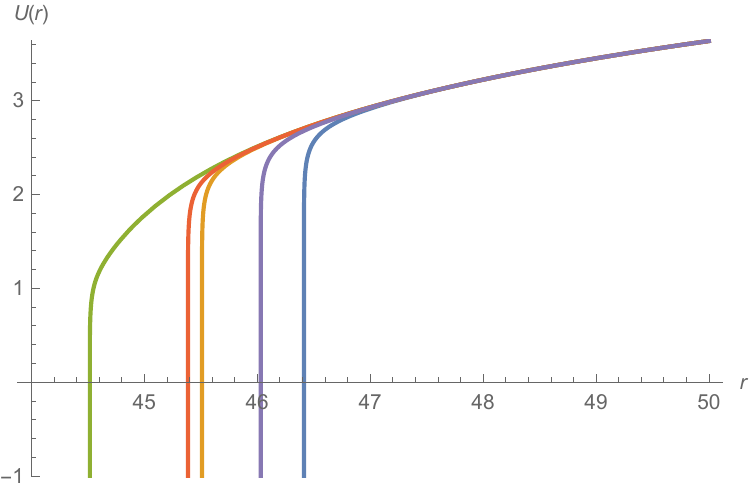}
\caption{$U$ solution}
  \end{subfigure}
  \begin{subfigure}[b]{0.32\linewidth}
    \includegraphics[width=\linewidth]{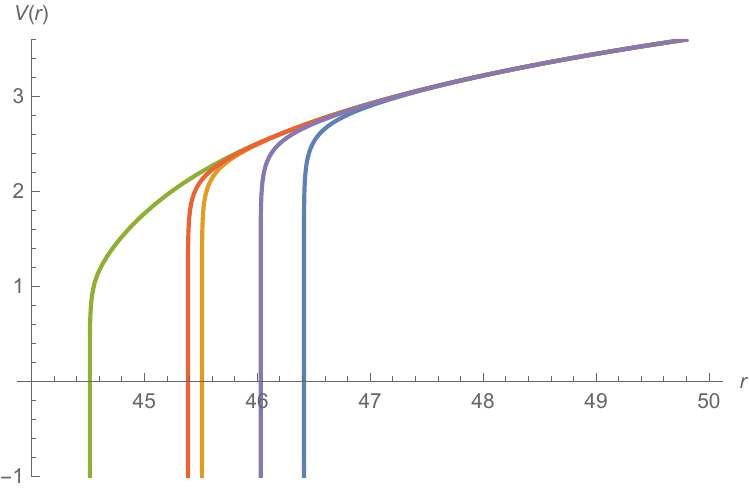}
\caption{$V$ solution}
  \end{subfigure}
  \begin{subfigure}[b]{0.32\linewidth}
    \includegraphics[width=\linewidth]{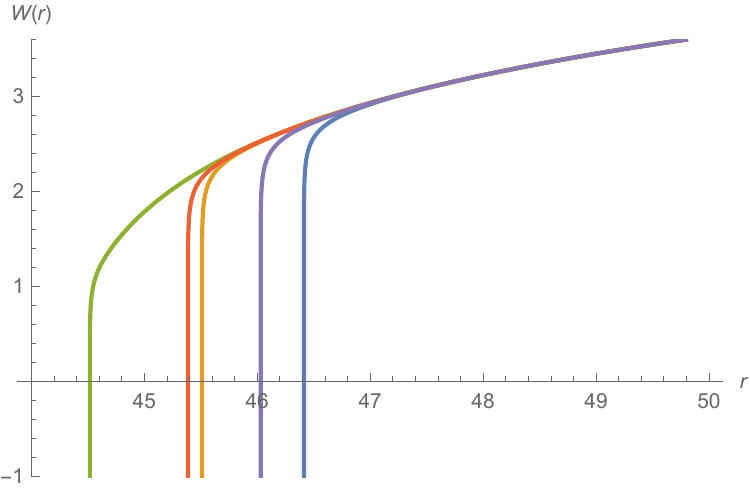}
\caption{$W$ solution}
  \end{subfigure}
  \begin{subfigure}[b]{0.32\linewidth}
    \includegraphics[width=\linewidth]{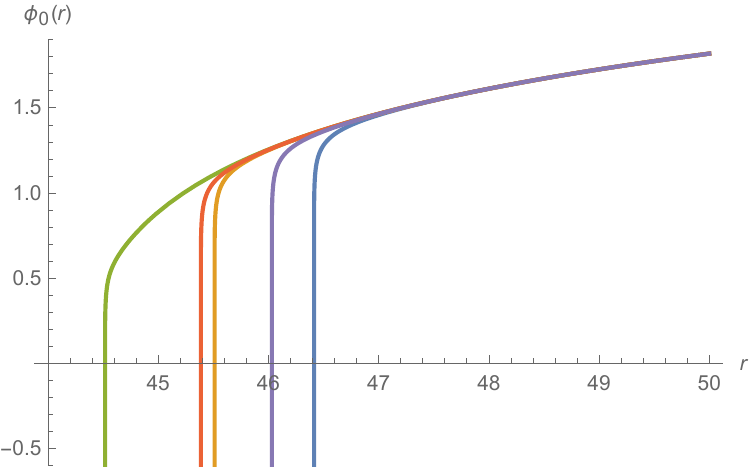}
\caption{$\phi_0$ solution}
  \end{subfigure}
  \begin{subfigure}[b]{0.32\linewidth}
    \includegraphics[width=\linewidth]{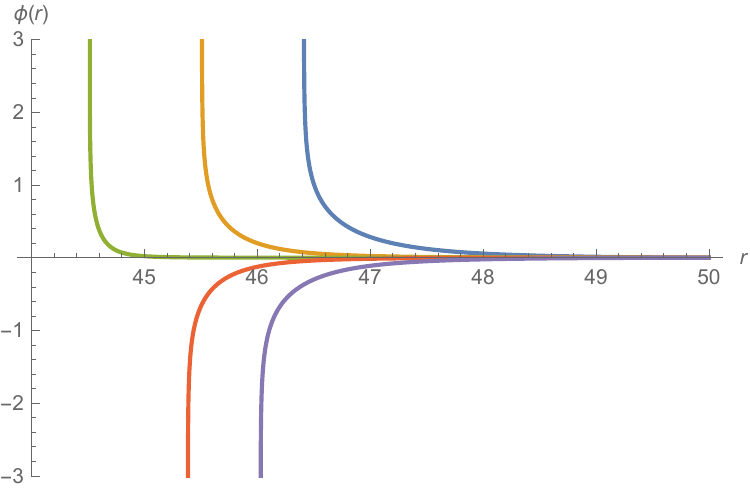}
\caption{$\phi$ solution}
  \end{subfigure}
\caption{Supersymmetric flows interpolating between asymptotically locally flat domain walls and $Mkw_2\times S^2\times H^2$-sliced curved domain walls in $SO(4)$ gauge group. The blue, orange, green, red and purple curves refer to $p_{21}=-0.5, -0.03, 0, 0.08, 0.25$, respectively.}
\label{Z_AdS3xS2xH2_SO2xSO2_SO4g_flow}
\end{figure}

\begin{figure}
  \centering
  \begin{subfigure}[b]{0.32\linewidth}
    \includegraphics[width=\linewidth]{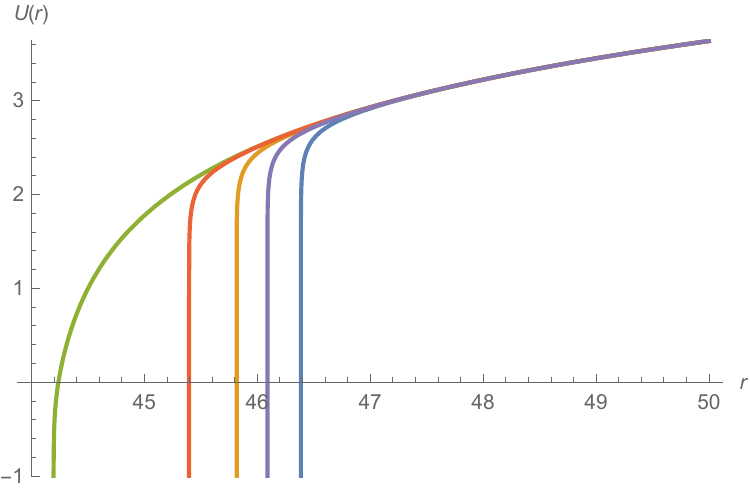}
\caption{$U$ solution}
  \end{subfigure}
  \begin{subfigure}[b]{0.32\linewidth}
    \includegraphics[width=\linewidth]{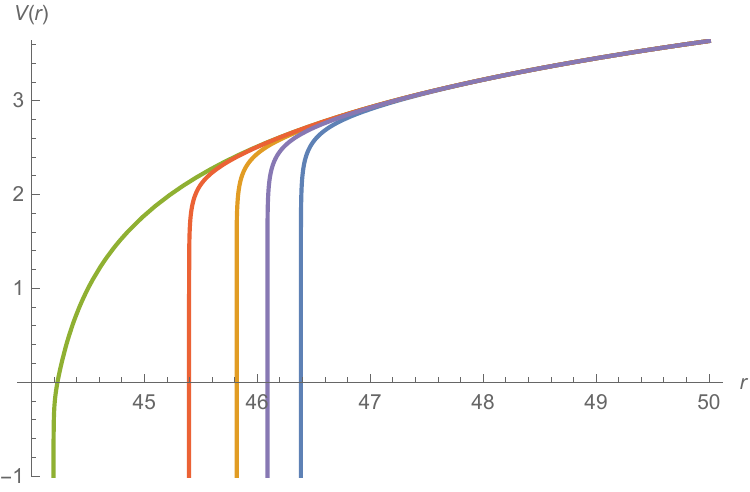}
\caption{$V$ solution}
  \end{subfigure}
  \begin{subfigure}[b]{0.32\linewidth}
    \includegraphics[width=\linewidth]{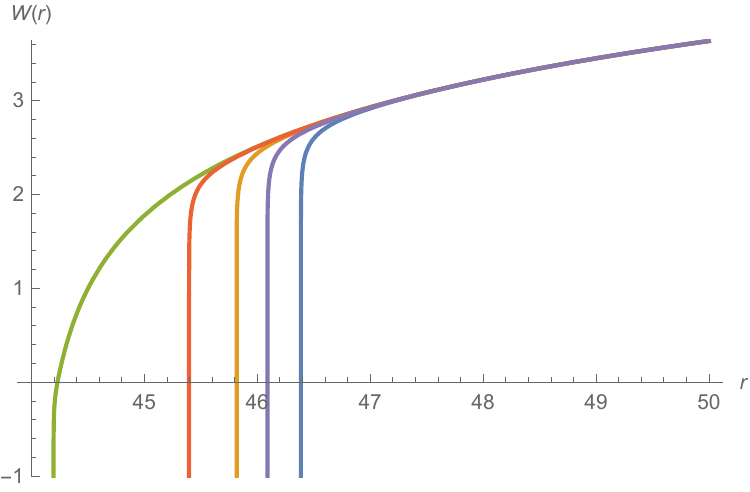}
\caption{$W$ solution}
  \end{subfigure}
  \begin{subfigure}[b]{0.32\linewidth}
    \includegraphics[width=\linewidth]{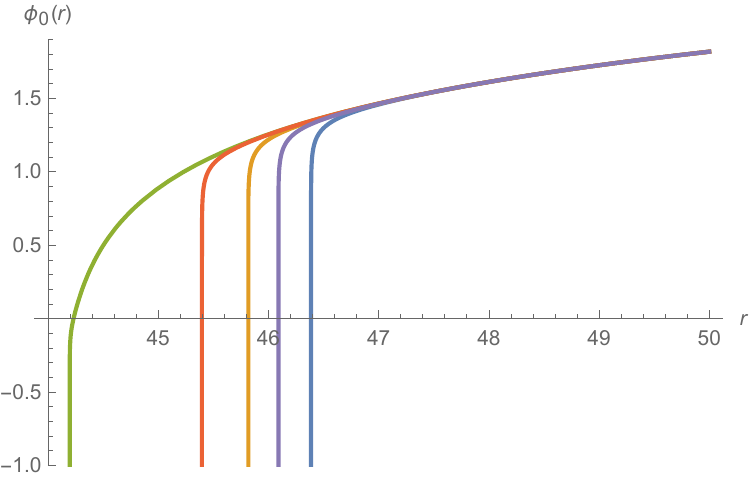}
\caption{$\phi_0$ solution}
  \end{subfigure}
  \begin{subfigure}[b]{0.32\linewidth}
    \includegraphics[width=\linewidth]{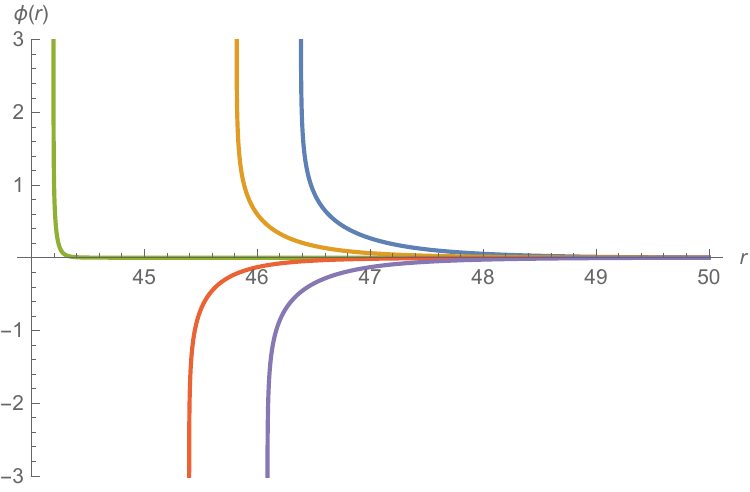}
\caption{$\phi$ solution}
  \end{subfigure}
\caption{Supersymmetric flows interpolating between asymptotically locally flat domain walls and $Mkw_2\times \mathbb{R}^2\times \mathbb{R}^2$-sliced domain walls in $SO(4)$ gauge group. The blue, orange, green, red and purple curves refer to $p_{21}=-0.5, -0.12, 0, 0.03, 0.25$, respectively.}
\label{Z_AdS3xR2xR2_SO2xSO2_SO4g_flow}
\end{figure}

\begin{figure}
  \centering
  \begin{subfigure}[b]{0.32\linewidth}
    \includegraphics[width=\linewidth]{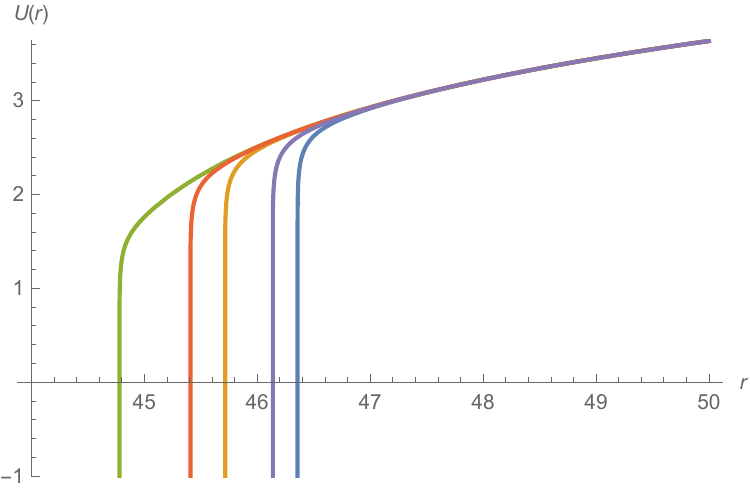}
\caption{$U$ solution}
  \end{subfigure}
  \begin{subfigure}[b]{0.32\linewidth}
    \includegraphics[width=\linewidth]{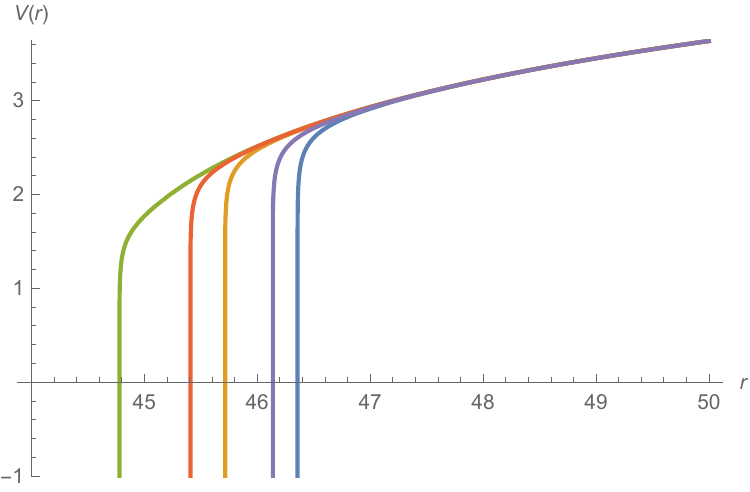}
\caption{$V$ solution}
  \end{subfigure}
  \begin{subfigure}[b]{0.32\linewidth}
    \includegraphics[width=\linewidth]{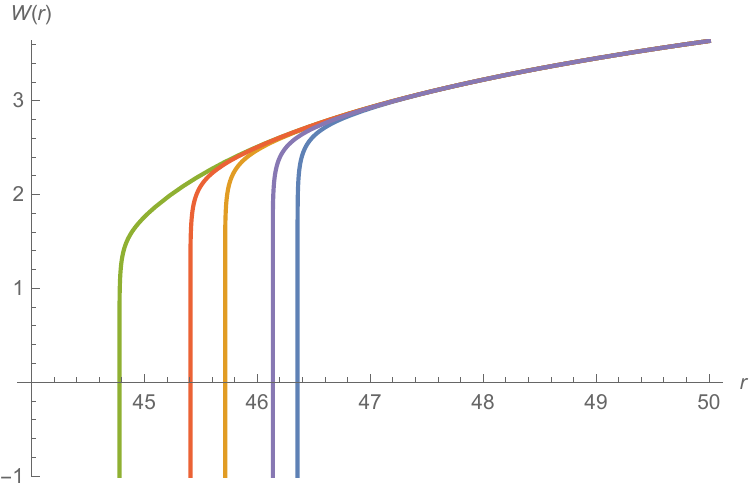}
\caption{$W$ solution}
  \end{subfigure}
  \begin{subfigure}[b]{0.32\linewidth}
    \includegraphics[width=\linewidth]{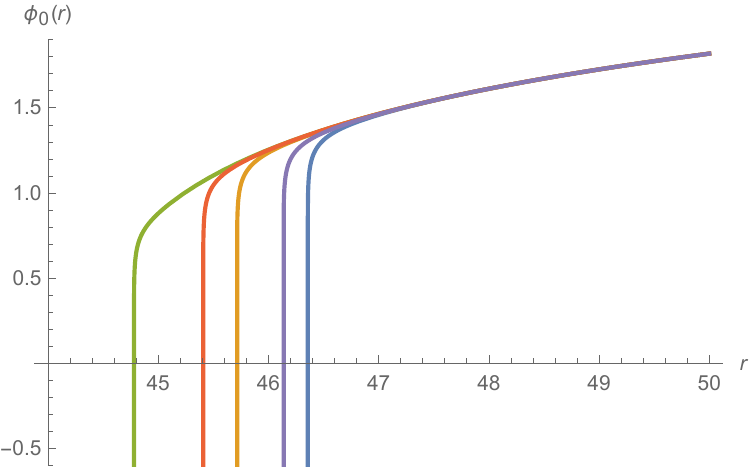}
\caption{$\phi_0$ solution}
  \end{subfigure}
  \begin{subfigure}[b]{0.32\linewidth}
    \includegraphics[width=\linewidth]{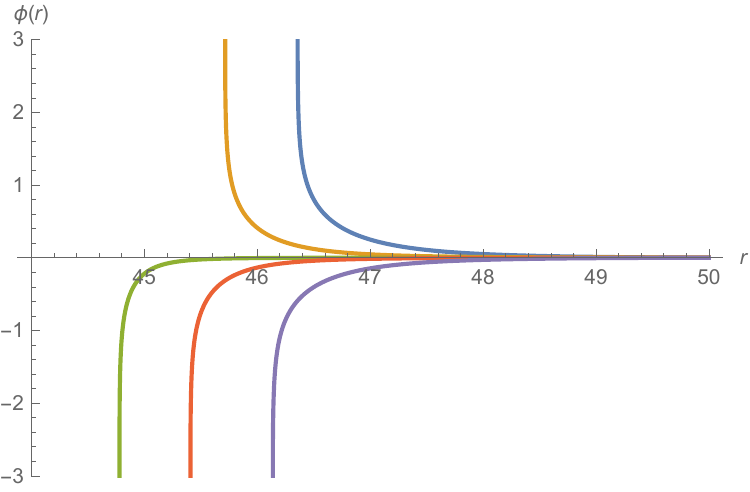}
\caption{$\phi$ solution}
  \end{subfigure}
\caption{Supersymmetric flows interpolating between asymptotically locally flat domain walls and $Mkw_2\times H^2\times \mathbb{R}^2$-sliced curved domain walls in $SO(4)$ gauge group. The blue, orange, green, red and purple curves refer to $-0.5, -0.12, -0.03, 0, 0.25$, respectively.}
\label{Z_AdS3xH2xR2_SO2xSO2_SO4g_flow}
\end{figure}

\begin{figure}
  \centering
  \begin{subfigure}[b]{0.32\linewidth}
    \includegraphics[width=\linewidth]{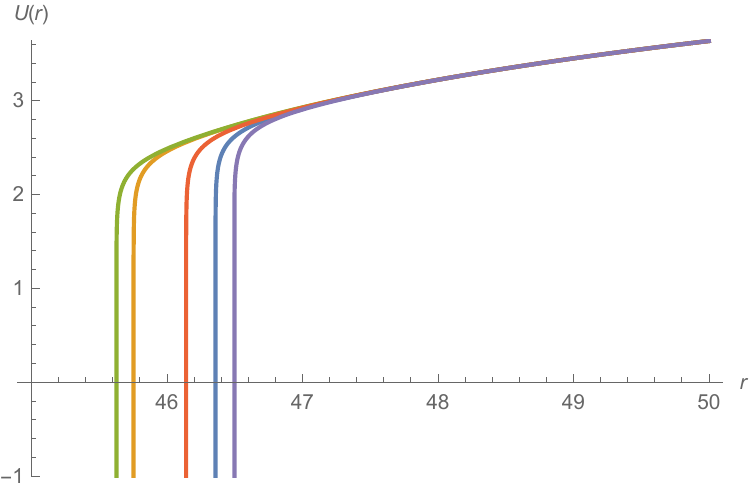}
\caption{$U$ solution}
  \end{subfigure}
  \begin{subfigure}[b]{0.32\linewidth}
    \includegraphics[width=\linewidth]{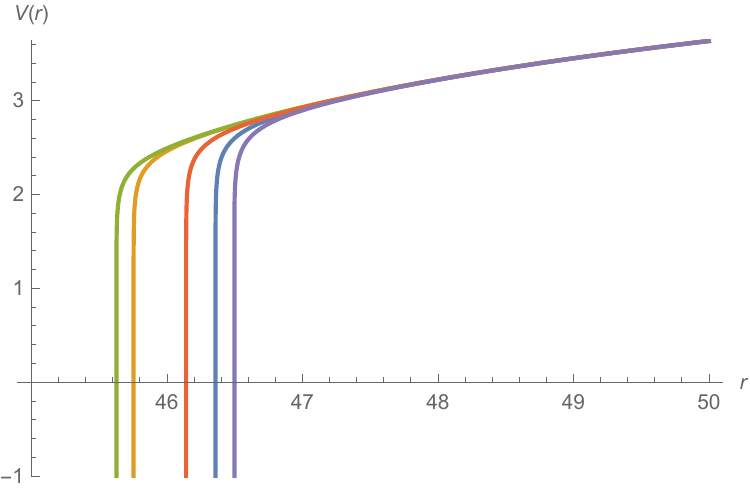}
\caption{$V$ solution}
  \end{subfigure}
  \begin{subfigure}[b]{0.32\linewidth}
    \includegraphics[width=\linewidth]{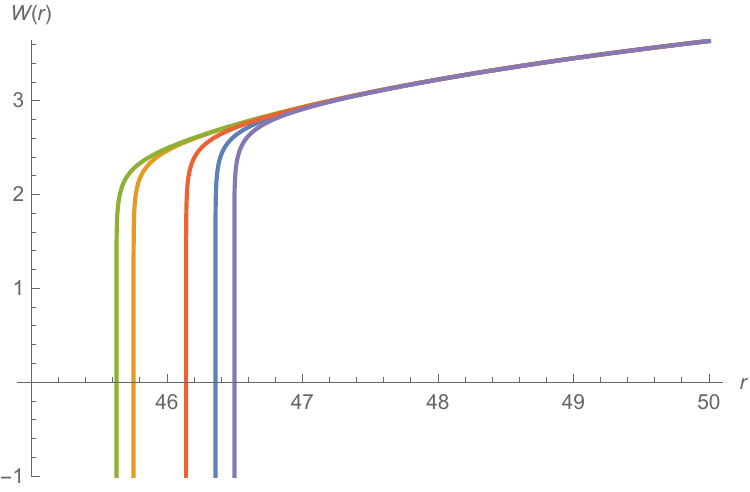}
\caption{$W$ solution}
  \end{subfigure}
  \begin{subfigure}[b]{0.32\linewidth}
    \includegraphics[width=\linewidth]{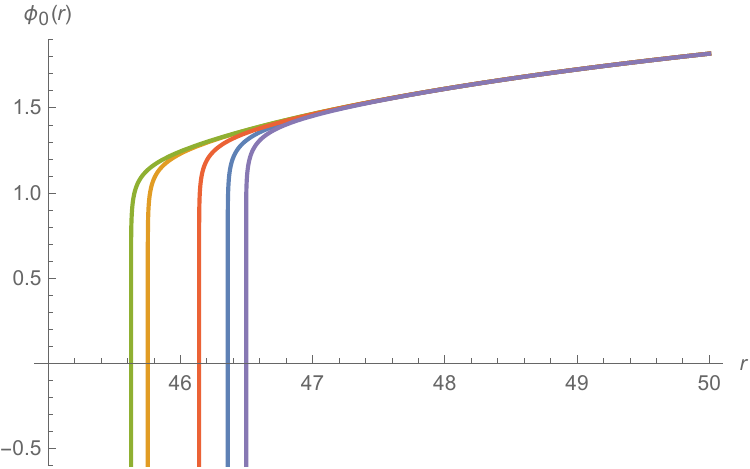}
\caption{$\phi_0$ solution}
  \end{subfigure}
  \begin{subfigure}[b]{0.32\linewidth}
    \includegraphics[width=\linewidth]{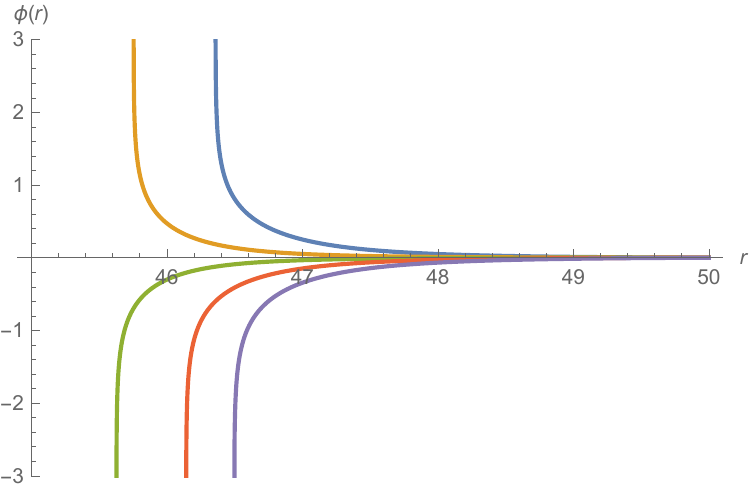}
\caption{$\phi$ solution}
  \end{subfigure}
\caption{Supersymmetric flows interpolating between asymptotically locally flat domain walls and $Mkw_2\times H^2\times H^2$-sliced curved domain walls in $SO(4)$ gauge group. The blue, orange, green, red and purple curves refer to $p_{21}=-0.5, -0.12, -0.01, 0.25, 0.6$, respectively.}
\label{Z_AdS3xH2xH2_SO2xSO2_SO4g_flow}
\end{figure}

\begin{figure}
  \centering
  \begin{subfigure}[b]{0.32\linewidth}
    \includegraphics[width=\linewidth]{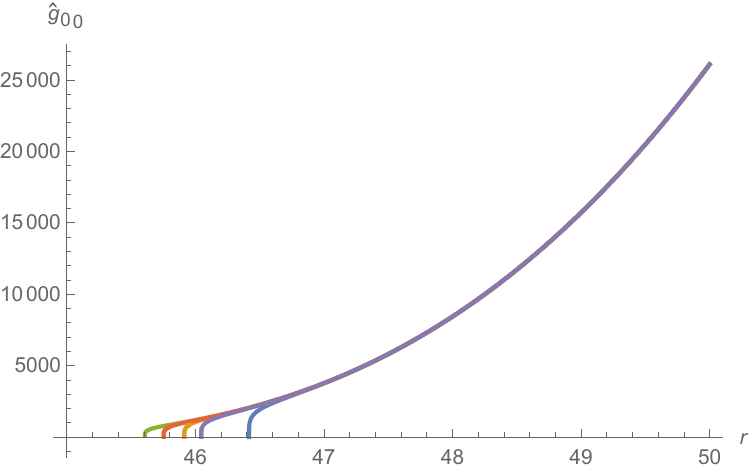}
\caption{$k_1=k_2=1$}
  \end{subfigure}
  \begin{subfigure}[b]{0.32\linewidth}
    \includegraphics[width=\linewidth]{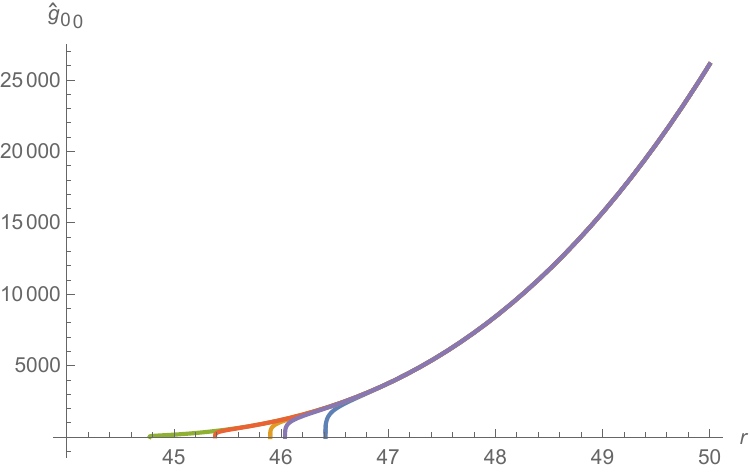}
\caption{$k_1=1, k_2=0$}
  \end{subfigure}
  \begin{subfigure}[b]{0.32\linewidth}
    \includegraphics[width=\linewidth]{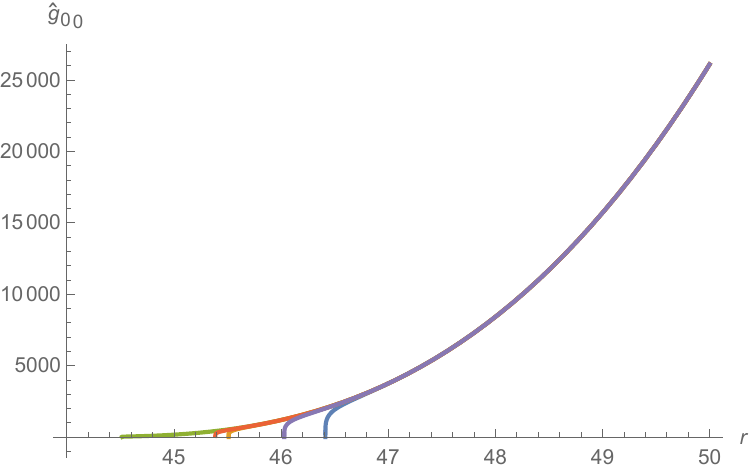}
\caption{$k_1=-k_2=1$}
  \end{subfigure}
  \begin{subfigure}[b]{0.32\linewidth}
    \includegraphics[width=\linewidth]{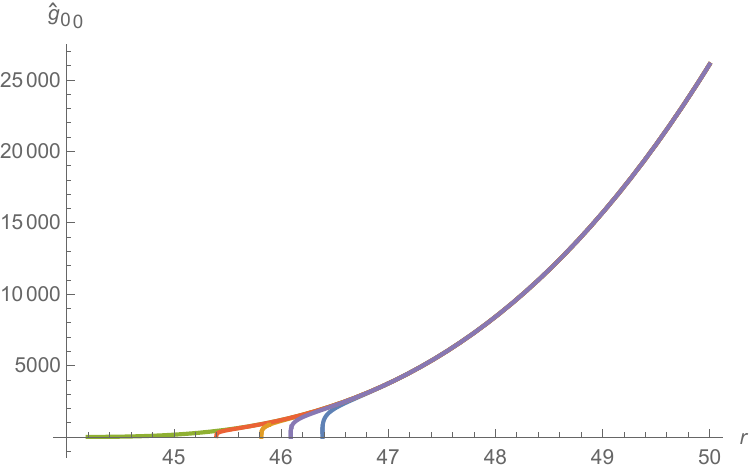}
\caption{$k_1=k_2=0$}
  \end{subfigure}
  \begin{subfigure}[b]{0.32\linewidth}
    \includegraphics[width=\linewidth]{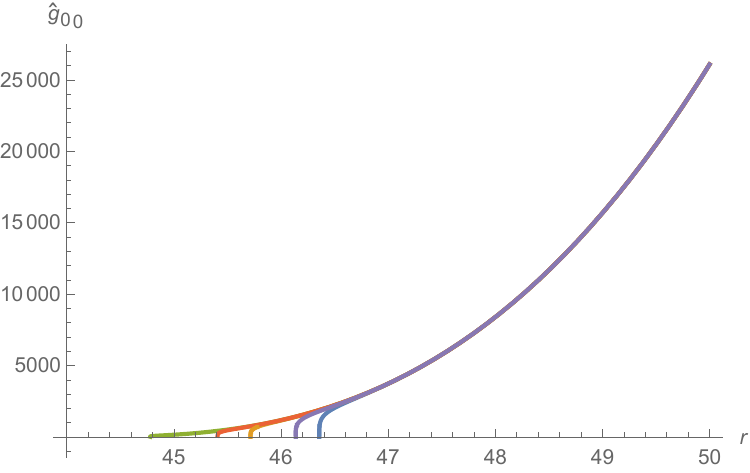}
\caption{$k_1=-1, k_2=0$}
  \end{subfigure}
  \begin{subfigure}[b]{0.32\linewidth}
    \includegraphics[width=\linewidth]{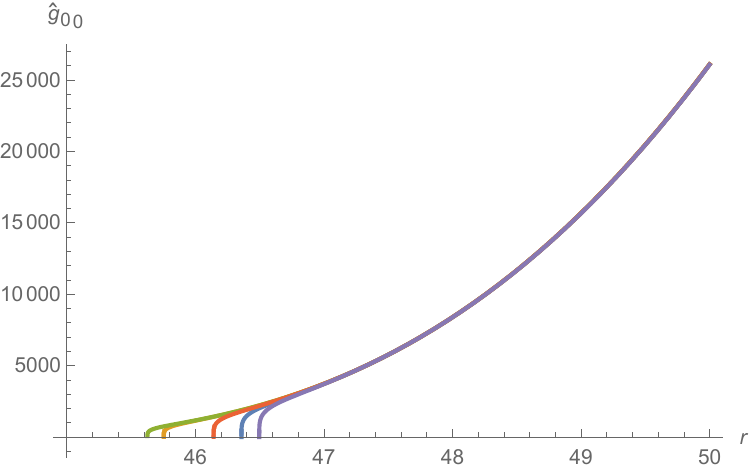}
\caption{$k_1=k_2=-1$}
  \end{subfigure}
\caption{Profiles of the ten-dimensional metric component $\hat{g}_{00}$ for the supersymmetric flows given in figures \ref{Z_AdS3xS2xS2_SO2xSO2_SO4g_flow} to \ref{Z_AdS3xH2xH2_SO2xSO2_SO4g_flow}.}
\label{Z_AdS3xSig2xSig2_SO2xSO2_SO4g_g00}
\end{figure}

For solutions with $SO(2)\times SO(2)$ twists, there are two gauge groups to consider, $SO(4)$ and $SO(2,2)$ with the embedding tensor \eqref{SO(2)xSO(2)wij}. The ansatz for the metric is given in \eqref{SO(2)xSO(2)7Dmetric}. To cancel the spin connection on $\Sigma^2_{k_1}\times\Sigma^2_{k_2}$, we turn on the following $SO(2)\times SO(2)$ gauge fields
\begin{eqnarray}
A^{12}_{(1)}&=&\frac{p_{11}}{4k_1}\frac{f_{k_1}'(\theta_1)}{f_{k_1}(\theta_1)}e^{-V}e^{\hat{\varphi}_1}+\frac{p_{12}}{4k_2}\frac{f_{k_2}'(\theta_2)}{f_{k_2}(\theta_2)}e^{-W}e^{\hat{\varphi}_2},\nonumber \\
A^{34}_{(1)}&=&\frac{p_{21}}{4k_1}\frac{f_{k_1}'(\theta_1)}{f_{k_1}(\theta_1)}e^{-V}e^{\hat{\varphi}_1}+\frac{p_{22}}{4k_2}\frac{f_{k_2}'(\theta_2)}{f_{k_2}(\theta_2)}e^{-W}e^{\hat{\varphi}_2} \label{ZSO(2)xSO(2)gaugeAnt}
\end{eqnarray} 
with the corresponding two-form field strengths
\begin{eqnarray}
F^{12}_{(2)}&=&-e^{-2V}\frac{p_{11}}{4}e^{\hat{\theta}_1}\wedge e^{\hat{\varphi}_1}-e^{-2W}\frac{p_{12}}{4}e^{\hat{\theta}_2}\wedge e^{\hat{\varphi}_2},\nonumber \\ 
F^{34}_{(2)}&=&-e^{-2V}\frac{p_{21}}{4}e^{\hat{\theta}_1}\wedge e^{\hat{\varphi}_1}-e^{-2W}\frac{p_{22}}{4}e^{\hat{\theta}_2}\wedge e^{\hat{\varphi}_2}\, .
\end{eqnarray}
\indent Following a similar analysis for gaugings in $\mathbf{15}$ representation, we also turn on the modified three-form field strength using the ansatz
\begin{equation}\label{ZAbSO(2)xSO(2)3form}
\mathcal{H}^{(3)}_{\hat{m}\hat{n}\hat{r} 5}= \beta e^{-2(V+W+4\phi_0)}\varepsilon_{\hat{m}\hat{n}}
\end{equation}
in which $\beta$ is a constant. We now impose the twist conditions
\begin{equation}\label{ZSO(2)xSO(2)QYM}
g(\sigma p_{11}+p_{21})=k_1\qquad \textrm{and} \qquad g(\sigma p_{12}+p_{22})=k_2
\end{equation}
and the projection conditions in \eqref{AddSO(2)xSO(2)ProjCon} together with \eqref{pureZProj}.
\\
\indent With all these, the resulting BPS equations are given by
\begin{eqnarray}
U'&=&-\frac{e^{2\phi_0}}{10}\left[e^{-2(V+\phi)}(e^{4\phi}p_{11}+p_{21})+e^{-2(W+\phi)}(e^{4\phi}p_{12}+p_{22})\right]\nonumber\\&&+\frac{g}{5}e^{-2(\phi_0+\phi)}(e^{4\phi}+\sigma)+\frac{3}{5}e^{-2(V+ W+2\phi_0)}\beta,\\
V'&=&-\frac{e^{2\phi_0}}{10}\left[4e^{-2(V+\phi)}(e^{4\phi}p_{11}+p_{21})-e^{-2(W+\phi)}(e^{4\phi}p_{12}+p_{22})\right]\nonumber\\&&+\frac{g}{5}e^{-2(\phi_0+\phi)}(e^{4\phi}+\sigma)-\frac{2}{5}e^{-2(V+ W+2\phi_0)}\beta,\\
W'&=&-\frac{e^{2\phi_0}}{10}\left[e^{-2(V+\phi)}(e^{4\phi}p_{11}+p_{21})-4e^{-2(W+\phi)}(e^{4\phi}p_{12}+p_{22})\right]\nonumber\\&&+\frac{g}{5}e^{-2(\phi_0+\phi)}(e^{4\phi}+\sigma)-\frac{2}{5}e^{-2(V+ W+2\phi_0)}\beta,\\
\phi_0'&=&-\frac{e^{2\phi_0}}{20}\left[e^{-2(V+\phi)}(e^{4\phi}p_{11}+p_{21})+e^{-2(W+\phi)}(e^{4\phi}p_{12}+p_{22})\right]\nonumber\\&&+\frac{g}{10}e^{-2(\phi_0+\phi)}(e^{4\phi}+\sigma)-\frac{1}{5}e^{-2(V+ W+2\phi_0)}\beta,\\
\phi'&=&-\frac{e^{2\phi_0}}{4}\left[e^{-2(V+\phi)}(e^{4\phi}p_{11}-p_{21})+e^{-2(W+\phi)}(e^{4\phi}p_{12}-p_{22})\right]\nonumber\\&&-\frac{g}{2}e^{-2(\phi_0+\phi)}(e^{4\phi}-\sigma).
\end{eqnarray}
Unlike the similar analysis for gaugings in $\mathbf{15}$ representation, it turns out that compatibility between these BPS equations and the second-ordered field equations requires 
\begin{equation}
p_{12} p_{21} + p_{11} p_{22}=0
\end{equation}
for any values of $\beta$. This implies that the constant $\beta$ is a free parameter in this case. However, we do not find any $AdS_3$ fixed points from the BPS equations.
\\
\indent For $SO(4)$ gauge group, examples of flow solutions between asymptotically locally flat domain walls and curved domain walls for various forms of $\Sigma^2\times \Sigma^2$ are shown in figures \ref{Z_AdS3xS2xS2_SO2xSO2_SO4g_flow} to \ref{Z_AdS3xH2xH2_SO2xSO2_SO4g_flow}. In these solutions, we have chosen particular values of $g=16$ and $\beta=2$. The green curve in figure \ref{Z_AdS3xR2xR2_SO2xSO2_SO4g_flow} is the flat domain wall solution given in \cite{our_7D_DW}. All of the IR singularities are physical as can be seen from the behavior of the ten-dimensional metric given in figure \ref{Z_AdS3xSig2xSig2_SO2xSO2_SO4g_g00}.

We have also considered $SO(2)$ twists on $\Sigma^2\times \Sigma^2$ by setting $p_{11}=p_{12}=0$ and obtain more complicated BPS equations. However, we do not find any $AdS_3$ fixed points either. Therefore, we will not give further detail on this analysis.

\subsubsection{Solutions with $SO(3)$ twists on $K^4$}
\begin{figure}
  \centering
  \begin{subfigure}[b]{0.38\linewidth}
    \includegraphics[width=\linewidth]{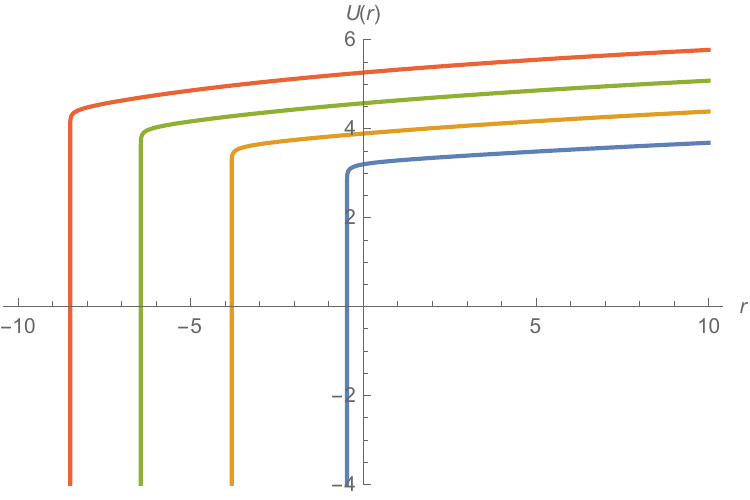}
\caption{$U$ solution}
  \end{subfigure}
  \begin{subfigure}[b]{0.38\linewidth}
    \includegraphics[width=\linewidth]{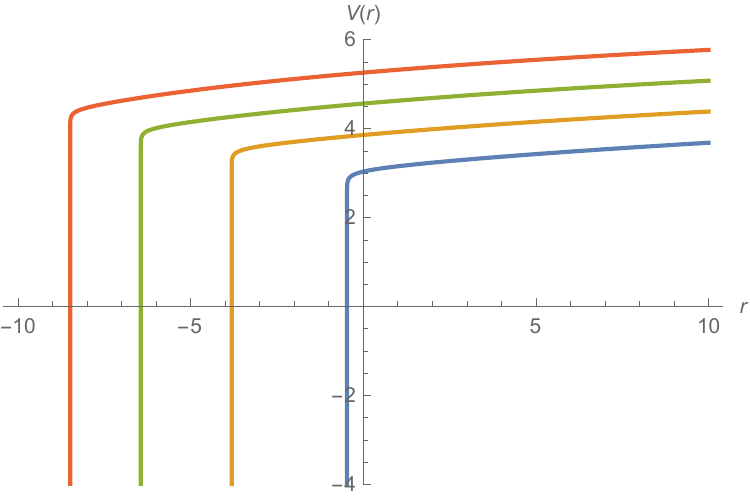}
\caption{$V$ solution}
  \end{subfigure}\\
  \begin{subfigure}[b]{0.38\linewidth}
    \includegraphics[width=\linewidth]{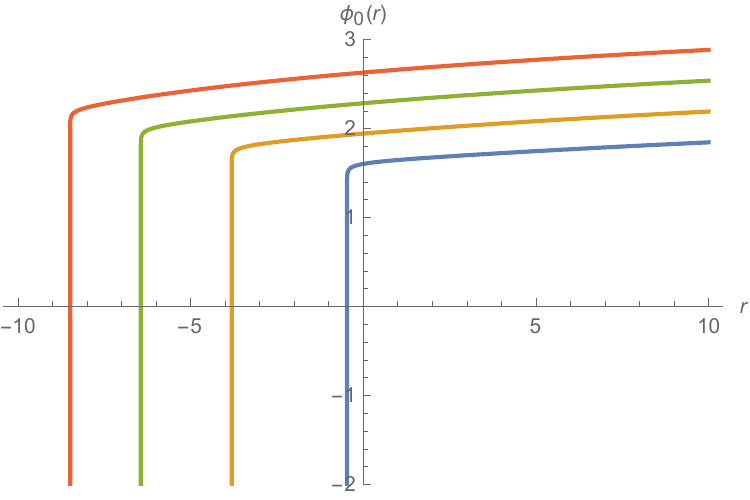}
\caption{$\phi_0$ solution}
  \end{subfigure}
  \begin{subfigure}[b]{0.38\linewidth}
    \includegraphics[width=\linewidth]{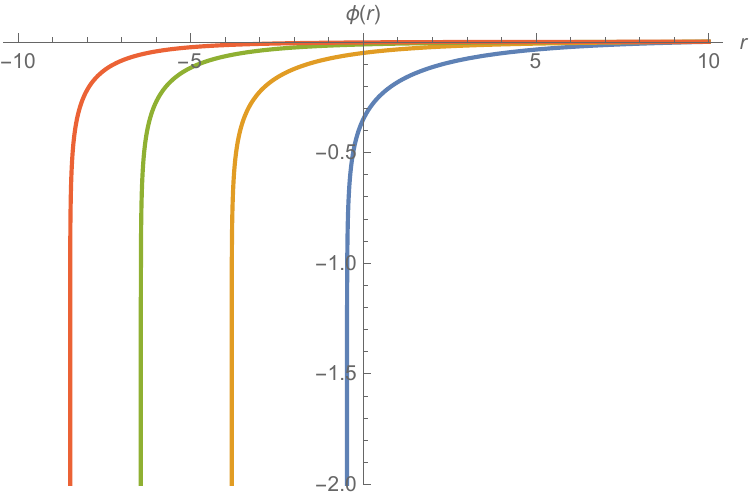}
\caption{$\phi$ solution}
  \end{subfigure}
\caption{Supersymmetric flows interpolating between asymptotically locally flat domain walls and $Mkw_2\times CP^2$-sliced curved domain walls in $SO(4)$ gauge group. The blue, orange, green and red curves refer to $g=4,8,16,32$, respectively.}
\label{Z_AdS3xCP2_SO3_SO4g_flow}
\end{figure}

\begin{figure}
  \centering
  \begin{subfigure}[b]{0.38\linewidth}
    \includegraphics[width=\linewidth]{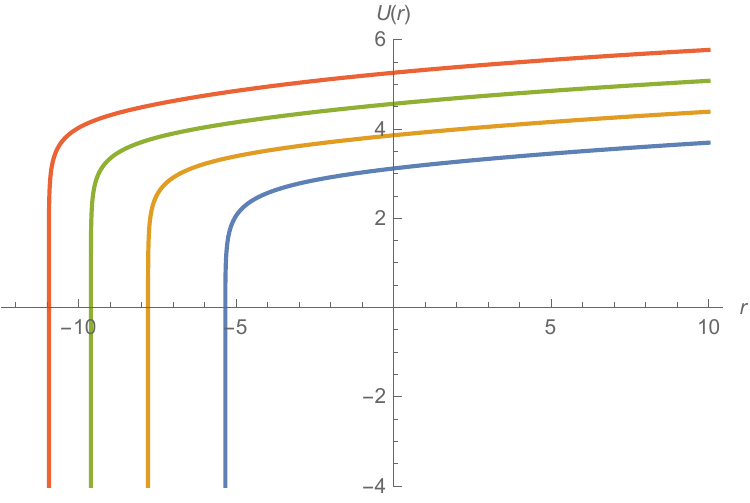}
\caption{$U$ solution}
  \end{subfigure}
  \begin{subfigure}[b]{0.38\linewidth}
    \includegraphics[width=\linewidth]{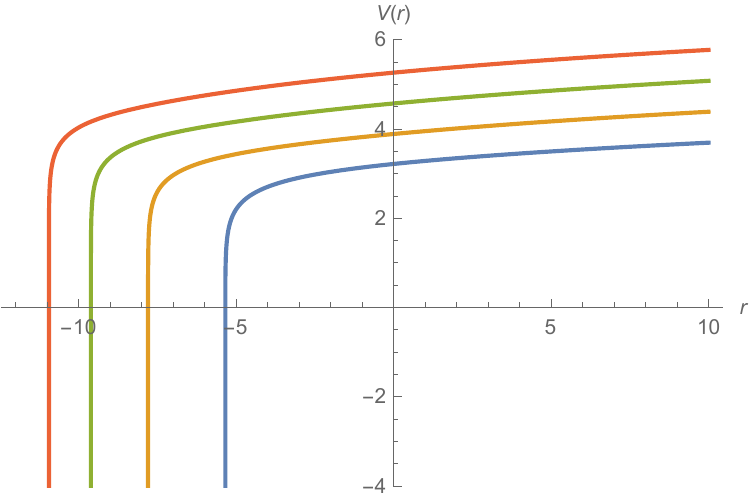}
\caption{$V$ solution}
  \end{subfigure}\\
  \begin{subfigure}[b]{0.38\linewidth}
    \includegraphics[width=\linewidth]{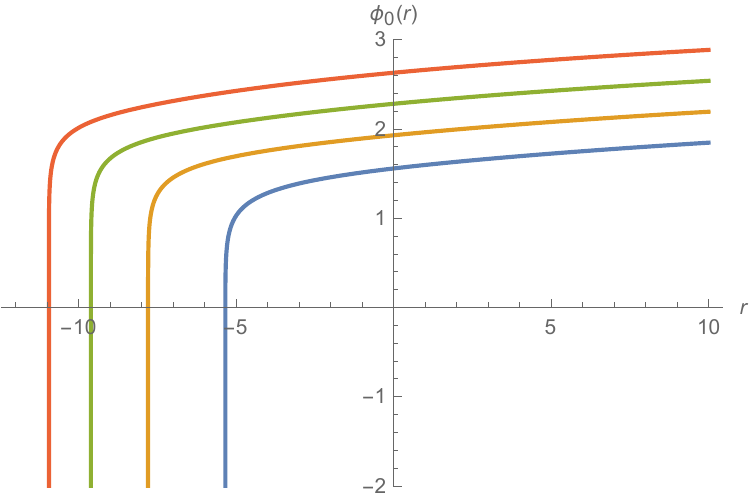}
\caption{$\phi_0$ solution}
  \end{subfigure}
  \begin{subfigure}[b]{0.38\linewidth}
    \includegraphics[width=\linewidth]{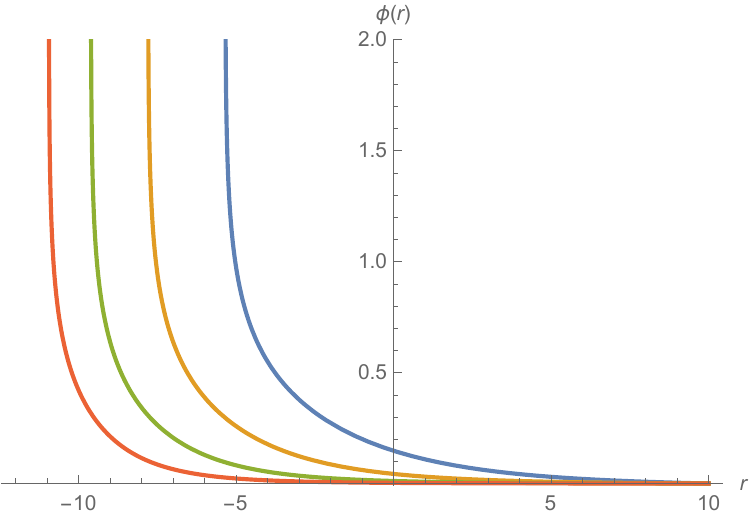}
\caption{$\phi$ solution}
  \end{subfigure}
\caption{Supersymmetric flows interpolating between asymptotically locally flat domain walls and $Mkw_2\times CH^2$-sliced curved domain walls in $SO(4)$ gauge group. The blue, orange, green and red curves refer to $g=4,8,16,32$, respectively.}
\label{Z_AdS3xCH2_SO3_SO4g_flow}
\end{figure}

\begin{figure}
  \centering
  \begin{subfigure}[b]{0.38\linewidth}
    \includegraphics[width=\linewidth]{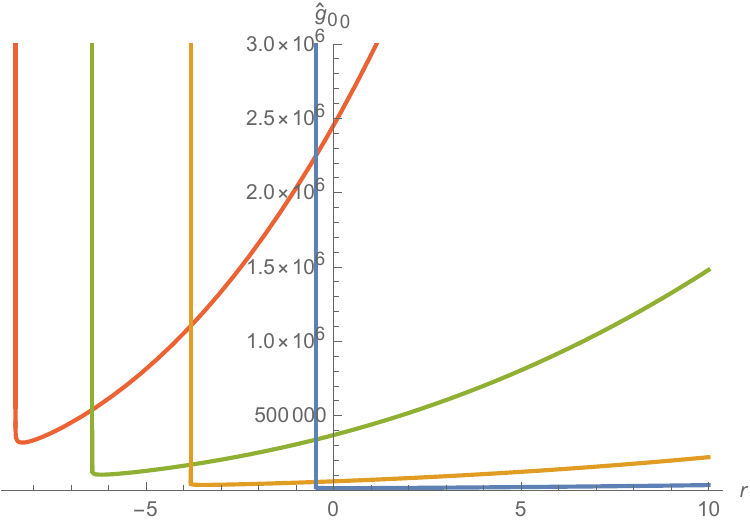}
\caption{$k=1$}
  \end{subfigure}
  \begin{subfigure}[b]{0.38\linewidth}
    \includegraphics[width=\linewidth]{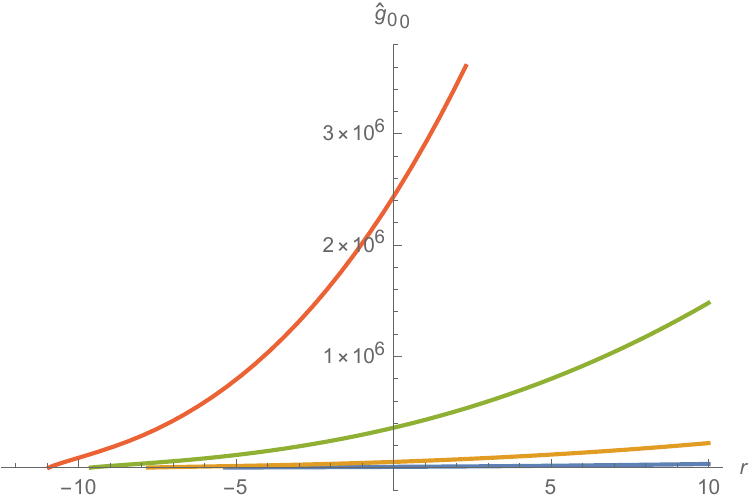}
\caption{$k=-1$}
  \end{subfigure}
\caption{Profiles of the ten-dimensional metric component $\hat{g}_{00}$ for the supersymmetric flows given in figures \ref{Z_AdS3xCP2_SO3_SO4g_flow} and \ref{Z_AdS3xCH2_SO3_SO4g_flow}.}
\label{Z_AdS3xK3_SO3_SO4g_g00}
\end{figure}

For $\Sigma^4$ being a Kahler four-cycle $K^4_k$, we perform an $SO(3)$ twist to cancel the $SU(2)$ part of the spin connection,  given in \eqref{AdS3xKahler4SpinCon}, by turning on the $SO(3)$ gauge fields
\begin{equation}\label{ZSO(3)KahlergaugeAnt}
A^{i4}_{(1)}=\frac{p}{4k}(f'_k(\psi)-1)\delta^{ij}\tau_j
\end{equation} 
with the two-form field strengths given by
\begin{eqnarray}
\mathcal{H}^{14}_{(2)}&=&-\frac{p}{4}e^{-2V}(e^{\hat{4}}\wedge e^{\hat{5}}-e^{\hat{3}}\wedge e^{\hat{6}}),\nonumber \\
\mathcal{H}^{24}_{(2)}&=&-\frac{p}{4}e^{-2V}(e^{\hat{5}}\wedge e^{\hat{3}}-e^{\hat{4}}\wedge e^{\hat{6}}),\nonumber \\
\mathcal{H}^{34}_{(2)}&=&-\frac{p}{4}e^{-2V}(e^{\hat{3}}\wedge e^{\hat{4}}-e^{\hat{5}}\wedge e^{\hat{6}})\, .
\end{eqnarray}
These field strengths do not lead to any problematic terms in the modified Bianchi's identity for the three-form field strengths. However, we can have a non-vanishing three-form field strength by using the following ansatz
\begin{equation}\label{ZSO(3)Kahler3form}
\mathcal{H}^{(3)}_{\hat{m}\hat{n}\hat{r} 5}= \beta e^{-4(V+\phi_0)}\varepsilon_{\hat{m}\hat{n}}.
\end{equation}
which is a manifestly closed three-form for a constant $\beta$.
\\
\indent With the $SL(4)/SO(4)$ coset representative and the embedding tensor given in \eqref{ZSO(3)coset} and \eqref{Z_SO3_embedding} together with the projections \eqref{YSO(3)KahlerProjCon} and \eqref{pureZProj}, we find the following BPS equations
\begin{eqnarray}
U'&\hspace{-0.3cm}=&\hspace{-0.3cm}\frac{g}{10}e^{-2(\phi_0+3\phi)}(3e^{8\phi}+\rho)-\frac{3}{5}e^{-2(V-\phi_0+\phi)}p-\frac{3}{5}e^{-4(V+\phi_0)}\beta,\\
V'&\hspace{-0.3cm}=&\hspace{-0.3cm}\frac{g}{10}e^{-2(\phi_0+3\phi)}(3e^{8\phi}+\rho)+\frac{9}{10}e^{-2(V-\phi_0+\phi)}p+\frac{2}{5}e^{-4(V+\phi_0)}\beta,\\
\phi_0'&\hspace{-0.3cm}=&\hspace{-0.3cm}\frac{g}{20}e^{-2(\phi_0+3\phi)}(3e^{8\phi}+\rho)-\frac{3}{10}e^{-2(V-\phi_0+\phi)}p+\frac{1}{5}e^{-4(V+\phi_0)}\beta,\\
\phi'&\hspace{-0.3cm}=&\hspace{-0.3cm}-\frac{g}{4}e^{-2(\phi_0+3\phi)}(e^{8\phi}-\rho)+\frac{1}{2}e^{-2(V-\phi_0+\phi)}p
\end{eqnarray}
in which we have used the twist condition \eqref{GenQYM}. We do not find any $AdS_3$ fixed points from these equations. Examples of supersymmetric flows for $\beta=-2$ are given in figures \ref{Z_AdS3xCP2_SO3_SO4g_flow} and \ref{Z_AdS3xCH2_SO3_SO4g_flow} for $k=1$ and $k=-1$, respectively. From the behavior of the ten-dimensional metric given in figure \ref{Z_AdS3xK3_SO3_SO4g_g00}, we find that the IR singularities for $k=-1$ are physical. In the case of flat $K^4_k$ with $k=0$, we have $p=0$ by the twist condition resulting in the standard flat domain wall solutions.

\subsubsection{Solutions with $SO(2)$ twists on $K^4$}\label{ZKahlerSO(2)section}
As the final case, we briefly consider the $SO(2)$ twist by turning on an $SO(2)$ gauge field to cancel the $U(1)$ part of the spin connection for the metric \eqref{U(1)Kahlermetric}. This gauge field is given by
\begin{equation}\label{ZKahlerSO(2)gaugeAnt}
A^{34}_{(1)}=-p\frac{3k\psi^2}{4\sqrt{f_k(\psi)}}\tau_3\, .
\end{equation}  
The $SO(2)$ singlet scalars from $SL(4)/SO(4)$ coset are described by the coset representative \eqref{ZSO(2)Ys}, and the embedding tensor is given in \eqref{ZSO2_theta}. We can also turn on the three-form field strength \eqref{ZSO(3)Kahler3form}. With the twist condition \eqref{GenQYM} and the projections \eqref{pureZProj} together with 
\begin{equation}\label{SO(2)KahlerProj}
\gamma^{\hat{3}\hat{4}}\epsilon^a=-\gamma^{\hat{5}\hat{6}}\epsilon^a=-{(\Gamma_{12})^a}_b\epsilon^b,
\end{equation}
the corresponding BPS equations are given by
\begin{eqnarray}
U'&=&\frac{g}{10}e^{-2(\phi_0+\phi_1)}\left[2e^{4\phi_1}-(\rho-\sigma)\sinh{2\phi_3}+(\rho+\sigma)\cosh{2\phi_3}\cosh{2\phi_2}\right]\nonumber\\&&-\frac{3}{5}e^{-2(V-\phi_0+\phi_1)}p-\frac{3}{5}e^{-4(V+\phi_0)}\beta,\\
V'&=&\frac{g}{10}e^{-2(\phi_0+\phi_1)}\left[2e^{4\phi_1}-(\rho-\sigma)\sinh{2\phi_3}+(\rho+\sigma)\cosh{2\phi_3}\cosh{2\phi_2}\right]\nonumber\\&&+\frac{9}{10}e^{-2(V-\phi_0+\phi_1)}p+\frac{2}{5}e^{-4(V+\phi_0)}\beta,\\
\phi_0'&=&\frac{g}{20}e^{-2(\phi_0+\phi_1)}\left[2e^{4\phi_1}-(\rho-\sigma)\sinh{2\phi_3}+(\rho+\sigma)\cosh{2\phi_2}\cosh{2\phi_3}\right]\nonumber\\&&-\frac{3}{10}e^{-2(V-\phi_0+\phi_1)}p+\frac{1}{5}e^{-4(V+\phi_0)}\beta,\\
\phi_1'&=&-\frac{g}{4}e^{-2(\phi_0+\phi_1)}\left[2e^{4\phi_1}+(\rho-\sigma)\sinh{2\phi_3}-(\rho+\sigma)\cosh{2\phi_2}\cosh{2\phi_3}\right]\nonumber\\&&+\frac{3}{2}e^{-2(V-\phi_0+\phi_1)}p,\\
\phi_2'&=&\ -\frac{g}{2}e^{-2(\phi_0+\phi_1)}(\rho+\sigma)\sinh{2\phi_2}\text{\ sech\ }{2\phi_3},\\
\phi_3'&=&\ \frac{g}{2}e^{-2(\phi_0+\phi_1)}\left[(\rho-\sigma)\cosh{2\phi_3}-(\rho+\sigma)\cosh{2\phi_2}\sinh{2\phi_3}\right].
\end{eqnarray}
As in all of the previous cases for gaugings in $\overline{\mathbf{40}}$ representation, there are no $AdS_3$ fixed points from these equations.

\section{Conclusions and discussions}\label{conclusion}
In this paper, we have extensively studied supersymmetric $AdS_n\times \Sigma^{7-n}$ solutions of the maximal gauged supergravity in seven dimensions with $CSO(p,q,5-p-q)$ and $CSO(p,q,4-p-q)$ gauge groups. These gauged supergravities can be embedded respectively in eleven-dimensional and type IIB supergravities on $H^{p,q}\circ T^{5-p-q}$ and $H^{p,q}\circ T^{4-p-q}$. Therefore, all the solutions given here have higher dimensional origins and could be interpreted as different brane configurations in string/M-theory. This makes applications of these solutions in the holographic context more interesting. Accordingly, we hope our results would be useful along this line of research.
\\
\indent For a particular case of $SO(5)$ gauge group, we have recovered all the previous results on $AdS_n\times \Sigma^{7-n}$ fixed points with $n=2,3,4,5$. We have provided numerical RG flows interpolating between the supersymmetric $N=4$ $AdS_7$ vacuum dual to $N=(2,0)$ SCFT in six dimensions and all these $AdS_n\times \Sigma^{7-n}$ fixed points. Some of these flows have not previously been discussed, so our results could complete the list of already known flow solutions. Furthermore, we have extended all these RG flows to singular geometries in the IR. These singularities take the form of curved domain walls with $Mkw_{n-1}\times \Sigma^{7-n}$ slices and can be interpreted as non-conformal field theories in $n-1$ dimensions. The flow solutions suggest that they describe non-conformal phases of the $(n-1)$-dimensional SCFTs obtained from twisted compactifications of $N=(2,0)$ SCFT in six dimensions.
\\
\indent We have also discovered novel classes of $AdS_5\times S^2$, $AdS_3\times S^2\times \Sigma^2$ and $AdS_3\times CP^2$ solutions in non-compact $SO(3,2)$ gauge group. Unlike in $SO(5)$ gauge group, there do not exist any supersymmetric $AdS_7$ fixed points in this gauge group. The maximally supersymmetric vacua are given by half-supersymmetric domain walls. In this case, we have studied RG flow solutions between these new fixed points and curved domain walls. We have also examined the behavior of the time component of the eleven-dimensional metric and found that many of the singularities are physically acceptable. The singular geometries identified here can be then interpreted as a holographic description of non-conformal field theories obtained from twisted compactifications of $N=(2,0)$ six-dimensional field theories. A similar study has been carried out for $SO(4,1)$ gauge group in which a new class of $AdS_3\times CP^2$ solutions has been found. 
\\
\indent Flow solutions for non-compact $SO(3,2)$ and $SO(4,1)$ gauge groups can also be interpreted as black $3$-brane and black strings in asymptotically curved domain wall space-time. These solutions are similar to four-dimensional black holes studied in \cite{Klemm_first_AdS4_BH}. In \cite{Hristov}, these black hole solutions have been shown to arise from a dimensional reduction of the $AdS_5$ black strings studied in \cite{2D_Bobev}. It has also been pointed out in \cite{Hristov} that the four-dimensional black holes, with curved domain wall asymptotics, should be seen from a higher-dimensional perspective as black strings in $AdS_5$. However, this is not the case for the solutions given in this paper. Our solutions cannot be related to any supersymmetric black objects in eight dimensions with asymptically $AdS_8$ space-time due to the absence of supersymmetric $AdS_d$ vacua for $d>7$.    
\\
\indent For $CSO(p,q,4-p-q)$ gauge group which is obtained from a truncation of type IIB theory, we have performed a similar analysis as in the $CSO(p,q,5-p-q)$ gauge group but have not found any $AdS_n$ fixed points. The resulting gauged supergravity admits half-supersymmetric domain walls as vacuum solutions which, upon uplifted to ten dimensions, describe $5$-branes in type IIB theory. We have given supersymmetric flow solutions interpolating between asymptotically locally flat domain walls, in which the effect of magnetic charges are small compared to the superpotential of the domain walls, and curved domain walls with $Mkw_{n-1}\times \Sigma^{7-n}$ worldvolume. By the standard DW/QFT correspondence, these solutions should be interpreted as RG flows across dimensions between non-conformal field theories in six and $n-1$ dimensions. It could be interesting to study these field theories on the worldvolume of the $5$-branes in type IIB theory. Our results suggest that these $N=(2,0)$ field theories have no conformal fixed points in lower dimensions. It could be interesting to have a definite conclusion whether this  is true in general. On the other hand, if this is not the case, it would also be interesting to extend the anylysis of this paper by using more general ansatz in particular with non-vanishing massive two-form fields and find new classes of $AdS_n\times \Sigma^{7-n}$ solutions of seven-dimensional gauged supergravity. 
{\large{\textbf{Acknowledgement}}} \\
This work is supported by The Thailand Research Fund (TRF) under grant RSA6280022.

\appendix
\section{Bosonic field equations}
For convenience, in this appendix, we collect all the bosonic field equations of the maximal gauged supergravity in seven dimensions. These equations are given by
\begin{eqnarray}
0&=&R_{\mu\nu}-\frac{1}{4}\mathcal{M}_{MP}\mathcal{M}_{NQ}(D_\mu\mathcal{M}^{MN})(D_\nu\mathcal{M}^{PQ})-\frac{2}{5}g_{\mu\nu}\mathbf{V}\nonumber\\&&-4\mathcal{M}_{MP}\mathcal{M}_{NQ}\left(\mathcal{H}_{\mu\rho}^{(2)MN}{{\mathcal{H}^{(2)PQ}}_{\nu}}^\rho-\frac{1}{10}g_{\mu\nu}\mathcal{H}_{\rho\sigma}^{(2)MN}\mathcal{H}^{(2)PQ\rho\sigma}\right)\nonumber\\
& &-\mathcal{M}^{MN}\left(\mathcal{H}_{\mu\rho\sigma M}^{(3)}{\mathcal{H}_\nu^{(3)\rho\sigma}}_N-\frac{2}{15}g_{\mu\nu}\mathcal{H}_{\rho\sigma\lambda M}^{(3)}{\mathcal{H}^{(3)\rho\sigma\lambda}}_N\right),
\end{eqnarray}
\begin{eqnarray}
0&=&D^\mu(\mathcal{M}_{MP}D_\mu\mathcal{M}^{PN})-\frac{g^2}{8}\mathcal{M}^{PQ}\mathcal{M}^{RN}\left(2Y_{RQ}Y_{PM}-Y_{PQ}Y_{RM}\right)\nonumber\\&&-\frac{4}{6}\mathcal{M}^{PN}\mathcal{H}_{\mu\nu\rho M}^{(3)}{\mathcal{H}^{(3)\mu\nu\rho}}_P-8\mathcal{M}_{MP}\mathcal{M}_{QR}\mathcal{H}_{\mu\nu}^{(2)PQ}
\mathcal{H}^{(2)RN\mu\nu}\nonumber\\&&+4g^2Z^{QT,P}Z^{NR,S}\mathcal{M}_{QM}(2\mathcal{M}_{TR}\mathcal{M}_{PS}-\mathcal{M}_{TP}\mathcal{M}_{RS})\nonumber \\
& &+4g^2Z^{QT,P}Z^{RS,N}\mathcal{M}_{QS}(2\mathcal{M}_{TP}\mathcal{M}_{RM}-\mathcal{M}_{TR}\mathcal{M}_{PM})\nonumber\\&&-4g^2\delta^N_MZ^{TU,P}Z^{QR,S}\mathcal{M}_{TQ}\left(\mathcal{M}_{UR}\mathcal{M}_{PS}
-\mathcal{M}_{UP}\mathcal{M}_{RS}\right) \nonumber\\&&+\frac{8}{5}\delta^N_M\left(\mathbf{V}+\mathcal{M}_{SP}\mathcal{M}_{QR}\mathcal{H}_{\mu\nu}^{(2)PQ}\mathcal{H}^{(2)RS\mu\nu }+\frac{1}{16}\mathcal{M}^{PQ}\mathcal{H}_{\mu\nu\rho P}^{(3)}{\mathcal{H}^{(3)\mu\nu\rho}}_Q\right),\nonumber \\
& &
\end{eqnarray}
\begin{eqnarray}
0&=&4D_\nu(\mathcal{M}_{MP}\mathcal{M}_{NQ}\mathcal{H}^{(2)PQ\nu\mu })-\frac{g}{2}{X_{MNP}}^Q\mathcal{M}_{QR}D^\mu\mathcal{M}^{PR}\nonumber\\&&-2\epsilon_{MNPQR}\mathcal{M}^{PS}{\mathcal{H}^{(3)\mu\nu\rho}}_S\mathcal{H}^{(2)QR}_{\nu\rho}+\frac{1}{9}e^{-1}\epsilon^{\mu\nu\rho\lambda\sigma\tau\kappa}\mathcal{H}^{(3)}_{\nu\rho\lambda M}\mathcal{H}^{(3)}_{\sigma\tau\kappa N},
\end{eqnarray}
\begin{eqnarray}
0&=&D_\rho\left(\mathcal{M}^{MN}{\mathcal{H}^{(3)\rho\mu\nu}}_N\right)-2gZ^{NP,M}\mathcal{M}_{NQ}\mathcal{M}_{PR}\mathcal{H}^{(2)QR\mu\nu}\nonumber\\&&-\frac{1}{3}e^{-1}\epsilon^{\mu\nu\rho\lambda\sigma\tau\kappa}\mathcal{H}^{(2)MN}_{\rho\lambda}\mathcal{H}^{(3)}_{\sigma\tau\kappa N},\label{3formfieldEQ}\\
0&=&e^{-1}\epsilon^{\mu\nu\rho\lambda\sigma\tau\kappa}Y_{MN}\mathcal{H}^{(4)N}_{\lambda\sigma\tau\kappa}-6Y_{MN}\mathcal{M}^{NP}{\mathcal{H}^{(3)\mu\nu\rho}}_P\, .\label{fullSD}
\end{eqnarray}


\end{document}